# Latent Dirichlet Allocation
# 漫游指南

马晨(sharpstill@163.com)

# 前言

LDA 算法是主题模型领域非常著名的算法，值得深入研究应用，该算法也有很深刻的数学背景和技术启发。曾经有哲人说：万物皆数。我个人是个十分喜欢数学，喜欢算法，热爱技术的人，非常想从算法中寻找人工智能的永恒之道。我尤其记得 19 世纪的数学家赫尔曼.汉克尔说的好：

*就大多数学科而言，一代人摧毁的正是另一代人所建造的，而他们所建立的也必将被另一代人所破坏。只有数学不同，每一代人都是在旧的建筑物上加进新的一层。*

所以说，数学的价值还具有一种永世不灭的恒久性，其他学科的时尚潮流往往随着时代的变迁被人遗忘，那些旨在改变世界的理想，最终往往变成思想垃圾。而只有数学和算法与此不同。

我们探究前人伟大的成果时，就能体会到奥利弗.亥维赛的精辟论说：“逻辑能够很有耐性，因为它是永恒的”。

我在选择分析 Latent Dirichlet Allocation(LDA)这个算法课题时，我考虑了很多因素，首先，该算法是已经被学术界和工业界广泛接受的；其次，该算法能带来更多的新技术启示；最后,该算法能为您的工作，您的研究带来最具实用性的技术启发。

LDA 算法恰好满足了这个条件。

虽然网上已经有许多分析 LDA 算法的博客文章，但是网上的博文相对零散不成体系，读者阅读起来有较大困难，我的目标是不放弃任何一位读者，只要读者有恒心和毅力，就一定可以从这部作品中受益，为什么需要这本书，因其有**独特的价值：**

1. **这部作品理论与实践并重**:网上的同类文章非常零散，理论推导部分也缺乏**关键细节**，这部作品的每一条公式都由作者手把手为您推理（每一条公式都有详细的解释和备注），并且按照初学者的思路娓娓道来，从逻辑链条上打通算法的整个环节，让用户有醍醐灌顶的认识。并且在**实践部分**，作者以多年的工作实践经验为基础，**精选**了 6 个实现简单但又有巨大应用价值的 LDA 的应用方法，这些精选的应用方法将成为读者未来工作实践不可多得的资料。

2. **这部作品饱含了作者的独到见解：**这部作品最大的特色是从理论分析开始就有包含着许多作者自己独到的理解和分析，从不同角度完美解释算法的整个流程。

**3. 读者可以在这部作品各取所需：**有的工程师对于算法推导不是很感兴趣，这种情况下可以跳过前几章，直接从第 4 章读 LDA 算法怎么具体实现。如若未来有兴趣研究 LDA 的来龙去脉时，可以再来看前几章的理论推导部分。如果读者对大数据环境下的 LDA 感兴趣，包括怎么在 Hadoop、Spark 上实现 LDA 算法可以直接看第 5 章。

**4. 这部作品首次将 LDA 引入大数据时代：**大数据时代最大的特色就是信息爆炸，各种文本数据，用户生成（UGC）数据也变得非常庞大，网上查阅到的 LDA 算法资料大部分都是不能应对大数据环境的，这部作品的第 5 章深入浅出地讲解了大数据环境下怎么实现并行化的 LDA 算法。

**5. 这部作品是国内关于 LDA 的变分推断技术讲解最细致的书。**

**章节安排：**

第 1 章为相关背景介绍，介绍了算法知识的来源：从 18 世纪的欧拉讲到剑桥大学的 David Blei。

第 2 章和第 3 章为算法的理论分析阶段：第 2 章为 LDA 算法的前置知识，为 LDA 做了理论工具上的准备。

在第 2 章力求做到关键证明不遗漏，这样就可以与后面第 3 章的 LDA 的算法推导构成一个完整的推理链条！这一章的有些证明需要一些简单的微积分知识，但如果读者忘记了所有基础的微积分知识的话，那么看不懂某条证明，就请姑且相信我的推理是对的吧，跳过去往后看，日后再复习。

第 3 章为 LDA 算法推导部分，用严谨的数学推导和清晰的讲解（每个公式都做了清晰的标注），让读者认识该推理方法。

第 4 章为实现和应用部分，用伪代码方式庖丁解牛，讲解代码实现的精髓，然后结合作者多年的工作实践，写了该算法的几个最具实用价值的应用，这些应用方法中有相当一部分是作者的亲身工作经验的总结，在任何其他书籍上找不到。

第 5 章为并行化，在大数据如火如荼的今天，要想大规模运行 LDA 算法，就要靠并行化技术了，这一章从 2 个算法的改进形式讲解了该技术的并行化，并且可以放在目前最流行的 spark 大数据引擎上运行。

第 6 章，第 7 章，第 8 章三个章节为 LDA 的变分 EM 技术的详细推导和 C 语言代码实现的一个详尽剖析，这个方法的来龙去脉都在这三个章节有所体现。

本文是一个指南指引性质的作品，仍未完善完美，本书还有些地方可能仍有

疏漏，本人水平有限，如果大家发现纰漏，请及时联系人民邮电出版社修正。

希望这部作品日臻完美。

作者：马晨　2016.4.19

# 目录

# 第1章 背景

LDA 算法使用的全部知识的渊源可以追溯到 18 世纪的欧拉，欧拉（Leonhard Euler ，1707 年 4 月 15 日～1783 年 9 月 18 日），瑞士数学家。欧拉一生贡献颇丰，1734 年，欧拉解决巴塞尔问题就立即出名了，巴塞尔问题就是问式(1-1)的值是多少。

$$\sum_{n=1}^{\infty}\frac{1}{n^2} = \lim_{n\to\infty}(\frac{1}{1^2}+\frac{1}{2^2}+...+\frac{1}{n^2}) = ? \tag{1-1}$$

这个问题困扰了几个世纪的数学家，当时的数学家只知道该级数的值小于 2，但不知道具体精确值，欧拉准确的推导出该式的值= $\pi^2 / 6$，欧拉的方法聪明而新颖，他创造性将有限多项式的观察推广到无穷级数，并假设相同的性质对于无穷级数也是成立的：

$$1-\frac{x^2}{3!}+\frac{x^4}{5!}-\frac{x^6}{7!}+\frac{x^8}{9!}=[1-\frac{x^2}{\pi^2}][1-\frac{x^2}{4\pi^2}][1-\frac{x^2}{9\pi^2}][1-\frac{x^2}{16\pi^2}]... \tag{1-2}$$

欧拉最后的发现是令人惊奇的，$\pi$这个数字在于圆周率无关的场合中出现了，这足以说明数学之中、自然之中、冥冥之中存在着某些神秘的联系。虽然以现代数学的眼光来看，欧拉的证明还不严密。但作为第一个（富有创造性的）证明，欧拉的这个证明永远有着其宝贵的价值。欧拉的另一个发现就是发现了 gamma 函数 $f(x)=\Gamma(x)$ ，该函数后被广泛应用于概率论，这个函数也是本文的主角之一。

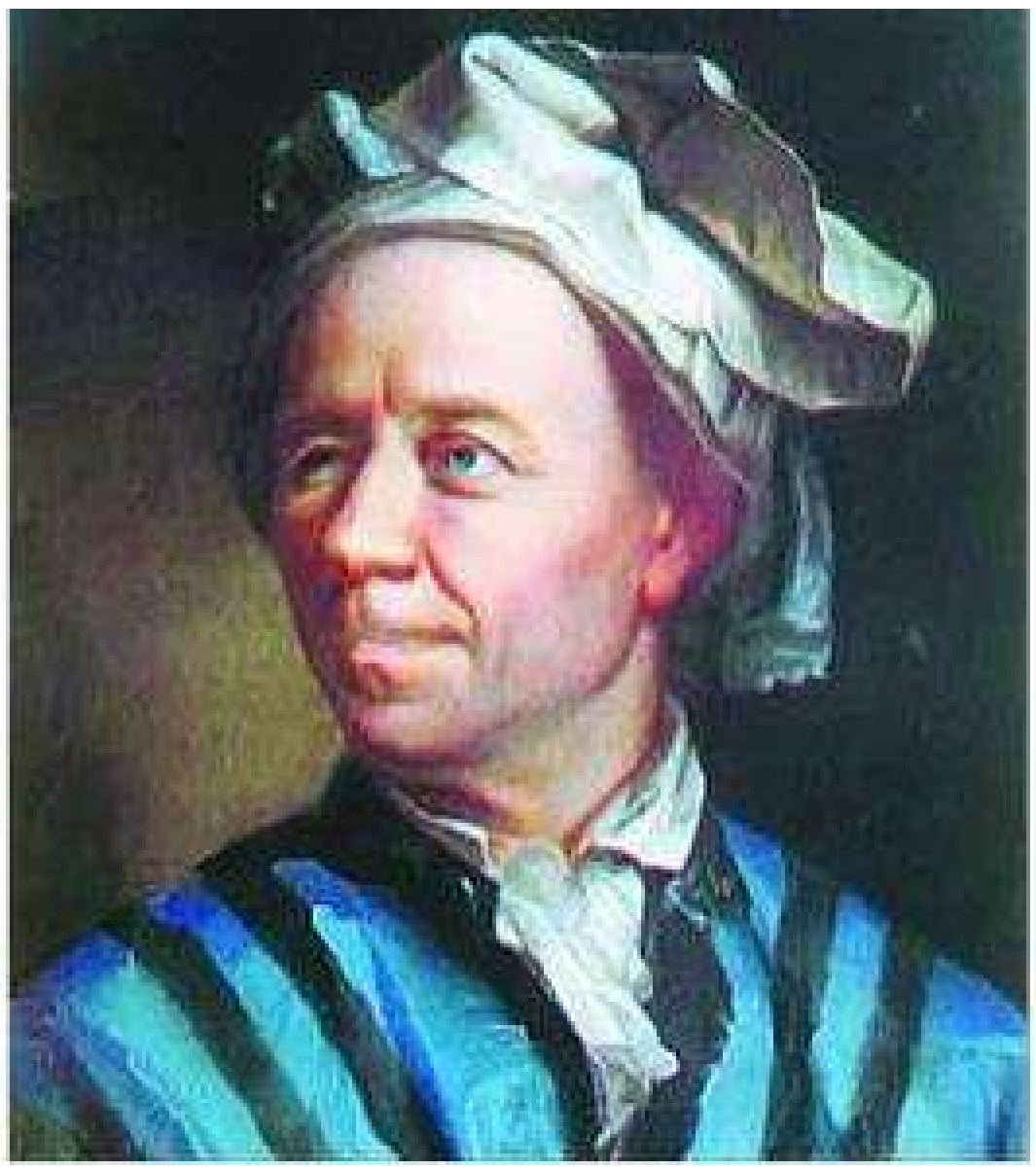

**图 1-1** Euler

作为算法标题之一的 Dirichlet， wiki 一下，一个 19 世纪的人映入了我们的眼帘，Dirichlet（1805～1859）德国数学家，生与现德国 Duren（当时属法国），卒于哥廷根。他是解析数论的奠基者，也是现代函数观念的定义者。在本文中该数学家的主要贡献是 Dirichlet 分布。

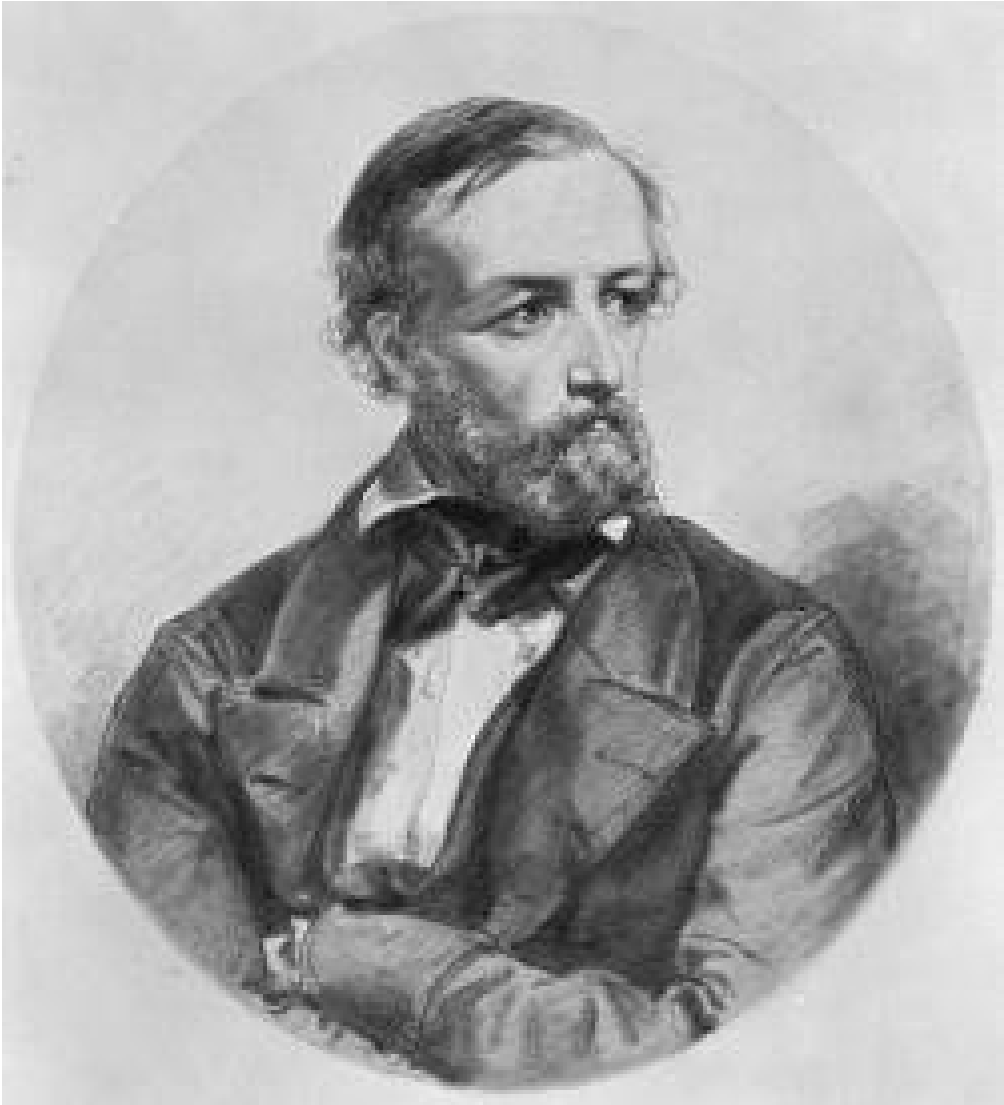

图 **1-2** **Peter Gustav Lejeune Dirichlet**

但是这还不是故事的全部，说到底 19 世纪的时候还没有发明计算机，LDA 应该不是这哥们发明的，于是继续 search，最后查明英国剑桥大学的 David M.Blei 是最初 LDA 论文的作者。Blei 同学借用了 Dirichlet Distribution，而创造了 Latent Dirichlet Allocation。

下面这张照片的 blei 以 PLSA（LDA 之前的另一个概率模型）为基础，加上

了贝叶斯先验，从而诞生了 LDA 算法，LDA 算法最初的论文使用的是变分 EM 方法训练（Variational Inference）。该方法较为复杂，而且最后训练出的 topic 主题非全局最优分布，而是局部最优分布。后期发明了 Collapsed Gibbs Sampling 方法，推导和使用都较为简洁。Blei 及其 LDA 算法正式的介绍如下：

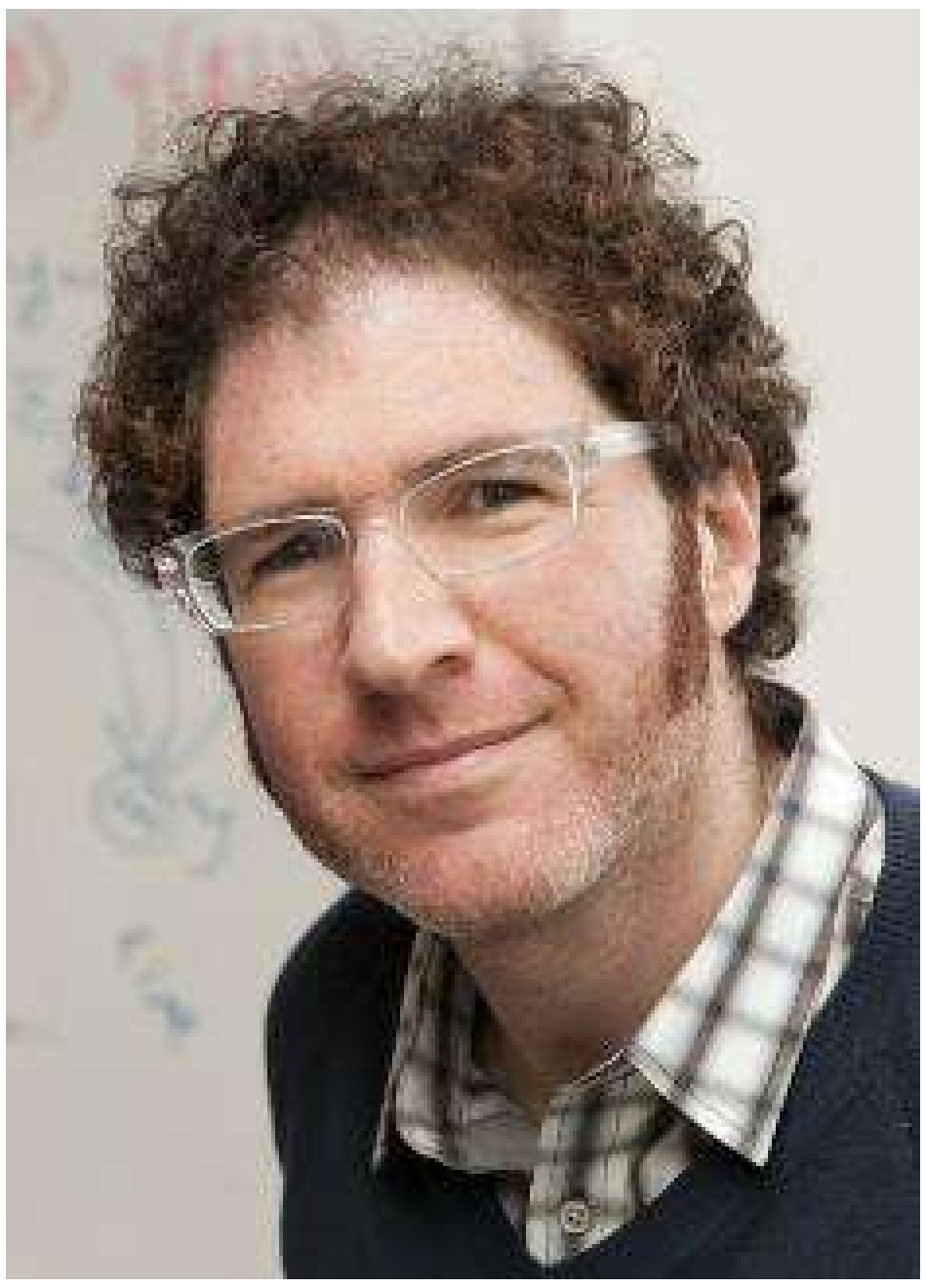

**图 1-3 David Blei**

Latent Dirichlet Allocation 是 Blei 等人于 2003 年提出的基于概率模型的主题模型算法，LDA 是一种无监督机器学习技术，可以用来识别大规模文档集或语料库中的潜在隐藏的主题信息。该方法假设每个词是由背后的一个潜在隐藏的主题中抽取出来。

对于语料库中的每篇文档，LDA 定义了如下生成过程（generative process）：

1. 对每一篇文档，从主题分布中抽取一个主题；
2. 从上述被抽到的主题所对应的单词分布中抽取一个单词；
3. 重复上述过程直至遍历文档中的每一个单词。

LDA 认为每篇文章是由多个主题 mix 混合而成的，而每个主题可以由多个词的概率表征。所以整个程序的输入和输出如表 1-1 所示：

表 1-1 LDA 算法的输入与输出

| | |
|---|---|
| **算法输入：** | 分词后的文章集（通常为一篇文章一行）<br>主题数 K，超参数α和β |
| **算法输出：** | 1.每篇文章的各个词被指定(assign)的主题编号：tassign-model.txt<br>2.每篇文章的主题概率分布θ：theta-model.txt<br>3.每个主题下的词概率分布φ：phi-model.txt<br>4.程序中词语 word 的 id 映射表：wordmap.txt<br>5.每个主题下φ概率排序从高到低 top n 特征词：twords.txt |

如果你想使用一下 LDA 算法，我建议你从 Gibbs LDA++代码起步使用。（http://gibbslda.sourceforge.net/），你用一下，就会发现该算法使用方式还算傻瓜，并且生成的结果文件也挺规则，根据 manual 一看便懂。输入分词后的文件，一个文章一行，输出其中看到每个主题规则文件.twords 如下格式所示：

| Topic 0th: | Topic 1th: |
|---|---|
| 食品 0.022937 | 农业 0.022957 |
| 不合格 0.006634 | 农村 0.022404 |
| 加多宝 0.006634 | 农民 0.014940 |
| 检测 0.006433 | 土地 0.013734 |
| 包装 0.004403 | 粮食 0.007909 |
| 儿童 0.004038 | 农产品 0.006644 |
| 抽检 0.004016 | 耕地 0.005823 |
| 王老吉 0.003759 | 集体 0.005067 |
| 批次 0.003483 | 农户 0.004852 |
| 检验 0.003379 | 种植 0.004674 |
| 红罐 0.003342 | 流转 0.004177 |
| 样品 0.003312 | 农业部 0.003693 |

# 第2章 前置知识

本章所描述的工具和线索在后期 LDA 算法的采样公式推导中会全部明了，关于为什么需要使用这些知识要素，这里面有很长的渊源和历史，比如在概率论和数理统计中，gamma 函数被广泛使用，而在最终的 LDA 采样公式中，你会发现，gamma 函数被神奇地约去消失了。我们在后面的章节中可以看到，LDA 算法精妙之处在于用令人屏息的洞察力作为纽带，将全部零散的部件组合在了一起。

## 2.1 gamma 函数

所谓的 gamma 函数其实就是**阶乘的函数形式**，上过小学我们都知道 n!=1·2·3⋯n，如果我问你 3 的阶乘是多少，你立即回答出 1·2·3=6，但是如果我问你 0.5 阶乘是什么，如果没有 gamma 函数就无法回答了，欧拉经过不懈努力，终于发现阶乘的更一般的函数形式 gamma 函数 $f(x)=\Gamma(x)$，直接给出：

$$\Gamma(x) = \int_0^{+\infty} e^{-t} t^{x-1} dt (x > 0) \tag{2-1}$$

可以直接算出 $\Gamma(1) = \int_0^{+\infty} e^{-t} dt = -e^{-t}\Big|_0^{+\infty} = 1$

- 也可以算出

$\Gamma(\frac{1}{2}) = \sqrt{\pi}$

由于 $\Gamma(x) = \int_0^{\infty} e^{-t} t^{x-1} dt$ 所以 $\Gamma(\frac{1}{2}) = \int_0^{\infty} e^{-t} t^{\frac{1}{2}-1} dt = \int_0^{\infty} e^{-t} t^{-\frac{1}{2}} dt$

设 $t=u^2$，则 $dt = 2udu$。当 $t = 0$ 时，$u = 0$；当 $t = +\infty$ 时，$u=+\infty$

则换元后 $=2\int_0^{\infty} e^{-u^2} u^{-1} u du = 2\int_0^{\infty} e^{-u^2} du = \int_{-\infty}^{\infty} e^{-u^2} du$

令 $I=\int_{-\infty}^{\infty} e^{-u^2} du$，则 $I^2 = \int_{-\infty}^{\infty} e^{-t^2} dt \bullet \int_{-\infty}^{\infty} e^{-u^2} du = \int_{-\infty}^{\infty}\int_{-\infty}^{\infty} e^{-t^2} e^{-u^2} dt du$

做二重积分换元法，做极坐标变换，令 $t = r\cos\theta$，$u = r\sin\theta$，

则当 t 和 u 的区域 D 都为 $-\infty$ 到 $+\infty$，即积分区域为整个坐标轴的时侯

r 半径的范围为 0 到 $+\infty$，而 $\theta$ 的范围为绕坐标轴一圈从 0 到 $2\pi$

使用雅可比行列式：

$$\mathrm{dtdu}=\begin{vmatrix}\frac{\partial t}{\partial r} & \frac{\partial t}{\partial \theta}\\ \frac{\partial u}{\partial r} & \frac{\partial u}{\partial \theta}\end{vmatrix} dr d\theta = \begin{vmatrix}\cos\theta & -r\sin\theta \\ \sin\theta & r\cos\theta\end{vmatrix} dr d\theta = r dr d\theta$$

故 $\mathrm{I}^2 = \int_{-\infty}^{\infty}\int_{-\infty}^{\infty} e^{-\mathrm{t}^2}e^{-\mathrm{u}^2}dtdu = \int_0^{2\pi}d\theta\int_0^{\infty}e^{-r^2}rdr = \pi$

则又因为 I 的被积函数大于 0，则 I>0，最后得 $\mathrm{I}=\sqrt{\pi}$ (2-2)

- 接下来验证 $\Gamma(x+1) = x\Gamma(x)$

$\Gamma(x+1) = \int_0^{+\infty} e^{-t}t^x dt$，做分部积分法（$\int_a^b udv = [uv]_a^b - \int_a^b vdu$）

$$\int_0^{+\infty} e^{-t}t^x dt = -\int_0^{+\infty} t^x d(e^{-t}) = -[t^x e^{-t}]_0^{+\infty} + x\int_0^{+\infty} e^{-t}t^{x-1}dt = x\Gamma(x)$$

也正因为如此，$\Gamma(n) = (n-1)!$ (2-3)

## 2.2 二项分布(binomial distribution)

在概率论中，二项分布即重复 n 次独立的伯努利试验。在每次试验中只有两种可能的结果（成功/失败），每次成功的概率为 p，而且两种结果发生与否互相对立，并且相互独立，与其它各次试验结果无关，事件发生与否的概率在每一次独立试验中都保持不变，则这一系列试验总称为 n 重伯努利实验，当试验次数为 1 时，二项分布就是伯努利分布。

在给出二项分布之前，我们来做一个例子，假设你在玩 CS 这个游戏，你拿着狙击枪，敌人出现你打中敌人的概率是 p，打不中敌人的概率是 1-p，那么敌人第一次出现你没打中而第二次出现你打中的概率是 (1-p)·p。如果敌人出现了 n 次，而你打中了其中的 k 次，而不确定具体在哪 k 次（第 1 次，还是第 4 次？），这样从 n 次中任取 k 次的次数是 $\mathrm{C}_n^k = \binom{n}{k}$，而这不确定的 k 次打中敌人的概率是：$\mathrm{C}_n^k p^k(1-p)^{n-k}$，通过这个例子我们便得知了二项分布的概率。

二项分布的概率密度函数是：

$$f(k;n,p)=\mathrm{P}(X=k)=\binom{n}{k}p^k(1-p)^{n-k}$$
$$for\ \text{k=0,1,2,...,n, where} \qquad (2\text{-}4)$$
$$\binom{n}{k}=\frac{n!}{k!(n-k)!}$$

## 2.3 beta 分布(beta distribution)

在概率论中，beta 分布是指一组定义在区间(0,1)的连续概率分布，有两个参数$\alpha$和$\beta$，且$\alpha,\beta$>0。

Beta 分布的概率密度函数是：

$$\begin{aligned}
&f(x;\alpha,\beta)=\frac{x^{\alpha-1}(1-x)^{\beta-1}}{\int_0^1 u^{\alpha-1}(1-u)^{\beta-1}du}\\
&=\frac{\Gamma(\alpha+\beta)}{\Gamma(\alpha)\Gamma(\beta)}x^{\alpha-1}(1-x)^{\beta-1}\\
&=\frac{1}{B(\alpha,\beta)}x^{\alpha-1}(1-x)^{\beta-1}
\end{aligned} \qquad (2\text{-}5)$$

随机变量 X 服从参数为的 beta 分布通常写作：$X\sim Beta(\alpha,\beta)$。

这个式子中分母的函数$B(\alpha,\beta)$称为β函数。

➢ 这里我们来证明一个重要的公式，该公式中的关系在 LDA 算法 Gibbs Sampling 采样公式中也有使用，这个关系也就是**β函数和 Gamma 函数的关系**（该公式也被称为第一型欧拉积分）：

$$\mathrm{B}(a,\mathrm{b})=\frac{\Gamma(a)\Gamma(b)}{\Gamma(a+b)}=\int_0^1\mu^{a-1}(1-\mu)^{\mathrm{b}-1}d\mu \qquad (2\text{-}6)$$

下面我们给出这个关系式的两种证明方法。

**证明方法 1：**

$$Beta(\mu\mid a,b)=\frac{\Gamma(a+b)}{\Gamma(a)\Gamma(b)}\mu^{a-1}(1-\mu)^{b-1}$$

$$\Gamma(a)\Gamma(\mathrm{b})=\int_0^\infty x^{a-1}e^{-x}dx\underbrace{\int_0^\infty y^{b-1}e^{-y}dy}_{\text{用dt换掉}dy}$$

设$t=x+y$,则$dy=d(t-x)=dt$

当$y=0$的时候，$t=x$；当$y=\infty$的时候，$t=\infty$

$\int_0^\infty y^{b-1}e^{-y}dy = \int_x^\infty (\mathrm{t}-\mathrm{x})^{b-1}e^{-(\mathrm{t}-\mathrm{x})}dt = \int_x^\infty (\mathrm{t}-\mathrm{x})^{b-1}e^{-t}e^{x}dt$

故原式=$\int_0^\infty x^{a-1}e^{-x}dx \bullet \int_x^\infty (\mathrm{t}-\mathrm{x})^{b-1}e^{-t}e^{x}dt$

$= \int_0^\infty x^{a-1}\int_x^\infty (t-x)^{b-1}e^{-t}dtdx$

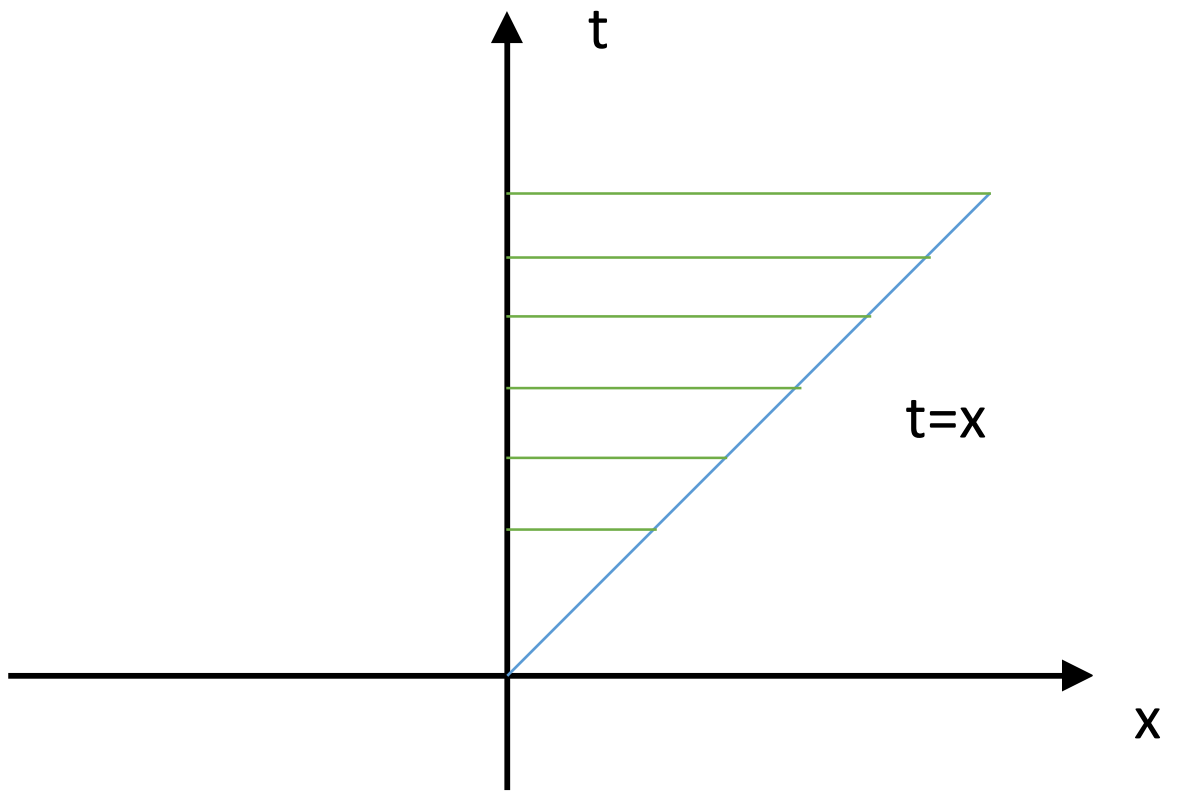

**图 2-1 积分区域**

在图 2-1 的上半部分（阴影部分）为积分区域，则交换积分次序：

$= \int_0^\infty e^{-t}\int_0^t x^{a-1}(\mathrm{t}-\mathrm{x})^{b-1}dxdt$

再令 $x = t\mu$，则 $x = t$ 的时候，$\mu = 1$；$x = 0$ 的时候，$\mu = 0$

按换元规则 $dx = td\mu;\ x^{a-1}(t-x)^{b-1} = (t\mu)^{a-1}(t-t\mu)^{b-1} = t^{a+b-2}\mu^{a-1}(1-\mu)^{b-1}$，则原式

$=\int_0^\infty e^{-t}\int_0^1 t^{a+b-2}\mu^{a-1}(1-\mu)^{b-1} \bullet td\mu dt$

$= \int_0^\infty e^{-t}t^{a+b-1}dt\int_0^1 \mu^{a-1}(1-\mu)^{b-1}d\mu$

$\therefore \Gamma(a)\Gamma(\mathrm{b}) = \Gamma(a+b)\int_0^1 \mu^{a-1}(1-\mu)^{b-1}d\mu$

$\int_0^1 \mu^{a-1}(1-\mu)^{b-1}d\mu=\frac{\Gamma(a)\Gamma(\mathrm{b})}{\Gamma(a+b)}$，

所以 $\int_0^1 \frac{\Gamma(a+b)}{\Gamma(a)\Gamma(\mathrm{b})}\mu^{a-1}(1-\mu)^{b-1}d\mu=1$

第二种证明方法不涉及多重积分，因此颇为简洁，如下：

$Beta(\mu \mid a,b) = \int_0^1 \mu^{a-1}(1-\mu)^{\mathrm{b}-1}d\mu$

这个方法的关键之处在于应用分部积分法，

此时设置 $u = \mu^{a-1}$，$dv = (1-\mu)^{b-1}d\mu$，然后便可得到

$du = (a-1)y^{a-2}dy$，$v = \frac{-(1-\mu)^b}{b}$，换元之后再使用分部积分法可得：

$$B(a,b)=\mu^{a-1}\left(\frac{-(1-\mu)^b}{b}\right)\Bigg|_0^1+\frac{a-1}{b}\int_0^1\mu^{a-2}\cdot(1-\mu)^b d\mu$$

这个加和式的前一部分等于 0，后一部分容易观察到这又是一个β函数：

$$=\frac{a-1}{b}B(a-1,b+1)$$

我们思如泉涌，如同多米诺骨牌，引发了连锁反应，很快便发现一个规律：

$$=\frac{(a-1)(a-2)\cdots 1}{b(b+1)(b+2)\cdots(b+a-2)}B(1,b+a-1)$$

$$=\frac{(a-1)(a-2)\cdots 1}{b(b+1)\cdots(b+a-2)}\int_0^1(1-\mu)^{a+b-2}d\mu$$

$$=\frac{(a-1)(a-2)\cdots 1}{b(b+1)\cdots(b+a-2)}\left[\frac{-(1-\mu)^{a+b-1}}{a+b-1}\right]\Bigg|_0^1$$

$$=\frac{(a-1)(a-2)\cdots 1}{b(b+1)\cdots(b+a-2)(b+a-1)}$$

$$=\frac{\Gamma(a)\Gamma(b)}{\Gamma(a+b)}$$

这个证明方法分别连续使用了分部积分法，在最后又利用了 gamma 函数的阶乘特性，显得非常巧妙和简洁。

➢ Beta 分布的期望

如果 $p \sim Beta(t \mid \alpha,\beta)$，则

$$E(p)=\int_0^1 t\bullet Beta(t\mid\alpha,\beta)dt$$

$$=\int_0^1 t\bullet\frac{\Gamma(\alpha+\beta)}{\Gamma(\alpha)\Gamma(\beta)}t^{\alpha-1}(1-t)^{\beta-1}dt=\frac{\Gamma(\alpha+\beta)}{\Gamma(\alpha)\Gamma(\beta)}\int_0^1 t^{\alpha}(1-t)^{\beta-1}dt$$

这个式子右边的概率对应到 $\mathrm{Beta}(\mathrm{t}\mid\alpha+1,\beta)$，

有 $\int_0^1\frac{\Gamma(\alpha+\beta+1)}{\Gamma(\alpha+1)\Gamma(\beta)}t^{\alpha}(1-t)^{\beta-1}dt=1$，带回到 E(p)式：

$$E(p)=\frac{\Gamma(\alpha+\beta)}{\Gamma(\alpha)\Gamma(\beta)}\bullet\frac{\Gamma(\alpha+1)\Gamma(\beta)}{\Gamma(\alpha+\beta+1)}$$，化简后

$$=\frac{\alpha}{\alpha+\beta}$$

这说明，beta 分布的均值可以用 $\frac{\alpha}{\alpha+\beta}$ 来估计

对于后面 2.6 节提到的 Dirichlet 分布也有相似结论：

如果 $\vec{p}\sim Dir(\vec{t}\mid\vec{\alpha})$，可以证明：$E(\vec{p})=\left(\frac{\alpha_1}{\sum_{i=1}^K\alpha_i},\frac{\alpha_2}{\sum_{i=1}^K\alpha_i},\frac{\alpha_3}{\sum_{i=1}^K\alpha_i}\ldots,\frac{\alpha_K}{\sum_{i=1}^K\alpha_i}\right)$ (2-7)

这个结论在 LDA 算法做完 Gibbs sampling 后，估计$\vec{\theta}$和$\vec{\varphi}$时用到。

## 2.4 多项分布(multinomial distribution)

多项分布[1]是二项分布的推广扩展，在 n 次独立试验中每次只输出 k 种结果中的一个，且每种结果都有一个确定的概率 p。多项分布给出了在多种输出状态的情况下，关于成功次数的各种组合的概率。

举个例子，投掷 n 次骰子，这个骰子共有 6 种结果输出（k=6），且 1 点出现概率为 $p_1$，2 点出现概率 $p_2$，…多项分布给出了在 n 次试验中，骰子 1 点出现 $x_1$ 次，2 点出现 $x_2$ 次, 3 点出现 $x_3$ 次，…，6 点出现 $x_6$ 次。这个结果组合的概率为：

$$f(x_1,...,x_k;n,p_1,...,p_k) = \mathrm{P}(X_1 = x_1 \text{ and } ... \text{ and } X_k = x_k)$$
$$= \begin{cases} \dfrac{n!}{x_1!...x_k!} p_1^{x_1} \bullet ... \bullet p_k^{x_k} & \text{when } \sum_{i=1}^{k} x_i = n \\ 0 & \text{otherwise} \end{cases} \tag{2-8}$$

公式(2-8)为多项分布的概率公式，注意在这个公式中，$x_i$为第 i 种状态的输出结果的频度，如果 k=2，只有两种情况，此公式将退化为二项分布，所以二项分布是特殊情况下的多项分布。

也可以用 gamma 函数表示（这个写法的形式和 Dirichlet 分布相似）：

$$f(x_1,...,x_k;n,p_1,...,p_k) = \frac{\Gamma(\sum_i x_i + 1)}{\prod_i \Gamma(x_i + 1)} \prod_{i=1}^{k} p_i^{x_i} \tag{2-9}$$

下面我们通过一个例题加深对多项分布的印象:

- 问题：同时投掷 5 枚骰子，出现两对点数一样的概率是多少？

解：现在先把问题简化成特定投掷到 2 个一点，2 个二点，1 个三点的概率是多大？

$X_1$ 到 $X_6$ 表示六个点的出现次数之和为 5，则：

$$\mathrm{P}(X_1 = 2, X_2 = 2, X_3 = 1, X_4 = 0, X_5 = 0, X_6 = 0) = \frac{5!}{2!2!1!0!0!0!}(\frac{1}{6})^2(\frac{1}{6})^2(\frac{1}{6})^1(\frac{1}{6})^0(\frac{1}{6})^0(\frac{1}{6})^0$$

这里 0!=1

[1] 陈希孺《概率论与数理统计》 p60, p67-例 2.7

$= \frac{5}{1296}$

再考虑，现在 $X_1$ 到 $X_6$ 其中 2 个取 2,1 个取 1 的种类有多少种？

| $X_1$ | $X_2$ | $X_3$ | $X_4$ | $X_5$ | $X_6$ |
|---|---|---|---|---|---|
| 2 | 2 | 1 | 0 | 0 | 0 |
| 2 | 2 | 0 | 1 | 0 | 0 |
| 2 | 0 | 2 | 1 | 0 | 0 |

.......

先不考虑 2,2,1 三者顺序时共有 $\binom{6}{3}$ 种取法;再考虑下 2,2,1 三者交换顺序有 3 种，因为两个 2 先后交换仍为 2,2。

所以 $X_1$ 到 $X_6$ 其中 2 个取 2,1 个取 1 的种类有 $3 \bullet \binom{6}{3} = 60$

最后答案概率=$60 \bullet \frac{5}{1296} = \frac{25}{108}$

**多项分布的极大似然估计：**

需要特别值得说明是 “多项分布的似然函数”容易让读者困惑。这里特别说明一下，我们将多项分布的概率公式(2-8)重新写下来：

$$p(x_1,\cdots,x_k \mid p_1,\cdots,p_k) = f(x_1,\ldots,x_k;n,p_1,\ldots,p_k) = \begin{cases} \frac{n!}{x_1!\ldots x_k!} p_1^{x_1} \cdot \ldots \cdot p_k^{x_k} & when \sum_{i=1}^{k} x_i = n \\ 0 & otherwise \end{cases}$$

注意这个公式中的 $x_i$ 为第 i 种状态的输出结果的频度，其出现在指数部分，每个状态的可能性为 $p_1,p_2,\cdots,p_k$，且 $\sum_{i=1}^{k} p_i = 1$。在极大似然估计中，由于使用 log 形式的似然函数（log-likelihood），随后对其求导，获取似然函数的极值。在这个过程中，多项式系数作为常数项通常被无情地忽略了，我们做如下分析：

根据极大似然估计的原理，对于确定的 n 次试验结果，多项分布的似然函数满足：

$$L(p_1,\cdots,p_k \mid n;x_1,\cdots,x_k) = p(x_1,\cdots,x_k \mid p_1,\cdots,p_k) = n!\prod_{i=1}^{k} \frac{p_i^{x_i}}{x_i!}$$

且 $\sum_{i=1}^{k} x_i = n$ 以及 $\sum_{i=1}^{k} p_i = 1$

接着使用 log-likelihood 技法:

$$\log\Big(L(p_1,...,p_k)\Big)=\log(n!)-\sum_{i=1}^{k}\log(x_i!)+\sum_{i=1}^{k}x_i\log(p_i)$$

s.t. $\sum_{i=1}^{k}p_i=1$

引入拉格朗日乘数法（如果不了解拉格朗日乘数法，去看本书 6.1.4 节），则：

$$Lagrange(\mathbf{p},\lambda)=\log(n!)-\sum_{i=1}^{k}\log(x_i!)+\sum_{i=1}^{k}x_i\log(p_i)+\lambda(\sum_{i=1}^{k}p_i-1)$$

紧接着对其按照参数 p 求导，前两项不含 p 求导得 0，被忽略，由此公式(2-8)多项式系数作为常数项就都被忽略了。

$$\frac{\partial\Big(Lagrange(\mathbf{p},\lambda)\Big)}{\partial p_i}=\frac{x_i}{p_i}+\lambda=0\Rightarrow p_i=-\frac{x_i}{\lambda}\quad(i=1,\cdots k)$$

$$\sum_{i=1}^{k}p_i=1\Rightarrow\sum_{i=1}^{k}(-\frac{x_i}{\lambda})=1$$

$$\sum_{i=1}^{k}(-\frac{x_i}{\lambda})=1\Rightarrow\lambda=-\sum_{i=1}^{k}x_i=-n\text{，}$$

所以再将 $\lambda$ 带入 $p_i=-\frac{x_i}{\lambda}$ 可得 $\hat{p}_i=\frac{x_i}{n}$。

直观思考一下多项分布的极大似然估计，其实可想而知，就是数数 $x_i$ 的个数然后算一下占整个样本中的比例就可以作为 $p_i$ 概率的估计了。所以通常在使用似然函数时，可以忽略其常数项—多项式系数。

## 2.5 狄利克雷分布(dirichlet distribution)

dirichlet 分布是 beta 分布在多项情况下的推广，也是多项分布的共轭先验分布（共轭先验分布在 2.6 节讲）。dirichlet 分布的概率密度函数如下：

$$f(p_1,...,p_{k-1};\alpha_1,...,\alpha_k)=\frac{1}{\Delta(\vec{\alpha})}\prod_{i=1}^{k}p_i^{\alpha_i-1}\text{，这里}\Delta(\vec{\alpha})=\frac{\prod_{i=1}^{k}\Gamma(\alpha_i)}{\Gamma(\sum_{i=1}^{k}\alpha_i)}\tag{2-10}$$

二项分布和多项分布很相似，Beta 分布和 Dirichlet 分布很相似，而至于“Beta 分布是二项式分布的共轭先验概率分布，而狄利克雷分布（Dirichlet 分

布）是多项式分布的共轭先验概率分布”这点在下文中说明。

另一个重要的公式是: $$\int \prod_{i=1}^{k} p_i^{\alpha_i - 1} d\vec{p} = \Delta(\vec{\alpha}) = \frac{\prod_{i=1}^{k}\Gamma(\alpha_i)}{\Gamma(\sum_{i=1}^{k}\alpha_i)}$$ (2-11)

为了简便表达，公式中引入了一个新的三角形符号的希腊字母Δ（念“德尔塔”Delte）代表B函数的多项版本，这个公式的结构和证明**相似于**上文中“β函数和 Gamma 函数的关系—公式(2-6)”，这个证明留给读者来完成。从此，公式中凡是出现积分中连乘时，就要像巴甫洛夫试验中流着口水的狗一样警觉，建立起“可以换成 gamma 函数”的条件反射。

## 2.6 共轭先验分布(conjugacy prior)

In Bayesian probability theory, if the posterior distributions p($\theta$|$x$) are in the same family as the prior probability distribution p($\theta$), the prior and posterior are then called conjugate distributions, and the prior is called a conjugate prior for the likelihood function. [2]

所谓的共轭，只是我们选取(choose)一个函数作为似然函数(likelihood function)的 prior probability distribution，使得后验分布函数[3](posterior distributions)和先验分布函数形式一致。比如 Beta 分布是二项式分布的共轭先验概率分布，而狄利克雷分布(Dirichlet 分布）是多项式分布的共轭先验概率分布。为什么要这样做呢？这得从贝叶斯估计[4]谈起：

根据贝叶斯规则，后验分布=似然函数*先验分布

$$p(\theta \mid x) = \frac{\overbrace{p(x \mid \theta)}^{likelihood} \overbrace{p(\theta)}^{prior\ belief}}{\underbrace{p(x)}_{evidence}} = \frac{p(x \mid \theta)p(\theta)}{\int p(x \mid \theta)p(\theta)d\theta} \propto p(x \mid \theta)p(\theta)$$ (2-12)

*参数估计是一个重要的话题。对于典型的离散型随机变量分布：二项式分布，多项式分布；典型的连续型随机变量分布：正态分布。他们都可以看着是参数分布，因为他们的函数形式都被一小部分的参数控制，比如正态分布的均值和方差，二项式分布事件发生的概率等。因此，给定一堆观测数据集（假定数据满*

[2] http://en.wikipedia.org/wiki/Conjugate_prior

[3] 后验之意为执果索因，2.6 节其实是贝叶斯估计，欲了解更多，参见陈希孺《概率论与数理统计》p155

*足独立同分布），我们需要有一个解决方案来确定这些参数值的大小，以便能够利用分布模型来做密度估计。这就是参数估计！*

*对于参数估计，一直存在两个学派的不同解决方案。一是频率学派解决方案：通过某些优化准则（比如似然函数）来选择特定参数值；二是贝叶斯学派解决方案：假定参数服从一个先验分布，通过观测到的数据，使用贝叶斯理论计算对应的后验分布。先验和后验的选择满足共轭，这些分布都是指数簇分布的例子。*[4]

---

简而言之，假设参数θ也是变量而非常量，而且在做试验前已经服从某个分布 p(θ)（来源于以前做试验数据计算得到，或来自于人们的主观经验），然后现在做新试验去更新这个分布假设。如果不知道最大似然估计(Maximum Likelihood)的概念，参见附录。

### 2.6.1 从二项分布到 beta 分布

注意二项分布概率密度函数为$f(k;n,p) = \mathrm{P}(X = k) = \binom{n}{k} p^k (1-p)^{n-k}$，将参数去掉，变成形式：$f(p) \propto p^k (1-p)^{n-k}$，再加上归一化因子 $B(\alpha,\beta)$（注意这个归一化因子含有 k 和 n，但绝不含 p），变为 beta 分布：$f(x;\alpha,\beta) = \frac{1}{B(\alpha,\beta)} x^{\alpha-1}(1-x)^{\beta-1}$，beta 分布前一项β函数是确保 beta 分布是归一化(normalized)。

➢ 求证：beta 分布确实是二项分布的共轭先验分布

证明：

1. 二项分布的似然函数：

$$L = \binom{s+f}{s} p^s (1-p)^f$$

这里 s 表示 n 次试验中成功的次数，f 表示 n 次伯努利试验中失败的次数。

2. 先验分布 beta 分布如下：

$P(p \mid \alpha,\beta) = \frac{p^{\alpha-1}(1-p)^{\beta-1}}{B(\alpha,\beta)}$，其中 $B(\alpha,\beta) = \frac{\Gamma(\alpha)\Gamma(\beta)}{\Gamma(\alpha+\beta)}$

---

[4] 采用贝叶斯法的出发点在于如果仅仅做了很少几次试验来估计参数（比如均值或概率 p 等），就会由于小样本数据造成估计不准。但利用之前已经拥有的经验（以前做过的试验数据）就可以令估计更为准确合理

3. 由于 prior distribution * likehood = post distribution

$$P(p \mid s,f,\alpha,\beta) = \frac{\binom{s+f}{s} \bullet p^s(1-p)^f \bullet p^{\alpha-1}(1-p)^{\beta-1} \Big/ B(\alpha,\beta)}{\int_{q=0}^{1} (\binom{s+f}{s} q^s(1-q)^f \bullet q^{\alpha-1}(1-q)^{\beta-1} \Big/ B(\alpha,\beta))dq}$$

$$= \frac{p^{s+\alpha-1}(1-p)^{f+\beta-1} \; / \; B(\alpha,\beta)}{\int_{q=0}^{1} (q^{s+\alpha-1}(1-q)^{f+\beta-1} \; / \; B(\alpha,\beta))dq}$$

$$= \frac{p^{s+\alpha-1}(1-p)^{f+\beta-1}}{B(s+\alpha, f+\beta)} \tag{2-13}$$

这就可以看到后验分布(post distribution)又变为 beta 分布，也就是和先验(prior distribution)一致了，因此我们称之为共轭(conjugacy)。

观察到后验和先验都是 beta 分布，但 $X \sim Beta(\alpha,\beta)$ 变为了 $X \sim Beta(\alpha+\mathrm{s},\beta+f)$，超参数变了。如果以后有新增的观测值，后验分布又可作为先验分布来进行计算。具体来讲，在某一个时间点，有一个观测值，此时可以得到后验，之后，每一个观测值的到来，都以之前的后验作为先验，乘以似然函数后，得到修正后的新后验。在这每一步中，其实我们不需要管什么似然函数，我们可以将后验分布看作是以代表 x=1 出现“次数”的参数 s 和代表 x=0 出现“次数”的参数 f 为参数的 beta 分布：当有一个新的 x=1 的观测量到来的时候，s=$\alpha$+1,f=$\beta$，即$\alpha$的值相应的加 1;否则 s=$\alpha$, f=$\beta$+ 1 即$\beta$的值加 1。所以这也就是超参数($\alpha$,$\beta$)又被称之为伪计数(pseudo count)的原因。

我们可以把上面性质所表示的方法看作为序列方法（sequential approach），该方法是贝叶斯观点中很自然得到的学习方法。它非常适合实时学习场景。在某一时刻，有一个观测数据，或是一小批量数据，在下一批观测数据到来之前我们就可以丢弃它们，因为我们可以在一开始的小批量数据中得到我们的后验分布模型，当有新的一批数据到来时，只需要更新这个模型就够了。一个实时学习应用场景是：在所有数据到来之前，预测就必须通过之前稳定到达的一部分数据流来做出预测。注意到，因为这种序列方法不需要将所有数据都载入内存，因此他在大数据的应用将非常有效。当然，之前频率学派所使用的最大化似然函数方法也可以转换成这种序列式方法的。

另外，我们做先验分布的目的是估计参数，比如投掷硬币试验，我们需要根据已有的观测数据，估计下一次试验硬币的正面结果概率是多少。

$$P(x=1 \mid D)=\int_0^1 p(x=1 \mid \mu) p(\mu \mid D) d\mu=\int_0^1 \mu p(\mu \mid D) d\mu=E(\mu \mid D) \tag{2-14}$$

这正是 beta 分布的均值，结合 beta 分布的均值公式可以得到：

$$P(x=1 \mid D)=\frac{s+\alpha}{s+\alpha+f+\beta} \tag{2-15}$$

### 2.6.2 从多项分布到 Dirichlet 分布

通过观察多项式分布的形式（$\prod_{k=1}^{K} p_k^{m_k}$），我们选取先验分布的形式为（保留带概率 $p_k$ 的项）：

$$P(\mu \mid \vec{\alpha}) \propto \prod_{k=1}^{K} p_k^{\alpha_k-1} \ (0 \leq p_k \leq 1 \text{ and } \sum_k p_k=1) \tag{2-16}$$

这里 $\alpha_1,\alpha_2,\alpha_3,\ldots,\alpha_k$ 代表这个分布的超参数（或伪计数），$\vec{\alpha}=(\alpha_1,\alpha_2,\alpha_3,\ldots,\alpha_k)^T$。由于有 $0 \leq \mu_k \leq 1$ and $\sum_k \mu_k=1$ 这两个条件的限制(用 $\mu_k$ 等价于 $p_k$)，因此 $\{\mu_k\}$ 之上的分布是 K-1 维度：

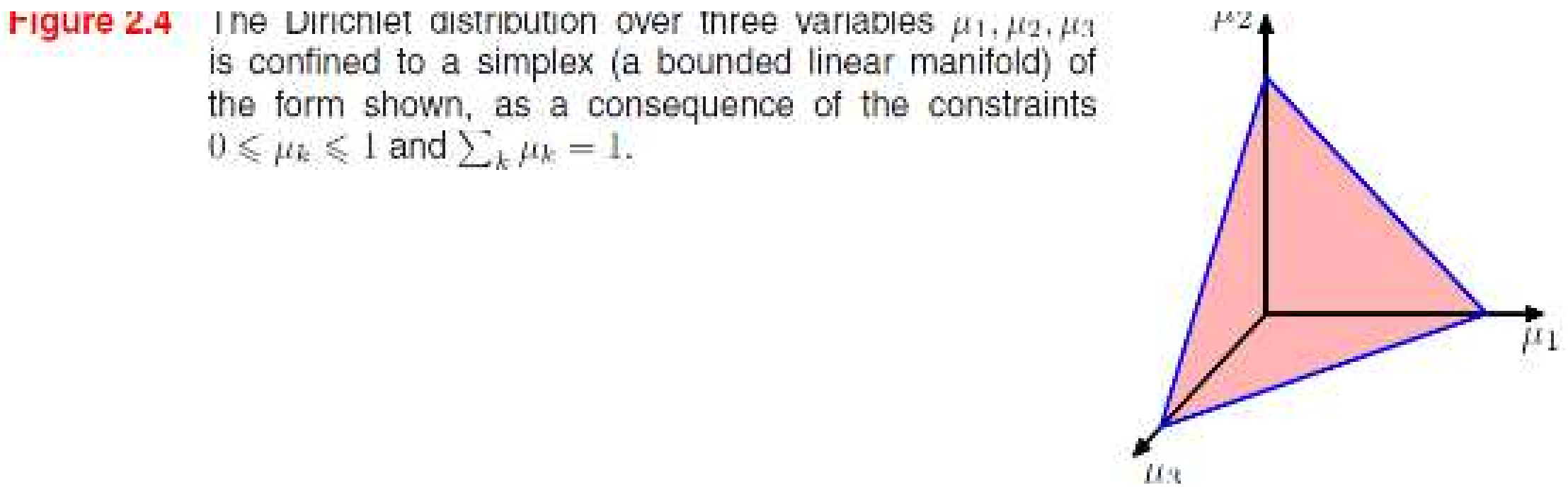

**图 2-2 节选自《Pattern Recognition and Machine Learning》[7] p77**

图 2-2 的含义是假设一共有 3 个概率，这 3 个概率$\mu$的范围被限制在红色框所围的坐标轴内。

将(2-16)式加入系数归一化(normalized),便可得到 Dirchlet 分布的概率密度表达式：

$$f(p_1,...,p_{k-1};\alpha_1,...,\alpha_k)=\frac{1}{\Delta(\vec{\alpha})}\prod_{i=1}^{k}p_i^{\alpha_i-1}\text{，这里 }\Delta(\vec{\alpha})=\frac{\prod_{i=1}^{k}\Gamma(\alpha_i)}{\Gamma(\sum_{i=1}^{k}\alpha_i)}$$

由于 $P(\vec{p}\mid D,\vec{\alpha})\propto\underbrace{p(D\mid\vec{p})}_{multinomial\ distribution}\bullet\underbrace{p(\vec{p}\mid\vec{\alpha})}_{dirichlet\ distribution}\propto\prod_{k=1}^{K}p_k^{\alpha_k+m_k-1}$，这里 $m_k$ 是:做 n 次实验输出 k 编号结果的次数，这个 $m_k$ 也叫充分统计量。跟上一节一样，这里的 $m_k$ 就是根据新试验去更新超参数。注意，f 函数内自变量的 $p_1$ 到 $p_{k-1}$ 而非 $p_k$，这是由于 $p_1$ 到 $p_{k-1}$ 一旦确定下来，而因为概率之和=1（$0\le p_k\le 1$ 以及 $\sum_k p_k=1$），因此前面这些“自由变量”一旦被确定，最后一个变量=1-($p_1+p_2+...+p_k$)固定下来了而不“自由”了。

- 求证：dirichlet 分布确实是多项分布的共轭先验分布

  证明:

  1. 多项分布的似然函数

$$L=\frac{n!}{x_1!...x_k!}\prod_{k=1}^{K}p_k^{x_k}\ \text{when}\ \sum_{i=1}^{k}x_i=n\ \text{and}\ \sum_{i=1}^{k}p_i=1$$

  2. 先验分布 dirichlet 分布

$$\mathrm{Dir}(\vec{\mathrm{p}}|\vec{\alpha})=\underbrace{\frac{\Gamma(\sum_{i=1}^{k}\alpha_i)}{\prod_{i=1}^{k}\Gamma(\alpha_i)}}_{1/\Delta(\vec{\alpha})}\prod_{k=1}^{K}p_k^{\alpha_k-1}$$

  3. 由于 $\text{后验}=\underbrace{\text{先验}}_{Dir}\bullet\underbrace{\text{似然函数}}_{Mult}$

$$=\frac{\overbrace{\frac{n!}{x_1!...x_k!}}^{\text{约去}}\prod_{k=1}^{K}p_k^{x_k}\bullet\frac{\prod_{k=1}^{K}p_k^{\alpha_k-1}}{\underbrace{\Delta(\vec{\alpha})}_{\text{约去}}}}{\int\underbrace{\frac{n!}{x_1!...x_k!}}_{\text{约去}}\prod_{k=1}^{K}p_k^{x_k}\bullet\frac{\prod_{k=1}^{K}p_k^{\alpha_k-1}}{\underbrace{\Delta(\vec{\alpha})}_{\text{约去}}}d\vec{p}}=\frac{\prod_{k=1}^{K}p_k^{x_k+\alpha_k-1}}{\int\prod_{k=1}^{K}p_k^{x_k+\alpha_k-1}d\vec{p}}$$

又因为 $\int_{\sum p_i=1}\prod_{i=1}^{k} p_i^{\alpha_i-1}d\vec{p} = \Delta(\vec{\alpha}) = \frac{\prod_{i=1}^{k}\Gamma(\alpha_i)}{\Gamma(\sum_{i=1}^{k}\alpha_i)}$，用这个式子替换掉分母：

$$\text{则原式} = \frac{\prod_{k=1}^{K} p_k^{x_k+\alpha_k-1}}{\Delta(\vec{\alpha}+\vec{x})} = Dir(\vec{p} \mid \vec{\alpha}+\vec{x}) \quad (2\text{-}17)$$

## 2.7 总结

1. 贝叶斯学派采用给参数赋予先验分布，并使得先验与后验共轭，通过求后验均值来得到参数的估计，频率学派通过某个优化准则比如最大化似然函数来求得参数的估计；不管是哪个学派思想，都要用到似然函数，注意到似然函数有所不同，这点在极大似然估计(MLE)和最大后验概率估计(MAP)体现得尤其明显。

2. 当拥有无限数据量时（beta 分布式中 s 和 f 都趋向于无穷，dirichlet 分布式中 m 趋向于无穷），贝叶斯方法和频率学派方法所得到的参数估计是一致的。当在有限的数据量下，贝叶斯学派的参数后验均值的大小介于先验均值和频率学派方法得到参数估计。比如在抛硬币实验中，当数据量有限时，先验均值为 0.5，后验均值将会比先验大，比频率学派得到参数估计小。

3. 随着观测数据的增多，后验分布曲线越来越陡峭（越来越集中），即方差越来越小（后验方差总比前验方差小），当数据量无穷大时，方差趋近于 0，即随着数据越来越多，后验的不确定性在减小。

## 参考文献

# 第3章 LDA 的 Gibbs Sampling 推导

## 3.1 unigram 假设

假设有 N 个服从 i.i.d(独立同分布)的单词从一个多项式分布(multinomial)中抽取，在 N 个词中，我们关注 $v_i$ 的发生次数为 n(t)，那么 $\vec{n} = (n_1, n_2, ... n_V)$ 则记为 w~Mult(w|p)，该文档生成的概率是

$$p(\vec{w}) = p(w_1)p(w_2)...p(w_n) = \prod_{t=1}^{V} p_t^{n_t} \quad \sum_{t=1}^{V} p_t = 1$$

(3-1)

其中 $n_t$ 为该单词在文章中出现的次数,V 为字典中的单词个数。在这里其中每个词的概率 $\hat{p}_i = \frac{n_i}{N}$。由于这种方法不考虑文章内单词间的顺序，因此被称之为词袋模型（bag of words）。可以注意到上面的(3-1)式子忽略了第 2 章的多项式公式(2-8)式中的多项式系数。得到了生成一篇文章的概率，可以将生成一篇文章的概率扩展到整个语料集。这里可以引入一个新的概念：概率图模型，来画出这种模型，正如图 3-1 所示，图中被涂色的 w 表示可观测变量，方框表示重复抽取的次数，N 表示一篇文档中总共 N 个单词，M 表示 M 篇文档。也就是说，重复抽取 M 篇文档，每个文档抽取 N 个单词，这样的生成模型生成了整个语料（corpus）。

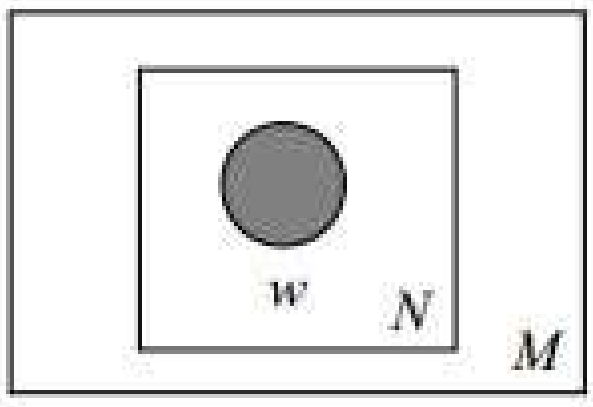

**图 3-1 基于 unigram 的图模型**

这个生成文档的概率可以用投掷骰子的游戏来模拟：上帝拿出一枚骰子，骰子有 V 个面，每个面代表一个词，每个面的概率是 $\vec{p} = (p_1, p_2, ..., p_V)$，投掷 N 次骰子每个面产生的次数分别是 $\vec{n} = (n_1, n_2, ... n_V)$ 次，然后计算生成语料库的概率，见图 3-2。

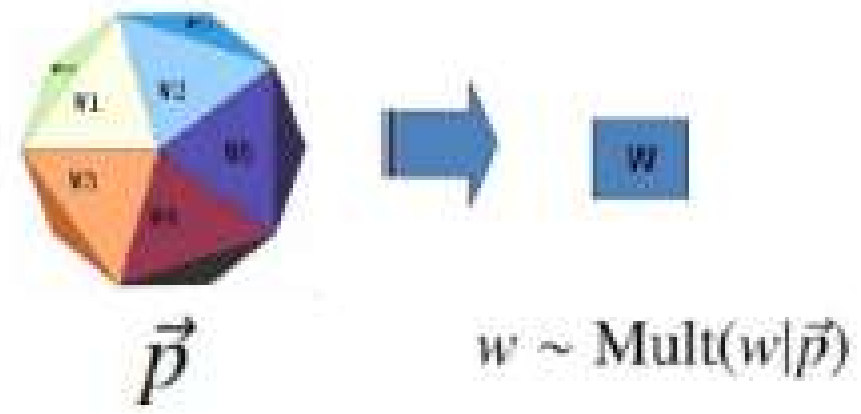

**图 3-2 上帝投掷 V 个面的骰子生成文档**

现在来引入 Dirichlet 分布作为多项分布的先验分布，则文本单词概率 $\vec{p} \sim \mathrm{Dir}(\vec{p}|\vec{\alpha})$。贝叶斯学派下，我们不知道上帝到底用哪个骰子来投掷，所以从一个服从 Dirichlet 分布的坛子中抽取一个骰子，然后投掷生成文档，图 3-3 形象地解释了这个过程。这也正是贝叶斯学派和频率学派对于待估计参数解释的差别所在，频率学派认为这个待估计的参数（概率 p）是定值，而只是我们不知道这个值是多少，需要用各种手段去估计（比如极大似然估计、矩估计等），而贝叶斯学派则认为该参数也是个随机变量，服从某种分布。

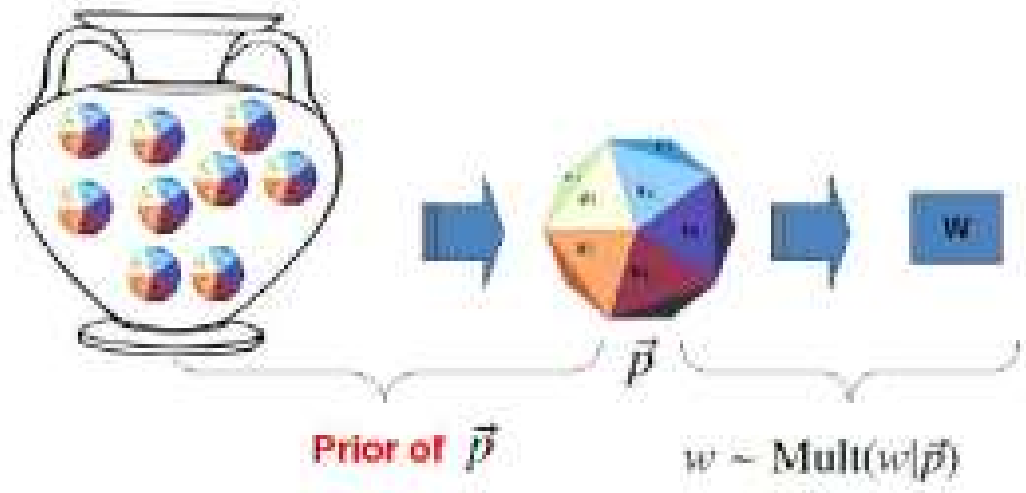

**图 3-3 贝叶斯观点下 Unigram Model**

根据 dirichlet 分布函数密度公式，我们得到了超参数(hyper parameter)（或伪计数(pseudo count)）$\alpha$的似然函数：

$$\mathrm{D}ir(\vec{p} \mid \vec{\alpha}) = \frac{1}{\Delta(\vec{\alpha})}\prod_{k=1}^{V} p_k^{\alpha_k - 1} \quad \vec{\alpha} = (\alpha_1, \alpha_2, ..., \alpha_V) \tag{3-2}$$

因为 $Dir(\vec{p} \mid \vec{\alpha}) + MultCount(\vec{n}) = Dir(\vec{p} \mid \vec{\alpha} + \vec{n})$，也即根据贝叶斯公式(2-12)，所以推出的后验分布：

$$p(\vec{p} \mid W, \vec{\alpha}) = \frac{\prod_{n=1}^{N} p(w_n \mid \vec{p}) p(\vec{p} \mid \vec{\alpha})}{\int_{\mathrm{P}} \prod_{n=1}^{N} p(w_n \mid \vec{p}) p(\vec{p} \mid \vec{\alpha}) d\vec{p}}$$
$$= \frac{\prod_{t=1}^{V} p^{\alpha_t + n(t) - 1}}{\Delta(\vec{\alpha} + \vec{n})} = Dir(\vec{p} \mid \vec{\alpha} + \vec{n}) \tag{3-3}$$

将这个贝叶斯观点下的 unigram model 画成概率图模型如图 3-4 所示。

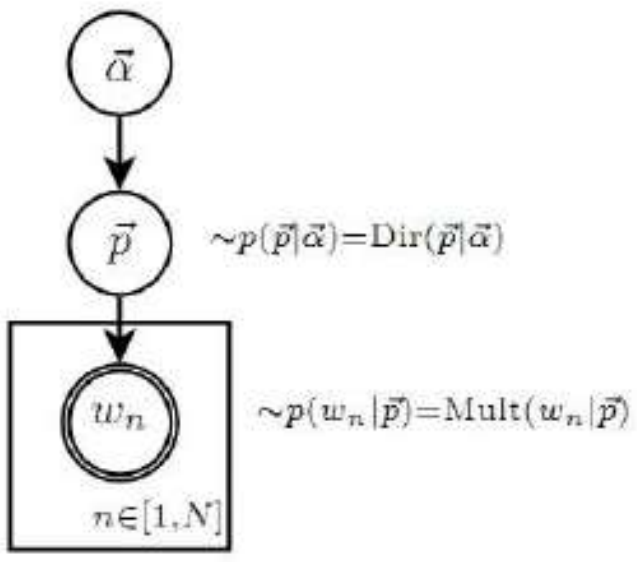

**图 3-4 Unigram Model 的概率图模型**[5]

在贝叶斯框架下，参数 **p** 可以得到估计，因为我们已经有了后验分布，根据第 2 章的公式(2-7)，于是：

$$E(\vec{p}) = \left(\frac{\alpha_1 + n_1}{\sum_{i=1}^{V}(\alpha_i + n_i)}, \frac{\alpha_2 + n_2}{\sum_{i=1}^{V}(\alpha_i + n_i)}, \frac{\alpha_3 + n_3}{\sum_{i=1}^{V}(\alpha_i + n_i)} \ldots, \frac{\alpha_V + n_V}{\sum_{i=1}^{V}(\alpha_i + n_i)}\right) \tag{3-4}$$

也就是说对于每一个 $p_i$，我们做如下参数估计即可：$\hat{p}_i = \frac{n_i + \alpha_i}{\sum_{i=1}^{V}(n_i + \alpha_i)}$

上述参数估计很直观：每个单词产生的概率估计值是对应事件的先验的伪计数和数据中的计数的和在整体计数中的比例。进一步可以产生整体文本语料概率为：[3]

$$p(W \mid \vec{\alpha}) = \int p(W \mid \vec{p}) p(\vec{p} \mid \vec{\alpha}) d\vec{p}$$

$$= \int \prod_{t=1}^{V} p_t^{n_t} Dir(\vec{p} \mid \vec{\alpha}) d\vec{p}$$

$$= \int \prod_{t=1}^{V} p_t^{n_t} \frac{1}{\Delta(\vec{\alpha})} \prod_{t=1}^{V} p_t^{\alpha_t - 1} d\vec{p}$$

$$= \frac{1}{\Delta(\vec{\alpha})} \int \prod_{t=1}^{V} p_t^{n_t + \alpha_t - 1} d\vec{p} \quad //\text{dirichlet} \int \text{ 换成}\Delta\text{(利用公式(2-11))}$$

$$= \frac{\Delta(\vec{n} + \vec{\alpha})}{\Delta(\vec{\alpha})} \tag{3-5}$$

## 3.2 Latent Dirichlet Allocation 介绍

Latent Dirchlet Allocation 是 Blei 等人于 2003 年提出的一种概率主题

[5] 图片来自于《LDA 数学八卦》

模型，LDA 是一种无监督机器学习模型，可以用来识别语料库中的潜在的主题信息。该方法假设每个词是由背后的一个潜在的主题中抽取出来。

LDA 的图模型如下：

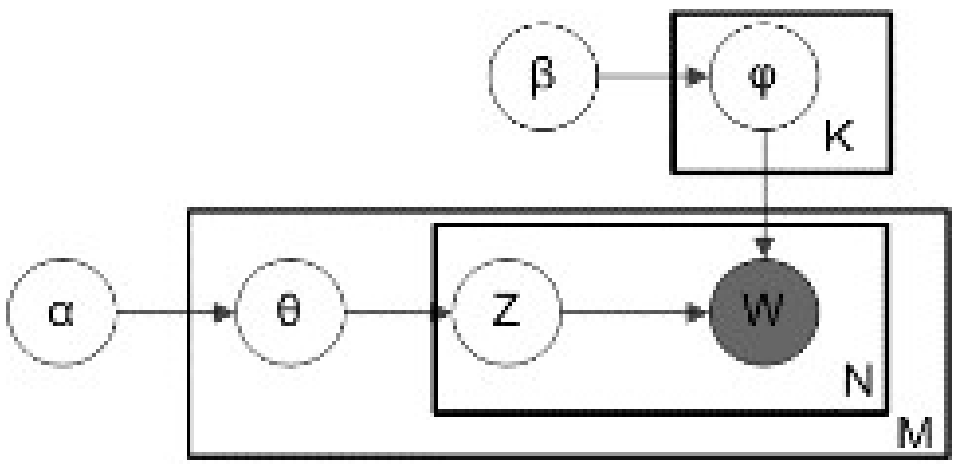

**图 3-5 LDA 图模型表示**

这个图模型表示法有时也称作“盘子表示法”（plate notation）。图中的阴影圆圈表示可观测变量（observed variable），非阴影圆圈表示潜在变量（latent variable），箭头表示两变量间的条件依赖性（conditional dependency），方框表示重复抽样，重复次数在方框的右下角。

M 代表训练语料中的文章数；

K 代表设置的主题数；

V 代表训练语料的词表单词数；

θ是一个 M*K 的矩阵，$\vec{\theta}_m$ 代表第 m 篇文章的主题分布；

φ是一个 K*V 的矩阵，$\vec{\varphi}_k$ 代表编号为 k 的主题之上的词分布；

α是每篇文档的主题分布的先验分布 Dirichlet 分布的参数（也被称为超参数），其中 $\vec{\theta}_i \sim Dir(\vec{\alpha})$；

β是每个主题的词分布的先验分布 Dirichlet 分布的参数（也被称为超参数），其中 $\vec{\varphi}_k \sim Dir(\vec{\beta})$；

w 是可被观测的词；

z 是每个对于被观测的词的潜在的主题分配。

对于语料库中的每篇文档，LDA 定义了一个生成过程（generative process）。

Smoothed 版本 LDA 模型的标准生成过程的描述如下：

1. 选取 $\vec{\theta}_{\mathrm{m}} \sim Dir(\vec{\alpha})$，这里 $\mathrm{m} \in \{1,...,M\}$；

2. 选取 $\vec{\varphi}_k \sim Dir(\vec{\beta})$，这里 $k \in \{1,...,K\}$；

3. 对于每个单词位置 $W_{i,j}$，这里 $j \in \{1,...,N_i\}$，$i \in \{1,...,M\}$；

选取一个 topic 主题从 $z_{i,j} \sim Multinomial(\theta_i)$

选取一个 word 词从 $w_{i,j} \sim Multinomial(\varphi_{z_{i,j}})$

如果没看懂就换一种比喻表述方法，爱因斯坦曾说：上帝不扔骰子。我们这

里假设我们的机器上帝是扔骰子的，所以我换一种投掷骰子的说法解释这个过程。

LDA 的逻辑与 3.1 节生成文档的逻辑类似，这里我引用《LDA 数学八卦》的一幅图如下说明 LDA 如何生成文档：

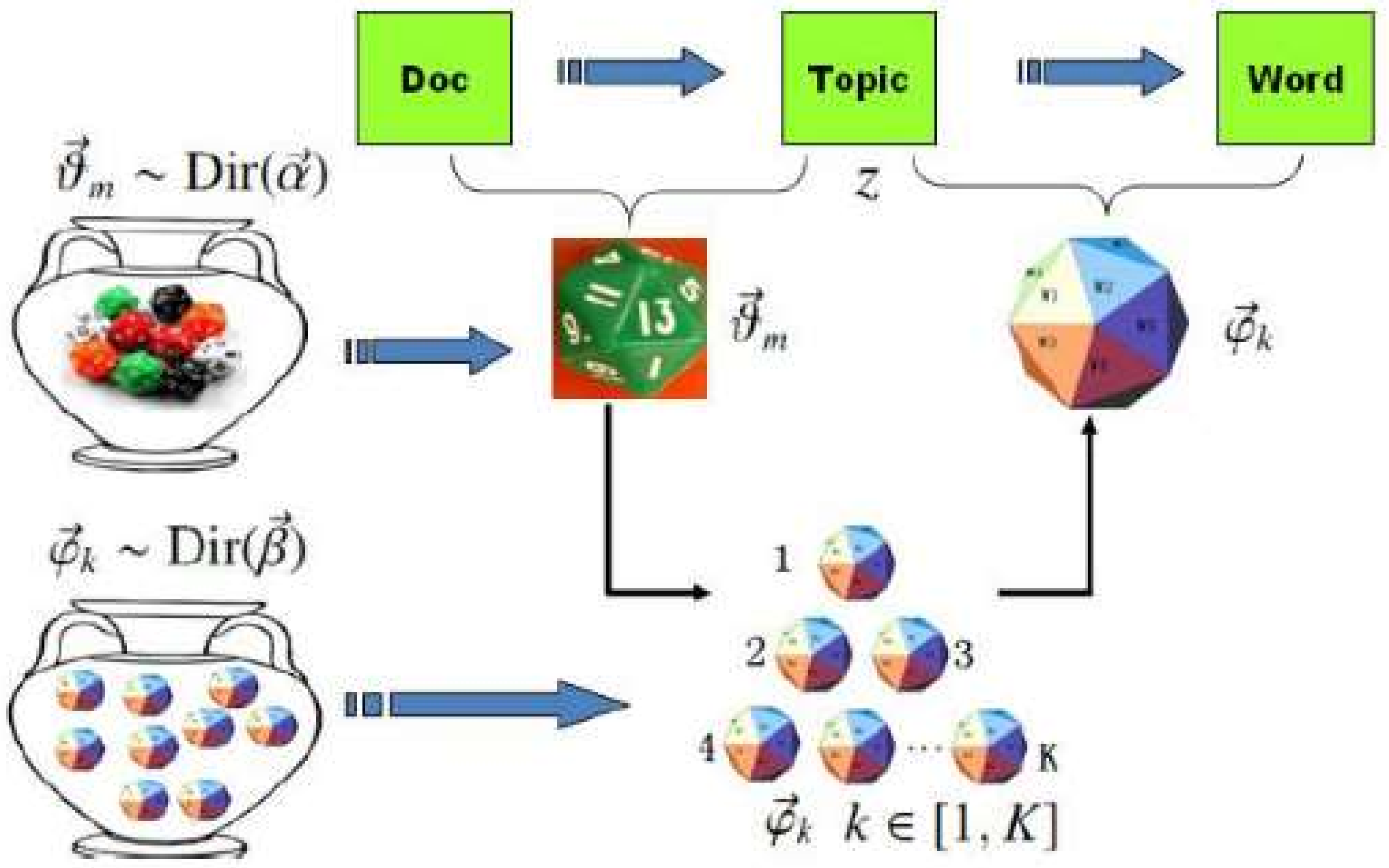

图 3-6 LDA 模型[6]

上帝先在服从 $\varphi_k \sim Dir(\vec{\beta})$ 分布的坛子抽取 $k \in (1,2,3,\ldots,K)$ 的共 K 个 $\varphi_k$ (topic→word) 骰子。每次生成一篇新的文档前，从服从 $\vec{\theta}_m \sim Dir(\vec{\alpha})$ 的坛子中抽取出一个 doc→topic 骰子，然后重复以下步骤：

i. 投掷这个 doc→topic 骰子 $\vec{\theta}_m$，得到一个 topic 编号 z。

ii. 从 $k \in (1,2,3,\ldots,K)$ 的共 K 个 $\varphi_k$ (topic→word) 骰子中选择编号为 z 的那个，投掷这枚骰子，于是得到一个词 w。

更佳的表述如下：

i. $\vec{\alpha} \rightarrow \vec{\theta} \rightarrow z_{m,n}$，这个过程表示在生成第 **m** 篇文档的时候，先从服从 $\vec{\theta}_m \sim Dir(\vec{\alpha})$ 的坛子中抽中一个 doc-topic 骰子 $\vec{\theta}_m$，然后投掷这枚骰子生成了文档中第 **n** 个词的主题(topic)编号 $z_{m,n}$。[7]此为 Dirichlet- Multinomial 共

[6] 图片引用自《LDA 数学八卦》
[7] 这里足见是主题分布的超参数

轭。

ii. $\vec{\beta} \to \vec{\varphi}_{\mathrm{k}} \to w_{m,n} \mid k = z_{m,n}$，这个过程表述了如下动作生成语料中第 m 篇文档的第 n 个词：从先前服从 $\vec{\varphi}_k \sim Dir(\vec{\beta})$ 分布的坛子已抽取好的 $\mathrm{k} \in (1,2,3,...,\mathrm{K})$ 的共 K 个 $\vec{\varphi}_{\mathrm{k}}$ (topic→word) 骰子之中，挑选编号为 $k = z_{m,n}$ 的那个骰子进行投掷，然后生成 word: $w_{m,n}$。[8]此为 Dirichlet- Multinomial 共轭。

简单来说，第 i 个过程就是 $p(\vec{z})$，第 ii 个过程就是 $p(\vec{w} \mid \vec{z})$，根据条件概率的基本公式可得：

$$p(\vec{w},\vec{z}) = p(\vec{w} \mid \vec{z})p(\vec{z}) \tag{3-6}$$

而 LDA 的目标是找出每个词后潜在的主题，所以为了达到这个目标，需要计算后验概率(3-7)。

$$p(\vec{z} \mid \vec{w}) = \frac{p(\vec{w},\vec{z})}{\sum_z p(\vec{w},\vec{z})} \tag{3-7}$$

难点在于分母，简单分析一下分母的情况便可得出：公式(3-7)**难以直接计算**。按照离散分布上边缘概率的处理方法：假设我们有概率 $p(a,b,c)$，可以计算在所有 c 的可能值上求和，就消去 c，而仅计算 $p(a,b) = \sum_c p(a,b,c)$。所以文档中一个单词 $w_i$ 的概率是：

$$p(w_i) = \sum_{k=1}^{K} p(w_i, z_i = k) = \sum_{k=1}^{K} p(w_i \mid z_i = k)p(z_i = k) \tag{3-8}$$

由此可得公式(3-7)的分母，也即整个语料集的所有单词的概率：

$$p(\vec{w}) = \sum_z p(\vec{w},\vec{z}) = \prod_{i=1}^{n}\sum_{k=1}^{K} p(w_i \mid z_i = k)p(z_i = k) \tag{3-9}$$

式(3-9)中的 n 是语料中所有单词实例的总数，因为计算分母陷入了 $K^n$ 项的难题，这个离散状态空间 （discrete state space）太大了以至于无法列举出来（enumerate）[3]。Thomas L.Griffiths 等人从统计物理学的 Potts Model 获得灵感，开发出了 LDA 的蒙特卡洛马尔可夫（Monte Carlo Markov Chain）求解方法。这就是下文 3.3 节和 3.4 节所介绍的内容。

[8]这里足见是词分布的超参数

## 3.3 马尔可夫链 → Metropolis-Hasting → Gibbs Sampling

在正式推导 LDA 的 Gibbs Sampling 采样公式之前，读者有必要了解为什么需要这样推导，做到知其然知其所以然。[9]

### 3.3.1 马尔可夫链(markov chain)

马尔可夫链条通俗说就是根据一个转移概率矩阵去转移的随机过程（马尔可夫过程）[10]，该随机过程在 Page Rank 算法中也有使用。如图 3-7 所示：

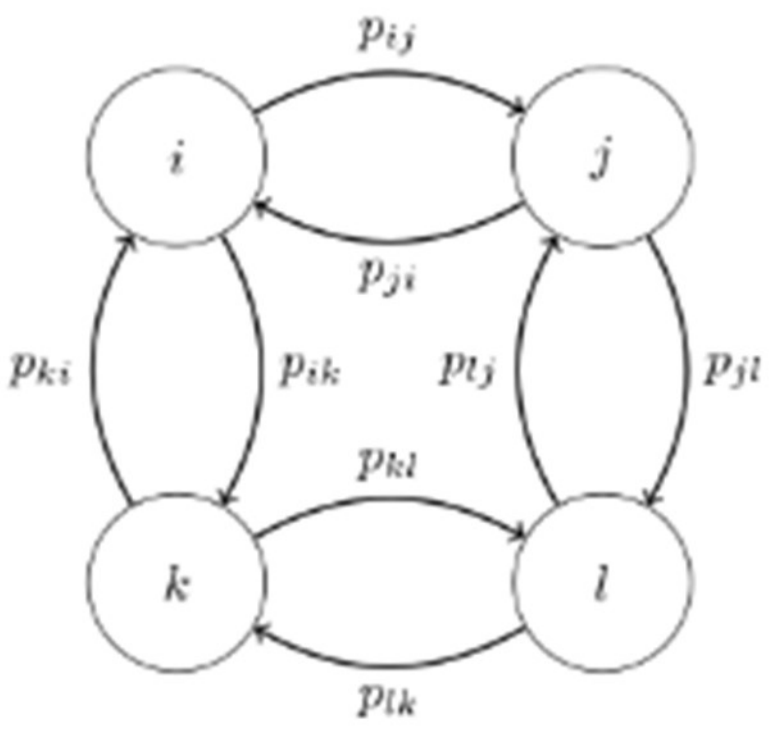

**图 3-7 马尔可夫转移图**

通俗理解：这里的每个圆环类似一个岛屿(状态)，比如 i 到 j 的概率是 $p_{ij}$，每个节点的出度概率之和=1。现在假设我要根据这个图去转移，那么首先需要把这张图“翻译”成如下的矩阵：

$$
P = \begin{array}{ccccc}
{}^{col}/_{row} & i & j & k & l \\
i & p_{ii} & p_{ij} & p_{ik} & p_{il} \\
j & p_{ji} & p_{jj} & p_{jk} & p_{jl} \\
k & p_{kl} & p_{kj} & p_{kk} & p_{kl} \\
l & p_{li} & p_{lj} & p_{lk} & p_{ll}
\end{array}
$$

即该矩阵的第 i 行第 j 列 $p_{ij}$ 表示从 i 岛屿走到 j 岛屿的概率。

这就是转移矩阵，我现在身处的位置用一个向量表示 $\pi_0$=(i,j,k,l)。

现在假设我站在 i 岛屿，我的位置向量 $\pi_0$=(1,0,0,0)，第一次转移，也就是相当于 $\pi_1 = \pi_0 \cdot P = [p_{ii}, p_{ij}, p_{ik}, p_{il}]$ ，相乘后向量 $\pi_1$ 每一项为小于 1 的小数。就是说我有 $P_{ii}$ 的机会比率留在原始 i 岛屿，有 $P_{ij}$ 的机会到达 j 岛屿，……；第二次转移时，以我刚才的位置向量为基础得到 $\pi_2 = \pi_1 \cdot P$ ；以此类推……

[9] LDA 推断方法有 MCMC 和变分 EM，这个版本讲的 MH 算法和 Gibbs Sampling 就是 MCMC 算法的一种

[10] 马尔可夫链是个很简单的概念，百度/google 一搜一大堆，我在这里用通俗说法

**有这么一种情况**，我的位置向量在若干次转移后会达到一个稳定状态，再转移π向量也不变化了，这个状态称之为平稳分布状态π*，这个情况需要满足一个条件：这就是 Detailed Balance。

下面通过一个例子深刻认识 Detailed Balance：

首先要明确：一个马尔可夫链条要成为 reversible markov chain 才有可能达到 Detailed Balance 条件，何为 reversible markov chain?也就是满足 Detailed Balance 的 马尔可夫链条。是不是有点鸡生蛋蛋生鸡的味道。换一种说法，reversible markov chain 也就是其转移概率满足 Kolmogorov’s criterion 的链条，这个 Kolmogorov’s criterion 即表示以下含义，在图 3-7 中：$p_{ij}p_{jl}p_{lk}p_{ki} = p_{ik}p_{kl}p_{lj}p_{ji}$。

也就说一个封闭的环中，一个方向的概率连乘积=反过来方向的概率连乘积。

例子：我们用假设构造这样一个转移矩阵：

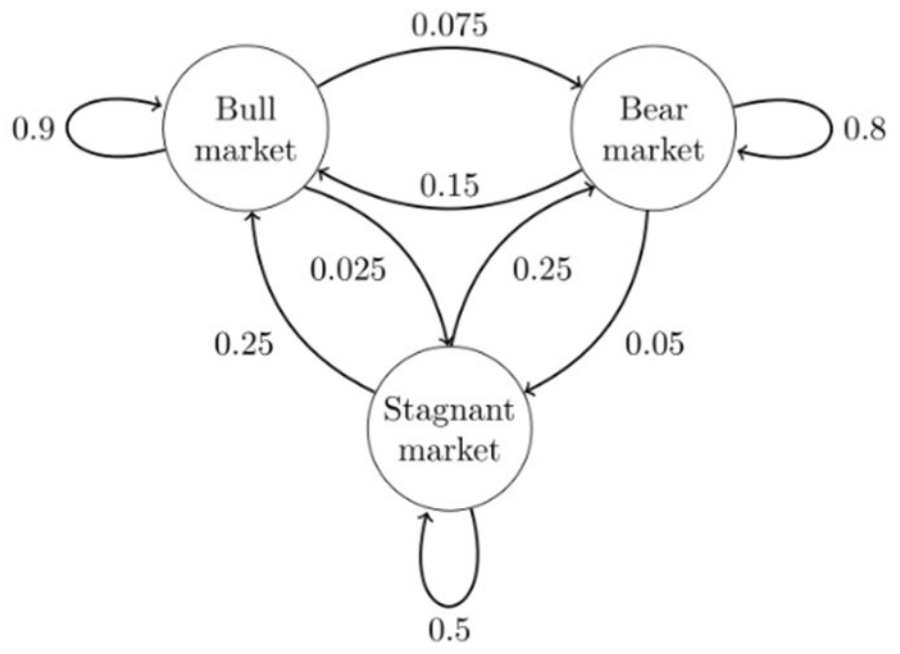

**图 3-8 detail balance 的转移矩阵图**

先构造出转移矩阵，$P =$

| col / row | Bull | Bear | Stagnant |
|---|---|---|---|
| Bull | 0.9 | 0.075 | 0.025 |
| Bear | 0.15 | 0.8 | 0.05 |
| Stagnant | 0.25 | 0.25 | 0.5 |

假设我们的初始向量为π=(1, 0, 0)，现在用计算机软件计算矩阵乘法得到 π × P 1000 次后的 stationary distribution = (0.625, 0.3125, 0.0625)。注意这个平稳分布有且只有一个（**是唯一的**）。

所谓的 detailed balance 就是：

$$\pi_i P_{ij} = \pi_j P_{ji} \tag{3-10}$$

这里的π是平稳分布(stationary distribution)的那个π*。代到这个例子里：

假设 $i = 1, j = 2$

$P_{ij} = 0.075,\ P_{ji} = 0.15$

则 $\pi_i = 0.625, \pi_j = 0.3125$

所以 $\pi_i \cdot P_{ij} = 0.625 \cdot 0.075 = 0.046875$

$\pi_j \cdot P_{ji} = 0.3125 \cdot 0.15 = 0.046875$

所以得出结论 $\pi_{\rm i} P_{ij} = \pi_j P_{ji}$，detailed balance 成立。

有了 detailed balance， 马尔可夫链就可以收敛[11]了，我们可以根据 detailed balance 去 sampling 生成一些点，使这些点收敛到 stationary distribution。因此这些点就是满足这个 stationary distribution 概率分布的点。收敛在这里称之为 burning-in，在 burning-in 之前的若干迭代步骤生成的点被抛弃掉。下面我们再来验证下：

由于 $\pi_{\rm i} P_{ij} = \pi_j P_{ji}$，因此两边同时对 i 的所有可能值求和，也就是：

$$\sum_{\rm i} \pi_{\rm i} P_{ij} = \sum_i \pi_j P_{ji} = \pi_j \underbrace{\sum_i P_{ji}}_{=1} = \pi_j$$

最后一步由于矩阵的每一行和=1（每个节点出度概率之和=1）。因此这个推导暗示了 $\pi^* = \pi^* \cdot P$

在这个例子里，我们取 j=2，也就是 $\pi_2 = 0.3125$，则

$$\sum_{\rm i} \pi_{\rm i} P_{i2} = 0.625 \bullet 0.075 + 0.3125 \bullet 0.8 + 0.0626 \bullet 0.25 = 0.3125 = \pi_2$$

另外这样的马尔可夫链还需要满足两个性质：1.irreducible， 马尔可夫链的所有状态节点需要可以彼此通信，不能有割裂的孤岛。2.aperiodic，非周期性，链条不会在特定的周期内在两个节点来回循环。

### 3.3.2 Metropolis-Hasting 算法

结合 LDA 算法，有了上述 detail balance 条件，受到这个平稳分布不再变化的启发，我们的**终极目标自然是：要使用一个马尔可夫链条，sample[12]出一系列的状态点，使其最终的平稳分布状态就是我们给定的那个联合概率分布（该联合概率就是 LDA 里的文档集被生成(generate)的概率）**。

**MH 算法的目的**：是根据一个需求的(desired distribution)概率分布 P(x)

[11] 关于 markov chain 收敛的更严格详细的证明，参见：http://users.aims.ac.za/~ioana/notes-ch4-2x1.pdf 4.4

[12] Sample 的意思也就是一个随机数生成器：按照某种概率分布来随机生成一系列数字

生成一系列样本状态点（因此，这个算法可以生成任意的概率分布）。为了达到这个目的，该算法使用马尔可夫过程去到达一个平稳分布π(x)(stationary distribution)，以使π(x)=P(x)。

**MH 算法推导：为了达到这个平稳分布，有两个条件需要满足：(1)满足 detail balance：** $\pi(x)P(x \to x') = \pi(x')P(x' \to x)$ （2）该平稳分布必须唯一（Ergodicity 可遍历[13]）。

MH 算法的方式是设计一个马尔可夫过程（通过构造转移概率）来满足上述两个条件，用 P(x)代替π(x)后(注意斜体字 *P* 为转移矩阵)，现在重点来分析 detail balance：

$$\mathrm{P}(x)P(x \to x') = \mathrm{P}(x')P(x' \to x)$$

$$\frac{P(x \to x')}{P(x' \to x)} = \frac{\mathrm{P}(x')}{\mathrm{P}(x)}$$

我们的方法是进一步将转移概率 $P(x \to x')$ 分解为两个子步骤：proposal distribution(建议概率) $\mathrm{g}(x \to x')$ 和 acceptance distribution（接受概率）$\mathrm{A}(x \to x')$。**建议概率**是说我们给出状态 x 后转移到 x’ 的条件概率，而**接受概率**是接受状态 x’ 的条件概率。

所以转移概率 $P(x \to x') = \mathrm{g}(x \to x')A(x \to x')$，代入公式(3-10)（detail balance)，并整理可得：$\frac{A(x \to x')}{A(x' \to x)} = \frac{\mathrm{P}(x')}{\mathrm{P}(x)} \frac{g(x' \to x)}{g(x \to x')}$，也就是说这样我们得到了一个接受比率：“从状态 $x$ 到 $x'$ 的接受概率”与“$x'$ 到 $x$ 的接受概率”的比率（接受率）。如果这个比率大于 1，则我们按照建议概率 $g(x \to x')$ 转移到 $x'$，否则停留在 $x$ 原地不动（拒绝接受建议）。Metropolis 的选择是进一步将上述接受率改为如下形式（比率大于 1 就=1，上限为 1）：

$$\alpha = A(x \to x') = \min\{1, \frac{\mathrm{P}(x')}{\mathrm{P}(x)} \frac{g(x' \to x)}{g(x \to x')}\} \tag{3-11}$$

接着就可以写出 MH 算法：

[13] https://en.wikipedia.org/wiki/Metropolis%E2%80%93Hastings_algorithm:第(2)个条件的 Ergodicity 可以进一步用两个条件来表示：(1) be aperiodic—the system does not return to the same state at fixed intervals; and (2) be positive recurrent—the expected number of steps for returning to the same state is finite.

**算法 3.1 Metropolis-Hasting 算法步骤**

初始化，随机选择一个初始状态点 x

for i=0 to N-1 do //第 i 个 sample 点为 $x_{(i)}$

  根据 $g(x \to x')$ 随机选择一个新状态 x’

  计算接受率 $\alpha = \min\{1, \frac{P(x')}{P(x)}\frac{g(x' \to x)}{g(x \to x')}\}$

  生成 0~1 之间随机数 $u$

  if($u$ < $\alpha$) then{ //若$\alpha$<1，按照$\alpha$为概率接受 x’

    $x_{(i+1)} = x'$

  }else{

    $x_{(i+1)} = x_{(i)}$

  }

MH 算法非常简单，但要小心设计建议概率 g，其实有时候考虑 MH 算法的特殊情况（Gibbs Sampling is a special case of MH）让问题变得更简单。

### 3.3.3 Gibbs Sampling

考虑我们目标要得到的是一个维度是 n 的多维的概率分布，如果采用设置多维概率分布 P 里的完全条件概率(full conditionals)作为建议概率(proposal)，那么接受率就会始终=1，一直接受$x'$。让我们来证明这一点[15]：

full conditionals 的公式如下：

$$p(x_j \mid x_{\neg j}) = p(x_j \mid x_1,\ldots,x_{j-1},x_{j+1},\ldots,x_n) = \frac{p(x_1,\ldots,x_n)}{p(x_1,\ldots,x_{j-1},x_{j+1},\ldots,x_n)} \tag{3-12}$$

第 i 次 sample 的点转移的 proposal distribution： $g(x \to x') = g(x' \mid x^{(i)})$

由于 *x* 是 n 维变量，当 j=1,2,…,n 每一维度轮流循环时

{

我们如果设置 $g(x' \mid x^{(i)}) = \begin{cases} p(x'_j \mid x^{(i)}_{\neg j}) & \text{if } x^{(i)}_{\neg j} = x'_{\neg j} \\ 0 & \text{otherwise} \end{cases}$

那么就可以造成$\alpha = 1$

$$\alpha = A(x^{(i)} \to x') = \min\{1, \frac{p(x')}{p(x^{(i)})}\frac{g(x^{(i)} \mid x')}{g(x' \mid x^{(i)})}\}$$

$$= \min\{1, \frac{p(x')}{p(x^{(i)})}\frac{p(x^{(i)}_j \mid x'_{\neg j})}{p(x'_j \mid x^{(i)}_{\neg j})}\}$$

$$\underset{x_{\neg j}^{(i)}=x'_{\neg j}}{=} \min\{1, \frac{\overbrace{p(x')}^{p(x'_j|x'_{\neg j})\cdot p(x'_{\neg j})}}{\underbrace{p(x^{(i)})}_{p(x_j^{(i)}|x_{\neg j}^{(i)})\cdot p(x_{\neg j}^{(i)})}} \cdot \frac{p(x_j^{(i)} \mid x_{\neg j}^{(i)})}{p(x'_j \mid x'_{\neg j})}\} = \min\{1, \frac{p(x'_j \mid x'_{\neg j}) \cdot p(x'_{\neg j})}{p(x_j^{(i)} \mid x_{\neg j}^{(i)}) \cdot p(x_{\neg j}^{(i)})} \frac{p(x_j^{(i)} \mid x_{\neg j}^{(i)})}{p(x'_j \mid x'_{\neg j})}\}$$

$$= \min\{1,1\} = 1$$

}

所以一旦在联合概率的 full conditionals 可用时，可以采用 n 维向量里轮流每一个维度循环的方式来迭代达到平稳状态。

而为了得到 full conditionals，就要先写出联合概率（这也就引出了下一节的文本生成的联合概率），再用公式(3-12)的方式求得。

因此通用的 Gibbs Sampling 迭代算法如算法 3.2 所示：

**算法 3.2 Gibbs Sampling 算法步骤**

初始化，随机选择一个初始状态点 $x_{1:n}^{(0)}$

for i=0 to N-1 do{ //第 i 个 sample 点为 $x^{(i)}$, x 的下标 j 为 x 的第 j 维

  sample $x_1^{(i+1)} \sim p(x_1 \mid x_2^{(i)}, x_3^{(i)} \dots, x_n^{(i)})$

  sample $x_2^{(i+1)} \sim p(x_2 \mid x_1^{(i+1)}, x_3^{(i)} \dots, x_n^{(i)})$

  …

  sample $x_j^{(i+1)} \sim p(x_j \mid x_1^{(i+1)}, x_2^{(i+1)} \dots, x_{j-1}^{(i+1)}, x_{j+1}^{(i)}, \dots, x_n^{(i)})$

  …

  sample $x_n^{(i+1)} \sim p(x_n \mid x_1^{(i+1)}, x_2^{(i+1)} \dots, x_{n-1}^{(i+1)})$

}

## 3.4 伟大的采样公式： Collapsed Gibbs Sampling 采样公式推导

鉴于 3.2 节的式(3-7)的 $p(\vec{z} \mid \vec{w})$ 难以求得，假如利用 Collpased Gibbs Sampling，就可以使用 3.3.3 节的 full conditional 公式（式(3-12)）写出：$p(z_i \mid \vec{z}_{\neg i}, \vec{w})$，用此公式来模拟 $p(\vec{z} \mid \vec{w})$。

现在我们想推导出 Gibbs Sampling 的采样公式，在做 Gibbs Sampling 公式推导之前最重要的一件事是写出整个文本训练集生成的联合概率（也就是按照 3.2 节的公式(3-6)）：

$$p(\vec{w}, \vec{z} \mid \vec{\alpha}, \vec{\beta}) = p(\vec{w} \mid \vec{z}, \vec{\beta}) p(\vec{z} \mid \vec{\alpha}) \quad (3\text{-}13)$$

注意到第一项不包含$\alpha$，第二项不包含$\beta$，所以两项可以分开处理。先不考虑β，先来看第一项：也就是说给定 topic 的情况下的多项分布的似然函数（参照式(3-1)）：

$$p(\vec{w} \mid \vec{z},\varphi) = \prod_{i=1}^{W} p(w_i \mid z_i) = \prod_{i=1}^{W} \varphi_{z_i,w_i} \tag{3-14}$$

这里的 W 代表语料中的所有词数，这些词在 topic 的 $z_i$=k 的条件下的多次独立“多项分布”试验中被产生，我们现在将这一项分解为两个连乘，一个 over topic，一个 over vocabulary[14]：

$$p(\vec{w} \mid \vec{z},\varphi) = \prod_{k=1}^{K} \prod_{\{i:z_i=k\}} p(w_i = t \mid z_i = k) = \prod_{k=1}^{K}\prod_{t=1}^{V} \varphi_{k,t}^{n_k^{(t)}} \tag{3-15}$$

这里我们用$n_k^{(t)}$代表 topic k 下单词 t 被观测到的次数，而最终目标函数

$p(\vec{w} \mid \vec{z},\vec{\beta})$由对$\varphi$求积分得到。

(3-15)式子的解释：首先不考虑超参数$\beta$，而是假设已知参数$\varphi$，这个$\varphi$就是那个 K*V 维的矩阵，表示从每一个 topic 产生词的概率，然后把$\varphi$积分掉，仿照公式(3-5)的推理，再利用公式(2-11)，可以求出第一部分表达式了[15]：

$$\begin{aligned}
p(\vec{w} \mid \vec{z},\vec{\beta}) &= \int \underbrace{p(\vec{w} \mid \vec{z},\varphi)}_{\prod_{k=1}^{K}\prod_{t=1}^{V}\varphi_{k,t}^{n_k^{(t)}}} \bullet \underbrace{p(\varphi \mid \vec{\beta})}_{Dir(\vec{\varphi}\mid\vec{\beta})=\prod_{k=1}^{K}\frac{1}{\Delta(\vec{\beta})}\prod_{t=1}^{V}\varphi_{k,t}^{\beta_t-1}} d\varphi \\
&= \int p(\vec{w} \mid \vec{z},\varphi) p(\varphi \mid \vec{\beta}) d\varphi \\
&= \int \prod_{k=1}^{K} \frac{1}{\Delta(\vec{\beta})} \prod_{t=1}^{V} \varphi_{k,t}^{n_k^{(t)}+\beta_t-1} d\vec{\varphi}_k \\
&= \prod_{k=1}^{K} \frac{\Delta(\vec{n}_k+\vec{\beta})}{\Delta(\vec{\beta})},\ \vec{n}_k = \{n_k^{(t)}\}_{t=1}^{V}
\end{aligned} \tag{3-16}$$

(3-16)中的$\vec{n}_k$向量即代表：第 k 个 topic 下单词的分布情况，$\vec{n}_k$=(该 topic 下第 1 个单词的个数,该 topic 下第 2 个单词的个数,...)。我们发现这里对$\varphi$的积分的处理：$\varphi$被积分掉（integrating out）而消失了[16]。这里积分的变量$\vec{\varphi}_k$是$\varphi$的所有可能值的集合，有点类似于边缘概率在离散分布上的处理，假设我们有概率

---

[14] 一个单词可同时被多个主题生成，$\varphi$就是从某个主题产生单词的概率，所以拆分成各主题的乘积

[15] 引自《parameter estimation for text analysis》 p21

[16] 这就是 Collapsed Gibbs Sampling 的 Collapsed 一词来历，这个 integrating out 出自《Gibbs Sampling for the Uninitiated》p15　这里可以积掉的深层原因是每个词上的 topic 采样都是独立的，而非《Uninitiated》一文里 p18 里的选定类标签 L 后一次性都将生成完一篇文章的所有 word。(《Uninitiated 中文版》的盘子图$\theta$一对多：一个源头θ箭头指向一篇文章的所有 word)

p(a,b,c)，可以计算在所有 c 的可能值上求和，就消去 c，而仅计算 $p(a,b)=\sum_c p(a,b,c)$。

这个过程属于 Dirichlet- Multinomial 共轭结构。可以仿照这个推理过程，$p(\vec{z}\mid\vec{\alpha})$ 也可以被推理得到，将他重写为两个的乘积：

$$p(\vec{z}\mid\theta)=\prod_{i=1}^{W}p(z_i\mid d_i)=\prod_{m=1}^{M}\prod_{k=1}^{K}p(z_i=k\mid d_i=m)=\prod_{m=1}^{M}\prod_{k=1}^{K}\theta_{m,k}^{n_m^{(k)}} \tag{3-17}$$

这里 $d_i$ 表示第 m 篇文档，$n_m^{(k)}$ 表示第 m 篇文档下第 k 号主题词数（该文档下 topic k 被指派给词的计数）。为了积分掉（integrating out）$\theta$，[17]仿照公式(3-16)的推理，同样利用公式(2-11)，我们做如下推导[18]：

$$p(\vec{z}\mid\vec{\alpha})=\int p(\vec{z}\mid\theta)p(\theta\mid\vec{\alpha})d\theta$$

$$=\int\prod_{m=1}^{M}\frac{1}{\Delta(\vec{\alpha})}\prod_{k=1}^{K}\theta_{m,k}^{n_m^{(k)}+\alpha_k-1}d\vec{\theta}_m$$

$$=\prod_{m=1}^{M}\frac{\Delta(\vec{n}_m+\vec{\alpha})}{\Delta(\vec{\alpha})},\quad \vec{n}_m=\{n_m^{(k)}\}_{k=1}^{K} \tag{3-18}$$

这里 $\vec{n}_m$ 即代表：第 m 篇文档中的主题分布情况，$\vec{n}_m$=(1 号主题词数,2 号主题词数,...)。这也属于 Dirichlet- Multinomial 共轭结构。

所以联合分布可以得到：

$$p(\vec{w},\vec{z}\mid\vec{\alpha},\vec{\beta})=p(\vec{w}\mid\vec{z},\vec{\beta})p(\vec{z}\mid\vec{\alpha})=\prod_{k=1}^{K}\frac{\Delta(\vec{n}_k+\vec{\beta})}{\Delta(\vec{\beta})}\bullet\prod_{m=1}^{M}\frac{\Delta(\vec{n}_m+\vec{\alpha})}{\Delta(\vec{\alpha})} \tag{3-19}$$

有了联合概率分布，紧接着在此基础上就可以根据公式(3-12)推导 full conditionals $p(z_i=k\mid\vec{z}_{\neg i},\vec{w})$，而现在我们已经讨论了推导最后的 Collapsed Gibbs Sampling 采样公式推导的全部要素，**唯有两点尚缺，这两点就是以下的事实：**

(1) Gamma 约去

利用 $\Gamma(n+1)=n\Gamma(n)=n!$

$$\frac{\Gamma(n)}{\Gamma(n+1)}=\frac{(n-1)!}{n!}=\frac{(n-1)\bullet(n-2)\bullet\ldots\bullet 1}{n\bullet(n-1)\bullet(n-2)\bullet\ldots\bullet 1}=\frac{1}{n}$$

---

[17] 上面的φ和θ “可以”被积分掉,原因在于为每个单词是被各 topic 独立生成，不互相影响φ和θ,因此可被积分掉.

[18] 引自《parameter estimation for text analysis》p21

也就是说分子和分母里的 gamma 函数只差 1 时，如果分母较大，分子 gamma 函数内数字为分母 gamma 内数字小 1，整个式子就等于 $\frac{1}{\text{分子gamma函数内数字}}$。

(2) 连乘号约去

由于 full conditionals 中包含分子分母，分子和分母的**唯一差别**只在与当前采样的第 m 篇文档第 i 个单词，所以其他无关乘积因子分子和分母皆因相等而约去。

当前第 m 篇文档的主题采样时：$\prod_{m=1}^{M}\frac{\Delta(\vec{n}_m+\vec{\alpha})}{\Delta(\vec{\alpha})}$ 约去后为 $\frac{\Delta(\vec{n}_m+\vec{\alpha})}{\Delta(\vec{\alpha})}$

当前第 k 号主题的主题采样时：$\prod_{k=1}^{K}\frac{\Delta(\vec{n}_k+\vec{\beta})}{\Delta(\vec{\beta})}$ 约去后为 $\frac{\Delta(\vec{n}_k+\vec{\beta})}{\Delta(\vec{\beta})}$

Collapsed Gibbs Sampling 采样公式推导如下文[19]：

[19] 进一步阅读材料：
普林斯顿大学材料：[Appendix D] Derivation of Gibbs sampling equations https://lists.cs.princeton.edu/pipermail/topic-models/attachments/20110210/89b1646c/attachment-0001.pdf
HP 实验室 Note：Gibbs Sampling Derivation for LDA and TOT http://home.in.tum.de/~xiaoh/pub/TOTGibbs.pdf

$$p(\vec{w},\vec{z} \mid \vec{\alpha},\vec{\beta}) = p(\vec{w} \mid \vec{z},\vec{\beta}) \bullet p(\vec{z} \mid \vec{\alpha})$$
$$= \prod_{k=1}^{K} \frac{\Delta(\vec{n}_k + \vec{\beta})}{\Delta(\vec{\beta})} \bullet \prod_{m=1}^{M} \frac{\Delta(\vec{n}_m + \vec{\alpha})}{\Delta(\vec{\alpha})}$$

解释：$\vec{n}_k = (\text{k}_th$主题$(topic)$下第1个单词个数，第2个单词个数，$...)$

$\vec{n}_m = (m_th$文章$(doc)$下第1号主题词数，第2号主题词数$,...)$

上面公式中$\Delta(\vec{\alpha}) = \dfrac{\prod_{k=1}^{K}\Gamma(\alpha_k)}{\Gamma(\sum_{k=1}^{K}\alpha_k)}$

$$p(z_i = k \mid \vec{z}_{\neg i},\vec{w}) = \frac{p(\vec{w},\vec{z})}{p(\vec{w},\vec{z}_{\neg i})} = \frac{p(\vec{w},\vec{z})}{p(w_i = t,\vec{w}_{\neg i},\vec{z}_{\neg i})} = \frac{p(\vec{w},\vec{z})}{\underbrace{p(\vec{w}_{\neg i},\vec{z}_{\neg i} \mid w_i = t)}_{p(\vec{w}_{\neg i},\vec{z}_{\neg i})}p(w_i = t)}$$

$$= \frac{p(\vec{w},\vec{z})}{p(\vec{w}_{\neg i},\vec{z}_{\neg i})} \cdot \frac{1}{p(w_i = t)}$$

$$\propto \frac{p(\vec{w},\vec{z})}{p(\vec{w}_{\neg i},\vec{z}_{\neg i})} = \frac{\prod_{k=1}^{K} \frac{\Delta(\vec{n}_k + \vec{\beta})}{\Delta(\vec{\beta})} \bullet \prod_{m=1}^{M} \frac{\Delta(\vec{n}_m + \vec{\alpha})}{\Delta(\vec{\alpha})}}{\prod_{k=1}^{K} \frac{\Delta(\vec{n}_{k,\neg \mathrm{i}} + \vec{\beta})}{\Delta(\vec{\beta})} \bullet \prod_{m=1}^{M} \frac{\Delta(\vec{n}_{\mathrm{m},\neg \mathrm{i}} + \vec{\alpha})}{\Delta(\vec{\alpha})}} = \frac{\Delta(\vec{n}_k + \vec{\beta})}{\Delta(\vec{n}_{k,\neg \mathrm{i}} + \vec{\beta})} \bullet \frac{\Delta(\vec{n}_m + \vec{\alpha})}{\Delta(\vec{n}_{\mathrm{m},\neg \mathrm{i}} + \vec{\alpha})}$$

($\vec{n}_{k,\neg \mathrm{i}}$表示除了当前该单词i以外topic_Kth的单词向量,

$\vec{n}_{\mathrm{m},\neg \mathrm{i}}$表示除了该单词外该文章的主题词向量)

(解释：由于只是推断单词$\mathrm{W}_{m,n}$的topic_Kth,与此topic_Kth和doc_mth无关的均视为常数被忽略，因此消去$\prod$ 符号)

(而分母的$\Delta(\vec{\beta})$和$\Delta(\vec{\alpha})$也在分式中由于上下都有相同而被消去)

现在看上面分母$\Delta(\vec{n}_{k,\neg \mathrm{i}} + \vec{\beta}) \underbrace{=}_{\substack{V\text{为字典单词数}\\ \text{每个主题下相同}}} \dfrac{\prod_{t=1}^{V}\Gamma(n_{t,\neg \mathrm{i}} + \beta_t)}{\Gamma(\sum_{t=1}^{V}(n_{t,\neg \mathrm{i}} + \beta_t))}$

$$\underbrace{=}_{\text{除了第}i\text{个单词需要减1}} \frac{\Gamma(n_1 + \beta_1)\Gamma(n_2 + \beta_2)...\Gamma(n_i - 1 + \beta_i)...\Gamma(n_V + \beta_V)}{\Gamma(\sum_{t=1}^{V}(n_{t,\neg \mathrm{i}} + \beta_t))}$$

看分子部分$\Delta(\vec{n}_k + \vec{\beta}) \underbrace{=}_{\substack{V\text{为字典单词数}\\ \text{每个主题下相同}}} \dfrac{\prod_{t=1}^{V}\Gamma(n_t + \beta_t)}{\Gamma(\sum_{t=1}^{V}(n_t + \beta_t))}$

$$= \frac{\Gamma(n_1 + \beta_1)\Gamma(n_2 + \beta_2)...\Gamma(n_i + \beta_i)...\Gamma(n_V + \beta_V)}{\Gamma(\sum_{t=1}^{V}(n_t + \beta_t))}$$

因此因式第一部分如下 $\propto \dfrac{\Gamma(n_i + \beta_i)}{\Gamma(n_i - 1 + \beta_i)} \bullet \dfrac{\Gamma(\sum_{t=1}^{V}(n_{t,\neg \mathrm{i}} + \beta_t))}{\Gamma(\sum_{t=1}^{V}(n_t + \beta_t))}$

同理因式第二部分如下 $\propto \overbrace{\frac{\Gamma(n_k+\alpha_k)}{\Gamma(n_k-1+\alpha_k)}}^{\text{小写k为}z_i=k\text{的那个}k\text{ 号主题}} \bullet \underbrace{\frac{\Gamma(\sum_{t=1}^{K}(n_{t,\neg i}+\alpha_t))}{\Gamma(\sum_{t=1}^{K}(n_t+\alpha_t))}}_{\text{此K是主题总数,}n_{t,\neg i}\text{是第m篇文章下第t号主题词数-1}}$

由于$\Gamma(x+1)=x\Gamma(x)=x!$

来看因式的这一部分 $\frac{\Gamma(n_i+\beta_i)}{\Gamma(n_i-1+\beta_i)}=\frac{1\bullet 2\bullet 3\bullet \ldots (n_i+\beta_i-2)\bullet \overbrace{(n_i+\beta_i-1)}^{\text{恰好是分母}\Gamma\text{里的}}}{1\bullet 2\bullet 3\bullet \ldots \bullet (n_i+\beta_i-2)}=n_i-1+\beta_i$

同理 $\frac{\overbrace{\Gamma(\sum_{t=1}^{V}(n_{t,\neg i}+\beta_t))}^{\text{小}}}{\underbrace{\Gamma(\sum_{t=1}^{V}(n_t+\beta_t))}_{\text{大}}} \underbrace{=}_{\text{按小的来}} \frac{1}{\sum_{t=1}^{V}(n_{t,\neg i}+\beta_t)}$

最后原式$p(z_i=k\mid \vec{z}_{\neg i},\vec{w})=\frac{p(\vec{w},\vec{z})}{p(\vec{w},\vec{z}_{\neg i})}=\underbrace{\frac{\overbrace{n_i-1}^{k_th\ \text{Topic第i个单词个数-1}}+\beta_i}{\sum_{t=1}^{V}(n_{t,\neg i}+\beta_t)}}_{W_i\text{这个词在}k_th\ Topic\text{下的概率}} \bullet \underbrace{\frac{\overbrace{n_k-1}^{m_th\ \text{Doc第k个主题词个数-1}}+\alpha_k}{\sum_{t=1}^{K}(n_{t,\neg i}+\alpha_t)}}_{topic_k_th\text{在m_th document里的概率}}$

为了将上面的注释中的“k_th Topic 的第 i 个单词个数-1”和“m_th Doc 的第 k 个主题词个数-1”融入公式符号中，并且为了引入“对称超参数”，也即公式中每个$\alpha_k$和$\beta_i$都按同一个$\alpha$和$\beta$处理，现在对公式做一些变换，得到：

$$\underbrace{=}_{\text{对称超参数}} \frac{n_{k,\neg i}^{(t)}+\beta}{\sum_{t=1}^{V}n_{k,\neg i}^{(t)}+V\beta} \bullet \frac{n_{m,\neg i}^{(t)}+\alpha}{\sum_{k=1}^{K}n_{m,\neg i}^{(k)}+K\alpha} \quad (3\text{-}20)$$

这就是 LDA 的 Collapsed Gibbs Sampling 采样公式了，其中的$\alpha$和$\beta$由于考虑了对称超参数，所以均使用同一个$\alpha$和$\beta$代替不同的$\alpha_i$和$\beta_i$代替。这一推导将永远是 MCMC 历史上的经典之作，它简洁、优雅，其关键之处还在于他利用了以下事实：Gamma 函数$\Gamma(x)$在最终的公式中神奇地消失了。

gamma 函数$\Gamma(x)$的出现有两个作用：

1. 换掉积分

2. 在分子分母同时出现 gamma 函数时利用 gamma 是阶乘的特性，而约去。

做完主题采样后，根据期望公式(3-4)就可以得到$\theta_{mat}$(doc->topic)和$\varphi_{mat}$(topic->word)两个重要的矩阵：

$\theta_{\text{mat}}=\left[\vec{\theta}_1 \quad \vec{\theta}_2 \quad \ldots \quad \vec{\theta}_m\right]$,其中每个$\vec{\theta}_i$向量里的每一项为

$$\theta_{m,k} \underbrace{=}_{\text{表示第m篇文章第k号主题}} \frac{n_{m,k} + \alpha_{\mathrm{k}}}{\sum_{i=1}^{\mathrm{K}}(n_{\mathrm{m},i} + \alpha_i)} \underbrace{=}_{\text{一般程序都只传入一个}\alpha} \frac{n_{m,k} + \alpha}{\sum_{i=1}^{\mathrm{K}} n_{\mathrm{m},i} + K\alpha} \tag{3-21}$$

$\varphi_{\text{mat}} = [\vec{\varphi}_1 \quad \vec{\varphi}_2 \quad \dots \quad \vec{\varphi}_K]$，其中每个$\vec{\varphi}_i$向量里的每一项为

$$\varphi_{k,w} \underbrace{=}_{\text{表示第k号主题第w号词}} \frac{n_{k,w} + \beta_w}{\sum_{i=1}^{V}(n_{k,i} + \beta_i)} \underbrace{=}_{\text{一般程序都只传入一个}\beta} \frac{n_{k,w} + \beta}{\sum_{i=1}^{V} n_{k,i} + V\beta} \tag{3-22}$$

## 3.5 总结

我们可以清楚的看到，在整个 LDA 迭代过程中，只是不断地对每个单词的 topic 编号重新指定，而$\theta_{\text{mat}}$(doc->topic)和$\varphi_{\text{mat}}$(topic->word)两个概率矩阵是迭代训练完成后通过期望公式计算出来的。

**训练后如何验证模型质量的好坏，验证结果的正确与否(或者验证训练是否已经收敛)呢？这里使用了一个 perplexity 的公式：**

$$b^{H(q)} = b^{-\frac{1}{N}\sum_{i=1}^{N}\log_b q(x_i)} \tag{3-23}$$

Perplexity 的意义：b 可以设置为 2 或 e，其中 H(q)就是该概率分布的熵。当概率 q 的 K 平均分布的时候，带入上式可以得到 q 的 perplexity 值=K。公式里的 $x_i$ 为测试文本，可以是句子或者文本，N 是测试集的大小（用来归一化），对于未知分布 q，**perplexity 的值越小，说明模型越好。**

$$perplexity(D_{test}) = \exp\{-\frac{\sum_{d=1}^{M}\log p(w_d)}{\sum_{d=1}^{M}N_d}\} = \exp\{-\frac{\sum_{d=1}^{M}\overbrace{\sum_{w_i \in d}\log\{\sum_{z\in d}(p(z \mid d)\cdot p(w_i \mid z))\}}^{\log p(w_d)}}{\sum_{d=1}^{M}N_d}\}$$

这里的$w_d$指的是第 d 篇文档中的所有词。

## 参考文献

# 第4章 实现与应用

根据前面的推导，LDA 的代码实现近在眼前，而 Gibbs Sampling 的 LDA 代码实现非常简洁，成为了机器学习界的典型案例。这一章，我分析 java 版本的 JGibbsLDA 代码(或 Gibbs LDA++)后，庖丁解牛，分析该代码实现。

## 4.1 实现

在分析 LDA 的代码实现之前，我们先来实现一个小算法（该小算法可以作为面试题），题目如下：

**投掷骰子程序**：写出一个函数，骰子有 K 个面，各面概率不一，给定各面的概率，**请写出投掷骰子后，骰子被投掷到的面编号的输出**。这里可将骰子抽象为一个 array 数组(double 类型数组)，这个数组内的元素为代表骰子各面的概率，函数逻辑为根据数组中各个 index 上的 double 浮点数概率，输出随机投掷到的那个 array index 编号：

**算法 4.1 Cumulative algorithm**

输入：double array[],array 每项代表概率

输出：随机投掷到的 array index

```
double p[] = new double[length];
copy array's element to p
for index from 1 to array.length do{
      p[index] += p[index-1];   //逐项累积
}
double u = random double in [0~ 1] * p[last element]
for index from 0 to p.length-1 do{
      if (p[index] > u){
            break
      }
return index
```

算法 4.1 在 gibbs sampling 中起到重要作用。该方法被称为“累积法”，将数组内概率先逐项累积，最后用该 p 数组范围内的均匀分布随机浮点数来随机投掷后，即可得到 index 值。

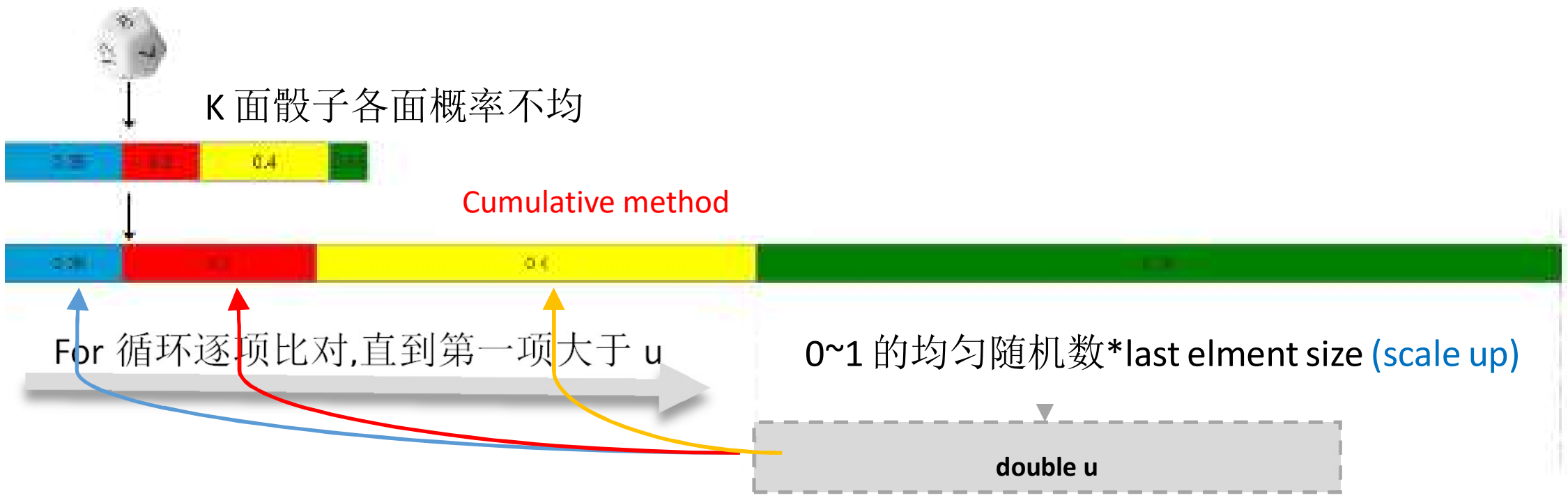

图 4-1 cumulative method

这个方法初看起来不是很明显，为何要累积？我们以一个直观的例子来说明这个问题，由于累积变换后数组的最后一项的长度就等于变换前的整体数组长度，在下面这个例子中假设我们要随机投掷到红色部分的面，这个面的机率只有 5%，而我们计算机程序的随机数生成器则是生成 0~1 之中每个小数机会都相等的均匀随机数。这个例子的图示如下：

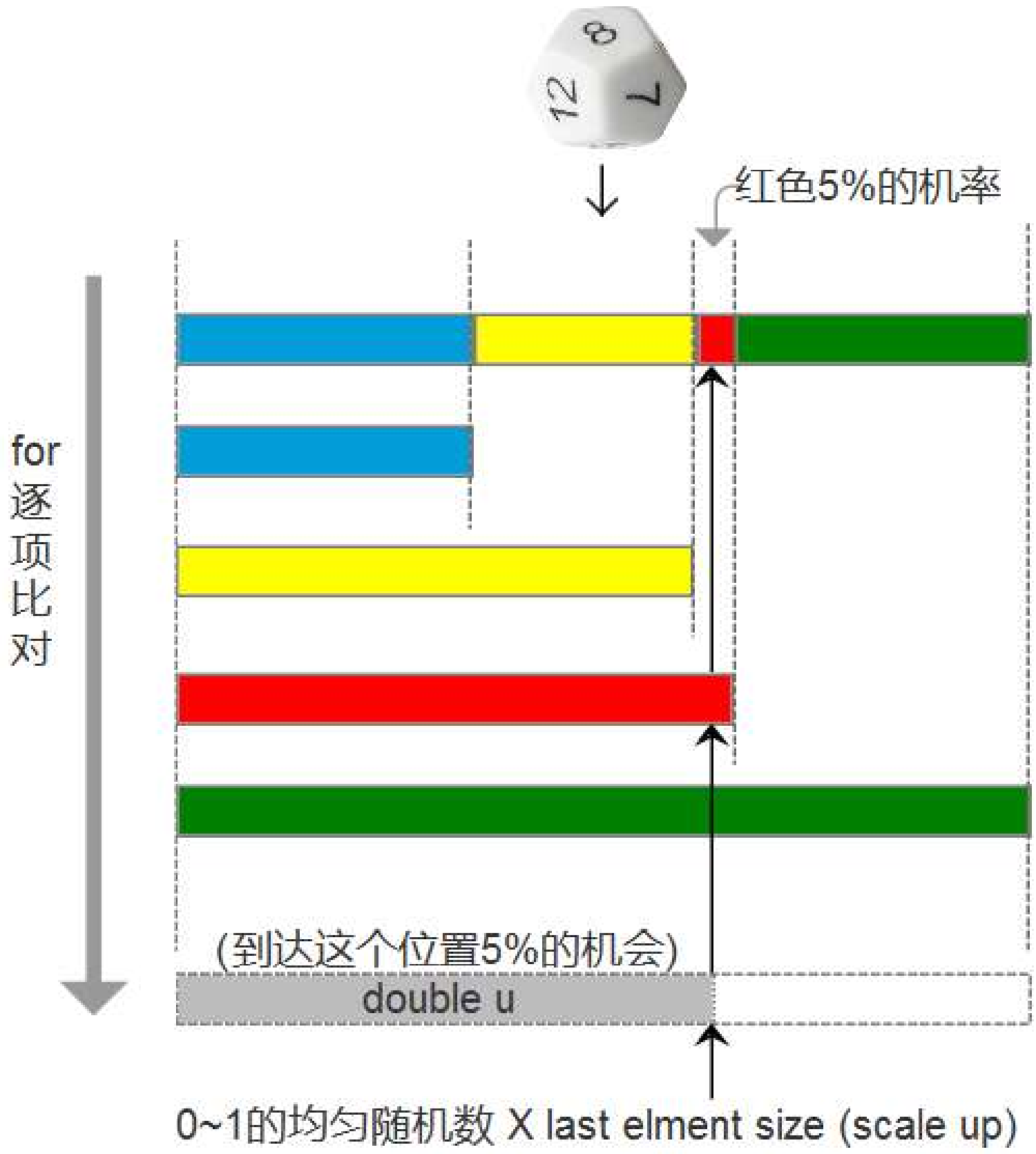

图 4-2 cumulative method 的一个直观例子

在上面这个例子中，蓝色和黄色的概率都在前面的 for 循环中被过滤了，因此在比对中我们如期到达了红色的面，此时判断 double u<累积后的红色概率而退出循环。我们紧接着来实现 Collapsed Gibbs sampling 的代码。[20]

根据前面的推导采样公式(3-20)可以写出[21]：

**算法 4.2 LDA Collapsed Gibbs Sampling**

输入：文档集(分词后)，K(主题数)，$\alpha$，$\beta$，iter_number(迭代次数)

输出：$\theta_{mat}$(doc->topic)和$\varphi_{mat}$(topic->word)、tassign 文件(topic assignment)

**申请几个统计量（int 数组（或二维数组））：**

```
nw[][]:number of instances of word/term i assigned to topic j，size V x K
nwsum[]:total number of words assigned to topic j，size K
nd[][]:第 i 篇文档里被指定第 j 个主题词的次数，  size M x K
ndsum[]:total number of words in document i，size M
z[][]: 是 int 二维数组[22]，topic assignment for each word，
      size M x per_doc_word_len 表示第 m 篇文档第 n 个 word 被指定的
topic index
```

**Initial 阶段：**

```
for (m, doc) in doc_set{   //m 是 doc 编号
   for word in doc{   //word 均有 wordid
       topic_index = random int from [0~ K-1]; 初始化阶段随机指定
       z[m][n] = topic_index; //将随机产生的主题存入 z 二维数组变量中
       nw[word_id][topic_index] ++; //为相应的统计量+1
       nwsum[topic_index]++;
       nd[m][ topic_index]++;
       ndsum[m]++;
   }
}
```

[20] 具体 java 版本代码请参考 JGibbs LDA 开源项目
[21] θ等价于 theta，φ等价于 phi
[22] 第二维每篇文档不一样（变长），第二维=per doc word count

//Initial 初始化结束，后面的迭代使用了-1→采样公式重新分配→+1 三重奏。

**Collasped Gibbs Sampling** 迭代阶段：（每一轮迭代有三重 **for** 循环）

for iter in iter_number{ //迭代 iter_number 次

 for (m, doc) in doc_set{ //m 是 doc 编号

   for word in doc{

    t=从 z[m][n]中取得当前 word 的主题编号(初始化来自随机)

    令 nw[word_id][t]、nwsum[t]、nd[m][ t]三个统计量均-1

    double p[] = new double p[K]

    for k in [0,1,2,…K-1]{  //从 0 到 K-1 号每个主题计算概率

$$p(z_i = k \mid \vec{z}_{\neg i}, \vec{w}) = \underbrace{\frac{nw[wordid][k] + \beta}{nwsum[k] + V \cdot \beta}}_{W_i\text{这个词在}k_th\ \ Topic\text{下的概率}} \bullet \underbrace{\frac{nd[m][k] + \alpha}{ndsum[m] + K \cdot \alpha}}_{topic_k_th\text{在m_th document里的概率}}$$

    //按照上述公式生成每个主题的概率存到临时概率数组 p 中

$$p[k] = p(z_i = k \mid \vec{z}_{\neg i}, \vec{w})$$

    }

    new_t = 输入数组 p 到算法 4.1 cumulative method 随机投掷

    令 nw[word_id][new_t]、nwsum[new_t]、nd[m][new_t]三个统计量均+1

   }

 }

}

**输出阶段（也可以在迭代一半时候即可输出）：**

根据 z 数组输出文件 tassign.txt

根据 nw、nwsum、nd、ndsum 套用公式(3-4)生成文件 theta.txt 和文件 phi.txt

文件 theta 格式:大矩阵文件，M 行 K 列

文件 phi 格式:大矩阵文件，K 行 V 列

下面讨论几个实现时的注意点：

**1. read old train file**:所谓的 model 文件就是算法 4.2 输出的这几个矩阵文件，如果输出了这几个文件，中断训练后，下次想再继续的时候：nw、nwsum、nd、ndsum 这四个统计量均从 tassign 文件中可以读取得到。

**2. predict**:另外，如果已经训练过一个 model 了，如果有一篇新文档，需要对其 predict（预测）新文档上的主题分布，可以利用已经训练的 trn_nw 和 trn_nwsum 用以下公式推断：

$$p(z_i = k \mid \vec{z}_{\neg i}, \vec{w}) = \underbrace{\frac{trn_nw_k^{(t)} + new_nw_{k,\neg i}^{(t)} + \beta_i}{\sum_{t=1}^{V}(trn_nw_k^{(t)} + new_nw_{k,\neg i}^{(t)} + \beta_t)}}_{W_i\text{这个词在}k_th\ Topic\text{下的概率}} \cdot \underbrace{\frac{n_{m,\neg i}^{(t)} + \alpha}{\sum_{k=1}^{K} n_{m,\neg i}^{(k)} + K\alpha}}_{topic_k_th\ \text{在}m_th\ document\text{里的概率}}$$

注意看这个公式，trn_nw 和 trn_nwsum 相当于起到了伪计数(pseudo count)的作用，公式后一项因子只与**当前文档**的主题词数计数有关，故不必加入已有 train model 的计数。此外，predict 在不管是初始化还是迭代时只修改 new_nw 和 new_nwsum 统计量，其余采样过程都一样。[23]

由于 trn_nw 二维数组变量已经很大，因此 new_nw 变量对其影响不是特别的大，所以可以减少一些对文档预测时候的采样次数，通常情况下，20 次采样足矣。假设对一篇 1000 字的文档（主题数 100）进行 20 次 Gibbs Sampling 采样进行 predict，大概需要花费多少时间复杂度呢？可以简单地做一个估算：

$$20(iter_num) \times 1000(N_d) \times 100(K) = 2000{,}000$$

这个数字大概是需要 200 万次循环，在笔者的机器上仍需花费 1 秒钟。

另外，-1 → 重新分配 topic → +1 三重奏我将其形象比喻为在采样每个单词前，从各个统计量中**抠掉**该单词（及其主题），然后重新分配。

再来用两张图，来说明：-1 → 重新分配 topic→ +1 三重奏：

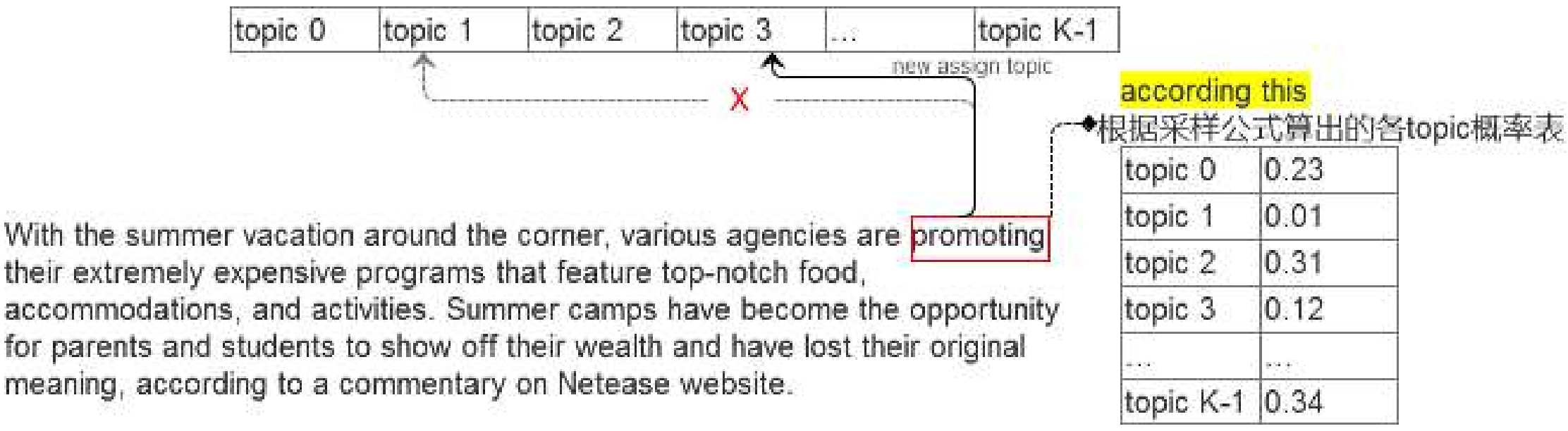

**图 4-3 -1 → reassign topic → +1**

23 https://lists.cs.princeton.edu/pipermail/topic-models/2012-June/001957.html

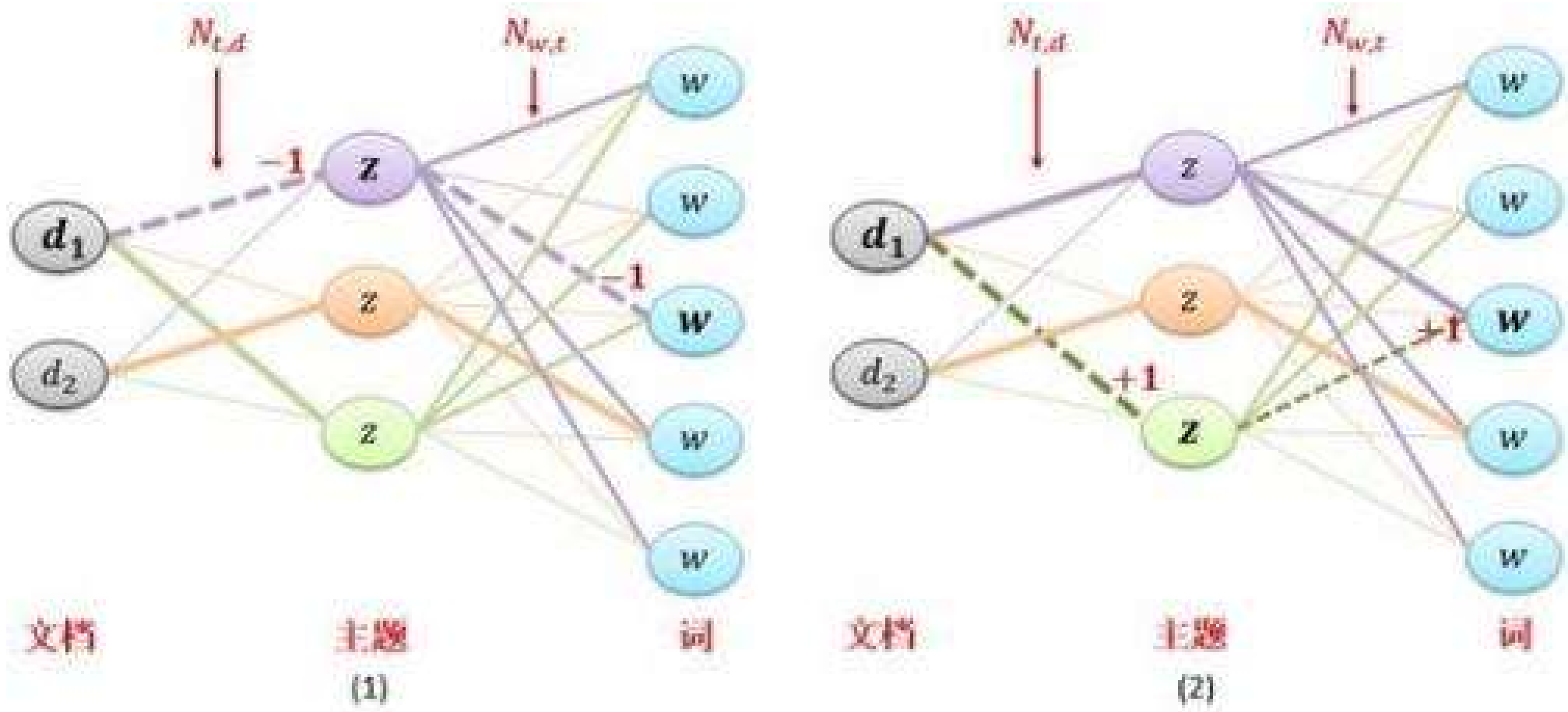

图 4-4 -1 → reassign topic → +1

JGibbs LDA 或 Gibbs LDA++输出的文件样例如下：

1. tassign 文件（一行一个 doc，冒号前是 wordid,冒号后是 topicid）

```
model-final.tassign
0:184 1:119 2:71 0:71 1:72 2:71 3:71 4:71 5:71 6:71 7:71 8:71 9:71 10:71 11:71 12:71 13:71 14:71 15:71 16:71 17:71 18:71 19:71 20:71 21:71 22:71 23:71 24:71 25:71 26:71 16:71 27:71 28:71 29:
140:62 141:71 7:71 142:71 143:71 144:64 140:71 141:71 7:71 142:117 143:64 144:64 140:36 145:117 146:117 147:71 140:71 141:71 148:117 149:117 150:64 151:117 152:117 153:64 7:71 154:7
329:184 330:64 331:184 332:99 333:184 329:184 330:72 331:184 332:50 333:184 330:64 331:184 334:50 335:117 336:184 337:184 338:184 339:50 340:184 341:184 342:184 226:113 343:50 34
624:64 625:64 626:117 627:29 624:64 625:117 626:29 627:29 628:117 156:64 629:64 630:147 486:64 490:117 631:64 632:117 7:71 4:64 633:64 634:117 635:71 147:147 624:64 636:64 637:71 63
148:117 655:117 240:50 226:50 720:117 362:50 148:117 655:117 240:50 226:50 720:50 362:50 721:50 4:117 722:50 723:50 724:50 725:117 726:50 171:117 727:117 728:117 606:117 729:117 2
962:117 240:117 655:117 962:117 240:117 655:117 356:117 683:117 963:117 136:117 964:189 965:147 592:117 966:117 251:117 151:64 967:117 968:117 969:117 137:50 964:64 970:24 251:11
1476:157 1477:50 1478:103 1479:149 1480:50 1481:139 1482:50 1476:147 1477:64 1478:80 1479:50 1480:50 1481:187 1482:50 1481:186 1483:64 1484:50 1485:147 1486:147 1487:147 1488:
7:64 1716:64 1717:71 78:71 1718:152 311:64 276:117 7:71 1716:80 1717:71 78:71 1718:10 311:64 276:117 16:71 1719:71 7:71 1720:71 1721:71 1041:64 1722:50 1723:71 1724:184 1725:71 7:7
626:117 834:50 7:50 1813:66 655:117 1802:50 1814:50 626:117 834:50 7:117 1813:50 655:117 1802:50 1814:50 673:50 524:50 265:117 226:117 1200:50 101:117 111:117 1012:117 7:117 1200:
1895:193 136:50 1639:64 1896:50 1895:64 136:64 1639:50 1896:50 171:117 1897:117 1193:117 1898:157 1617:29 1366:184 171:117 1895:191 1899:64 237:29 136:117 1540:50 356:3 246:50 1
1949:112 1950:43 158:64 1843:43 1951:64 155:29 215:64 914:64 1949:61 1950:29 158:64 1843:43 1951:64 155:64 215:71 914:64 1952:29 1953:43 1954:43 1955:29 1956:29 1617:29 1366:29 1
2087:64 389:184 2088:184 397:184 2087:182 389:184 2088:64 397:184 409:147 2089:145 718:184 2090:178 2091:43 1846:64 389:184 2092:108 2093:64 2094:78 17:184 389:184 2088:184 209
78:117 2124:71 2125:119 38:117 78:117 2124:117 2125:117 38:117 171:117 1480:117 4:64 1317:117 2126:117 583:117 732:117 733:117 1480:117 2127:117 151:117 2128:117 149:117 2129:11
2350:184 1344:147 2351:43 2352:49 2353:43 2354:86 2350:117 1344:50 2351:152 2352:26 2353:43 2354:43 148:117 2355:49 2356:199 2357:183 1617:49 171:117 2252:43 1295:117 2358:24 2
2404:49 2405:49 2406:50 1137:99 2407:101 950:147 2408:96 2404:49 2405:49 2406:49 1137:49 2407:35 950:19 2408:50 267:50 2409:50 2410:137 950:50 2411:178 2412:49 2413:49 2414:49 2
156:64 133:64 2487:64 2488:119 100:50 2419:147 2489:64 2490:174 156:64 133:64 2487:64 2488:89 100:64 2419:71 2489:50 2490:37 2491:64 156:147 1193:147 2492:167 2493:147 2494:152
2636:147 2637:147 2593:80 2638:106 2639:89 2593:147 1510:50 2636:147 2637:147 2593:50 2638:34 2639:147 2593:50 1510:50 2640:10 2641:182 1617:43 681:147 2636:147 2637:50 367:147
1007:117 626:117 929:79 2685:50 2686:49 2687:184 1007:117 626:117 929:50 2685:79 2686:9 2687:147 251:117 252:117 1832:147 1833:184 1837:50 115:50 2688:117 1837:50 2689:50 2690:1
1950:117 1183:117 2804:117 177:117 1950:117 1183:117 2804:117 177:117 2805:117 2806:117 147:117 206:117 2807:117 1950:117 177:117 181:117 2808:117 2043:50 2809:117 177:117 215
644:64 1846:117 307:117 626:117 2994:117 2995:117 1480:117 2996:64 644:71 1846:71 307:117 626:29 2994:29 2995:64 1480:64 2996:71 2997:172 2998:14 2999:122 3000:64 409:64 3001:29
```

冒号前是wordid，冒号后是分配的topicid

一行一篇文章

2. theta 文件（矩阵文件：行号=doc idx，列号=topic idx）

Doc1

Doc2

Doc3

…

```
topic0 topic1 topic2 topic3 topic4 topic5 topic6 topic7 topic8 topic9 topic10 topic11 topic12 topic13 topic14 ...
0.002339 0.002339 0.002339 0.005263 0.002339 0.002339 0.002339 0.002339 0.002339 0.002339 0.002339 0.002339 0.002339 0.002339 0.002339 0.002339 0.002339 0.002339 0
0.003309 0.001471 0.001471 0.001471 0.001471 0.001471 0.001471 0.001471 0.001471 0.001471 0.001471 0.001471 0.001471 0.001471 0.003309 0.001471 0.001471 0.001471 0
0.001057 0.001057 0.001057 0.001057 0.001057 0.001057 0.002378 0.001057 0.001057 0.001057 0.001057 0.002378 0.002378 0.001057 0.001057 0.001057 0.001057 0.001057 0
0.002254 0.002254 0.002254 0.002254 0.002254 0.002254 0.002254 0.005070 0.005070 0.002254 0.002254 0.002254 0.002254 0.002254 0.002254 0.002254 0.002254 0.002254 0
0.001170 0.001170 0.001170 0.001170 0.002632 0.001170 0.001170 0.001170 0.001170 0.001170 0.001170 0.001170 0.002632 0.001170 0.001170 0.001170 0.002632 0.001170 0
0.000390 0.000390 0.000390 0.000390 0.000390 0.000390 0.000878 0.000878 0.000390 0.000390 0.000390 0.000390 0.000390 0.000390 0.000390 0.000390 0.000390 0.000878 0
0.001192 0.007154 0.002683 0.001192 0.001192 0.001192 0.004173 0.001192 0.001192 0.001192 0.001192 0.002683 0.001192 0.001192 0.001192 0.002683 0.004173 0.001192 0
0.001743 0.001743 0.001743 0.001743 0.001743 0.001743 0.001743 0.001743 0.001743 0.001743 0.003922 0.001743 0.001743 0.001743 0.001743 0.001743 0.001743 0.001743 0
0.001713 0.001713 0.001713 0.001713 0.001713 0.001713 0.001713 0.001713 0.001713 0.001713 0.001713 0.001713 0.001713 0.001713 0.001713 0.001713 0.005996 0.001713 0
0.002235 0.002235 0.002235 0.005028 0.002235 0.002235 0.005028 0.002235 0.002235 0.002235 0.002235 0.002235 0.002235 0.002235 0.002235 0.002235 0.002235 0.002235 0
0.001544 0.001544 0.001544 0.003475 0.001544 0.001544 0.003475 0.005405 0.001544 0.001544 0.003475 0.001544 0.003475 0.001544 0.003475 0.001544 0.001544 0.001544 0
0.002996 0.006742 0.002996 0.002996 0.002996 0.002996 0.002996 0.002996 0.002996 0.002996 0.002996 0.002996 0.002996 0.002996 0.002996 0.002996 0.002996 0.002996 0
0.001787 0.000794 0.001787 0.003774 0.000794 0.001787 0.000794 0.002781 0.000794 0.000794 0.000794 0.000794 0.000794 0.001787 0.000794 0.002781 0.000794 0.000794 0
0.002632 0.002632 0.002632 0.002632 0.002632 0.002632 0.002632 0.002632 0.002632 0.002632 0.002632 0.002632 0.002632 0.002632 0.002632 0.002632 0.002632 0.002632 0
0.002532 0.002532 0.008861 0.002532 0.002532 0.002532 0.002532 0.002532 0.008861 0.002532 0.002532 0.002532 0.002532 0.005696 0.002532 0.002532 0.002532 0.005696 0
0.001416 0.001416 0.001416 0.001416 0.006726 0.001416 0.001416 0.001416 0.001416 0.001416 0.003186 0.001416 0.001416 0.001416 0.004956 0.001416 0.003186 0.003186 0
0.002439 0.002439 0.002439 0.002439 0.002439 0.002439 0.005488 0.002439 0.002439 0.002439 0.008537 0.002439 0.002439 0.002439 0.008537 0.002439 0.002439 0.002439 0
0.001081 0.001081 0.001081 0.001081 0.003784 0.001081 0.001081 0.001081 0.002432 0.003784 0.002432 0.001081 0.001081 0.002432 0.001081 0.002432 0.001081 0.001081 0
0.000500 0.000500 0.001749 0.000500 0.000500 0.002374 0.000500 0.000500 0.000500 0.000500 0.000500 0.000500 0.000500 0.000500 0.000500 0.001124 0.000500 0.001124 0
0.002996 0.002996 0.002996 0.002996 0.010487 0.002996 0.002996 0.002996 0.002996 0.002996 0.002996 0.002996 0.002996 0.002996 0.006742 0.002996 0.002996 0.002996 0
0.001663 0.001663 0.001663 0.001663 0.001663 0.001663 0.001663 0.001663 0.001663 0.001663 0.003742 0.001663 0.001663 0.001663 0.001663 0.001663 0.001663 0.001663 0
0.001136 0.003977 0.001136 0.001136 0.001136 0.001136 0.001136 0.001136 0.001136 0.001136 0.001136 0.001136 0.002557 0.002557 0.001136 0.001136 0.002557 0.001136 0
```

4. phi 文件（矩阵文件：行号=topic idx，列号=word idx）

| | word0 | word1 | word2 | word3 | word4 | word5 | word6 | word7 | word8 | word9 | word10 | word11 | word12 | … | | | | | |
|---|---|---|---|---|---|---|---|---|---|---|---|---|---|---|---|---|---|---|---|
| topic0 | 0.002339 | 0.002339 | 0.002339 | 0.005263 | 0.002339 | 0.002339 | 0.002339 | 0.002339 | 0.002339 | 0.002339 | 0.002339 | 0.002339 | 0.002339 | 0.002339 | 0.002339 | 0.002339 | 0.002339 | 0.002339 | 0 |
| | 0.003309 | 0.001471 | 0.001471 | 0.001471 | 0.001471 | 0.001471 | 0.001471 | 0.001471 | 0.001471 | 0.001471 | 0.001471 | 0.001471 | 0.001471 | 0.001471 | 0.003309 | 0.001471 | 0.001471 | 0.001471 | 0 |
| | 0.001057 | 0.001057 | 0.001057 | 0.001057 | 0.001057 | 0.001057 | 0.002378 | 0.001057 | 0.001057 | 0.001057 | 0.001057 | 0.002378 | 0.002378 | 0.001057 | 0.001057 | 0.001057 | 0.001057 | 0.001057 | 0 |
| topic1 | 0.002254 | 0.002254 | 0.002254 | 0.002254 | 0.002254 | 0.002254 | 0.002254 | 0.005070 | 0.005070 | 0.002254 | 0.002254 | 0.002254 | 0.002254 | 0.002254 | 0.002254 | 0.002254 | 0.002254 | 0.002254 | 0 |
| | 0.001170 | 0.001170 | 0.001170 | 0.001170 | 0.002632 | 0.001170 | 0.001170 | 0.001170 | 0.001170 | 0.001170 | 0.001170 | 0.001170 | 0.002632 | 0.001170 | 0.001170 | 0.001170 | 0.002632 | 0.001170 | 0 |
| topic2 | 0.000390 | 0.000390 | 0.000390 | 0.000390 | 0.000390 | 0.000390 | 0.000878 | 0.000878 | 0.000390 | 0.000390 | 0.000390 | 0.000390 | 0.000390 | 0.000390 | 0.000390 | 0.000390 | 0.000390 | 0.000878 | 0 |
| | 0.001192 | 0.007154 | 0.002683 | 0.001192 | 0.001192 | 0.001192 | 0.004173 | 0.001192 | 0.001192 | 0.001192 | 0.001192 | 0.002683 | 0.001192 | 0.001192 | 0.001192 | 0.002683 | 0.004173 | 0.001192 | 0 |
| … | 0.001743 | 0.001743 | 0.001743 | 0.001743 | 0.001743 | 0.001743 | 0.001743 | 0.001743 | 0.001743 | 0.001743 | 0.003922 | 0.001743 | 0.001743 | 0.001743 | 0.001743 | 0.001743 | 0.001743 | 0.001743 | 0 |
| | 0.001713 | 0.001713 | 0.001713 | 0.001713 | 0.001713 | 0.001713 | 0.001713 | 0.001713 | 0.001713 | 0.001713 | 0.001713 | 0.001713 | 0.001713 | 0.001713 | 0.001713 | 0.001713 | 0.005996 | 0.001713 | 0 |
| | 0.002235 | 0.002235 | 0.002235 | 0.005028 | 0.002235 | 0.002235 | 0.005028 | 0.002235 | 0.002235 | 0.002235 | 0.002235 | 0.002235 | 0.002235 | 0.002235 | 0.002235 | 0.002235 | 0.002235 | 0.002235 | 0 |
| | 0.001544 | 0.001544 | 0.001544 | 0.003475 | 0.001544 | 0.001544 | 0.003475 | 0.005405 | 0.001544 | 0.001544 | 0.003475 | 0.001544 | 0.003475 | 0.001544 | 0.003475 | 0.001544 | 0.001544 | 0.001544 | 0 |
| | 0.002996 | 0.006742 | 0.002996 | 0.002996 | 0.002996 | 0.002996 | 0.002996 | 0.002996 | 0.002996 | 0.002996 | 0.002996 | 0.002996 | 0.002996 | 0.002996 | 0.002996 | 0.002996 | 0.002996 | 0.002996 | 0 |
| | 0.001787 | 0.000794 | 0.001787 | 0.003774 | 0.000794 | 0.001787 | 0.000794 | 0.002781 | 0.000794 | 0.000794 | 0.000794 | 0.000794 | 0.000794 | 0.001787 | 0.000794 | 0.002781 | 0.000794 | 0.000794 | 0 |
| | 0.002632 | 0.002632 | 0.002632 | 0.002632 | 0.002632 | 0.002632 | 0.002632 | 0.002632 | 0.002632 | 0.002632 | 0.002632 | 0.002632 | 0.002632 | 0.002632 | 0.002632 | 0.002632 | 0.002632 | 0.002632 | 0 |
| | 0.002532 | 0.002532 | 0.008861 | 0.002532 | 0.002532 | 0.002532 | 0.002532 | 0.002532 | 0.008861 | 0.002532 | 0.002532 | 0.002532 | 0.002532 | 0.005696 | 0.002532 | 0.002532 | 0.002532 | 0.005696 | 0 |
| | 0.001416 | 0.001416 | 0.001416 | 0.001416 | 0.006726 | 0.001416 | 0.001416 | 0.001416 | 0.001416 | 0.001416 | 0.003186 | 0.001416 | 0.001416 | 0.001416 | 0.004956 | 0.001416 | 0.003186 | 0.003186 | 0 |
| | 0.002439 | 0.002439 | 0.002439 | 0.002439 | 0.002439 | 0.002439 | 0.005488 | 0.002439 | 0.002439 | 0.002439 | 0.008537 | 0.002439 | 0.002439 | 0.002439 | 0.008537 | 0.002439 | 0.002439 | 0.002439 | 0 |
| | 0.001081 | 0.001081 | 0.001081 | 0.001081 | 0.003784 | 0.001081 | 0.001081 | 0.001081 | 0.002432 | 0.003784 | 0.002432 | 0.001081 | 0.001081 | 0.002432 | 0.001081 | 0.002432 | 0.001081 | 0.001081 | 0 |
| | 0.000500 | 0.000500 | 0.001749 | 0.000500 | 0.000500 | 0.002374 | 0.000500 | 0.000500 | 0.000500 | 0.000500 | 0.000500 | 0.000500 | 0.000500 | 0.000500 | 0.000500 | 0.001124 | 0.000500 | 0.001124 | 0 |
| | 0.002996 | 0.002996 | 0.002996 | 0.002996 | 0.010487 | 0.002996 | 0.002996 | 0.002996 | 0.002996 | 0.002996 | 0.002996 | 0.002996 | 0.002996 | 0.002996 | 0.006742 | 0.002996 | 0.002996 | 0.002996 | 0 |
| | 0.001663 | 0.001663 | 0.001663 | 0.001663 | 0.001663 | 0.001663 | 0.001663 | 0.001663 | 0.001663 | 0.001663 | 0.003742 | 0.001663 | 0.001663 | 0.001663 | 0.001663 | 0.001663 | 0.001663 | 0.001663 | 0 |
| | 0.001136 | 0.003977 | 0.001136 | 0.001136 | 0.001136 | 0.001136 | 0.001136 | 0.001136 | 0.001136 | 0.001136 | 0.001136 | 0.001136 | 0.002557 | 0.002557 | 0.001136 | 0.001136 | 0.002557 | 0.001136 | 0 |

4. wordmap 文件（word→wordid）

法规 1134
法规处 14599
法警 95871
法语 5523
法语节 86971
法语系 94238
法诺夫 78519
法赫德 70171
法辛 20246
法郎 32960
法里德 136597
法门 93723
法院 8762
法雷尔 80298
法餐 38716
法餐厅 112467
法马 105507
泗水 124802
泗洪 125832
泗洪县 125816
泗阳 99157

**算法复杂度分析：**

1. **时间复杂度**：Collapsed Gibbs Sampling LDA 最大的特点是每次迭代内部要经历三重 for 循环，因此每轮迭代经历的循环数是 doc_num * per_word_num * K = corpus_word_count *K，所以最终的时间复杂度 =O(iter_num * corpus_word_count * K)，这就说明了你启动时设置的主题数 K 越多，算法运行就越耗时间。

2. **空间复杂度**：算法运行中需要几个统计量，耗费空间最多的有两个 nw 和 nd，其中 nw 占用内存空间最大=主题数 K * vocabulary_size(词典中 unique 单词数)，所以算法的空间复杂度=O(K * V)。例如，考虑 K=300，V=10 万词，且都用 int（4 字节）来存储，二维数组 nw 需要多大的内存呢？

$$\frac{300个主题,10万word的统计量nw占用大小}{1MB内存占用字节数} = \frac{300 \times 100000 \times 4}{2^{10} \times 2^{10}} = 114.44MB$$

**参数设置：**

topic number K：许多读者问，如何设置主题个数，其实现在没有特别好的办法[24]，目前只有交叉验证（cross validation）：通过设置不同的 K 值训练后验证比较求得最佳值。我的建议是一开始不要设置太大而逐步增大实验，Blei 在论文《Latent Dirichlet Allocation》提出过一个方法，采用设置不同的 topic 数量，画出 topic_number-perplexity 曲线；Thomas L. Griffiths 等人在《Finding scientific topics》也提出过一个验证方法，画出 topic_number-logP(w|T)曲线，然后找到曲线中的纵轴最高点便是 topic 数量的最佳值。有兴趣的读者可以去读读这两篇论文原文的相应部分。这个参数同时也跟文章数量有关，可以通过一个思想实验来验证：设想两个极端情况：如果仅有一篇文章做训练，则设置几百个 topic 不合适，如果将好几亿篇文章拿来做 topic model，则仅仅设置很少 topic 也是不合适的。

**$\alpha$、$\beta$超参数：**这个超参数其实作为伪计数，不是很影响算法的结果。

**迭代次数 iter_number：**gibbs sampling 前面 1000 次甚至更多可以抛弃，这样会达到更好的效果。

## 4.2 应用

算法实用才更有价值，我将 LDA 称之为万能工具，在许多文本分析的角度都可以应用（甚至推荐领域），下面我们通过几个实用例子来谈谈如何应用 LDA 算法。选取应用的原则是：利用基本的 LDA 算法的输出即可做出、无需改造原有 LDA 代码的，这也是对读者要求较小的应用方式。

### 4.2.1 相似文档发现

这个方法可以被用作新闻推荐中，正文详情页的“相关推荐”，该方法所述的相似文档是指的“主题层面”上的相似，这就比其他的基于 word 来挖掘的相似度更有意义，整个方法非常简单：

**(1) Hellinger distance**

[24] HDP 等较为复杂的模型可以自动确定这个参数，但是模型复杂，计算复杂

该距离是度量两个概率分布差异的距离，如下所述（当然你也可以用 KL 距离等）：

$$H(P,Q)=\frac{1}{\sqrt{2}}\sqrt{\sum_{i=1}^{K}(\sqrt{p_i}-\sqrt{q_i})^2} \tag{4-1}$$

由于前面的 $\frac{1}{\sqrt{2}}$ 是常数，不影响排序，因此两个文章的相似度可以用下面的距离做出：

$$doc_similarity(d,f)=\sum_{i=0}^{K-1}(\sqrt{\theta_{d,\mathrm{i}}}-\sqrt{\theta_{f,i}})^2 \tag{4-2}$$

(2)算法步骤

由于 LDA 可以将文章表征成多主题分布的混合，所以利用上述公式，这样仅仅用了一个文件：theta 文件，就可以分析出主题层面上的距离。

输入：LDA 训练后的 theta 文件

**步骤：**1. 读取 LDA 训练后的 theta 文件

2. 为每个 doc 分别计算其他 doc 与该 doc 的主题距离（用上面的公式），再按照距离从小到大排序，输出每个 doc 最相关的前 n 个主题文章。

(3) 结果演示

比如 ISIS 和极端组织字面上并不相等，但表达意思一致，在这种情况下，依靠基于 LDA 的办法就可以成功为其推荐出极端组织的文章，部分结果如下所示（截图左边为“正文页主文章”，右边一系列为与之主题相似的文章）：

| | | |
|---|---|---|
| 美国可能向越南出售什么武器？（图）[1]- 中国日报网 | 美媒关注中国隐形战机歼-31亮相珠海航展 - 中文国际 - 中 | http://news.sina.com.cn/w/2014-11-13/16303113 |
| | 未能取到标题 | http://news.sina.com.cn/w/2014-09-18/0913308 |
| | 美国欲加强在澳洲军事存在 增加驻军及战机轮防 | http://news.sina.com.cn/w/2014-08-11/1236306 |
| | 未能取到标题 | http://news.sina.com.cn/w/2014-01-08/0851291 |
| | 美国解除武器禁运将助越南海军提升远海监察能力[1]- 中国 | http://news.sina.com.cn/w/2014-10-24/1643310 |
| | 未能取到标题 | http://news.sina.com.cn/w/2014-03-13/0950296 |
| | 美国计划打造本土宇宙飞船取代俄联盟号 | http://news.sina.com.cn/w/2014-08-22/1702307 |
| | 未能取到标题 | http://news.sina.com.cn/w/2014-03-06/0942296 |
| | 未能取到标题 | http://news.sina.com.cn/w/2014-03-12/0921296 |
| | 澳可能与日联手造潜艇 | http://news.sina.com.cn/w/2014-07-29/0419305 |
| | 俄媒：2025年前俄卫星集群组成将达150个 - 中文国际 - 中 | http://news.sina.com.cn/w/2014-11-17/1936311 |
| | 美欲摆脱依赖俄罗斯 巨资发展“太空巴士” | http://news.sina.com.cn/w/2014-09-18/0939308 |
| | 未能取到标题 | http://news.sina.com.cn/w/2014-09-23/1307309 |
| | 美国军工巨头将与澳企联手 追踪太空垃圾(图) | http://news.sina.com.cn/w/p/2014-08-28/09233 |
| | 未能取到标题 | http://news.sina.com.cn/w/2014-04-27/1006300 |
| | 美拟建最快超级计算机赶超中国 | http://news.sina.com.cn/w/2014-11-16/0320311 |
| 索马里首都汽车炸弹袭击致5人死亡 | 尼日利亚北部村庄教堂遭袭至少30人死亡 | http://news.sina.com.cn/w/2014-06-30/0409304 |
| | 伊拉克巴格达警方检查站遇袭 至少6人死亡 | http://news.sina.com.cn/w/2014-11-01/2302310 |
| | 伊拉克发生多起爆炸袭击致10人死亡 | http://news.sina.com.cn/w/2014-11-07/0615311 |
| | 未能取到标题 | http://news.sina.com.cn/w/2014-01-10/1027292 |
| | 阿富汗遭自杀性袭击致多人死 塔利班声称负责 | http://news.sina.com.cn/w/2014-07-02/1344304 |
| | 伊拉克首都北部城区遭炮击致16人受伤 | http://news.sina.com.cn/w/2014-11-04/0712310 |
| | 未能取到标题 | http://news.sina.com.cn/w/2014-01-02/0949291 |
| | 未能取到标题 | http://news.sina.com.cn/w/2014-03-11/1602296 |
| | 索马里首都汽车炸弹袭击致10人死亡 | http://news.sina.com.cn/w/2014-10-13/0644309 |
| | 阿富汗南部自杀式袭击致4人死亡18人受伤 | http://news.sina.com.cn/w/2014-10-08/1738309 |
| | 未能取到标题 | http://news.sina.com.cn/w/2014-01-13/1356292 |
| | 阿富汗首都发生爆炸袭击致4死12伤 塔利班称负责 | http://news.sina.com.cn/w/2014-10-21/1324310 |
| | 也门路边炸弹爆炸导致14人死亡 | http://news.sina.com.cn/w/2014-08-14/0841306 |
| | 未能取到标题 | http://news.sina.com.cn/w/2014-09-05/0529307 |
| | 伊拉克首都发生3起爆炸袭击致16死54伤 | http://news.sina.com.cn/w/2014-05-23/0119302 |
| | 尼日利亚约贝州居民遭袭已致8人死亡 | http://news.sina.com.cn/w/2014-07-30/0001305 |
| 土耳其将允许伊拉克库尔德人越境对IS作战 | 伊拉克极端组织宣称已“斩首”一名美国记者 | http://news.sina.com.cn/w/2014-08-20/07513071 |
| | 未能取到标题 | http://news.sina.com.cn/w/2014-04-22/05582998 |
| | 未能取到标题 | http://news.sina.com.cn/w/2014-01-08/19062919 |
| | 杀害美国记者是对美国的恐怖袭击 | http://news.sina.com.cn/w/2014-08-24/22003073 |
| | 美国将增兵伊拉克 帮助剿灭极端武装 | http://news.sina.com.cn/w/2014-11-09/02103111 |
| | 卡塔尔否认援助叙极端组织:仅帮助温和民兵武装 | http://news.sina.com.cn/w/2014-10-27/13293105 |
| | 美军或派遣军事顾问协助伊拉克打击伊斯兰国 | http://news.sina.com.cn/w/2014-11-14/15083114 |
| | 未能取到标题 | http://news.sina.com.cn/w/2014-01-05/17112916 |
| | 未能取到标题 | http://news.sina.com.cn/w/2014-10-09/05583096 |
| | 伊拉克军队持续向武装分子据点展开进攻(图) | http://news.sina.com.cn/w/p/2014-09-02/134130 |
| | 叙利亚反对派表示支持美国打击极端组织计划 | http://news.sina.com.cn/w/2014-09-11/14453082 |
| | 尼日利亚极端组织发被绑女学生视频 开获释条件 | http://news.sina.com.cn/w/2014-05-12/17363011 |
| | 奥巴马称部分美军即将撤离伊拉克 | http://news.sina.com.cn/w/2014-08-16/07093069 |
| | “伊斯兰国”再发处死美记者视频 | http://news.sina.com.cn/w/2014-09-04/03593079 |
| 泰国船只相撞致23名中国人受伤2名韩国人失踪 | 未能取到标题 | http://news.sina.com.cn/w/2014-10-20/06193101 |
| | 泰国船只相撞 中国游客23伤 | http://news.sina.com.cn/w/2014-10-21/04593101 |
| | 加拿大首都议会大厦附近发生枪击事件 | http://news.sina.com.cn/w/2014-10-22/22503102 |
| | 韩国一客运列车与旅游列车相撞 致十余人受伤 | http://news.sina.com.cn/w/2014-07-22/17453056 |
| | 尼加拉瓜金矿坍塌事故20人获救 5名旷工仍失踪 | http://news.sina.com.cn/w/2014-08-30/15273076 |
| | 几内亚沉船事故已致18人死亡 仍有20人下落不明 | http://news.sina.com.cn/w/2014-10-13/09133098 |
| | 中国游客在普吉岛受伤 | http://news.sina.com.cn/w/2014-10-20/02293101 |
| | 中国一渔船太平洋起火沉没 | http://news.sina.com.cn/w/2014-05-06/10103006 |
| | 孟加拉国一艘客轮沉没造成12人死亡 | http://news.sina.com.cn/w/2014-05-04/17583005 |
| | 秘鲁外海两渔船相撞致9人死亡3人失踪 3人获救 | http://news.sina.com.cn/w/2014-06-25/09143041 |
| | 几内亚沉船事故已致18人死亡 仍有20人下落不明 | http://news.sina.com.cn/w/2014-10-13/09133098 |
| | 中国籍油轮在济州海域发生空调爆炸 2名人身亡 | http://news.sina.com.cn/w/2014-05-22/14303019 |
| | 未能取到标题 | http://news.sina.com.cn/w/2014-04-21/16002997 |
| | 孟加拉国载有数百人渡轮沉没已致6人死亡 | http://news.sina.com.cn/w/2014-05-15/20193014 |
| | 一辆非法移民船在多米尼加海域倾覆6人死亡 | http://news.sina.com.cn/w/2014-05-02/11493004 |
| | 几内亚载69人木船因超载沉没 9人死亡数人失踪 | http://news.sina.com.cn/w/2014-10-11/08343097 |

**图 4-5 主题相似的文章推荐**

### 4.2.2 自动打标签

利用 LDA 的训练结果做文章的简易自动打标签同样可以得到不错的结果，该方法的实现也异常简单：

算法输入：theta 文件（doc→topic），phi 文件（topic→word）

算法步骤：

1. 分词（可以考虑使用 stanford segment）

2. 词性标注（可以考虑使用 stanford chinese POS tagging）

3. 去掉副词、介词等非实意词。

4. LDA 训练

5. 通过读取 theta 文件找出每篇文章概率最大那个主题，获得主题编号。

6. 读取 phi 文件里每个主题下 word 概率，读入内存。

7. 根据该文章最大主题编号找出该文章下该概率最大主题编号下的概率最大 n 个 word 词（max top n），（换句话说：该文章最大主题下的最大概率的 n 个词）作为该文章标签输出。

输出结果样例如下所示，可以看到，由于是同一个主题下最大概率的 n 个词，因此仅从标签表达意义就可以看出文章大体是讲什么的。

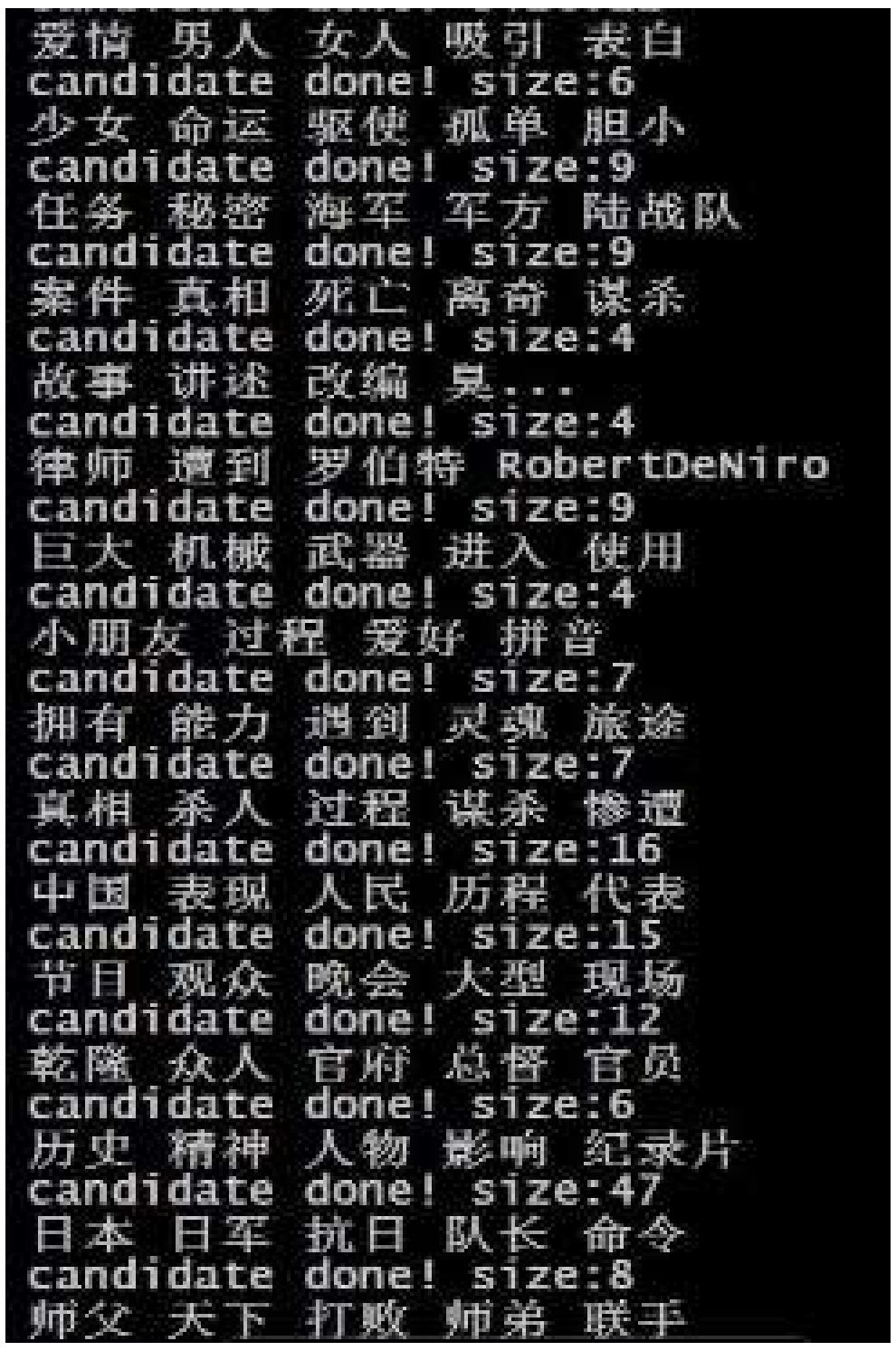

图 4-6 基于 LDA 打标签

#### 4.2.3 LDA 与 LR（逻辑斯蒂回归）结合做新闻个性化推荐系统

常见的个性化新闻推荐手段有基于关键词的，基于新闻频道栏目的等，但是基于关键词的由于粒度较细，不太容易满足推荐系统的“多样性”，而基于喜爱栏目（比如 NBA）的个性化推荐粒度又太粗，所以就比较有必要研究基于主题的推荐系统了。

下面我分别介绍两种方法为用户推荐用户喜爱主题的新闻文章。

**方法 1：LDA + LR（逻辑斯蒂回归）**

这个方法基于这样的假设：用户从新闻列表页点击的新闻时，肯定是对该新闻较为喜欢，而上下相邻的几条新闻未点击，肯定用户的眼睛看到了而未点：表明不喜欢。

**算法步骤：**

1.收集点击日志：

用户在一个新闻网站上（比如一个门户）通常的点击情况图 4-7 所示，我们需要收集下用户在列表页点击的文章和用户未点击的前几个和后几个文章（收集时需要做标识区别）。

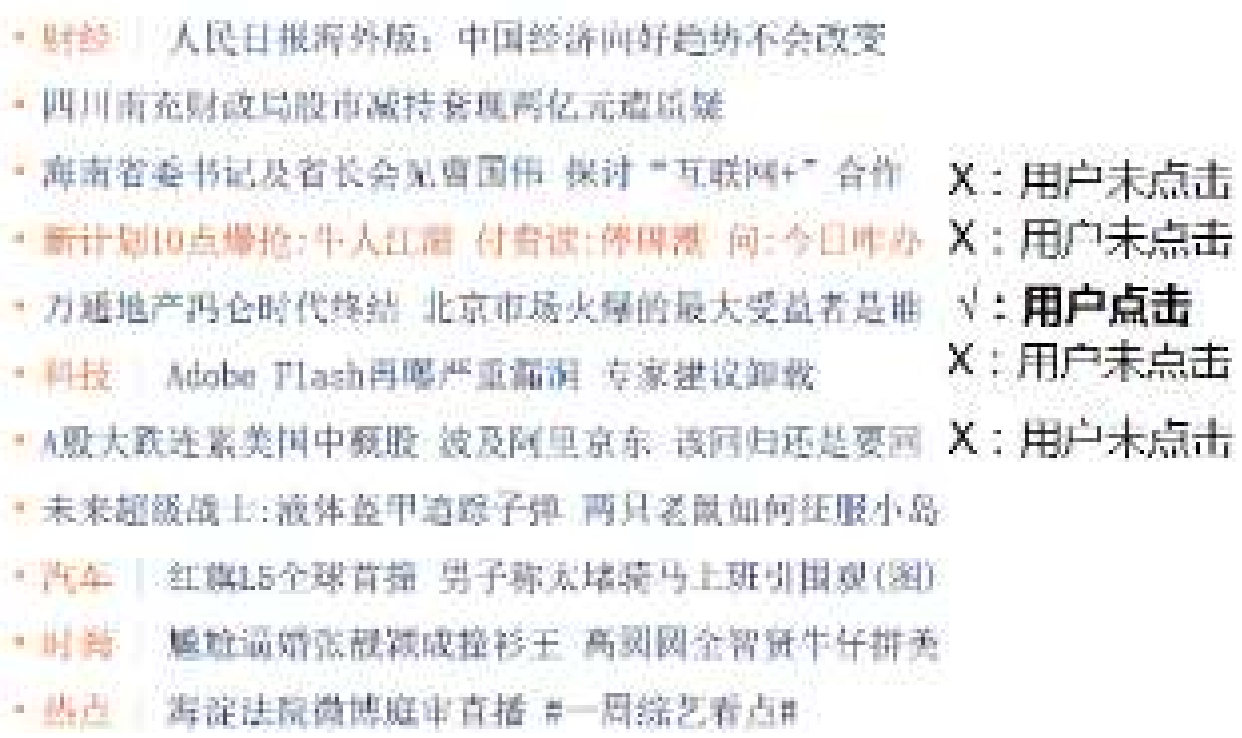

图 4-7 需要收集的用户点击的 item 条目和上下几条未点击

2. 使用 LDA 为新闻文章做主题维度特征提取

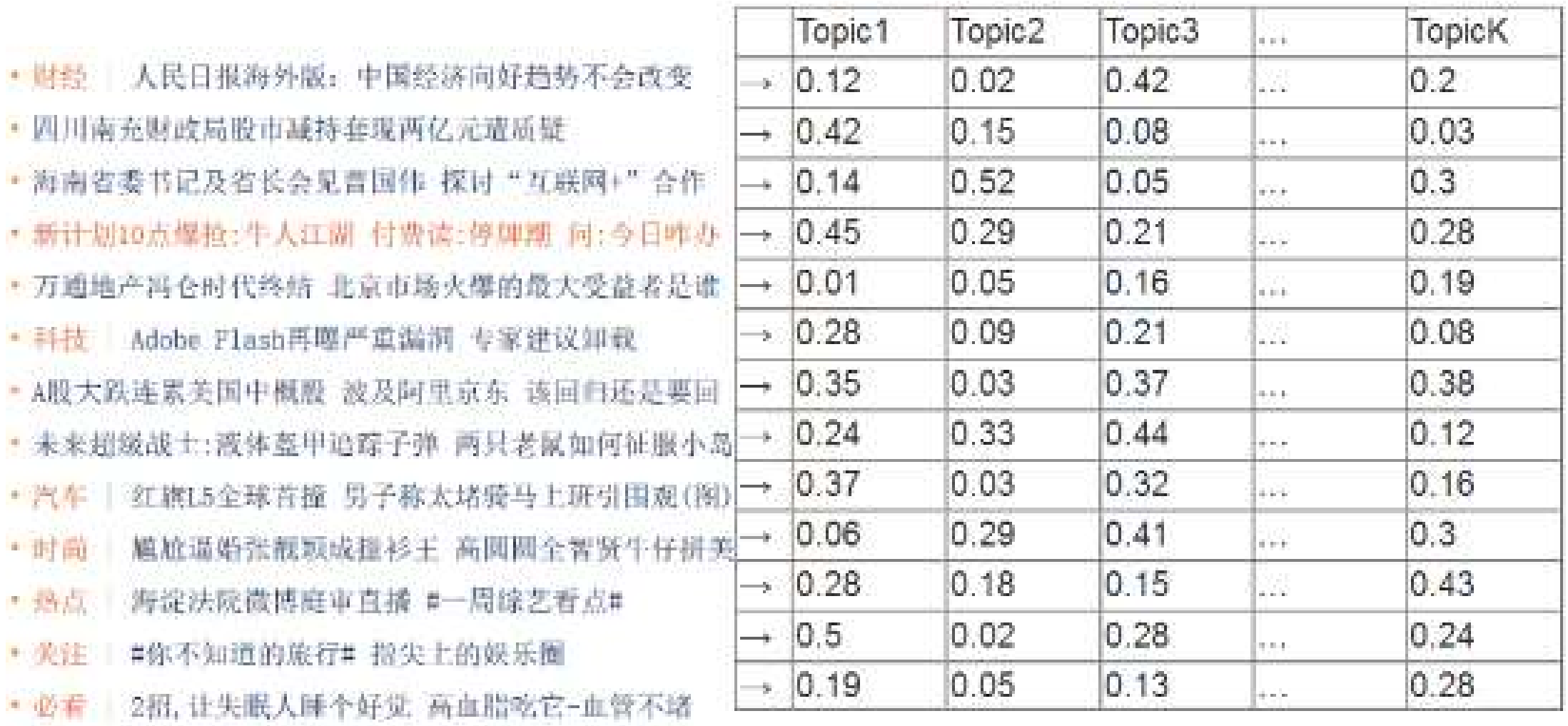

图 4-8 将文章用 LDA 特征提取(theta 文件)

3. LR 算法使用 log 点击数据做训练

将文章通过 LDA 训练后降维为 K 个主题（比如 200 个主题后），每个主题的概率为一个小于 1 的浮点小数。在利用前面的点击 log 数据，就可以将每个用户点击标为正样本，用户未点击为负样本。

| docid | 用户点击 | Topic1 | Topic2 | Topic3 | ... | TopicK |
|---|---|---|---|---|---|---|
| 842 | 0 | 0.12 | 0.02 | 0.42 | ... | 0.2 |
| 886 | 0 | 0.42 | 0.15 | 0.08 | ... | 0.03 |
| 822 | 0 | 0.14 | 0.52 | 0.05 | ... | 0.26 |
| 981 | 1 | 0.05 | 0.2 | 0.17 | ... | 0.47 |
| 417 | 0 | 0.49 | 0.3 | 0.17 | ... | 0.23 |

表 4-1 某一个用户的 LR 训练数据(1 为点击，0 为未点击)

紧接着就可以使用 LR 训练出该用户对 K 个 topic 里每个 topic 上的喜好权重 weight 了，再用这个 weight 向量对**每篇新的新闻文章**使用线性加权公式：doc_score = w1 * topic1 + w2 * topic2 + …，从大到小按照 score 排序后，

可以立即为用户提供个性化推荐。

**注意**：考虑到通常很多网站用户粘性没有那么大，一个用户的点击记录不可能很多，可以先使用用户聚类算法将用户聚类成一组一组，然后**一组用户作为训练集**训练 topic 喜好程度的 weight 权重。

**方法 2：user profile 记录喜好 topic 法**

算法步骤：

1.**提取 topic**：文章 LDA 训练后的 theta 文件，提取每篇文章概率最大的前 3 个 topic 主题

2.**save topic→user profile**：当用户访问文章 A 后，就把文章 A 的 top 3 的 topic 贴到用户兴趣档案里，并将概率分值相加，如果过了 1 天用户没有访问网站或访问这个 topic 的文章，就按照该 topic 乘以 0.8 衰减，直至衰减到 0。

3.**产生推荐**：下次用户再次访问网站时，从用户兴趣档案找出用户最感兴趣（分值最大）的 k 个 topic，然后选取这几个 topic 下热门的文章为用户推荐。

### 4.2.4 topic rank[10]

由于 LDA 输出的主题编号是随机的，主题也是未经排序的，因此可以开发一个为每个主题打分的算法，将显著的、有特色的主题打分高往前排，而将垃圾的、意义不明的主题往后排，这种排序算法也是很有意义的。

算法步骤：

该算法总体而言核心思路就是计算当前 topic 的概率分布与**垃圾分布**的距离。

**1.垃圾 topic 分布分为两种，准备 2 个距离：**

(1) topic→doc：由于 LDA 算法运行出的 theta 文件是 doc→topic 的概率分布，我们需要转换成 topic→doc 的概率分布，所以 k 号 topic 的第 m 篇文章的概率为 $\hat{\theta}_{\mathrm{k,m}} = \frac{nd[m][k] + \alpha}{\sum_{\mathrm{i=1}}^{D} nd[i][k] + D \cdot \alpha}$，(这里的 D 是训练文章数)Topic 在 doc 上的概率向量 $\vec{\theta}_{topic_k} = [\theta_{k,1}, \theta_{k,2}, \theta_{k,3}, \ldots, \theta_{k,D}]$（其中 D 是语料集的文章总数），衡量其与下面这个背景噪音向量 $\vec{\theta}_{background} \underbrace{=}_{\text{共D项}} [1/D, 1/D, 1/D, \ldots, 1/D]$ 的距离（使用 KL divergence、1-余弦、1-皮尔逊相关度等），距离越远分值越高，主题越有特色。

(2) topic→word：使用 phi 文件的 topic→word 的概率分布，计算当前 topic 的 phi 向量 $\vec{\phi}=[\phi_{k,w_0},\phi_{k,w_1},\phi_{k,w_2},...,\phi_{k,w_V}]$ 与向量 $\vec{\phi}_{uniform}=[1/V,1/V,1/V,...,1/V]$ 的距离。

**2. 加权排序**

将上述几个当前 topic 向量与垃圾向量的加权求和后排序。final_score=a * $score_{topic->doc}$ + b*$score_{topic->word}$ 即可。

为了验证这个算法，我专门设计了一个页面，并且为每个主题人工标注了一个名字，如下截图：

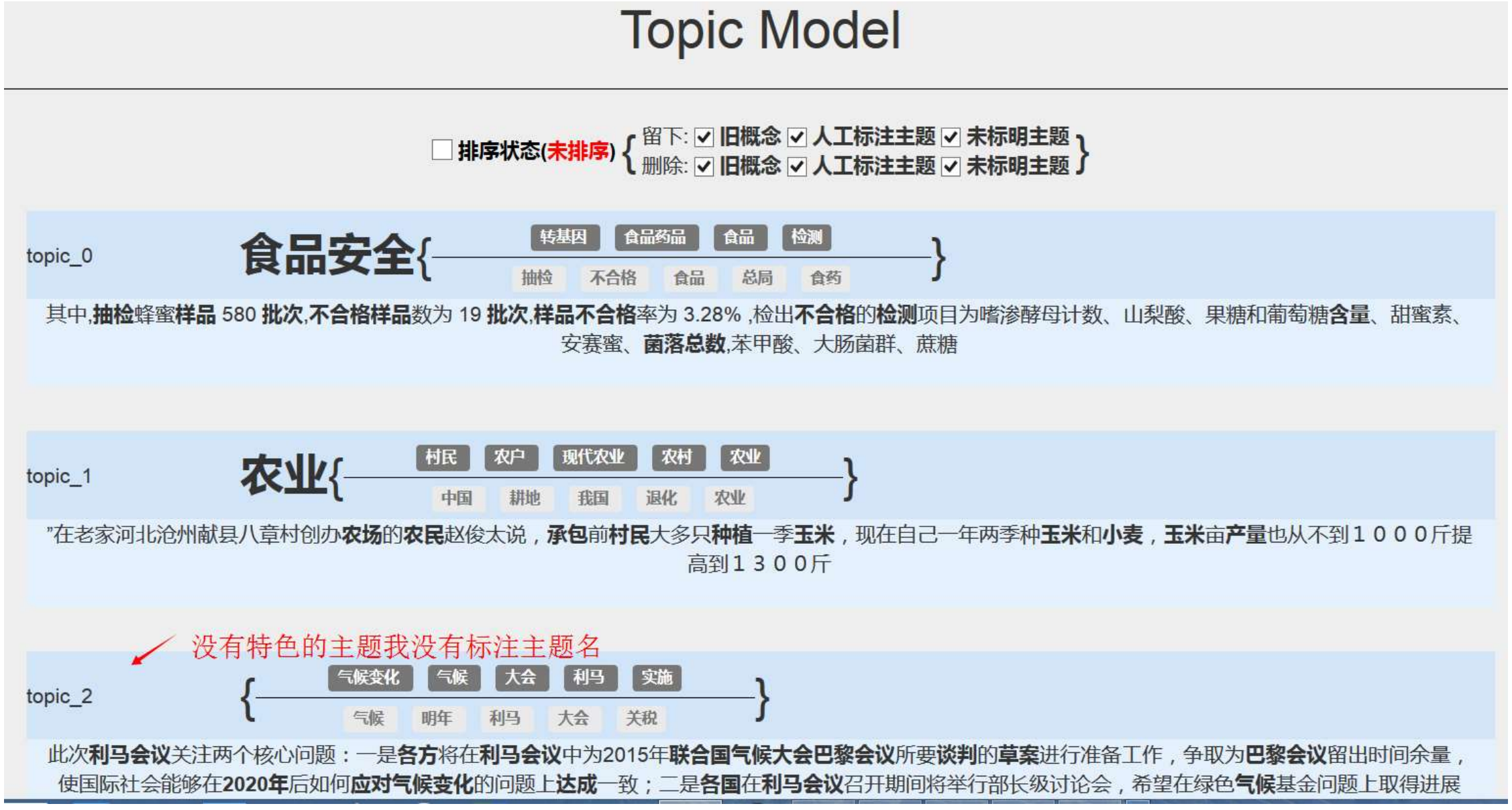

图 4-9 rank 排序前的主题

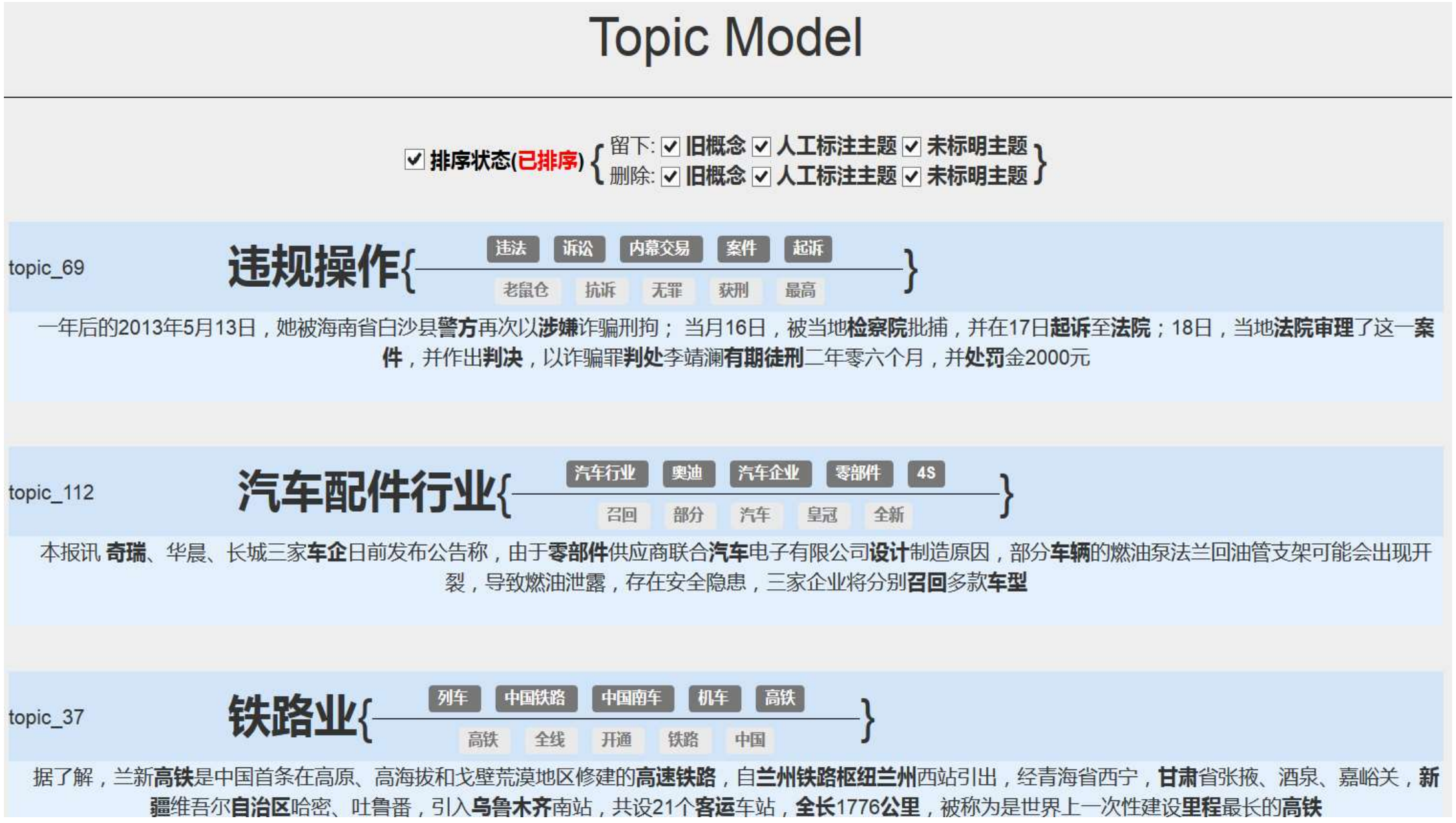

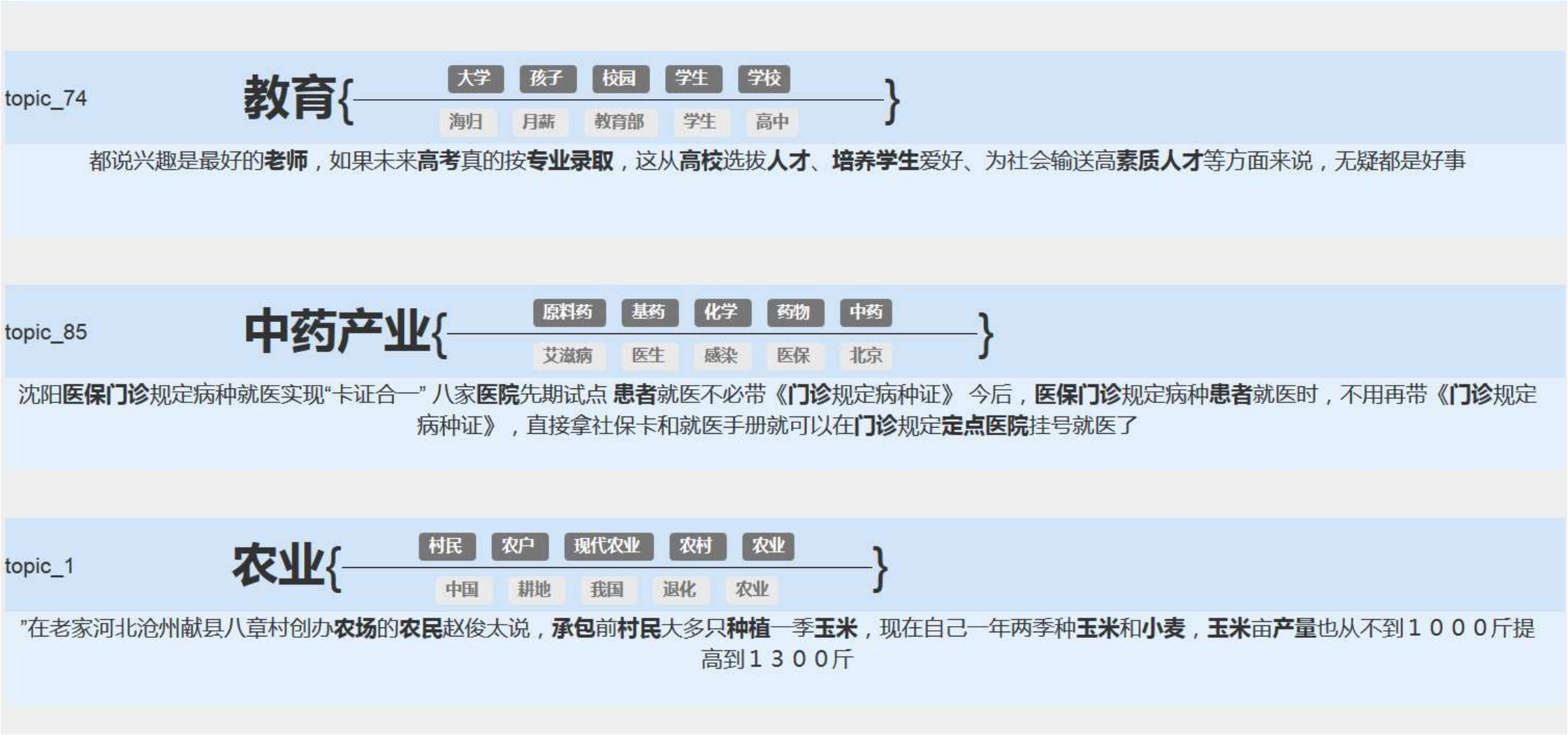

图 4-10 排序后的前几位高分 topic

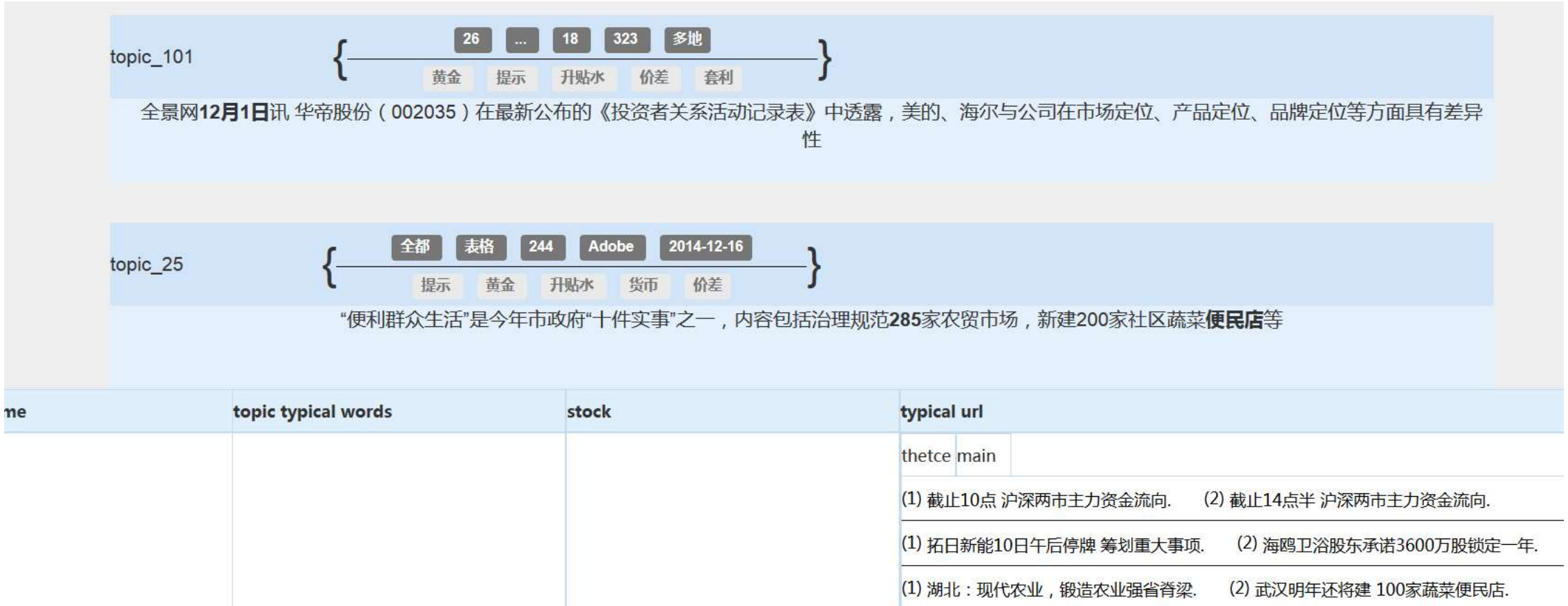

图 4-11 排序后低分的垃圾 topic

可以看到在排序后，出人意料地将人工标注的名称的特色 topic 打分较高（排序后的前几位均有名称标注），而排序后分值最低的垃圾 topic 则普遍没有标注名称，这说明人工挑选的 topic 与算法打分结果相一致。

### 4.2.5 word rank

仿照 topic rank 的思路，就可以做出 word rank，首先考虑下 word rank 有什么用，在哪些场景有使用，再给出基于 LDA 的 word rank 算法

word rank 的应用场景之一：**协助文本分类器**

比如在淘宝网搜索时，经常需要“分类搜索”的功能，如下截图：

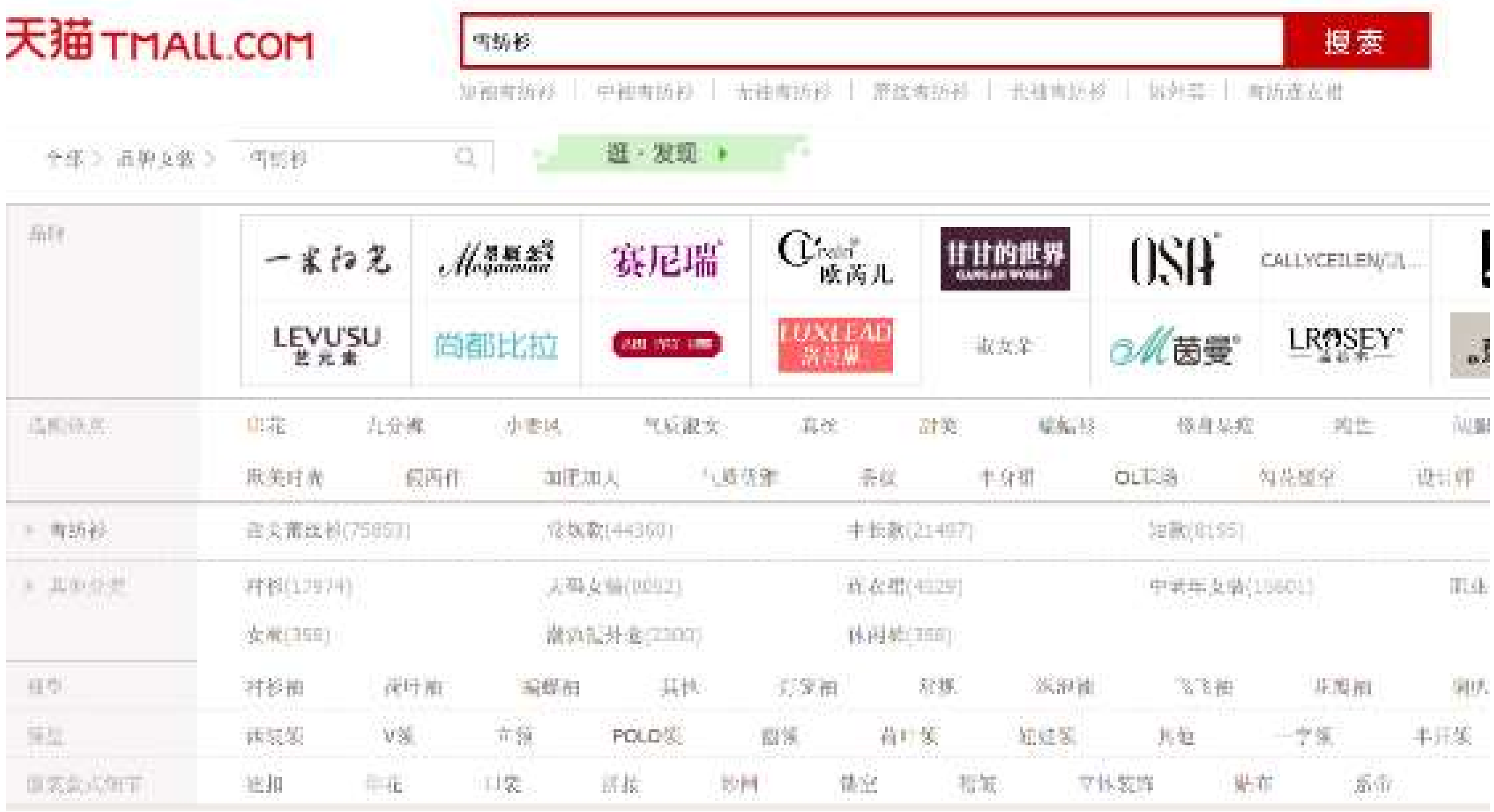

图 4-12 分类搜索

现在我们需要按照店家给出的商品标题描述分类，但是，如果你仔细观察店家给出的商品标题，会发现如下情况：

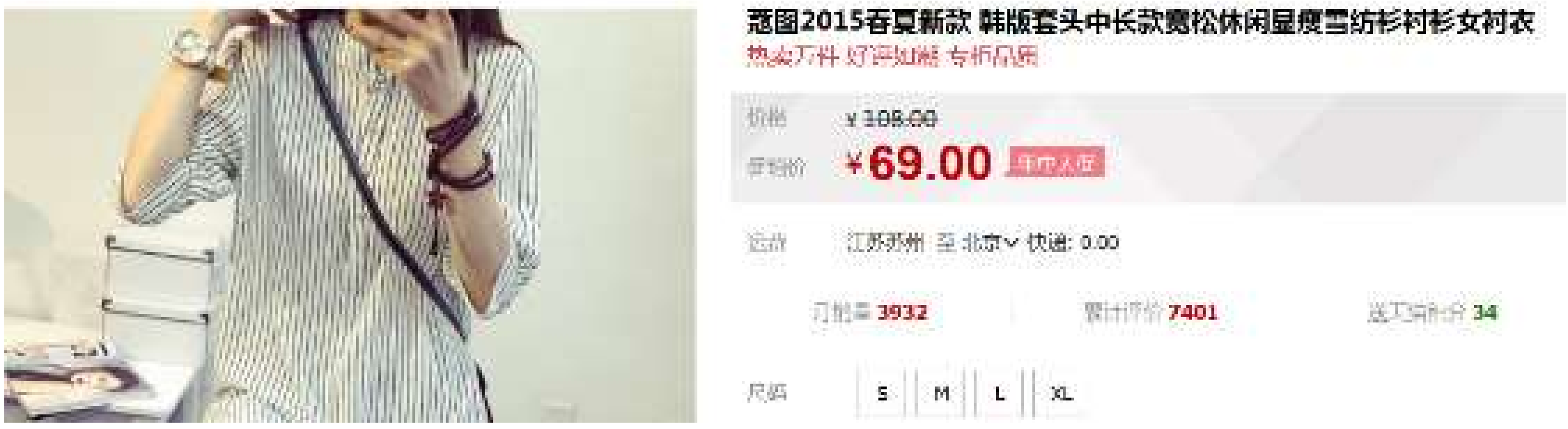

图 4-13 商品的标题描述会误导搜索分类

如果我按照传统的分词后 TF-IDF 提取 word 特征的方式分类，就会出问题，比如上面这个例子中，店家为了增加他们被搜索命中的机会，通常在标题上填写很多重复冗余无用的信息，比如图上的标题中“套头”这个词的意思是：没有扣子或者拉链的，必须从头上套着穿的。但是这个词是不能用作分类的依据的，搜索时我只想按照商品的主要特征词来分类，而非“套头”。

如果有一个方法可以获取到每个分类下的主要特征词，而忽略每个分类下冗余词，那么就可以正确的分类了。所以步骤是：1.根据印花、真丝等类别词匹配(match)找出相应的商品的文本；2.每个分类执行 word rank 算法，为每个词打分，过滤掉分值最低的垃圾词；3.word rank 的分值作为 SVM/LR 等分类器的特征依据。

分类器在另一个情况下也可以使用 word rank 作为辅助：仅有正样本。比如

考虑这样一种情形，希望对新闻分类，找出“奇闻异事”类型的新闻作为推荐，预计可提高一些点击量。所谓的“奇闻异事”类型的新闻是指类似“连体婴儿”、“怪人怪事”、“外星人 UFO”之类的报道，困难在于这种分类训练集仅可以找到一些“奇闻异事”的报道作为正样本（例如用爬虫去猎奇网站抓取），但不太容易获得“非奇闻异事”即负样本作为分类器的训练集。此时可以对这些“奇闻”的语料集进行 word rank 找出 rank 排名靠前具有代表特征的词，譬如“金字塔”、“连体人”等，不含这些词的文章可被视为分类器的负样本进行训练。

下面的算法步骤详述了 word rank 算法：

1. 计算 word 在每个主题上的概率分布，得到向量

读取 phi 文件，由于 phi 中的概率值是 topic→word 的，而我们需要的是 word→topic 的反向关系，因此使用 $\vec{w}_c=[w_{c,0},w_{c,1},w_{c,2},\ldots,w_{c,K-1}]$，其中每一项的

$$w_{c,k}=\frac{nw[c][k]+\beta}{\sum_{t\in K}nw[c][t]+K\cdot\beta}$$，（这里的大写 K 是主题总数）。

2. 计算垃圾/噪音 word 向量

$\vec{w}_{uniform}=[1/K,1/K,1/K,\ldots,1/K]$，每一项都是 1/K 的平均值。

3. 使用 KL 距离度量两者距离

可以采用 1-余弦相似度，或 KL 距离，或几种距离度量的平均值度量两者距离，最后均一化，距离越大的分值越高。

我以汽车新闻类型的文章为案例，用该算法输出样例：

| word rank 得分较高的 word | word rank 得分较低的 word |
|---|---|
| 瑞虎 0.919845109474 | 悍马 0.0163962476521 |
| 专车 0.919889049627 | 加倍 0.0181429299963 |
| 奔腾 0.920182327547 | 韦伯 0.0183433086821 |
| 吉利 0.920315761625 | 充充电 0.0185650744705 |
| 菲翔 0.920482668896 | 假定 0.0187398834494 |
| 凯迪拉克 0.920863099045 | 程惊雷 0.0188122305783 |
| 长城 0.920909053849 | 六百 0.019123897586 |
| 本田 0.920911274423 | 投运 0.0191464581692 |
| 福克斯 0.921148048364 | 一臂之力 0.0191952400808 |
| 宝骏 0.921434482516 | 欧阳明 0.0194425138855 |
| 荣威 0.921925139295 | 发卡 0.0194874355798 |
| 路虎 0.922897265021 | 教育机构 0.0195078269565 |
| 瑞风 0.922900276469 | 交卷 0.0195195159147 |
| 捷豹 0.923057646905 | 污染天气 0.0195711744352 |
| 景逸 0.923665401422 | 于海霞 0.0195859600484 |
| 海马 0.924589687164 | 源点 0.0195859600484 |
| 奔驰 0.924722328098 | 暂住证 0.0195920128674 |
| 沃尔沃 0.92505266906 | 一概而论 0.0196311823389 |
| 南通 0.92513072146 | 刘理萌 0.0196311823389 |
| 东风标致 0.925149629944 | 长信 0.0196311823389 |
| 哈弗 0.925476645704 | 语气 0.0196637204583 |
| 雷克萨斯 0.926134979376 | 娄格嘉 0.0197160360731 |
| 马自达 0.926283860082 | 购物网站 0.0197221613537 |
| 起亚 0.926606816176 | 私企 0.0197550919891 |
| 奥迪 0.926743050409 | 文德 0.0197550919891 |
| 昆山 0.927461877372 | 奸商 0.0197611161447 |

**表 4-2 word rank 汽车文章例子**

可以看到得分较高的 word 大部分均为汽车品牌，大部分的得分就较为准确，经过测试，平均可以达到 80%的准确度，但得分低的里面“悍马”的分就太低了，这就表明了此算法仍有优化空间（留给读者，可以结合 topic rank 的 score 进行优化）。

### 4.2.6 文章质量评分算法

这是迄今为止，由我发现的基于原始的 LDA 算法训练结果就能产出的最强大的技术之一。说的通俗点，这个文章质量评分算法就好比“计算机程序自动为高考作文判分”。其实将 topic rank 思路扩展开，便可以延伸出此技术。

比如博文推荐方面，由于博文质量由用户自主撰写而成，良莠不齐，因此需要一个博文质量自动评分系统。在经过 LDA 算法对文本的训练后，可以根据 topic 主题在当前文章的分布情况以及文章中各个词 word 在主题 topic 上的概率分布信息开发出一个新的基于 LDA 的文本质量评估算法。这个算法对比以往的基于 SVM 分类器等方法有**几个关键优势**：1.该算法是无监督的，不需为每篇文章标注垃圾分类数据，2.该算法可以为每篇文章产生一个分值，而这是二分类的分类器做不到的。

该算法的步骤是：

1.先找出该文档的单词在特定主题下生成的概率分布即如下公式：

$$\vec{p}_d = [p(z_{w_1} \mid d)p(w_1 \mid z_{w_1}), p(z_{w_2} \mid d)p(w_2 \mid z_{w_2}), ..., p(z_{w_N} \mid d)p(w_N \mid z_{w_N})] \tag{4-3}$$

在这个公式中等式左边的向量是指该文档的生成概率向量，其中每一项指的是这个文档的一个词在该主题上产生的概率 * 文档中该主题的概率。这个式子里的 $p(w_1 \mid z_{w_1})$ 即 $\varphi_{z,w_1}$，而 $p(z_{w_1} \mid d)$ 即 $\theta_{\mathrm{d},z_{w_i}}$，其中 $z_{w_i}$ 来自于 Gibbs Sampling 采样时为每个单词分配的主题编号，注意到 $w_1$~$w_N$ 是指从词典中编号 1 直到 N 的所有词，如果该文档 d 中不含单词编号 i 的词，则向量中编号 i 位置的概率为 0。

2. 生成这个向量后，与“众文档平均概率向量[25]”（Mass Article probability vector）的概率进行距离度量，其中，“众文档平均概率向量”这个向量由如下公式得出：

$$\vec{p}_{mass} = [\sum_{k\in K} p(k)p(w_1 \mid z_k), \sum_{k\in K} p(k)p(w_2 \mid z_k), ..., \sum_{k\in K} p(k)p(w_N \mid z_k)] \tag{4-4}$$

其中 p(k)表示 k 编号主题在整个语料库（corpus）中的概率，可以由如下公式得到：

$$p(\mathrm{k}) = \frac{\sum_{d\in Corpus} \theta_{d,k}}{\sum_{d\in Corpus} (\sum_{z\in K} \theta_{d,z})} \tag{4-5}$$

其中，$\theta_{\mathrm{d,k}}$ 表示每个文档中主题 k 的概率值。分子是指语料库(corpus)的所有文字的 k 编号主题，而分母的部分相当于是语料库所有文章的所有 $\theta$ 之和。

3. 使用 KL 距离(4-6)、1-cosine 相似度(4-7)、1-pearson correlation coefficient(4-8)三种相似度/距离度量标准公式（criterion）衡量上面这两个向量之间的距离。

$$D_{KL}(P \parallel Q) = \sum_i \ln(\frac{P(i)}{Q(i)})\mathrm{P(i)} \tag{4-6}$$

$$D_{\cos ine}(\vec{P},\vec{Q}) = 1 - \frac{\vec{P} \bullet \vec{Q}}{\parallel \vec{P} \parallel \bullet \parallel \vec{\mathrm{Q}} \parallel} = 1 - \frac{\sum_{i=1}^{n} P_i \bullet Q_i}{\sqrt{\sum_{i=1}^{n}(A_i)^2} \bullet \sqrt{\sum_{i=1}^{n}(B_i)^2}} \tag{4-7}$$

$$D_{Pearson's\ correlation\ coefficient}(\vec{P},\vec{Q}) = 1 - \frac{\mathrm{cov}(\vec{P},\vec{Q})}{\sigma_{\vec{P}} \bullet \sigma_{\vec{Q}}} \tag{4-8}$$

距离算出来后，再使用公式

[25] 有时我也称之为上帝向量或完美向量

$$D_{rescale}(\text{criterion}) = \frac{D_{\vec{P},\vec{Q}}(\text{criterion}) - Min_D(\text{criterion})}{Max_\text{D}(\text{criterion}) - Min_D(\text{criterion})} \quad (4\text{-}9)$$

来将三种不同度量标准的距离重新归一化到 0~1 之间。越接近“众文档平均概率向量”的文章质量就越高，反之质量就越垃圾。最后使用

$$score = \frac{\sum_{c \in criterion} \text{D}_{rescale}(\text{c})}{3} \quad (4\text{-}10)$$

返回这三种距离的平均值作为对文档的评分分值（区间 0~1 之间）。

先以旅游博客作为打分测试集，整个质量打分算法运行结果如下文件所示：

| 旅游博文评分较高的 10 篇文章 | | | |
|---|---|---|---|
| **网址 url** | **评分** | **阅读量** | **文章标题** |
| http://blog.sina.com.cn/s/blog_53bd0d1701015f4l.html | 0.998066008917 | 61 | 山西徒步路线 |
| http://blog.sina.com.cn/s/blog_69cb346901017r65.html | 0.991291044954 | 697 | 关于海洋航行者号“神户”“长崎”“济州岛”上岸观光的攻略之毛毛雨 |
| http://blog.sina.com.cn/s/blog_65376d0501013ikj.html | 0.981205704494 | 17 | 最全全国旅游景点逃票大攻略 |
| http://blog.sina.com.cn/s/blog_459154f301017xz3.html | 0.979939083754 | 189 | 【心无羁，行不止】独家奉送新鲜出炉西藏 10DAYS 跟团+拉萨自由行攻略-近 3 万字吐血整理你读完绝不会失望 |
| http://blog.sina.com.cn/s/blog_6d4f6b220101b8cx.html | 0.979059053598 | 216 | 东欧德奥匈捷斯五国十日游记[无图] |
| http://blog.sina.com.cn/s/blog_45e5ae5a01014mry.html | 0.978861793044 | 110 | 《康藏朝圣之旅》 第三篇 魂牵拉姆纳错  （转载） |
| http://blog.sina.com.cn/s/blog_3d4ceed401017a48.html | 0.976334174062 | 278 | 2012 莫斯科、圣彼得堡七日游之流水账 |
| http://blog.sina.com.cn/s/blog_4a0ab71d01014per.html | 0.975317724569 | 526 | 美国两周游游记（纽约、新泽西、堪萨斯、拉斯维加斯、洛杉矶） |
| http://blog.sina.com.cn/s/blog_51688d58010168sk.html | 0.974945350427 | 69 | 2012 年十月金秋，走入绚彩额济纳,绝美胡杨林 |
| http://blog.sina.com.cn/s/blog_3e1b470f010179ve.html | 0.972611234889 | 191 | 泰国-大马一月游 |

**表 4-3 评分较高的 10 篇文章**

| 旅游博文评分较低的 10 篇文章 | | | |
|---|---|---|---|
| **网址 url** | **评分** | **阅读量** | **文章标题** |
| http://blog.sina.com.cn/s/blog_a2621cf20100yri6.html | 0.0825872 | 0 | 米径 3 公分 30 元/株 |
| http://blog.sina.com.cn/s/blog_a259d83e01014lkt.html | 0.0818101 | 2 | 米径 2-3-4-5-6-8-10-12 公分 |
| http://blog.sina.com.cn/s/blog_a2d9fcda01012twf.html | 0.0817654 | 0 | 16-60-190-260-360-760-1200 元/株 |
| http://blog.sina.com.cn/s/blog_7308354b01018uzp.html | 0.08163 | 437 | 神道八百万神 |
| http://blog.sina.com.cn/s/blog_9e05cefd01013gch.html | 0.0816215 | 5 | 2-3 分枝地径 2 公分高 180 公分 4 元/株 |
| http://blog.sina.com.cn/s/blog_a50c721201015smj.html | 0.0683244 | 3 | [原创]www.pc-ge.com 湖南阳光板批发 阳光板价格阳光板价格 PC 阳光板 |
| http://blog.sina.com.cn/s/blog_71663f9201016cr7.html | 0.06791 | 474 | (23) 四 川 律 师 通 信 录  Szechwan Lawyers Contacts |
| http://blog.sina.com.cn/s/blog_4d56fe5b01012ypz.html | 0.0612046 | 59 | 解放前测绘资料目录-甘肃 |
| http://blog.sina.com.cn/s/blog_a54beb4601015n17.html | 0.0595885 | 4 | [转载]www.pc-ge.com 湖南阳光板批发 2 |
| http://blog.sina.com.cn/s/blog_a574e2c801013dgd.html | 0.0220331 | 20 | 一色的青瓦，飞檐硐兽，古香古色，看起来如同泼墨山水 |

**表 4-4 评分较低的 10 篇文章**

可以看到打分最低的“垃圾文章”有以下若干种类型：

1. 批发货物的牛皮癣广告，如《[原创]www.pc-ge.com 湖南阳光板批发 阳光板价格阳光板价格 PC 阳光板》；

2. 各种名词列举的博文，如 http://blog.sina.com.cn/s/blog_4a708d1301014gvx.html，《互为近义词的成语大全》；还有《河南花木报价表，各种花木价格表》；

3. 简单列举的景点，如 http://blog.sina.com.cn/s/blog_577b695a01017bbd.html，《中国国家级自然景区名录》；

4. 航班和汽车等时刻表，如 http://blog.sina.com.cn/s/blog_5776fab60100zeue.html，《成都双流国际机场完整航班时刻表》等。

**实时接口（online learning）：**

由于 LDA 训练需要时间，如果我们需要一个实时接口，实时为新文章打分，可以采用以下策略：先将 LDA 的 model 训练完成后，将 nw（**稀疏矩阵）、平均概率向量、wordmap、距离最大值和最小值（用于均一化）存于类似 redis 等的数据库 cache 下来，此为 train 模型的记忆**。在线上文章分词后实时传送给接口后，连接读取 redis 中的模型记忆，使用前面介绍的 predict 方法为新文章预测主题分布后，再按照上面批量打分的方法衡量距离打分即可。

另外可以进一步挖掘一下文章的“阅读点击量”和“质量评分”之间是否存在关系，但是博客文章的点击阅读量(PV)经常与作者知名度、作者粉丝多少有关，因此可以按照区间平均点击阅读量(PV)来看两者的关系：

| 分值区间 | 平均 pv | 文章数 |
|---|---|---|
| 0.9~1.0 | 449.841 | 390 |
| 0.8~0.9 | 588.2538 | 2975 |
| 0.7~0.8 | 399.3124 | 3073 |
| 0.6~0.7 | 341.1599 | 2277 |
| 0.5~0.6 | 167.3782 | 3038 |
| 0.4~0.5 | 149.2586 | 1655 |
| 0.3~0.4 | 137.2095 | 797 |
| 0.2~0.3 | 71.75649 | 501 |
| 0.1~0.2 | 76.86979 | 192 |
| 0~0.1 | 58 | 20 |

表 4-5 区间平均 PV 与文章质量打分的关系

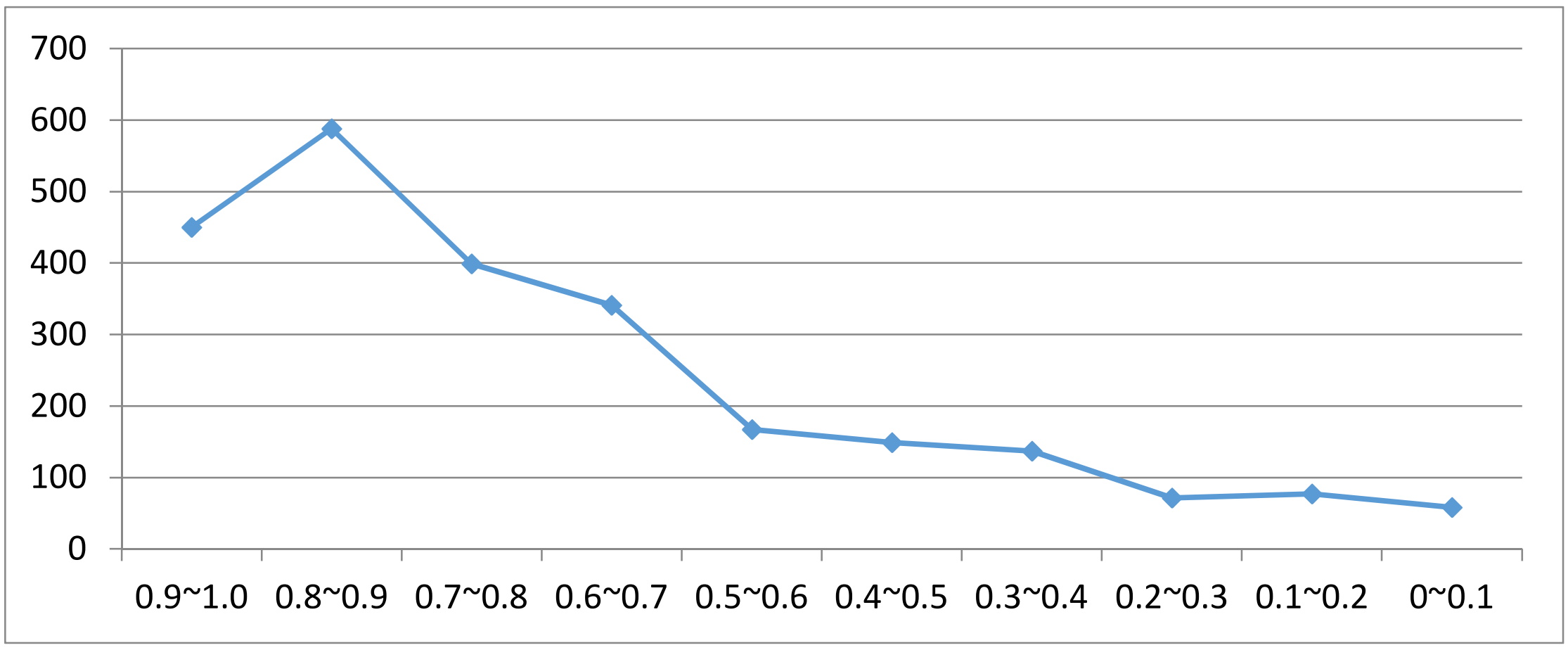

图 4-14 区间平均 PV 与文章质量打分的关系

可以从上图看到，按照每 0.1 分为一个区间进行统计文章的平均 pv，会发现存在一定关系的，即：文章打分高的点击阅读量相对来说更高，文章打分最低的点击阅读量相对来说更低。基于这一点，可以有理由将文章质量打分算法融入推荐系统，过滤掉垃圾文章，提高点击率。

当然该方法也可用于“没有标注训练集的文本垃圾过滤”等场景中。

### 4.2.7 总结

除了上面提到的应用外，LDA 还有许许多多其他的应用，限于篇幅有限，就不一一展示了，算法优化无止境，上面提到的应用都有很多优化余地，这些就留给读者完成。

## 参考文献

# 第5章 并行化

通过上一章的实现可以看到，LDA 算法运行的时间复杂度还是比较高的，跟主题数 K 和训练语料数据量都有关，如果主题数过多，更是会造成训练时间的大幅上涨。因此如果训练文本语料数据或主题数过多时，有必要开发并行计算版本的 LDA。

使用 Gibbs Sampling 版本的 LDA 开发并行化，存在着一个**主要难题**，就是 nw、nwsum、nd、ndsum 这几个统计量的并行更新问题，更具体地说，如果简单地将文章集(doc set)拆分成若干份并行计算，A 进程和 B 进程同时启动时，由于每次重新指定主题(reassign topic)后，都会修改掉统计量（-1 → 重新分配 topic → +1），因此就会造成修改读写冲突，破坏统计量的一致性。**3 个依赖**：(1)从列上看：不同文章并行化时，某文章的单词 a“依赖于”另一篇文章相同单词 a 采样后修改的 nw 和 nwsum；(2)从行上看：同一篇文章的后一个单词“依赖于”前一个单词采样后修改的 nd 和 ndsum。(3)从主题层面看，同一个主题的后一次采样“依赖于”同一个主题前一次采样的 nwsum。

## 5.1 AD-LDA

为了解决这个问题，David Newman 等人提出了 AD-LDA 算法[12]，这种方法将文章分散到不同的机器上（**按行拆分**），该算法由于可以被转换成 map-reduce 操作，因此早期的 LDA 并行化实现都采用 AD-LDA，该算法的核心是提出了 global update 操作（解决依赖 1），整个步骤如下所示：

算法 5.1 AD-LDA 算法

算法步骤：

1. 将训练集中的 D 篇文章分散到 P 个机器上。并按照以下规则将统计量分散到各个机器上：

| |
|---|
| nd、ndsum：这两个统计量仅与文章有关，拆分到各个机器上；<br>nw、nwsum：这两个统计量被完整 copy 到各个机器上，而不是拆分。<br>我们用 nw_p、nwsum_p 代表各个机器上的 copy 版本，注意到这个数组与全局占用内存一致。 |

2. 在各个机器 p 上，使用 nw_p,nwsum_p,nd_p,ndsum_p 执行原始单机 LDA 算法（各个机器互不知晓对方存在），注意几个统计量都被替换为了 copy 版本，即(nw_p=nw，nwsum_p=nwsum)，各机器中的统计量（nw_p 等）独立修改，其余过程与单机 LDA 算法（算法 4.2）完全一致。
3. merge back:在各个机器分布式 Gibbs Sampling 运行结束后，z_p 数组已经被指定 topic 完成。执行 global update 操作：

$$nw[w][k] \leftarrow nw[w][k] + \sum_{p\in P}(nw_p[w][k] - nw[w][k]) \tag{5-1}$$

nw 是所有进程在 Gibbs Sampling 之前的统计量（全局 nw），在各个机器上迭代结束后，nw_p 被合并回了全局统计量 nw，紧接着，全局 nwsum(topic’s word count)也可以被计算出来[26]：

$$nwsum[k] = \sum_{w\in V} nw[w][k]$$

4. 一轮迭代结束，转第 2 步

注意到这个算法被看作单机版本的 Gibbs Sampling 的近似(approximate)，因为在各个机器上不同进程启动 Gibbs Sampling 时，nw_p 和 nwsum_p 互不知晓其他机器的存在而进行采样，就会造成前面提到的修改冲突，破坏统计量一致性，所以说这一步其实已经产生了误差，不能完全等同于单机版的 Gibbs sampling，但其后的 global update 在每一轮迭代后都将这个问题尽可能地修复了。

**问题：**内存问题——AD-LDA 算法仍然未能解决耗费内存最大空间的统计量 nw 的问题，因此如果词典 vocabulary 和 topic 数量过多时，仍然在每个机器上

[26] 注意到公式(5-1)正确地反映了全局 topic assignment z，因为更新后的 nw 如实地反映了各个机器上 topic assignment z，由于各机器 p 启动 Gibbs Sampling 时都是从全局 nw 开始修改:

都会耗费较大内存，而由于单台机器的内存有限，尤其在 Hadoop 等对每个 job 内存限制时更不可接受。

## 5.2 spark-LDA

由于存在上述问题，邱卓林研究了一个基于 spark 的 LDA 分布式方法[27]，该方法虽然题目是基于 spark 的，但其实是个 LDA 的分布式 Gibbs Sampling 的一般性的原理，可以被实现在任意并行框架中（例如 MPI），该方法克服了 AD-LDA 占用内存大的问题，并且误差比 AD-LDA 更小。算法用到的变量如表 5-1。

| |
|---|
| X：数据集 |
| $X_P$ ：切分后的第 p 份子数据集 |
| nd：全局 document→topic count 统计量 |
| nd_i：nd 的第 i 部分 |
| nw_j：nw 的第 j 部分 |
| nwsum：topic k 被指定的单词总数 |
| nwsum_p：在第 p 份子块中 sampling 后重新统计的 topic k 的单词总数 |

**表 5-1 spark-LDA 变量含义**

算法步骤如下：

### 5.2.1 切分块

Spark-LDA 类似于 AD-LDA，也是将原始训练数据进行了切分后分散到各个机器上去执行，但是进一步考虑了列切分，由于 LDA 的采样中可以**不考虑**同一篇文章中单词出现的顺序，所以只要没有统计量更新依赖，就可以把文章像切蛋糕一样切分开并行执行。故这个算法的**关键思想**在于训练语料文本矩阵中**同一行(同一篇文章)**或**同一列(同一个单词)**的 word 都会产生**统计量更新依赖**，所以**同一行或同一列**的数据不能被同时 sampling 采样。（这里的行列是指每篇文章排序后的 wordid）如表 5-2 所示：

[27] 详见《Collapsed Gibbs Sampling for Latent Dirichlet Allocation on Spark》，该方法思想类似腾讯 peacock

| doc 编号 | 文章原始内容 | 转换成 wordid 后 | 对 wordid 排序 |
|---|---|---|---|
| doc m | He published several books, including Chromic Salts Technology, Rejuvenating Chemical Industry through Science and Technology, Soft Science and Reform, Large Linear Target Programming and Application, Research in China's Economic Development and Reform and Economic Reform and Development in China. | 306 310 482 326 497 403 400 323 375 394 392 371 415 343 417 316 329 440 417 484 436 305 444 392 463 312 315 425 | 305 306 310 312 315 316 323 326 329 343 371 375 392 392 394 400 403 415 417 417 425 436 440 444 463 482 484 497 |
| doc n | In recent years, he devoted himself to the use of complexity science to study issues relating to the development and reform of China, made enormous efforts to explore and explain the characteristics and law of development of the fictitious economy and actively studied and promoted the development of venture capital in China. | 444 312 384 306 458 372 471 424 466 359 406 415 490 476 496 312 343 315 379 393 454 454 447 361 397 325 336 482 374 495 338 492 393 425 | 306 312 312 315 325 336 338 343 359 361 372 374 379 384 393 393 397 406 415 424 425 444 447 454 454 458 466 471 476 482 490 492 495 496 |

表 5-2 先将原始训练文章转换为排序后的 **wordid**

| Doc_m | 305 | 306 | 310 | 312 | 315 | 316 | 323 | 326 | 329 | 343 | 371 | 375 | 391 | 392 | 394 | 400 | 403 | 415 | 417 |
|---|---|---|---|---|---|---|---|---|---|---|---|---|---|---|---|---|---|---|---|
| Doc_p | | 306 | 310 | 312 | | | 323 | | 329 | 343 | | | | 392 | | | 403 | | ... |
| Doc_q | | 306 | 310 | | | | 323 | | | | | 375 | 391 | | | 400 | 403 | 415 | ... |

表 **5-3** 每篇文章排序后 **wordid**，原训练语料→矩阵

上表是将原始语料转换为排序后(sorted)的矩阵后（并且对齐后），同一行和同一列的 wordid 在 sampling 时会产生统计量依赖。空的格子表示该文章没有这个词而被对齐留空。以上表中 323 为例，**画十字线**的上的单词均不可被同时并行执行 sampling。

经过分析，整个算法的步骤：可以先将原始数据 X 划分为 P*P 的份，这里的 P 是我们设置并发数，如图 5-1 所示的 P=3，每一份数据都相应地标注了坐标，注意到由于十字线上(同一行列)不可以被同时执行，因此图 5-1 的执行分为几个阶段执行，对角线上的(0,0)、(1,1)、(2,2)这 3 块数据先被同时执行，而每一块内部执行单机版的 Gibbs Sampling，这三块并行执行结束后，再执行另外不在十字线上的 3 块(0,1)、(1,2)、(2,0)，这三块执行结束后最后执行(0,2)、(1,0)、(2,1)三块（如图 5-1 的 shuffle 步骤之后的三行小组，三行行内并行计算，行间串行计算)：

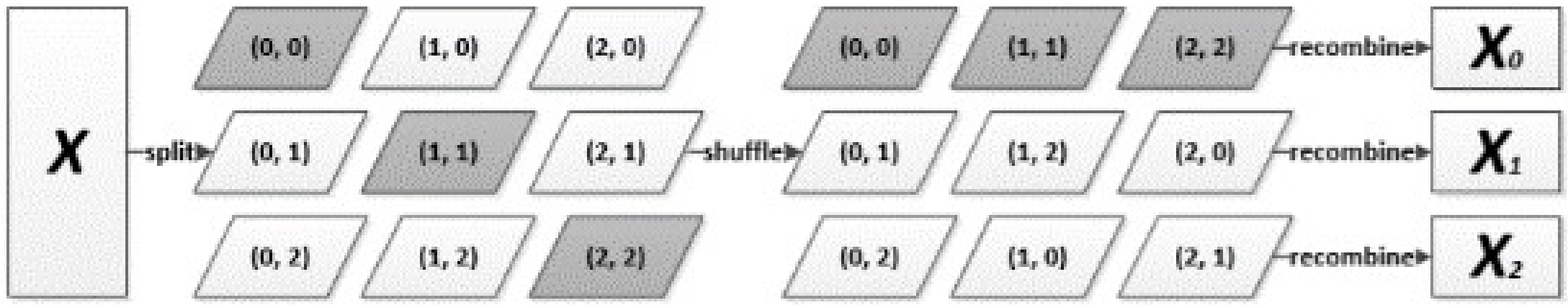

**图 5-1 spark-LDA 切分流程，并发度 3**

这里的“一块”的概念是包括多行和多列的，我们希望每块的数量都均等，这样在每台机器上才能负载均衡。但在切分时候由于“对齐”会造成空的 word 格子的原因，所以，简单地按照列数均分和行数均分来切分是不行的。为了尽可能每块均等，可以分割多次，然后挑选出各个块之间**差距的最大值最小**的那一次。

**切分策略：**(1)首先将数据集 X 按照列切分成均等的 P 份，如果包含的单词数不均等，可以随机交换列，再切分来保证均等。（2）在将切好的 P 份列，切成 P 份行，如果不相等，多次**随机交换行**，最后挑选出各个块之间**差距的最大值最小**的那一次。

### 5.2.2 选择

切分成 P ＊ P 块这个步骤后，就是挑选出哪些无冲突的 P 个块可以同时被并发执行（组内并行，组间串行）。通用的策略有 2 种：

(1)八皇后法

八皇后问题是一个经典的问题，在 8X8 格的国际象棋上摆放八个皇后，使其不能互相攻击，即任意两个皇后都不能处于同一行、同一列或同一斜线上，问有多少种摆法。所以这里的挑选无冲突块类似于八皇后摆棋子，我们这里的情况是弱化的八皇后问题，即挑选的块可以出现在同一个斜线上。八皇后的经典解法是回溯法[28]。

[28] http://fdcwqmst.blog.163.com/blog/static/16406145520109294346965/

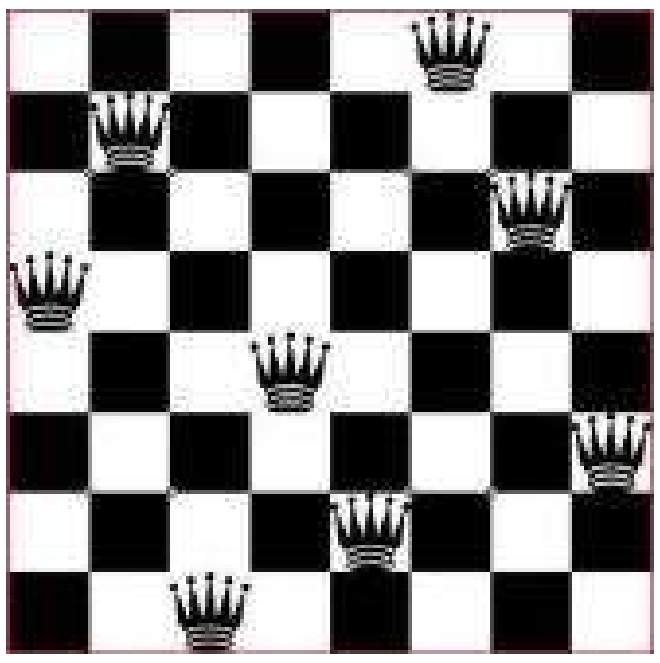

图 5-2 八皇后问题

(2)对角线法

腾讯的 peacock 系统[17]采用的是更简单的对角线法，由于同一行或同一列的块不能被同时选择，因此选择对角线肯定是没有问题的。沿着一条条对角线进行采样计算，一旦一条对角线计算完成后，就并行计算下一条对角线，同一对角线内并行计算，不同对角线间串行计算。

### 5.2.3 计算和合并

由于上述算法过程中的方法是：组内并行，组间串行，所以小组并行执行的一次迭代完毕后，小组的 nd、nw 等统计量需要同步(sync)到下一个小组内。而组内各个块内 sampling 计算过程与**单机版 gibbs sampling 完全一致**，只是将统计量切换：nd、ndsum 和 nw 切分成 P 份，比如第 j 份是 nd_j 和 nw_j，由于在公式(3-20)中，nwsum(topic k 被指定的单词的总数)无论如何都会产生冲突(依赖 3)，所以整个算法的误差也就在这个统计量了。因此在第 p 台机器上的 nwsum 被修改为 nwsum_p 后（并行计算完结后）都需要用 global update 方法 merge back。

$$nwsum[k] \leftarrow nwsum[k] + \sum_{p \in P} (nwsum_p[k] - nwsum[k]) \tag{5-2}$$

除了 nwsum 需要 merge back 之外，其余几个统计量，nd_j 和 nw_j 都需要合并到下个小组的文档 d 和单词 w。(这种合并过程可以用 spark 的 broadcast）整个过程见图 5-3。

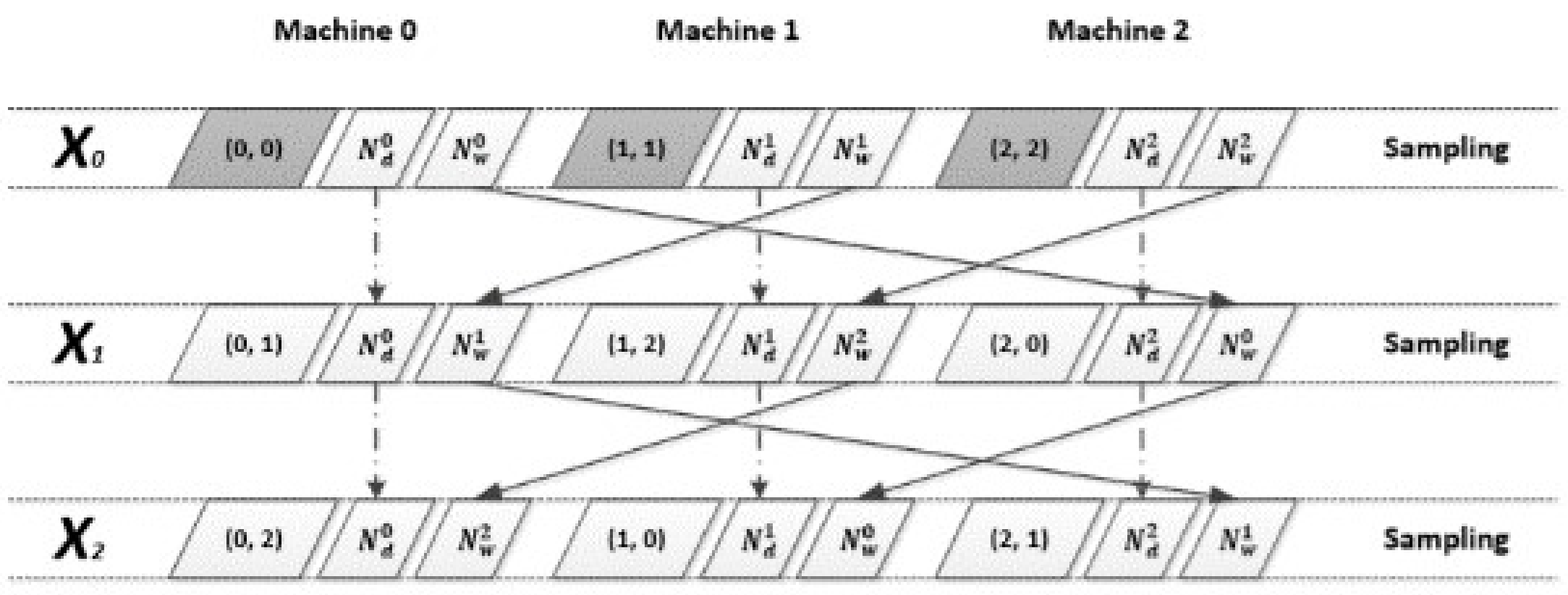

**图 5-3 计算后合并，同一个单词的 nw 和同一个文档的 nd 统计量被合并**

### 5.2.4 总结

**实验证明：**这种算法产生的混淆度(perlexity)与单机版 Gibbs Sampling 一致，因此误差非常小，而且分布式计算中的耗费内存也可以接受。但是实现复杂度稍高，需要比较好的 spark 编程功底。

(1)整个训练文章集(doc set)不仅以行切分(行指文章)，也以列切分。

(2)各个小块被切分尽可能数据量均等。

(3)统计量 nw 和 nd 因此也被切分到各个机器进程上。

(4)执行流程为：每轮迭代组内并行，组间串行。

(5)同一行或同一列的数据不能在一个小组内的不同块上同时并行执行。

(6)各个独立进程运行完毕后，误差仅在 nwsum，通过 global update 合并 nwsum。

## 参考文献

# 第6章 变分贝叶斯的启蒙

在我们的故事叙述到这里的时候，读者已经发现，Gibbs Sampling 技术虽然实现简单且输出效果不错，但是其采样的效率犹如“大力神海格力斯式的笨重劳动”（欧拉语）。因此有必要看看从另一种思路出发而产生的技术，变分贝叶斯法（Variational Bayes）推导是 Blei 的 LDA 论文原作中的经典方法，如果说学习 LDA 模型，但没有学过原作中的 VB 法推导，仍然称不上学会了 LDA。这种方法运行速度比 Gibbs Sampling 方法快，而且这个方法可以不经过采样 z 直接推断出隐变量和参数。虽然有读者会争论说这种方法得到的参数陷入了局部最优，不是很精确，但是在后期我们可以看到这个变分思想作为新技术的启示，结合其他技术（包括 Gibbs Sampling 等）可以产生巨大威力。这一章主要讲述变分贝叶斯(Variational Bayes)技术的启蒙知识。

## 6.1 前置知识

在正式讨论这种技术之前，在第 2 章的一些前置知识基础上，有必要需要补充一些简单的基础知识。记住：真理的碎片有可能散落在不起眼的地方或特别的地方，所以我们需要将它们捡拾收集起来，以达到我们的技术目标。

### 6.1.1 指数分布族(exponential family)

指数，也就是数学上的次方之意，比如平方的指数=2，立方的指数=3，随着越来越多的分布被发现，人们逐渐发现有一些特定的分布可以被总结为“指数分布族”，也就是利用特定的方法，可以将这些分布转换为某一种相同的**形式，这些形式的期望等性质都比较好求解，具有统一的形式**。下面直接给出指数分布族的密度函数：

$$p(x \mid \eta) = h(x)g(\eta)\exp\{\eta^T u(x)\} \tag{6-1}$$

你可以特别注意到这里的指数指的是**自然常数 e 的指数**，公式中 exp{x}也就是 e 的 x 次方，这个公式的$\eta$、g($\eta$)、u(x)都不是随便定义，均有意义，解释如下：

x ：有可能是标量（譬如一个实数）或是一个向量

$\eta$ ：自然参数(natural parameter)，更详细的解释在下文例子之后

u(x)：充分统计量(sufficient statistics)(后文会进一步解释这个概念)

g($\eta$) : 归一化参数，由于 p(x| $\eta$)的概率之和=1，这个 g($\eta$)是用于将其归一化的（还有个名称叫 partition function）。

即： $g(\eta)\int h(x)\exp\{\eta^T u(x)\}dx = 1$ ，所以 $g(\eta) = \dfrac{1}{\int h(x)\exp\{\eta^T u(x)\}dx}$

通常公式(6-1)也可以写为：

$$p(x \mid \eta) = h(x)\exp\{\eta^T u(x) - a(\eta)\} \tag{6-2}$$

注意到这个公式与前面式(6-1)的区别仅仅在于形式上从 g($\eta$)被转化为了 a($\eta$)，这个 a($\eta$)被称为-log(归一化参数)（log-normalizer or log-partition function），见公式(6-3)。

$$a(\eta) = \log(\frac{1}{g(\eta)}) = \log(\int h(x)\exp\{\eta^T u(x)\}dx) \tag{6-3}$$

将一些普通分布转化为指数分布族的形式方法也非常简单：

Step 1：将 p(x|$\mu$)凑成指数形式 exp{ln(p (x|$\mu$))}

Step 2：整理为公式(6-1)或公式(6-2)的形式。

通过下面几个例子可以加深理解这一点：

**例 6.1 Bernoulli 分布的指数分布族形式**

伯努利分布与前文所述的二项分布相似，只是没有了二项式系数。

$$p(x \mid \mu) = Bern(x \mid \mu) = \mu^x(1-\mu)^x$$

先对这个式子进行指数变换，则原式变为

$p(x \mid \mu) = \exp\{x \cdot \ln\mu + (1-x)\ln(1-\mu)\}$

第二步，对这个式子整理:

$p(x \mid \mu) = \exp\{x \cdot \ln\mu + (1-x)\ln(1-\mu)\}$

$= \exp\{x \cdot \ln\dfrac{\mu}{1-\mu} + \ln(1-\mu)\}$

$= (1-\mu)\cdot\exp\{x \cdot \underbrace{\ln\dfrac{\mu}{1-\mu}}_{\eta}\}$ //利用 $e^{a+b} = e^b e^a$

现在已经凑出了$\eta$，剩下的目标是凑出 g($\eta$)的形式，所以目标转变为将(1-$\mu$)转换为$\eta$的函数形式 g($\eta$)

$\eta = \ln\dfrac{\mu}{1-\mu} \underbrace{\Rightarrow}_{\text{写出反函数}\sigma(\eta)} \mathrm{e}^{\eta} = \dfrac{\mu}{1-\mu} \Rightarrow \sigma(\eta) = \mu = \dfrac{1}{1+\mathrm{e}^{-\eta}}$

所以$1-\mu = 1 - \dfrac{1}{1+\mathrm{e}^{-\eta}} = \dfrac{\mathrm{e}^{-\eta}}{1+\mathrm{e}^{-\eta}} = \dfrac{1}{1+\mathrm{e}^{\eta}} = \sigma(-\eta)$

最终 $p(x \mid \mu) = \underbrace{(1-\mu)}_{\sigma(-\eta)} \cdot \exp\{x \cdot \underbrace{\ln \frac{\mu}{1-\mu}}_{\eta}\}$

上述式子中归一化参数 g($\eta$)= σ(-$\eta$); h(x)=1

经历了这个例子，相信读者也掌握了这个将其他分布（normal, exponential, log-normal, gamma, chi-squared, beta, Dirichlet, Bernoulli, categorical, Poisson, geometric, inverse Gaussian, von Mises and von Mises-Fisher distributions）转换为指数分布族的方法。那我们再来个例子：

**例 6.2 狄利克雷分布 (Dirichlet Distribution)的指数分布族形式**

我们又来到了我们熟悉的 Dirichlet Distribution，找到第 2 章我们学过的概率密度函数公式(2-10)，将概率 p 字母用$\theta$代替后如公式(6-4)所示。

$$p(\vec{\theta} \mid \vec{\alpha}) = \frac{\Gamma(\sum_{i=1}^{k} \alpha_i)}{\prod_{i=1}^{k} \Gamma(\alpha_i)} \prod_{i=1}^{k} \theta_i^{\alpha_i - 1} \tag{6-4}$$

先对这个式子进行指数变换，则原式变为:

$$\begin{aligned} p(\vec{\theta} \mid \vec{\alpha}) &= \exp\{\ln(\frac{\Gamma(\sum_{i=1}^{k} \alpha_i)}{\prod_{i=1}^{k} \Gamma(\alpha_i)} \prod_{i=1}^{k} \theta_i^{\alpha_i - 1})\} \\ &= \exp\{\ln(\Gamma(\sum_{i=1}^{k} \alpha_i)) - \sum_{i=1}^{k} \ln(\Gamma(\alpha_i)) + \sum_{i=1}^{k} \ln(\theta_i^{\alpha_i - 1})\} \quad \text{//使用}\ln(A \bullet B) = \ln(A) + \ln(B) \\ &= \exp\{\sum_{i=1}^{k} \underbrace{(\alpha_i - 1)}_{\eta_i} \underbrace{\ln(\theta_i)}_{u(x_i)} \underbrace{- \sum_{i=1}^{k} \ln(\Gamma(\alpha_i)) + \ln(\Gamma(\sum_{i=1}^{k} \alpha_i))}_{-a(\eta)}\} \quad \text{//使用}\ln(A^{\alpha}) = \alpha \ln(A) \end{aligned} \tag{6-5}$$

在 Dirichlet Distribution 的指数分布族表达式形式中，h(x)=1

有人也许会想，只不过把分布函数的形式换了种写法罢了，这有什么用啊？其实读者大可不必如此着急，一时不明了的概念在后期会柳暗花明，指数分布族的用处在于与其他方法相结合（比如贝叶斯估计等)，在后文中我们可看到这一点。

很多初学者会疑惑，为何 LDA 要选 Dirichlet 分布？在这里也可以解释，Dirichlet 分布是指数分布族，且具有有限维度的充分统计量（sufficient statistics)，也是多项分布（multinomial distribution）的共轭先验分布。因此在第 7 章的推断和参数估计中会利用到 Dirichlet 分布的这些特性。

### 6.1.2 再谈数学期望

读者注意了，这里的期望是指概率论与数理统计中的期望，和我们日常口语

的期望不是一回事，统计学中期望这个概念究竟是什么意思呢？其实应该有个更平易近人的名字：随机变量之平均值，为什么要起这么个不够形象易懂的名字？由于概率论最早从研究赌博开始发展，当初期望主要是用于计算赌局的赢钱金额数，所以后来也就把这个名词沿袭下来了。

我通过一些例子谈谈数学期望，我们从平均数开始出发，假如要计算一个班级的平均成绩：E(X) = $(x_1+ x_2+ x_3+\cdots+ x_N)/N$；让我们再加上概率这个权重，也就是说，每个变量 x 的概率权重不等，所以 E(X)期望等于“x 的可能值与其概率之积的累加”。

$$E(X) = x_1 p_1 + x_2 p_2 + x_3 p_3 + \ldots + x_n p_n$$

这个式子对初学者而言可能看起来不是很明显，那换种说法再来解释一下，假设我做 N 次试验，每次把 X 的取值记下来，在这 N 次试验中，有 $N_1$ 次取到 $x_1$，有 $N_2$ 次取到 $x_2$，…，而这 N 次试验中 X 的总共取值为 $x_1N_1+ x_2N_2+\cdots+ x_NN_N$，而**平均**每次试验中 X 的取值，记为 E(X)

$$\begin{aligned} E(X) &= (x_1N_1+ x_2N_2+ x_3N_3+\cdots+ x_NN_N)/N \\ &= x_1(N_1/N) + x_2(N_2/N) +\cdots+ x_N(N_N/N) \end{aligned}$$

$N_1/N$ 是事件 $x_1$ 发生的频率，按照概率统计的定义，当 N 很大时候，频率接近概率，也即 $N_1/N$ 接近 $p_1$，因此就得到了下面这个公式：

$$E(X) = \sum_{i=1}^{\infty} x_i p_i \tag{6-6}$$

一个关于期望的性质是：若干个随机变量之和的期望等于各变量的期望之和。

$$E(X_1 + X_2 + X_3 + \ldots + X_n) = E(X_1) + E(X_2) + \ldots + E(X_n) \tag{6-7}$$

读者可以想想为什么该性质能成立，这个性质的证明留给读者来完成。

拥有了期望的公式(6-6)，我们还不满足，想进一步探究随机变量**函数**的期望 $E_f(g(x))$，也就是下面的定理。

**求证：**（随机变量函数的期望），设随机变量为离散型，有分布 $P(X=x_i)=p_i(i=1,2\cdots)$；或者为连续型，有概率密度函数 f(x)，则:

(1)随机变量离散型时：

$$E(g(X)) = \sum_i g(x_i)p_i \quad (\text{当}\sum_i |g(x_i)|\, p_i < \infty\text{时}) \tag{6-8}$$

(2)随机变量连续型时：

$$E_f(g(X)) = \int_{-\infty}^{\infty} g(x)f(x)dx \quad (\text{当}\int_{-\infty}^{\infty} |g(x)|\, f(x)dx < \infty\text{时}) \tag{6-9}$$

证明：

(1)先来看第一种离散的情况，这比较好证明，因为 $P(X=x_i)=p_i$, 所以 $P(g(X)=g(x_i))=p_i$, 由此立即可以得证(1)中等式。

(2)再来看第二种连续型随机变量的情况，我们下面来证明等式：

在连续情况下： $E_f(g(X))=\int_{-\infty}^{\infty}g(x)f(x)dx$ （当 $\int_{-\infty}^{\infty}|g(x)|f(x)dx<\infty$ 时）

我们先假设 g(X)函数为严格上升并可导，**这个证明一旦想清楚换元就很简单**，下面给出证明：

1. 先来观察 Y=g(x)，由于 g(x)属于严格上升函数，因此分布函数换成反函数 X=h(y)表达：

$$P(Y\le y)=P(g(X)\le y)\overbrace{=}^{\text{换成X表达}}P(X\le \mathrm{h}(y))=\int_{-\infty}^{h(y)}f(t)dt$$

P 关于 y 求导数，得到 y 的密度函数 $l(y)=f(h(y))h'(y)$

2. 又因为 h 是 g 的反函数，则 h(g(x))=x(此式左右两边同时对 x 求导)，

推理出=> $h'(g(x))\cdot g'(x)=1$ 即 $h'(g(x))=\dfrac{1}{g'(x)}$

因此 $E(g(X))=\int_{-\infty}^{\infty}y\cdot l(y)dy=\int_{-\infty}^{\infty}y\cdot f(h(y))h'(y)dy$ ，

把 y 换成 g(x), 则 $f(h(g(x)))=f(x)$; $h'(g(x))=\dfrac{1}{g'(x)}$; $dy=g'(x)dx$

最后 $E_f(g(X))=\int_{-\infty}^{\infty}g(x)\cdot\dfrac{f(x)}{g'(x)}\cdot g'(x)dx=\int_{-\infty}^{\infty}g(x)f(x)dx$

关于 g(X)非单调函数的证明超出了本书范围，留给读者来完成。

证明方法 2：

我们再来给出第二种证明方法，我们在这里仅就证明 $g(X)\ge 0$ （非负）情况下式(6-9)成立。

要证明式(6-9)，还需要一个另一个结论：对于**非负随机变量 Y**，有

$$E(Y)=\int_0^{\infty}P\{Y>y\}dy \qquad (6\text{-}10)$$

先来看看如何得到这个结论，我们假定 Y 是一个连续型随机变量，其密度函数为 $f_Y$ ，此时得到

$$\int_0^{\infty}P\{Y>y\}dy=\int_0^{\infty}\int_y^{\infty}f_Y(x)dxdy$$

此处利用了事实$P\{Y>y\}=\int_y^{\infty}f_Y(x)dx$，交换上述的积分次序，可以得到

$$\int_0^{\infty}P\{Y>y\}dy=\int_0^{\infty}\int_0^{x}f_Y(x)dydx=\int_0^{\infty}(\int_0^{x}dy)f_Y(x)dx=\int_0^{\infty}xf_Y(x)dx=E[Y]$$

得到结论(6-10)之后，我们接着来证明(6-9)式：

对于非负函数$g(x)\geq 0$，根据式(6-10)，有

$$E_f\left(g(X)\right)=\int_0^{\infty}P\{g(X)>y\}dy=\int_0^{\infty}\int_{x:g(x)>y}f(x)dxdy$$

交换积分次序，可得：

$$E_f\left(g(X)\right)=\int_{x:g(x)>0}(\int_0^{g(x)}dy)f(x)dx=\int_{x:g(x)>0}g(x)f(x)dx$$

这样，式(6-9)就得到了证明。$g(X)$为一般条件下的证明留给读者来完成。

当然你不关心上述的证明也可以，只需要记住这个公式，这个公式中 E 的下标 f 表示是关于概率密度函数 f(x)来求导的，通常也可以写作$\langle g(X)\rangle_f$，整体表达见公式(6-11)。

$$\left\langle g(X)\right\rangle_f=E_f(g(X))=\int_{-\infty}^{\infty}g(x)f(x)dx \tag{6-11}$$

公式(6-11)是一个众所周知的基本的概率论公式，在后文的变分推断技术推导中，这个公式会被反复用到。

最后来看一下二维(多维)随机变量函数的数学期望：

**求证：**二维(多维)随机变量函数的数学期望有以下命题:

(1)设随机变量离散型随机变量(X,Y)有联合分布列为$p(X=x_i,Y=y_j)=p_{ij}$，

$g(x,y)$为二维函数，则$Z=g(X,Y)$的期望为：

$$E\left(g(X,Y)\right)=\sum_i\sum_j g(x_i,y_j)p_{ij} \tag{6-12}$$

(2)设连续型随机变量(X,Y)的联合分布密度为$f(x,y)$，$g(x,y)$为二维函数，则$Z=g(X,Y)$的期望为：

$$E_{f(x,y)}\left(g(X,Y)\right)=\int_{-\infty}^{\infty}\int_{-\infty}^{\infty}f(x,y)g(x,y)dxdy \tag{6-13}$$

**证明：**这里仅就$g(X,Y)\geq 0$（非负）证明此命题。

要证明式(6-13)，就要利用前面已经得到的结论公式(6-10)：

$$E_{f(x,y)}\left(g(X,Y)\right)=\int_0^{\infty}P\{g(X,Y)>t\}dt$$

将概率

$$P\{g(X,Y)>t\}=\iint_{(x,y):g(x,y)>t}f(x,y)dxdy$$

代入期望公式可得：

$$E_{f(x,y)}\big(g(X,Y)\big)=\int_0^\infty\iint_{(x,y):g(x,y)>t}f(x,y)dxdydt$$

将上述三重积分交换积分次序，得到：

$$E_{f(x,y)}\big(g(X,Y)\big)=\int_{-\infty}^\infty\int_{-\infty}^\infty\int_{t=0}^{g(x,y)}f(x,y)dtdxdy=\int_{-\infty}^\infty\int_{-\infty}^\infty f(x,y)g(x,y)dxdy$$

这样便证明了式(6-13)。

了解了数学期望的相关知识，自然可以对指数分布族再做一番探究。

### 6.1.3 进一步观察指数分布族

**观察** 1. 先来观察一下公式(6-3)中的 a($\eta$)，如果对 a($\eta$)关于$\eta$求导，会得到什么呢？

$$\frac{d}{d\eta}a(\eta)=\frac{d}{d\eta}(\log(\int h(x)\exp\{\eta^T u(x)\}dx))$$

这里的 log 其实是 ln（以 e 为底的对数），考虑到导数的性质，$(\ln x)'=\frac{1}{x}$，应用**复合函数的链式求导法则**$\big(f(g(x))\big)'=f'(g(x))g'(x)$，在这次的求导中 f 函数就是 log，因此：

$$\begin{aligned}\frac{d}{d\eta}a(\eta)&=\frac{d}{d\eta}\Big(\log(\int h(x)\exp\{\eta^T u(x)\}dx)\Big)\\&=\frac{\int u(x)h(x)\exp\{\eta^T u(x)\}dx}{\underbrace{\int h(x)\exp\{\eta^T u(x)\}dx}_{\exp(a(\eta))}}\quad \text{//链式求导法则}\\&=\int u(x)\underbrace{h(x)\exp\{\eta^T u(x)-a(\eta)\}}_{p(x|\eta)}dx\\&=E(u(x))\end{aligned}$$

由此可见，a($\eta$)关于$\eta$求导得到了 E(u(x))。

现在我们将这个结论用到我们最熟悉的 dirichlet distribution 中，由于按照(6-5)式中的狄利克雷分布的指数分布族表达式，u(x)=ln($\theta$)，又由于

$$\frac{d}{d\eta}a(\eta)\underbrace{=}_{\eta\text{换为}\alpha}\frac{d}{d\alpha_i}(\sum_{i=1}^k\ln(\Gamma(\alpha_i))-\ln(\Gamma(\sum_{i=1}^k\alpha_i)))$$

这里由于看到要对 log-gamma 函数求导，因此引入一个新的数学符号，下面公式中这个长得像鱼叉一样的符号ψ是希腊字母 Psi（普赛），该函数就是 Digamma 函数：

$$\Psi(x)=\frac{d}{dx}\ln(\Gamma(x))=\frac{\Gamma'(x)}{\Gamma(x)}$$

则可以轻松得到以下公式：

$$E_{p(\theta|\alpha)}(\ln(\theta_i))=\frac{d}{d\alpha_i}\left(\sum_{i=1}^{k}\ln(\Gamma(\alpha_i))-\ln(\Gamma(\sum_{i=1}^{k}\alpha_i))\right)=\Psi(\alpha_i)-\Psi(\sum_{i=1}^{k}\alpha_i) \qquad (6\text{-}14)$$

这个狄利克雷分布的关于 $\ln(\theta_i)$ 的期望表达式在 Blei 用变分推断法推导 LDA 时被反复用到，是个**十分重要**的式子。

**观察 2.** u(x)被称为充分统计量（sufficient statistics)，何为充分之意，为何叫充分统计量，可以这么理解：对于要估计的参数 $\eta$ 来说，$\eta$ 的似然函数**仅仅依赖**于 u(x)。为了清晰地看出“充分”这一点，下面我们使用极大似然估计（MLE）简单地估计一下参数 $\eta$:

$$\text{likehood：}l(\eta;x_1,x_2,...,x_n)=\log(\prod_{i=1}^{n}p(x_i\mid\eta))$$

$$=\sum_{i=1}^{n}(\log(h(x_i))+\eta^T u(x_i)-a(\eta))$$

$$=\sum_{i=1}^{n}\log(h(x_i))+\eta^T\sum_{i=1}^{n}u(x_i)-n\cdot a(\eta)$$

紧接着按照极大似然估计的标准做法对 c 求导数使其等于 0：

$$\frac{\partial}{\partial\eta}l(\eta;x_1,x_2,...,x_n)=\sum_i u(x_i)-n\cdot(\frac{\partial}{\partial\eta}a(\eta))=0 \quad //\text{h 被看作常数忽略}$$

由于 $\frac{\partial}{\partial\eta}a(\eta)=E_\eta(u(x))$，为了使上述求导等于 0，则将此式子的期望替换代入上式，所以整理可得 $E_{\eta_{MLE}}(u(x))=\frac{1}{n}\sum_{i=1}^{n}u(x_i)$，也就是说最后所求 $\eta_{MLE}$ **必须满足**使“充分统计量 u(x)关于极大似然估计值 $\eta_{MLE}$ 的期望”=“充分估计量 $u(x_i)$ 关于所有样本点 $x_i$ 的平均值”。这样就得到了最终 $\eta$ 的极大似然估计 $\eta_{MLE}$。所以可以清晰地看到充分统计量概念的来历，这里我们不需要存储每个数据点 x，而仅仅需要存储 $\frac{1}{n}\sum_{i=1}^{n}u(x_i)$ 就可以估计出参数 $\eta$，这便是充分统计量之要义。

### 6.1.4 拉格朗日的杰作——拉格朗日乘数法

拉格朗日(全名为约瑟夫·路易斯·拉格朗日)，法国著名的 3L 人物之一，（法国 18 世纪后期到 19 世纪初数学界著名的三个人物：勒让德（adrien-marielegendre)、拉格朗日（joseph louis lagrange)和拉普拉斯（pierre-simonlaplace)三个人的姓氏的第一个字母为“L”，又生活在同一时代，所以人们称他们为“三 L”。）拉格朗日是数学界无畏的高手，在数学，力学，天文学都做出过杰出的贡献，另外拉格朗日与欧拉合作开启了变分法这一全新的学科。后来变分法又与贝叶斯方法所结合产生出了强大的威力，为机器学习打开了一扇新的窗户。本节所讲的是另一个拉格朗日广为人知的成果：拉格朗日乘数法。

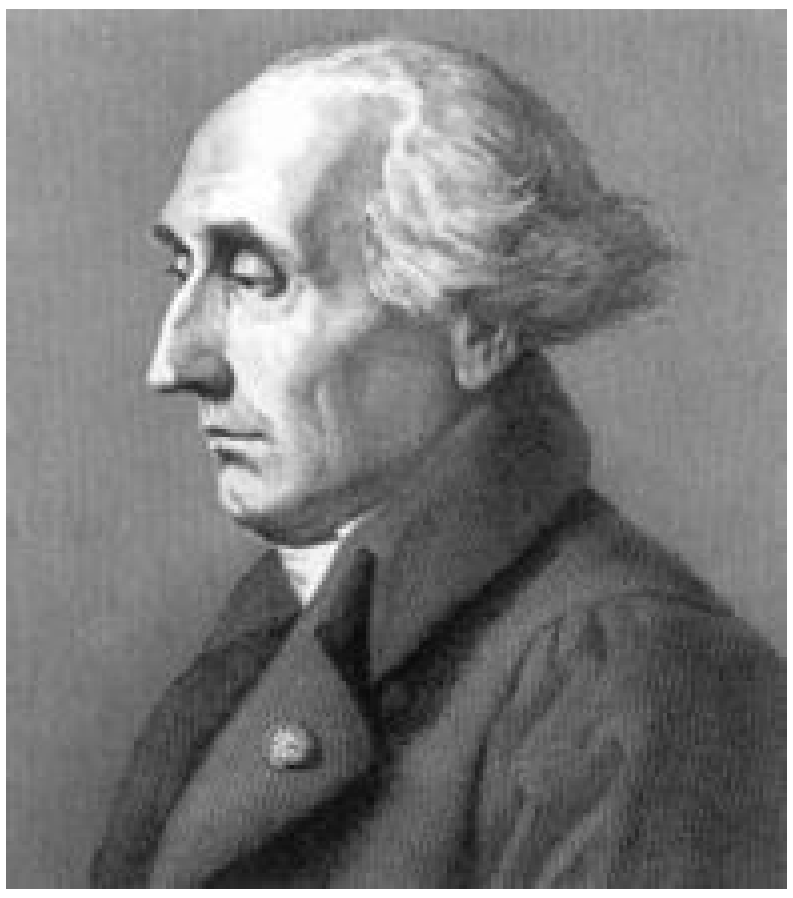

**图 6-1 约瑟夫·路易斯·拉格朗日**

任何学过高等数学或微积分的人都接触过拉格朗日乘数法，什么是拉格朗日乘数法？简而言之：条件极值。

要 z=f(x,y)在条件 g(x,y)=0 下的求极值的自变量 x，y 的情况，譬如：要用**指定表面积**的铁皮围城一个立方体，体积最大为多少？

所谓的条件极值，有个更正式的名字的叫：约束最优化（constraint optimization)，是指在一个或多个条件的约束下，求一个函数的最佳极值。举例来说，我们需要得到 z=f(x,y)的最小值，但前提是 x,y 必须满足方程 g(x,y)=0。最直截了当的做法是解出条件方程，用 x=h(y)。将此式代入 z=f(x,y)即可得到单一变量的函数 z=f(h(y),y)，求出导数 dz/dy=0 即可得到 y 的极值。但这个方法有其不足，首先明确看到的一点就是不一定能得到 x=h(y)。

天才拉格朗日发明的乘数法，优雅而简洁地解决了这个问题。拉格朗日乘数法的步骤只有 2 个部分，非常简洁，我们先来给出拉格朗日乘数法的解法步骤，再来通过一个直观的例子解释这个解法。

设给定函数 z=f(x, y)和附加条件 g(x, y)=0，为寻找 z=f(x, y)在附加条件下的极值点，先做拉格朗日函数 L(x, y)=f(x, y)+$\lambda$g(x, y),其中$\lambda$为新引入的一个乘法因子参数，求 L(x,y)对 x 和 y 的一阶偏导数，令它们=0，并与附加条件联立：

$$\begin{cases} \frac{\partial}{\partial x}\mathrm{L(x,y)} = \frac{\partial}{\partial x}f(x,y) + \lambda \cdot \frac{\partial}{\partial x}g(x,y) = 0 \\ \frac{\partial}{\partial y}\mathrm{L(x,y)} = \frac{\partial}{\partial y}f(x,y) + \lambda \cdot \frac{\partial}{\partial y}g(x,y) = 0 \\ g(x,y)=0 \end{cases}$$

由上述方程组解出 x，y 及$\lambda$，如此求得的(x, y)，就是函数 z=$f$(x, y)在附加条件 g(x, y)=0 下的可能极值点。

在多维(多元)函数情况下，也是一样的，轮流对各个变元求偏导即可，我们可以将符号变一下，使用$\nabla f$和$\nabla g$表示梯度，所谓的梯度，其实就是 f 和 g 对各个变元(x, y, z, …)的偏导数，如果将上面的方程组中的式子移项就会发现，上述方程即可写为$\nabla f = \lambda \nabla \mathrm{g}$（λ可正可负）和 g(x, y, z, …)=0

**例 6.3** 求函数 f(x, y)=3x+4y 在圆 $x^2+y^2=1$ 的极大值和极小值

解：此题的解答如果使用拉格朗日乘数法非常简单，首先，构造出一个拉格朗日函数 L(x, y)= 3x+4y−$\lambda$($x^2+y^2-1$)

$$\frac{\partial}{\partial x}\mathrm{L(x,y)} = \frac{\partial}{\partial x}f(x,y) + \lambda \cdot \frac{\partial}{\partial x}g(x,y) = 0$$

$$\Rightarrow \ 3 - \lambda \cdot (2x) = 0 \ \Rightarrow \ \mathrm{x}=\frac{3}{2\lambda}$$

$$\frac{\partial}{\partial y}\mathrm{L(x,y)} = \frac{\partial}{\partial y}f(x,y) + \lambda \cdot \frac{\partial}{\partial y}g(x,y) = 0$$

$$\Rightarrow \ 4 - \lambda \cdot (2y) = 0 \ \Rightarrow \ \mathrm{y}=\frac{2}{\lambda}$$

$$g(x,y) = 0 \Rightarrow \ x^2 + \mathrm{y}^2 - 1 = 0 \ \Rightarrow \ (\frac{3}{2\lambda})^2 + (\frac{2}{\lambda})^2 - 1 = 0 \ \Rightarrow \ \lambda = \pm\frac{5}{2}$$

所以$\mathrm{x}=\pm\frac{3}{5}, y = \pm\frac{4}{5}$

最后将 x 和 y 代入 f(x, y)得到极大值为 5，极小值为−5。

下面我们通过图 6-2，更深刻理解拉格朗日乘数法。

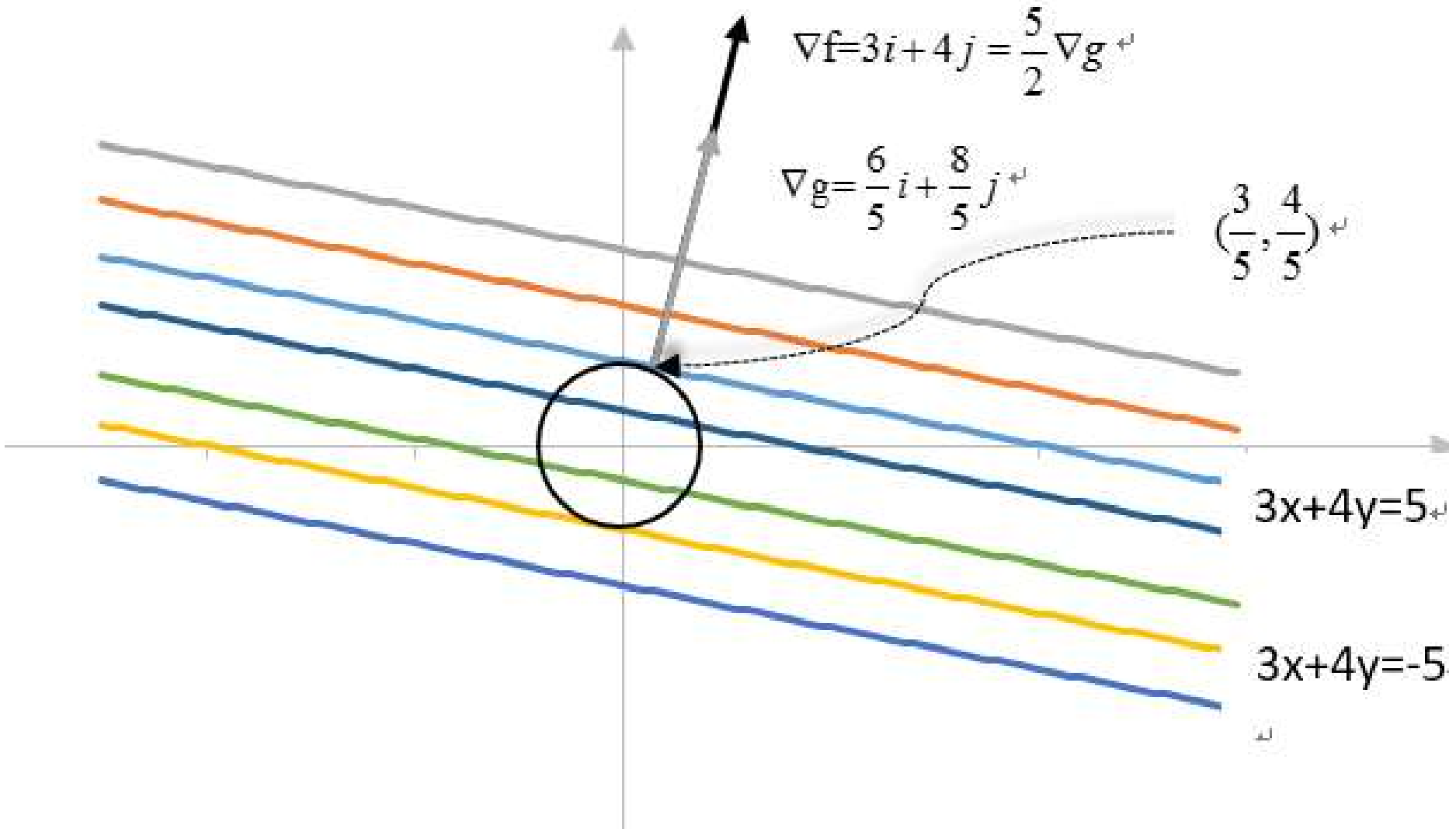

图 6-2 拉格朗日乘数法小例子图

这张图中的一排平行直线就是我们要寻求极大值的 f(x,y)，而圆形就是限制条件 $x^2+y^2=1$，此题需要求的就是离原点最远的点是**哪一条**（并且与圆相交）？**同圆相切的直线是离原点最远的直线！**还记得我们前面提到过的 $\nabla f = \lambda \nabla g$（$\lambda$可正可负）吗？这个意义也就明白了，在切点，任何同这个直线正交（垂直）的向量也是同圆正交的向量。因此，f 的梯度是 g 的倍数$\lambda$。

这个方法遵循了数学优秀方法的一贯特征：简洁优雅。拉格朗日也将自己的名字永远地刻在了这个方法上。当然，拉格朗日乘数法也可以扩展延伸限制条件为多个方程的情况。

多个约束条件下的拉格朗日乘数法：

$$\nabla f = \lambda_1 \cdot \nabla g_1 + \lambda_2 \cdot \nabla g_2 + \lambda_3 \cdot \nabla g_3 + \ldots = \sum \lambda_i \cdot \nabla g_i$$

也即构造拉格朗日函数：

$$L = f(\vec{x}) + \sum_{i=1}^{m} \lambda_i \cdot g_i(\vec{x}) \tag{6-15}$$

后续步骤和单个限制条件时相同。

当然拉格朗日乘数法的学问远不止于此，弱水三千只取一瓢，我的原则是，只取到需要足够使用的部分就可以了，未来如果需要用到其更深的部分，再取之即可。

### 6.1.5 指数分布族的一点深入的思考

在了解了拉格朗日乘数法之后，我们可以重返指数分布族了，对其进行更深

的理解，这一小节主要是从理论上对指数分布族进行一些起源的发掘，与后面 LDA 的变分法推导关系并不大，这一小节是解决部分读者这样一个疑问：为什么指数分布族的形如 (6-1)式？这里我引入一个概念，最大熵（Maximum Entropy)。

别被“熵”这个概念吓到，熵用来代表“随机变量不确定性的程度”，也就是信息无序的程度，变量的不确定性越大，熵也就越大，搞清楚他所需要的信息量也就越大。熵最早来自于热力学，后来被伟大的信息论之父香农用到了信息论中，香农当时想出这个熵的计算方法后，还没有想清楚该起什么名字，于是去问冯诺依曼，冯诺依曼建议叫“熵”，并说：“没有人知道熵是什么东西，因此你在争论中可以无往而不胜”。

熵的计算公式非常简单：

$$H(x) = -\sum_{x} P(x) \cdot \log[P(x)] \tag{6-16}$$

为了简便起见，这里我们使用自然常数 e 作为对数 log 的底[29]，特别注意到公式中的负号，因为概率 P 是一个小于 1 的数字，因此 log 对数是小于 0 的数字，所以求和后为负数，前面再加一个负号，就使得熵为正数了。

现在假设我们想找到一个特殊的概率分布 p(X)，我们所知道的所有信息只是关于函数的期望等于常数：$E_{p(x)}[f_k(x)] = C_k$，其中 $f_1, f_2, f_3, \ldots, f_k$ 是一系列的关于 x 的随机变量的函数，目标只有一个：在满足这些条件的情况下，概率分布 p(X) 的**最佳选取**是怎样的？

这里就引入了最大熵的选取原则，最大熵，就是在已知部分知识的前提下，关于未知分布最合理的推断就是符合已知知识不确定性最大或最随机的推断，这是我们可以作出的唯一不偏不倚的选择，任何其它的选择都意味着我们增加了其它的约束和假设，这些约束和假设根据我们掌握的信息无法作出。换而言之，引用吴军的《数学之美》中的一个通俗解释就是：当我们需要对一个随机事件的概率分布进行预测时，我们的预测应当满足全部已知的条件，而对未知的情况不要做任何主观假设。（不做主观假设这点很重要）在这种情况下，概率分布最均匀，预测的风险最小。因为这时概率分布的信息熵最大，所以人们称这种模型叫"最大熵模型"。

巴拉巴拉，说了这么多，其实也可以死记住一点：熵最大的选择就是最优选择。现在我们的任务很简单了，可以看到选取的 p(X)概率分布是 $[p(x_1), p(x_2), \cdots p(x_n)]$等等的一个向量的函数，只需要最大化(maximize)公式

[29] 因为 $\frac{d}{dx}(\ln(x)) = \frac{1}{x}$，但信息论中由于考虑到要利用的数值都是二进制的，所以熵通常为以 2 为底。

(6-16)，现在再加上优化的限制条件(constraint condition)。

$$P^*(x) = \arg\max_{p(x)} H(x)$$

$$\text{s.t.} \quad \sum_{x=1}^{n} P(x) = 1 \text{ 和 } \sum_{x=1}^{n} P(x) \cdot f_k(x) = C_k$$

其中 $f_k(x)$ 为一组特征函数，而优化中约束的意义就在于这一组特征函数在模型 $P(X)$ 下的均值等于其数据上的均值（ $\tilde{P}(X)$ 为数据分布），也可以将约束条件写为：

$$\text{s.t.} \quad E_P[f_k(x)] = E_{\tilde{P}}[f_k(x)], \quad k = 1,2,...,D$$

这种条件极值又一次激起了我们的条件反射：拉格朗日乘数法。

定义拉格朗日函数：

$$L = -\sum_{x=1}^{n} P(x) \cdot \log[P(x)] + \lambda_0 (\sum_{x=1}^{n} P(x) - 1) + \sum_{k} \lambda_k \cdot (\sum_{x=1}^{n} P(x) \cdot f_k(x) - C_k) \tag{6-17}$$

紧接着就是烂熟于心的求导了：

$$0 = \frac{\partial L}{\partial P(x)} = -\log[P(x)] - 1 + \lambda_0 + \sum_{k} \lambda_k \cdot f_k(x)$$

$$0 = \frac{\partial L}{\partial \lambda_0} = \sum_{x=1}^{n} P(x) - 1$$

$$0 = \frac{\partial L}{\partial \lambda_k} = \sum_{x=1}^{n} P(x) \cdot f_k(x) - C_k$$

方程组中仅仅使用第一个方程，则可以见到：

$$P(x) = \exp\{\sum_{k} \lambda_k \cdot f_k(x) + \lambda_0 - 1\} \propto \exp\{\sum_{k} \lambda_k \cdot f_k(x)\} = \exp\{\lambda^T \cdot u(x)\} \tag{6-18}$$

公式中的 $\lambda=(\lambda_1,\lambda_2,\lambda_3,...,\lambda_k)^T$，$\phi(x)=(f_1(x),f_2(x),f_3(x),...,f_K(x))^T$，这就昭示了这样一个事实：P(X)属于指数分布族。

### 6.1.6 Jensen 不等式

Jensen 不等式在求解变分贝叶斯时得到变分下界时有重要作用，这个不等式也非常简单。首先来看看凸函数(convex function)的定义：

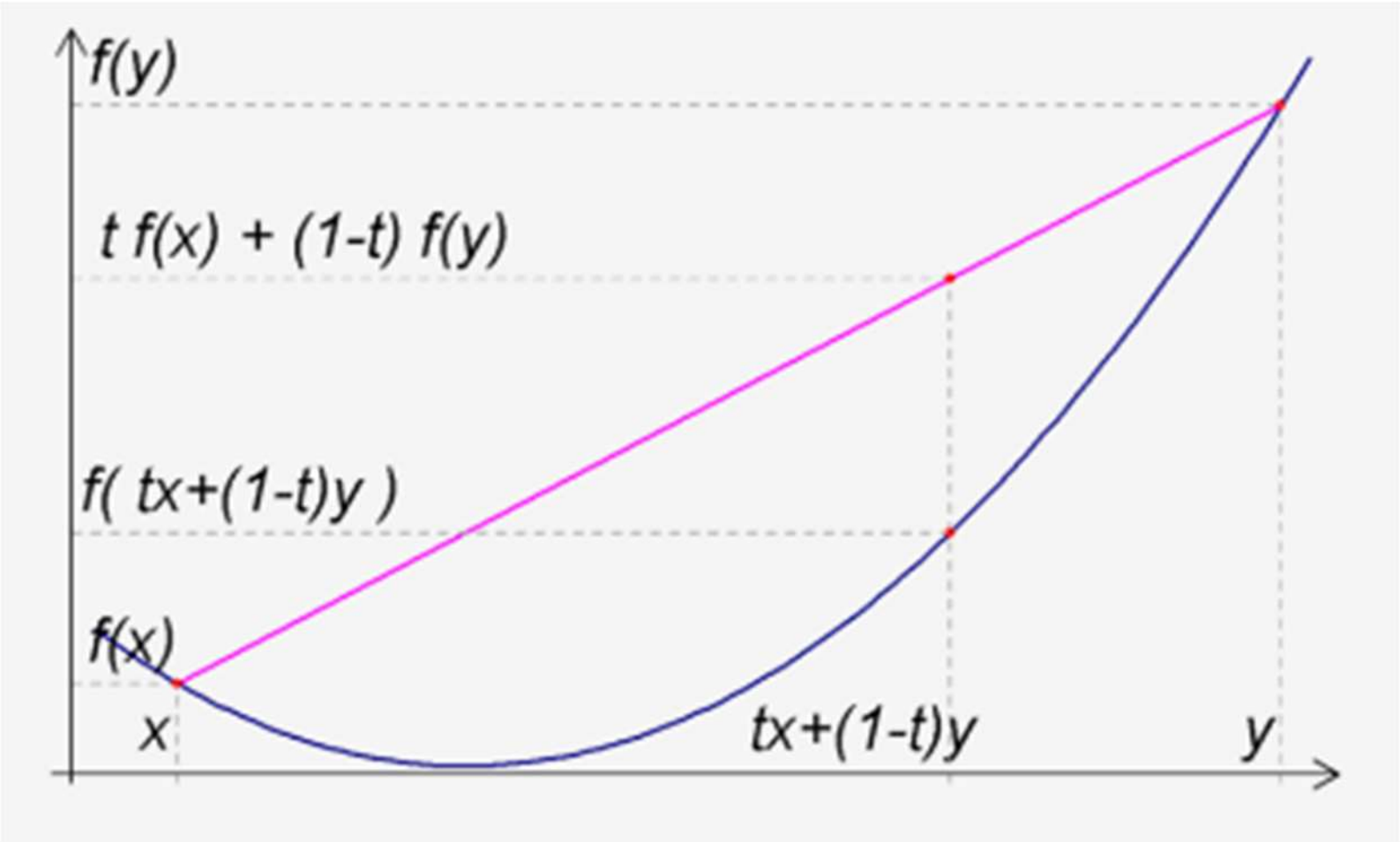

**图 6-3 凸函数（convex function）**

凸函数是定义在某个向量空间的凸子集 C（区间）上的实值函数 f，如果定义在其定义域 C 上任意两点 x，y，以及 $t \in [0,1]$，有：

$$f(tx+(1-t)y) \le tf(x)+(1-t)f(y) \tag{6-19}$$

凸函数的函数图像如图 6-3 所示，这个图像有个很形象的解释，(x, f(x)) 和 (y, f(y)) 两点的连线位于凸函数曲线之上。

了解了凸函数的概念之后，这里首先给出 Jensen 不等式的定义，再给出其证明。

定义 6.1 Jensen 不等式

令 f(x) 是在一个定义在区间 I 上的凸函数。如果 $x_1, x_2, \ldots, x_N \in I$ 且

$\lambda_1, \lambda_2, \lambda_3, \ldots, \lambda_N \ge N$ 且 $\sum_{i=1}^{N} \lambda_i = 1$，

$$f\left(\sum_{i=1}^{N} \lambda_i x_i\right) \le \sum_{i=1}^{N} \lambda_i f(x_i) \tag{6-20}$$

或者简化一点，可以使用期望公式表示（$\lambda_i$ 换用 $P(x_i)$ 表示）：

$$f[E(x)] \le E[f(X)] \tag{6-21}$$

$$f\left(\sum_{i=1}^{N} x_i P(x_i)\right) \le \sum_{i=1}^{N} f(x_i) P(x_i) \tag{6-22}$$

证明：

这个证明的主要思路是数学归纳法，先证明 N=1 的结论成立，再假设 N=k-1 时已经成立，证明 N=k 成立，那么得出结论：整个不等式就证明是成立的。

先来看 N=1 的情况，此时 $f(P(x_1) \cdot x_1) \le P(x_1) f(x_1)$，由于只有一项，那么 $P(x_1)=1$，所以 $f(1 \cdot x_1) = 1 \cdot f(x_1)$

再来看 N=2 的情况，如果$\lambda_1$和$\lambda_2$是两个任意的非负实数，且$\lambda_1+\lambda_2=1$，f 函数的凸性质（也即凸函数公式 (6-19)）就保证了：

$$\forall x_1, x_2: \quad f(\lambda_1 x_1 + \lambda_2 x_2) \le \lambda_1 f(x_1) + \lambda_2 f(x_2)$$

现在假设 N=k-1 时已经成立，则设$P'(x_i) = P(x_i) / (1 - P(x_k))$，其中$i = 1,2,...k-1$，则：

$$\sum_{i=1}^{k} f(x_i)P(x_i) = (1 - P(x_k))\sum_{i=1}^{k-1} f(x_i)P'(x_i) + f(x_k)P(x_k)$$

$$\ge (1 - P(x_k))f\left(\sum_{i=1}^{k-1} x_i P'(x_i)\right) + f(x_k)P(x_k) \qquad \text{//数学归纳假设}$$

$$\ge f\left((1 - P(x_k))\sum_{i=1}^{k-1} x_i P'(x_i) + x_k P(x_k)\right) \qquad \text{//利用 N=2 的情况}$$

$$= f\left(\sum_{i=1}^{k-1} x_i P(x_i) + x_k P(x_k)\right)$$

$$= f\left(\sum_{i=1}^{k} x_i P(x_i)\right)$$

所以得证。

现在来看一个利用该性质的例子，假设$f(x) = \ln(x)$，由于该函数属于凹函数（concave）[30]，因此不等式符号与(6-22)式相反，所以：

$$\ln\left(\sum_{i=1}^{N} x_i P(x_i)\right) \ge \sum_{i=1}^{N} \ln(x_i) P(x_i)$$

这个结论在标准 EM 算法的推导中被用到。

## 6.2 补充材料：变分法的启蒙

6.2 节以及本章之后的小节与 LDA 变分贝叶斯法的关系不大，这是由于 blei 使用了简化版的变分 EM 方法。所以 6.2 节以及本章之后的小节是为了启发其他新的技术思路而设置，读者倘若较为着急掌握 LDA 的变分方法，可跳过这些内容直接去看第 7 章。

变分法的源头可以追溯到牛顿时代，从最经典的最速降线问题开始，几位当时顶级的数学高手为变分法做出了开创性的思考，直到欧拉-拉格朗日方程（E-L

[30] 从$f(x) = \ln(x)$的函数图像也可看出这一点

方程）的确立，才将变分法置于一个不可置疑的基础上。

### 6.2.1 改变世界的方程：欧拉-拉格朗日方程

欧拉-拉格朗日方程(euler-lagrange equation)确实可称得上是改变世界的方程，变分法从一个 17 世纪末简单的竞赛挑战谜题线索开始，一步一步深刻影响了物理学和整个世界，而奠基变分法不可动摇地位的就是欧拉-拉格朗日方程。这两位才华横溢的数学家将自己的名字刻入了这个不可置疑的方程，这个方程在物理学的经典定律“最小作用力原理[31]”中也得到应用，这个原理后面不可缺少强大武器是泛函分析的变分法。变分法自然也是微积分发展到泛函时的必然产物（**所谓泛函，就是函数的函数**），如果说变分法和泛函是一座建筑大厦，那么欧拉-拉格朗日方程就像这个建筑大门的钥匙，因此我们首先从这个方程开始起步。

首先直接给出欧拉-拉格朗日方程（E-L 方程），接着证明他：

$$\frac{\partial F}{\partial f} - \frac{d}{dx}\frac{\partial F}{\partial f'} = 0 \tag{6-23}$$

这个方程其实是个微分方程（因为其中出现了导数），凡是**满足这个微分方程的解**的函数 f 必为一系列函数（泛函）中令 F 的积分到达极值的那个函数。简而言之：泛函极值。

泛函的英文单词是 functional，恰好与函数式编程（functional program）是同一个单词，泛函极值是以积分的面目示人，我们接下来证明这个方程：

**证明：**

我们首先说明我们的目标，我们希望能在区间 f(a)=A 和 f(b)=B 之间找到一个函数 f，这个函数 f 能令泛函（functional）的积分 $J$ 达到极大值：

$$J = \int_a^b F(x, f(x), f'(x))dx \tag{6-24}$$

这里的 F 就是函数的函数，我们假设 F 是一阶可导的函数，利用这个假设，我们勇敢地将我们的推理进行下去。

**第一个观察：**F 的两个端点分别是 a 和 b，这两个端点是固定死的，而泛函变化的部分在中间，一系列函数中哪个函数让 J 逼近极值？（如图 6-4）

[31] 最小作用量原理是一个极其重要的物理定律

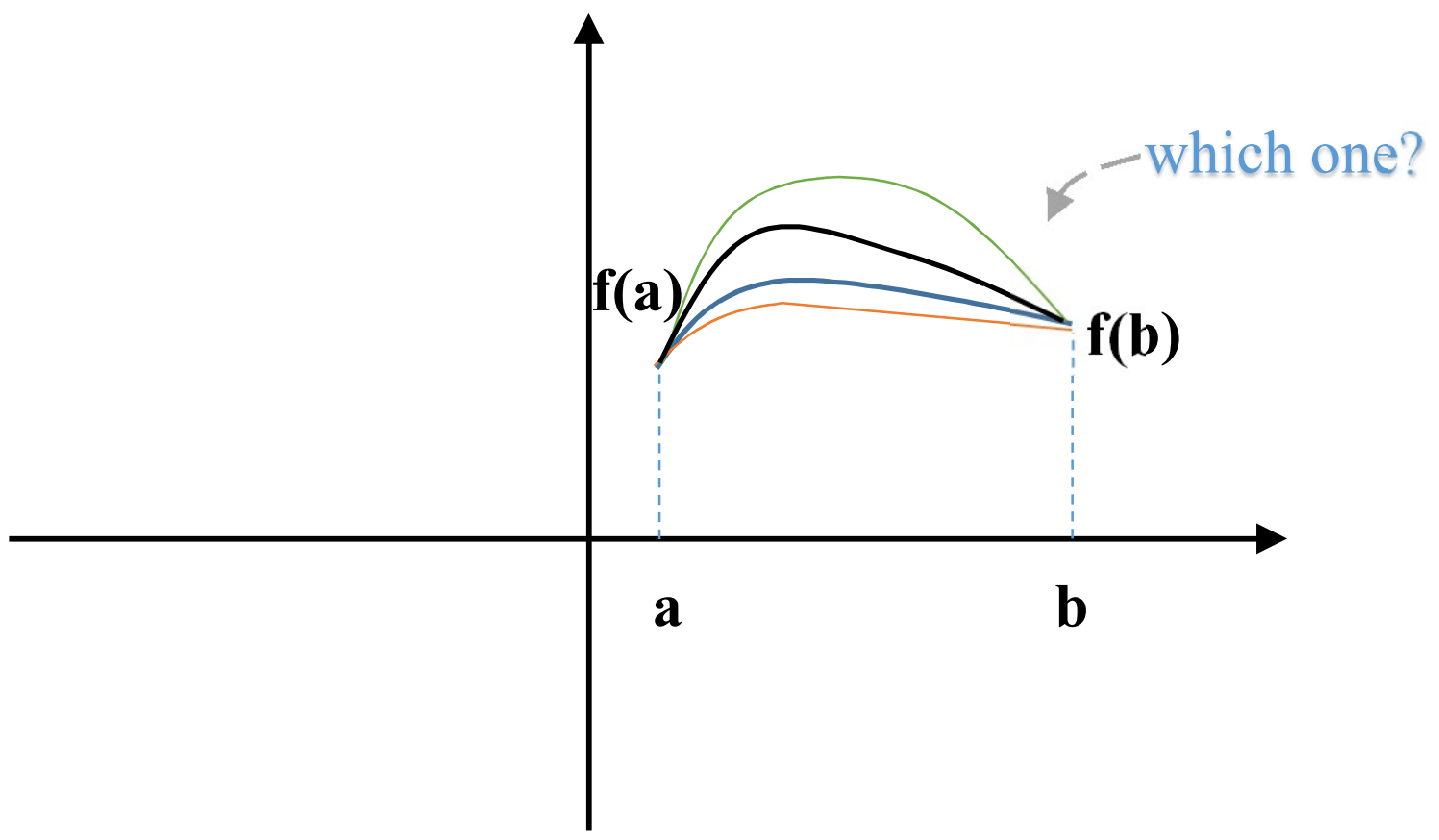

图 6-4 泛函积分的函数图

**第二个观察**：现在我们假设其中有一个函数恰好达到极大值！这个函数就是 f(x)，那么意味着这样一个事实：只要有一点点**轻微的扰动，便可让该泛函积分 J 缩小！（或者在 f(x)为极小值时，有轻微扰动，就会让 J 增大）**所以将这个说法用数学的方式表达出来，便是假设一个轻微扰动后的函数 $g_\varepsilon(x)$：

$$g_\varepsilon(x) = f(x) + \varepsilon \cdot \eta(x) \tag{6-25}$$

这里的ε是一个很小的正数，而 $\eta(x)$ 是一个新定义的关于 x 的函数，这里的 $\varepsilon \cdot \eta(x)$ 即代表 $f(x)$ 之上轻微的扰动（这便是 E-L 方程的神来之笔，这里的 $\varepsilon \cdot \eta(x)$ 还可用一个希腊符号简写：$\delta f$ 表示 $f$ 的变分）。这里的 $g_\varepsilon(x)$ 就可看成是一系列实验的可能曲线，而 $f(x)$ 代表取到极值的那个最佳曲线！结合第一个观察，可以看到两个端点是固死的，也即 $\eta(a) = \eta(b) = 0$。

现在我们要把这个"轻微扰动"的思想代回到我们的泛函公式(6-24)：

$$J_\varepsilon = \int_a^b F(x, g_\varepsilon(x), g_\varepsilon'(x))dx = \int_a^b F_\varepsilon dx \tag{6-26}$$

读者应该敏锐地观察到上面公式中的 J 的下标是ε，这是因为定积分在区间 a 和 b 上积分之后的最终结果中 x 会消失掉，而变成了ε的函数。

工作进行到这里，出现了最后一个也是最重要的一个观察。

**第三个观察**：$\left.\frac{dJ_\varepsilon}{d\varepsilon}\right|_{\varepsilon=0} = 0$，这个观察的思想是：**如果 $f(x)$ 恰好是令 J 达到极值的那一个函数，那么泛函积分 J 对ε求导就等于0**。（这是众所周知的道理：极值求导=0），而 $g_\varepsilon(x) = f(x) + \varepsilon \cdot \eta(x)$ 在ε=0的时候恰好 $g_\varepsilon(x) = f(x)$，所以就得出了这个观察的结论！

第三个观察也可换一种更为通俗易懂的说法：$g_\varepsilon(x)$ 代表一系列实验的可能曲线，当ε=0的时候所有的曲线都会收敛于最佳曲线 $f(x)$，而此时 J 对ε求导就等

于0。为什么偏偏要对ε求导呢？如前文所言，这是因为 J 是一个定积分，该定积分的积分变量是 x，积分后所有的 x 都会消失，而变为关于ε的函数。这便是变分法的精髓所在，用一个ε来应对形形色色不同的函数。

有了这三个重要的观察，后面的工作就回到我们熟悉的求导了：

$$\frac{dJ_\varepsilon}{d\varepsilon}=\frac{d}{d\varepsilon}\int_a^b F_\varepsilon dx=\int_a^b \frac{dF_\varepsilon}{d\varepsilon}dx \tag{6-27}$$

这就将我们的求导从 J 对ε求导转移到了 F 对ε求导：

$$\frac{F_\varepsilon}{d\varepsilon}=\frac{dx}{d\varepsilon}\frac{\partial F_\varepsilon}{\partial x}+\frac{dg_\varepsilon}{d\varepsilon}\frac{\partial F_\varepsilon}{\partial g_\varepsilon}+\frac{dg'_\varepsilon}{d\varepsilon}\frac{\partial F_\varepsilon}{\partial g'_\varepsilon}$$

$$=\frac{dg_\varepsilon}{d\varepsilon}\frac{\partial F_\varepsilon}{\partial g_\varepsilon}+\frac{dg'_\varepsilon}{d\varepsilon}\frac{\partial F_\varepsilon}{\partial g'_\varepsilon}$$

$$=\eta(x)\frac{\partial F_\varepsilon}{\partial g_\varepsilon}+\eta'(x)\frac{\partial F_\varepsilon}{\partial g'_\varepsilon}$$

所以，

$$\frac{dJ_\varepsilon}{d\varepsilon}=\int_a^b \eta(x)\frac{\partial F_\varepsilon}{\partial g_\varepsilon}+\eta'(x)\frac{\partial F_\varepsilon}{\partial g'_\varepsilon}dx$$

当ε=0时， $g_\varepsilon(x)=f(x)$ ， $F_\varepsilon=F(x,f(x),f'(x))$，恰恰在此时，J 有极值：

$$\left.\frac{dJ_\varepsilon}{d\varepsilon}\right|_{\varepsilon=0}=\int_a^b \eta(x)\frac{\partial F}{\partial f(x)}+\eta'(x)\frac{\partial F}{\partial f'(x)}dx=0$$

此时对上面积分内的第二部分使用利用分部积分公式$\int_a^b udv=[uv]_a^b-\int_a^b vdu$

$$\int_a^b \eta'(x)\frac{\partial F}{\partial f'(x)}dx=\int_a^b \frac{\partial F}{\partial f'(x)}d\eta(x)$$
$$=\left[\frac{\partial F}{\partial f'(x)}\eta(x)\right]_a^b-\int_a^b \eta(x)d\frac{\partial F}{\partial f'(x)}$$

等式的第一项使用边界固定点上$\eta(a)=\eta(b)=0$，所以第一项=0。

等式的第二项则为：

$$-\int_a^b \eta(x)d\frac{\partial F}{\partial f'(x)}=-\int_a^b \eta(x)\cdot[\frac{d}{dx}\frac{\partial F}{\partial f'(x)}]dx$$

该部分代回原式：

$$\left.\frac{dJ_\varepsilon}{d\varepsilon}\right|_{\varepsilon=0}=\int_a^b \eta(x)\cdot[\frac{\partial F}{\partial f(x)}-\frac{d}{dx}\frac{\partial F}{\partial f'(x)}]dx=0 \tag{6-28}$$

最后的一步很简单，因为要满足这个积分式子=0，而 $\eta(x)$ 是一个函数，取值不一定为 0，所以为了保证整个积分=0，必然就使得方括号内的式子=0，这就导致了 E-L 方程：

$$\frac{\partial F}{\partial f} - \frac{d}{dx}\frac{\partial F}{\partial f'} = 0$$

得证（Q.E.D.）。

正如众所周知的对函数求导=0 就可以求得函数极值一样，泛函极值就通过这个 E-L 的微分方程求得。

### 6.2.2 E-L 方程的两种降阶形式

上面的这个公式是 E-L 方程的最基本公式形式，在向前一步，便可得到平时使用时更为方便的公式降阶形式：

为了方便，我们换一下符号，设 $y = f(x)$，因此泛函变为：$F(x,y,y')$，原方程(6-23)变为：

$$\frac{\partial F}{\partial y} - \frac{d}{dx}\frac{\partial F}{\partial y'} = 0 \tag{6-29}$$

(1)当函数 $F(x,y,y')$ 不显含 $y$ 时，函数 $F$ 变为 $F(x,y')$。方程(6-29)变为：

$$\frac{d}{dx}\frac{\partial F}{\partial y'} = 0 \quad => \quad \frac{\partial F}{\partial y'} = C \tag{6-30}$$

(2)当函数 $F(x,y,y')$ 不显含 $x$ 时，函数 $F$ 变为 $F(y,y')$。泛函积分(6-24)变为：

$$\int_a^b F(y,y')dx = \int_a^b F(y,\frac{1}{dx/dy})\frac{dx}{dy}dy$$

设 $\frac{dx}{dy} = x'$，则 $F(y,\frac{1}{dx/dy})\frac{dx}{dy} = x'F(y,\frac{1}{x'}) = L(y,x')$

$L$ 是我们定义的新函数，注意到 $L(y,x')$ 不显含 $x$，因此：

$$\int_a^b F(y,y')dx = \int_a^b L(y,x')dy$$

特别注意到上式子中 $y$ 成了新的积分变量。

类比第一种情况，就会发现：这里的 L 与第一种情况的 F 类似（只是 $x$ 和 $y$ 符号交换了），因此：

$$\frac{\partial L}{\partial x'} = C$$

$$\frac{\partial L}{\partial x'}=\frac{\partial}{\partial x'}(x'F(y,\frac{1}{x'}))\underbrace{=}_{\text{复合函数求导}}F(y,\frac{1}{x'})+\underbrace{x'\bullet(-\frac{1}{x'^2})\bullet\frac{\partial F}{\partial(\frac{1}{x'})}}_{-y'\frac{\partial F}{\partial(\mathrm{y}')}}=F-y'\frac{\partial F}{\partial(\mathrm{y}')}=C$$

这样就得到了不显含$x$时，E-L 方程非常方便的形式：

$$F-y'\frac{\partial F}{\partial(y')}=C \tag{6-31}$$

这个形式非常简便易行。已经了解到了欧拉-拉格朗日方程(euler-lagrange equation)公式之后，我们需要将步子迈得大一点，更近一步，在实际情况中，往往碰到的是有限制条件的泛函优化问题，这就需要我们将拉格朗日乘数法和 E-L 方程进行配合使用。

### 6.2.3 E-L 方程与拉格朗日乘数法的联姻

E-L 方程和拉格朗日乘数法联合使用可以发挥更强大的能力，该方法备受我国的钱三强推崇，这种方法后来在有限元方法中也有广泛应用。我们从 6.2.1 节所述的第三种观察来推导与拉格朗日乘数法的联姻，我们需要分成两种情况讨论，一种是限制条件 G 函数中自变量不包含导数的推导；另一种是通用的简洁解决方法。

(1)限制条件自变量不包含导数的情况

假设我有两个函数需要优化，y 和 z，自然而然写出下列公式：

$$\frac{dJ}{d\varepsilon}=\int_{x_1}^{x_2}[(\frac{\partial F}{\partial y}-\frac{d}{dx}\frac{\partial F}{\partial y'})\frac{\partial y}{\partial\varepsilon}+(\frac{\partial F}{\partial z}-\frac{d}{dx}\frac{\partial F}{\partial z'})\frac{\partial z}{\partial\varepsilon}]dx \tag{6-32}$$

在条件$G(y,z,x)=0$的条件写的泛函极值。

由于函数 G 等于常数 0，常数求导必然等于 0，另外 x 不含ε，因此：

$$\frac{dG}{d\varepsilon}=0\ \Rightarrow\ \frac{\partial G}{\partial y}\frac{dy}{d\varepsilon}+\frac{\partial G}{\partial z}\frac{dz}{d\varepsilon}=0 \tag{6-33}$$

由于$\frac{\partial y}{\partial\varepsilon}=\eta_1(x)$，$\frac{\partial z}{\partial\varepsilon}=\eta_2(x)$，再代回上一步的公式，就很自然得到两者的比例关系（限制条件的作用如此展现)：

$$\frac{\partial M}{\partial y}\eta_1(x)+\frac{\partial G}{\partial z}\eta_2(x)=0\ \underbrace{\Rightarrow}_{G_y=\frac{\partial G}{\partial y}}\ \eta_2(x)=-\frac{G_y}{G_z}\eta_1(x)$$

将这个比例关系式代回到(6-32)式子中，取消$\eta_1(x)$，那么就得到：

$$\frac{dJ}{d\varepsilon}=\int_{x_1}^{x_2}\left[\left((\frac{\partial F}{\partial y}-\frac{d}{dx}\frac{\partial F}{\partial y'})-(\frac{\partial F}{\partial z}-\frac{d}{dx}\frac{\partial F}{\partial z'})\frac{G_y}{G_z}\right)\eta_1(x)\right]dx=0$$

与前面E–L 方程的推导一样，这一部由于必须要让整个积分结果=0，所以自然而然：

$$(\frac{\partial F}{\partial y}-\frac{d}{dx}\frac{\partial F}{\partial y'})-(\frac{\partial F}{\partial z}-\frac{d}{dx}\frac{\partial F}{\partial z'})\frac{G_y}{G_z}=0$$

再整理一下（移项可得），可以想到 F 和 y，z 均是只含 x 的函数，因此设计一个 $\lambda(x)$ 就得到：

$$-\lambda(x)=\frac{(\frac{\partial F}{\partial y}-\frac{d}{dx}\frac{\partial F}{\partial y'})}{\partial G/\partial y}=\frac{(\frac{\partial F}{\partial z}-\frac{d}{dx}\frac{\partial F}{\partial z'})}{\partial G/\partial z}$$

这里的 $\lambda(x)$ 为何前面要加负号，下一步马上明了：

$$-\lambda(x)=\frac{(\frac{\partial F}{\partial y}-\frac{d}{dx}\frac{\partial F}{\partial y'})}{\partial M/\partial y}\Rightarrow\frac{\partial F}{\partial y}-\frac{d}{dx}\frac{\partial F}{\partial y'}+\frac{\partial G}{\partial y}\lambda(x)=0$$

第一条方程式已经出现。

z 也是同理，再加上限制条件，因此得到三个方程式：

$$\begin{cases}\frac{\partial F}{\partial y}-\frac{d}{dx}\frac{\partial F}{\partial y'}+\frac{\partial G}{\partial y}\lambda(x)=0\\ \frac{\partial F}{\partial z}-\frac{d}{dx}\frac{\partial F}{\partial z'}+\frac{\partial G}{\partial z}\lambda(x)=0\\ G(y,z,x)=0\end{cases}\qquad(6\text{-}34)$$

三个方程，三个未知数：$y,z,\lambda(x)$，因此方程组理论上可解出。

(2)通用简洁方法的解释

有人或许会问，为何这里的λ是一个关于 x 的函数，而非像前面介绍的拉格朗日乘数法那样是一个实数？[32]那我们现在再从另一个角度（更简洁的角度）看待这个问题，还记得拉格朗日乘数法的写作方式吗？又会回忆起这些如同咒语般的死步骤：

[32] 相关讨论请看：http://math.stackexchange.com/questions/279518/constrained-variational-problems-intuition

设给定函数 z=f(x,y)和附加条件 g(x,y)=0，为寻找 z=f(x,y)在附加条件下的极值点，先做拉格朗日函数 L(x,y)=f(x,y)+$\lambda$g(x,y)，然后依次对各个变量求导令其等于 0 联立方程组求解。

现在情况来到了泛函的条件极值，其实方法也是一样的，我们这次将限制条件改为带导数的限制条件 G，那么这次的题目就可以如下：

题目：$J(y)=\int_D F(x,y,y')dx$ 在限制条件 $G(x,y,y')=0$ 中的最优化的泛函极值。

求解：

仿照一般的拉格朗日乘数法，写出拉格朗日函数 $L=J(y)+\lambda\cdot G(x,y,y')$，也就是：$L=\int_D F(x,y,y')dx+\lambda\cdot G(x,y,y')$

$\lambda$ 就是拉格朗日乘子，请注意到这个式子中 F 和 G 函数都仅是关于 x 的函数，而 J 是关于 y 的函数。富有怀疑精神的人会争论，这里的 G 函数=0 啊，因此这个函数与原始的泛函 J 没有区别！没错，但是后一项 $\lambda\cdot G(x,y,y')$ 仍旧有其意义，主要在于如果做此偏导 $\frac{\partial L}{\partial\lambda}=G(x,y,y')=0$，就会发现 $\frac{\partial L}{\partial\lambda}\neq\frac{\partial J}{\partial\lambda}$ ，对 L 的偏导恰好等于 $G(x,y,y')=0$，也就是限制条件从拉格朗日函数求导中出现了。

为了解决这个拉格朗日函数问题，我们需要将其写成一般的变分函数问题，也就是将其转变为能用到 E-L 方程的形式，现在的问题就在于要将 $\lambda\cdot G(x,y,y')$ 弄进积分符号中，因为 $G(x,y,y')=0$，所以你可以直接塞到积分符号里面即可！（而需要注意的是积分中需要对各个 x 都加上 $\lambda$,因此 $\lambda$ 变成了 $\lambda(x)$）

那么囊括了 $\lambda\cdot G(x,y,y')$ 因素的 $F$ 函数就变成了新的 $\tilde{F}$，原来的泛函变为了新的泛函：

$$\tilde{J}(y)=\int_D \tilde{F}(x,y,y')dx \tag{6-35}$$

现在只需要简单地写出根据这个新泛函的 E-L 方程即可：

$$\frac{\partial\tilde{F}}{\partial y}-\frac{d}{dx}\frac{\partial\tilde{F}}{\partial y'}=0 \tag{6-36}$$

拿到了这个公式，你现在再来观察上文中的方程组(6-34)，就会很明白了，看看方程组中的第一个是如何得到的：

$$\frac{\partial F}{\partial y}-\frac{d}{dx}\frac{\partial F}{\partial y'}+\frac{\partial G}{\partial y}\lambda(x)=0 \quad (6\text{-}37)$$

如果按照这个通用方法来解释，就会如此：

$$\tilde{J}(y)=\int_D F(x,y,y')dx+\lambda\bullet G(y,x)$$

$$=\int_D\left(F(x,y,y')+\lambda(x)\cdot G(y,x)\right)dx$$

由于 G 函数=0，可以被凑成积分形式（0 的积分还是 0），而λ也被凑入积分，因为针对每个 x 点都要做限制而变成了 $\lambda(x)$ 函数形式，这一点与(6-15)式相似。

这时，就会本能察觉到：$\tilde{F}(x,y,y')=F(x,y,y')+\lambda(x)\cdot G(y,x)$，因此应用 E-L 方程(6-36)，由于 G 中不含：$y'$，因此 G 对 $y'$ 求导=0，那么可以得到方程(6-37)了。

我们之后通过一个例题加深理解这个技术。

**例 6.4** 一块钢板围成什么曲面做成的半壁容器其容积最大？如果只考虑底面，就可化成平面问题，定长直线，围成什么曲线使其所围面积最大。（见图 6-5）

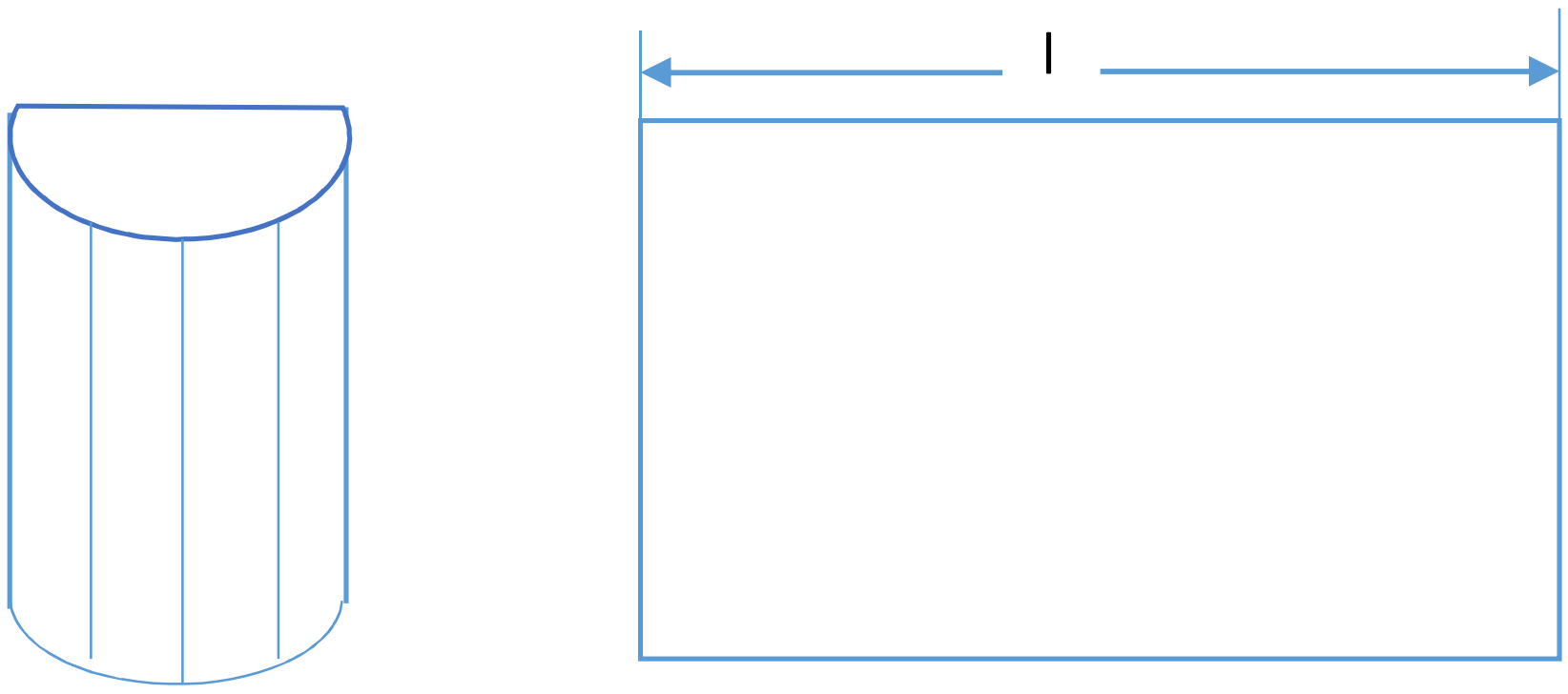

**图 6-5 例 6.4 的示意图**

解：我们很快发现这个是有条件限制（长方形中的长=l）下的泛函极值问题。这就要求我们同时**配合使用 E-L 方程和拉格朗日乘数法**。

回忆起“弧微分”的公式 $ds=\sqrt{1+y'^2}dx$，就可以将题目中的条件写下：

上图中长方形中长的条件 $\int_0^a\sqrt{1+y'^2}dx=l$ ，要优化泛函 $\int_0^a F_l(x,y,y')dx=\int_0^a ydx$，注意到限制条件可写为 $\int_0^a\sqrt{1+y'^2}dx=0$，（我们先忽略 $l$，因为其位置处在积分符号外），如果连积分符号带被积函数整个直接塞入

(6-35)，由于限制条件本身就带有积分符号（而且积分上下限相同），那这么做未免生硬，其实可以直接如下处理更简单：

$$\int_D F(x,y,y')dx + \lambda \cdot G(x,y,y') = \int_0^a y dx + \lambda\left(\int_0^a \sqrt{1+y'^2}dx\right)$$

$$= \int_0^a y + \lambda\sqrt{1+y'^2}dx \quad ，所以 L = \tilde{F}(x,y,y') = y + \lambda\sqrt{1+y'^2}$$

这就得到了在均包含积分符号时候的简便的拉格朗日函数方法的构造方法，即直接提取出上面 2 个公式中的被积函数，其中的$\lambda$便是拉格朗日乘子，特别注意到这个$\lambda$是一个实数而非函数，因为限制是加于整个限制条件之上而非限制在每一个 x 点。后续的处理与标准的拉格朗日乘数法有所不同，因为泛函不需要分别对拉格朗日函数中每个变量求导（求导得 0 解方程那是普通函数求极值的做法！），但替换为直接将这个函数代入 E-L 方程即可。

注意到这个方程不显含$x$，根据降阶的欧拉-拉格朗日公式(6-31)，则可以顺利写出下面的公式：

$$y + \lambda\sqrt{1+y'^2} - \lambda y'\frac{2y'}{2\sqrt{1+y'^2}} = C$$

$$y + \lambda\frac{(1+y'^2) - y'^2}{\sqrt{1+y'^2}} = C$$

$$y = -\frac{\lambda}{\sqrt{1+y'^2}} + C$$

我们画出图 6-6。

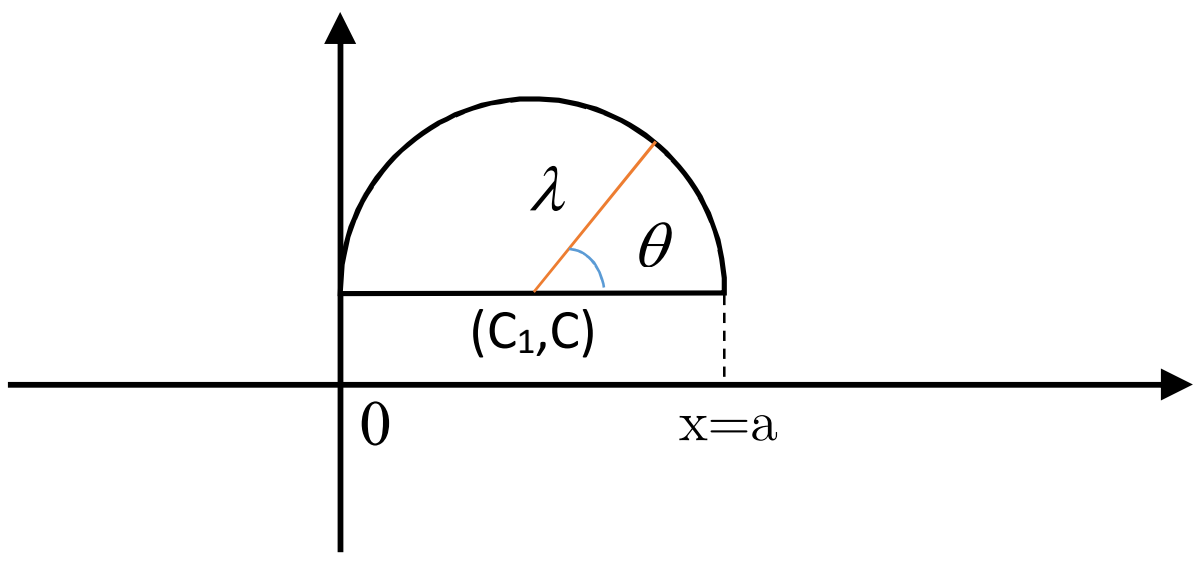

**图 6-6 解出的圆的图像**

回忆起三角函数的世界，抛却那些复杂而灵巧的三角函数公式变换，弱水三千只取一瓢，只需要回忆起 1 个公式即可：$1+\tan^2\theta = \sec^2\theta$

由于我们看到这个公式中包含$\sqrt{1+y'^2}$，这就触发了我们引入三角函数换元的灵感：

设 $y' = \frac{dy}{dx} = \tan\theta$，则原公式变为：（你应该知道 $\sec\theta = 1/\cos\theta$）

$$y = -\frac{\lambda}{\sqrt{1+\tan^2\theta}} + C = -\frac{\lambda}{\sec\theta} + C = -\lambda\cos\theta + C$$

所以 $\frac{dy}{d\theta} = \lambda\sin\theta$

由于 $y' = \frac{dy}{dx} = \tan\theta \ \Rightarrow \ dx = \frac{dy}{\tan\theta} = \frac{\lambda\sin\theta d\theta}{\sin\theta/\cos\theta} = \lambda\cos\theta d\theta$

两边同时积分 $\int dx = \int \lambda\cos\theta d\theta \ \Rightarrow \ x = \lambda\sin\theta + C_1$

这样就解出了 x 和 y 的参数方程组：

$$\begin{cases} x = \lambda\sin\theta + C_1 \\ y = -\lambda\cos\theta + C \end{cases}$$

如果对中学的圆的参数方程还有印象，回想起：$\sin^2\theta + \cos^2\theta = 1$，这就把上面的方程组化为下面的形式：

$$(x - C_1)^2 + (y - C)^2 = \lambda^2$$

这就恰好是圆的标准方程，圆心点在 $(C_1, C)$，半径是 $\lambda$，通过 x=0 和 x=a 两点。最后因为限制条件内的半圆周长=l，所以 $\pi\cdot\lambda = l \ \Rightarrow \ \lambda = l/\pi$。

**例 6.5** 上一个例子的限制条件和泛函交换一下呢？

优化目标：$J(y) = \int_0^a \sqrt{1+y'^2}dx$ ，限制条件：$\int_0^a ydx = A$

也就是说：什么样的曲线下包围给定面积的情况下最短？

跟上一例一样，提取出被积函数，构造拉格朗日函数：

$$L = \tilde{F}(x,y,y') = \sqrt{1+y'^2} + \lambda y$$

然后代入降阶的欧拉-拉格朗日公式(6-31)：

$$\sqrt{1+y'^2} + \lambda y - y'\frac{y'}{\sqrt{1+y'^2}} = C$$

$$y = -\frac{1}{\lambda}\cdot\frac{1}{\sqrt{1+y'^2}} + \frac{C}{\lambda}$$

设 $y' = \frac{dy}{dx} = \tan\theta$，则原公式变为：（你应该知道 $\sec\theta = 1/\cos\theta$）

$$y = -\frac{1}{\lambda}\cos\theta + \frac{C}{\lambda} = -\frac{1}{\lambda}\cos\theta + C_1$$

跟上一例子一样，得到 x 如下：

$$x = \frac{1}{\lambda}\sin\theta + C_2$$

最后同样得到圆的标准方程：

$$(x - C_2)^2 + (y - C_1)^2 = \frac{1}{\lambda^2}$$

还是圆弧！也就是说：如果曲线下方包围的面积给定了，这个曲线在圆弧时候达到曲线长度的最小值！因为可以想象，这条圆弧划出了足够的高度，而又圆满地包围了这个面积。

如果读者足够细心，会发现上例和本例有个区别，一个求极大值，另一个求极小值，没错，E-L 方程与普通函数判定极值时所用的求导得 0 方法一样，无法判定求得是极大还是极小，而仅能判断求得的是极值。

从这道例题中学到的一点便是：如何将欧拉-拉格朗日方程（E-L 方程）与拉格朗日乘数法配合起来使用，这招在后期的变分贝叶斯推导中也有相应使用。

更多变分法的例子请看：

http://www.exampleproblems.com/wiki/index.php/Calculus_of_Variations

## 参考文献

# 第7章 LDA 的变分贝叶斯法

如果说 Collapsed Gibbs Sampling 版本的 LDA 犹如《新约圣经》，那么 Variational Bayes 法推导的 LDA 就像《旧约圣经》一样，如果说一个人学习 LDA 却不知道 LDA 的变分推断方法，就像一个基督徒不读《圣经》一样，这仍然称不上了解 LDA，溯本求源是我们应有的态度。LDA 的变分贝叶斯法既是 LDA 模型的起点，也是一个启发新技术的契机。

## 7.1 Latent Dirichlet Allocation 的另一个视角

"A wise man never knows all, only fools know everything."

--非洲谚语

如果你已经学过 Gibbs Sampling 版的 LDA，从这一章开始，我需要你洗净脑袋，不要以为知道关于 LDA 所有的事情了。在这一章里，以 David M.Blei 的 LDA 原始论文《Latent Dirichlet Allocation》为依据重新组织撰写[33]，但我会补充进许多原始论文里没有的技术细节，以帮助读者理解。

尽管在前面第 3 章的 3.2 节已经详尽介绍过了 LDA 算法的生成规则，这里需要说明的是，历史上第一次出现的 LDA 算法，在 Blei 原始经典论文中一开始的 LDA 的概率模型图不是这个样子(basic 版本)。图 7-1 给出 smoothed 和 basic 两种版本的概率模型图，应该特别注意到β节点的不同，在 basic LDA 的模型图中，β节点直接被用来重复采样生成每个词。

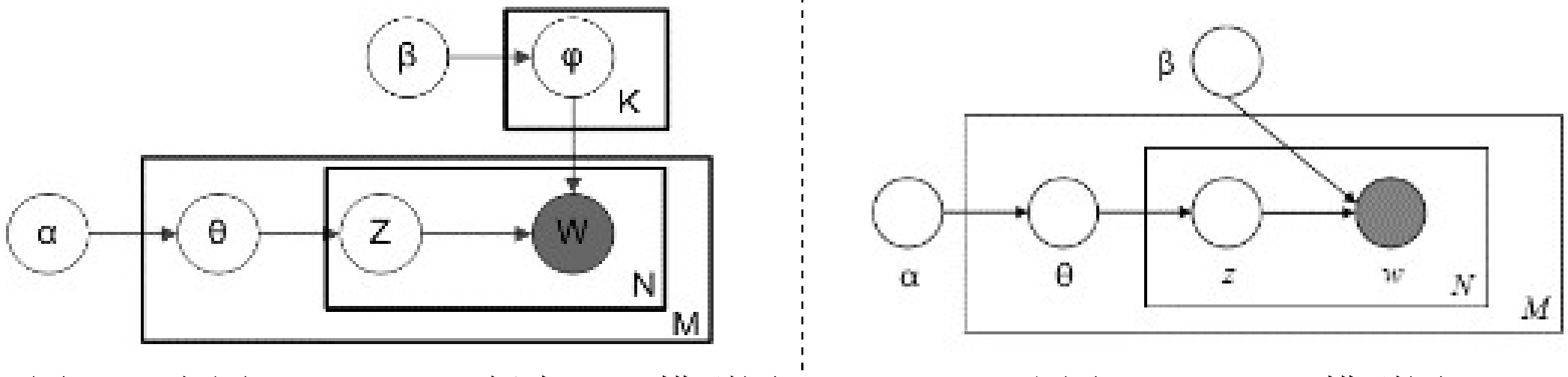

**图 7-1** 左图：**smoothed** 版本 **LDA** 模型图　　右图：**basic LDA** 模型图

在这个故事里，LDA 的生成咒语如下：

1. 选取文档长度（文档中的单词个数） $N \sim possion(\xi)$
2. 选取文档中的主题分布 $\vec{\theta} \sim Dir(\vec{\alpha})$

---

[33] 主要推导部分在第 5 节和附录(Appendix)部分

如前面章节所述，这里的α是 Dirichlet 分布的超参数，一共K维（而K是设置的主题总数），而 $\vec{\theta}=(\theta_0,\theta_1,...,\theta_{K-1})^T$ 是产生该文章主题的多项分布的概率参数，其中 $\theta_i$ 是第 i 个主题被选择的概率，从 Dirichlet 产生参数 $\vec{\theta}$ 后，再用 $\theta_i$ 去产生具体主题 z。

3. 上面两步完成后，进而产生一篇文章 N 个词里的每个词：

For each word $w_n$ in $1\sim N$ words:

(a) 选取一个 topic $z\sim Mult(\vec{\theta})$

(b) 选取一个 word $w_n$ from $p(w_n\mid z_n,\vec{\beta})$

请读者格外注意：这里的β的含义与前文的 Gibbs 版本**有所不同**，它是一个多项分布的参数，是一个 $K\times V$ 的矩阵，表示从 $Z^i$ 到 $W^j$ 的产生概率即：

$$\beta_{ij}=p(w^j=1\mid z^i=1) \tag{7-1}$$

这与 Gibbs 版本的φ矩阵含义相同，**但是，这里的β不是从某个分布（在 Gibbs 版本中，$\vec{\varphi}_k\sim Dir(\vec{\varphi}\mid\vec{\beta})$ ）中抽取的随机量，而是一个确定的量（fixed quantity），这个确定的量就是我们模型中将要估计的**。另外，在推导之前还有些简化模型的工作要做，第1步只是生成文档的长度，Possion 分布假设不是很重要[34]，所以可以使用真实的文档长度分布（document length distribution）来代替。还有，注意到 N（文档的长度即单词个数 N）与其他变量（θ和 z）无关，所以只是一个辅助变量，我们会在后续的推导中忽略这个变量的随机性。

这个推导故事中的 z 指定的使用与 Gibbs 版本略有不同，所以如果熟悉 Gibbs 版本的读者需要稍微重新认识一下 LDA，所有的变量的具体含义见表 7-1。

**表 7-1 basic LDA 各个变量的解释**

| |
|---|
| K：主题个数(用户设置) |
| M：训练预料中的文章数 |
| N：一篇文章内的单词数 |
| V：语料中所有使用到的单词构成的词典单词数 |
| w：word 单词，这个版本的 w 代表文章中的具体单词，具体使用时是这样的：$\vec{w}_d=(w^v_{d,1},w^v_{d,2},...,w^v_{d,N})$ 表示第 d 篇文档里的第 n 个单词[35]，因此是个三维矩阵（三维数组）形式存储<d, n, v>（第 d 篇文档的第 n 个单词是单词 v（单词表 |

[34] 关于 Possion 分布为何物，请看本章后记

[35] 上标表示词表中的单词 v，下标表示来源于文章中的哪个位置

的 word_id=v))。你应该注意到公式(7-1)里的$w^j$里的 j 也是上标，这是因为在这个版本里我们将 word 使用 one-hot 的向量表示（比如在{1,2,3,…,V}中只有用到的单词 v 为 1，其他皆为 0，因此$w^v = 1$，而$w^u = 0$，$u \neq v$）

z：对每个的单词的主题的分配指定。如同 w，这个版本里 z 也与前面介绍的 Gibbs 版本略有不同，主要区别在于这个 z 是一个三维的矩阵，存储方式为 <d,n,i>，表示第 d 篇文档里的第 n 个单词是否是编号为 i 的主题。

θ：与 Gibbs 版本一样，文章的主题分布

α：与 Gibbs 版本一样，θ的超参数，$\theta \sim Dir(\vec{\theta} \mid \vec{\alpha})$

β：表示从$z^i$主题到$w^j$单词的产生概率。它是一个多项分布的参数，是一个$K \times V$的矩阵，表示从$z^i$到$w^j$的产生概率即$\beta_{ij} = p(w^j = 1 \mid z^i = 1)$，这与 Gibbs 版本的φ矩阵含义类似，但有所区别的是**β是确定的量（fixed quantity）**。

Doc：文档使用$\vec{w} = (w_1, w_2, ..., w_N)$，这里的$w_n$是第 n 个单词（正如前文所述，每个 word 是个向量）

Corpus：训练语料库由 M 篇文档组成，$D = \{\vec{w}_1, \vec{w}_2, \vec{w}_3, ..., \vec{w}_M\}$

## 7.2 分析模型

解决这个模型的关键之处在于求解给定一篇文档时的隐变量(latent variable)的后验分布（posterior distribution）：

$$p(\vec{\theta}, \vec{z} \mid \vec{w}, \vec{\alpha}, \vec{\beta}) = \frac{p(\vec{\theta}, \vec{z}, \vec{w} \mid \vec{\alpha}, \vec{\beta})}{p(\vec{w} \mid \vec{\alpha}, \vec{\beta})} \tag{7-2}$$

这个公式是 LDA 变分推导的症结所在，我们这里的做法是先分析每一篇独立的文档，整个语料集的概率就是将一个个独立的概率乘起来，这也就意味着文档之间的概率相互独立。

首先我们观察公式(7-2)的分子，根据模型图 7-1 右图将分子分解开：

$$p(\vec{\theta}, \vec{z}, \vec{w} \mid \vec{\alpha}, \vec{\beta}) = p(\vec{w} \mid \vec{z}, \vec{\beta}) p(\vec{z} \mid \vec{\theta}) p(\vec{\theta} \mid \vec{\alpha}) \tag{7-3}$$

我们一项一项来看，这里的$p(\vec{w} \mid \vec{z}, \vec{\beta})$代表了给定一篇文章的长度为 N（单词个数为 N），其中每个词从$K \times V$的概率矩阵β中按照每个词的主题分配 z 选取，

在该情况下的该文章的概率。所以我们可以将其分解成每一个单词(word n)的概率相乘起来：

$$p(\vec{w} \mid \vec{z}, \vec{\beta}) = \prod_{n=1}^{N} \beta_{z_n, w_n} \tag{7-4}$$

接下来，$p(z_n \mid \vec{\theta})$ 就是 $\vec{\theta}$ 向量中第 i 个 topic 主题编号分量的概率 $\theta_i$，该概率使得当前文章第 n 个单词的第 i 个 topic 的主题分配 $z_n^i = 1$。最后 $p(\vec{\theta} \mid \vec{\alpha})$ 来源于 Dirichlet 分布的公式(2-10)。

综上所述，现在已经可以推导出给定参数α和β时，文档的主题分布 $\vec{\theta}$ 、topic 分配的集合 z 、和一篇文档内的单词集 $\vec{w}$ 的联合概率公式：

$$p(\vec{\theta}, \vec{z}, \vec{w} \mid \vec{\alpha}, \vec{\beta}) = p(\vec{\theta} \mid \vec{\alpha}) \prod_{n=1}^{N} p(z_n \mid \vec{\theta}) p(w_n \mid z_n, \vec{\beta}) \tag{7-5}$$

$$= \left( \frac{\Gamma(\sum_{i=1}^{K} \alpha_i)}{\prod_{i=1}^{K} \Gamma(\alpha_i)} \prod_{i=1}^{K} \theta_i^{\alpha_i - 1} \right) \prod_{n=1}^{N} \theta_{z_n} \beta_{z_n, w_n} \tag{7-6}$$

这里的 $\theta_{z_n}$ 代表 $\vec{\theta}$ 向量中的第 $z_n$ 个分量，我们可以重新将其表达为指数上标的表示法：利用整个字典长度 V，由于 $p(z_n \mid \vec{\theta}) = \prod_{j=1}^{V} (\theta_{z_n})^{w_n^j} = \prod_{i=1}^{K} \prod_{j=1}^{V} (\theta_i)^{w_n^j z_n^i}$，（对语料词典中所有单词 $j \in \{1, \ldots, V\}$ 连乘中只有当前文章第 n 个单词 $w_n^j = 1$），同理可得 $p(w_n \mid z_n, \vec{\beta}) = \beta_{z_n, w_n} = \prod_{i=1}^{K} \prod_{j=1}^{V} (\beta_{i,j})^{w_n^j z_n^i}$，因此公式(7-6)也可以写为：

$$p(\vec{\theta}, \vec{z}, \vec{w} \mid \vec{\alpha}, \vec{\beta}) = \left( \frac{\Gamma(\sum_{i=1}^{K} \alpha_i)}{\prod_{i=1}^{K} \Gamma(\alpha_i)} \prod_{i=1}^{K} \theta_i^{\alpha_i - 1} \right) \prod_{n=1}^{N} \prod_{i=1}^{K} \prod_{j=1}^{V} (\theta_i \beta_{i,j})^{w_n^j z_n^i} \tag{7-7}$$

为了加深理解，图 7-2 进一步做了解释：

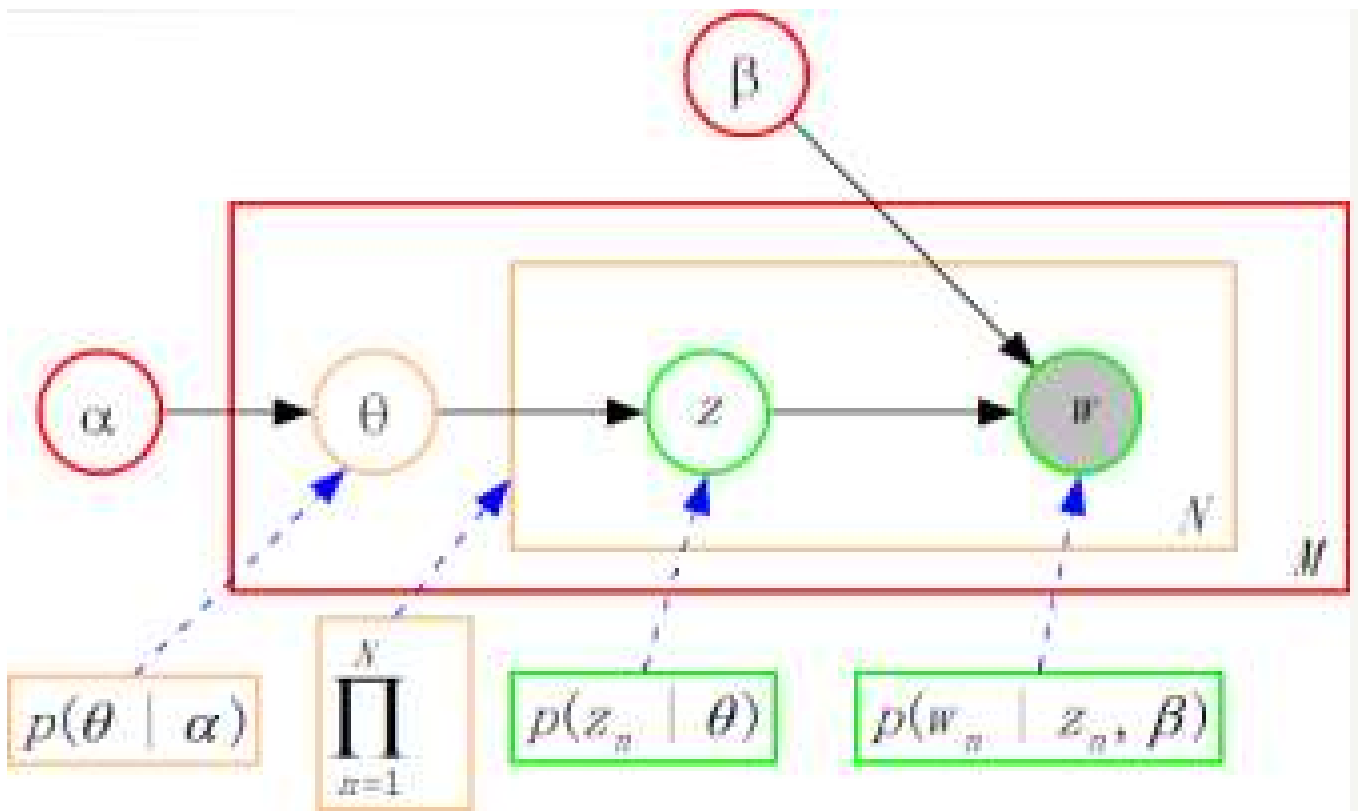

**图 7-2 生成一篇文档概率的解释图**

在图 7-2 中，LDA 的三个表示层被用三种颜色表示了出来：

1. corpus-level （红色）：α和β是语料级别的参数，也就是说对于每个文档都是一样的，因此在生成整个文档集(corpus)过程中只需要采样(sample)一次。

2. document-level （橙色）：θ是文档级别的参数，意即每个文档的θ参数是不一样的，所以对于每个文档都要采样(sample)一次θ。

3. word-level （绿色）：最后 z 和 w 都是单词级别的变量，单词级别的变量 w 和 z 为每篇文档的每个单词采样(sample)一次。

现在我们来看(7-2)的分母：对公式(7-5)中的θ求积分，对 z 的可能值求和，消去θ和 z 之后，就可以得到**一篇文档**的边缘分布。

$$p(\vec{w} \mid \vec{\alpha}, \vec{\beta}) = \int p(\vec{\theta} \mid \vec{\alpha}) \left( \prod_{n=1}^{N} \sum_{z_n} p(z_n \mid \vec{\theta}) p(w_n \mid z_n, \vec{\beta}) \right) d\vec{\theta} \tag{7-8}$$

公式(7-8)中的 $z_n$ 是指**一篇文档**上的第 n 个单词的主题分配，N 表示该文档的单词总数。

此处可以进一步理解下公式(7-8)，公式(7-8)中将 z 的所有可能值求和就可以消去 z，这实际上就构成了给定一篇文档中单词（word） $w_i$ 的概率，因此求和符号 sigma 里那一部分便可以写成：

$$p(w \mid \vec{\theta}, \vec{\beta}) = \sum_{z} p(w \mid z, \vec{\beta}) p(z \mid \vec{\theta}) \tag{7-9}$$

注意到这个概率是一个随机变量因其依赖于θ，我们现在可以定义下面所述的对于一篇文档的生成过程（generative process）：

1. 选取文档的主题分布 $\vec{\theta} \sim Dir(\vec{\alpha})$
2. For each word $w_n$ of N words:
   a) 从概率 $p(w \mid \vec{\theta}, \vec{\beta})$ 中选取一个单词 $w_n$

这个过程也就将文档边缘分布以连续混合分布（continuous mixture distribution）定义出来[36]：

$$p(\vec{w} \mid \vec{\alpha}, \vec{\beta}) = \int p(\vec{\theta} \mid \vec{\alpha}) \left( \prod_{n=1}^{N} p(w_n \mid \vec{\theta}, \vec{\beta}) \right) d\vec{\theta} \tag{7-10}$$

可以发现公式(7-8)实际就是公式(7-9)代入公式(7-10)之后的结果。公式(7-10)中 $p(w_n \mid \vec{\theta}, \vec{\beta})$ 可看成是连续混合分布中的混合成分（mixture components）， $p(\vec{\theta} \mid \vec{\alpha})$ 可看成是混合权重（mixture weights）。

[36] 与高斯混合模型类似

最后，将各个单文档的边缘分布乘起来（共 M 篇文章），就可得到整个语料集的概率：

$$p(D \mid \vec{\alpha},\vec{\beta}) = \prod_{d=1}^{M} \int p(\vec{\theta}_d \mid \vec{\alpha}) \left( \prod_{n=1}^{N} \sum_{z_{dn}} p(z_n \mid \vec{\theta}_d) p(w_{dn} \mid z_{dn},\vec{\beta}) \right) d\theta_d \tag{7-11}$$

公式(7-11)中的 $z_{dn}$ 指的是第 d 篇文章中的第 n 个词的主题分配。

如前所述，解决这个模型的关键之处在于求解(7-2)，但不幸的是，这个公式非常难以求解（intractable）：分子是公式(7-5)，而分母是公式(7-8)，问题就在于这个分母，仔细观察分母，可以将公式(7-8)替换为具体的模型参数，便将公式(7-8)写成：

$$p(\vec{w} \mid \vec{\alpha},\vec{\beta}) = \frac{\Gamma(\sum_i \alpha_i)}{\prod_i \Gamma(\alpha_i)} \int \left( \prod_{i=1}^{K} \theta_i^{\alpha_i - 1} \right) \left( \prod_{n=1}^{N} \sum_{i=1}^{K} \prod_{j=1}^{V} (\theta_i \beta_{ij})^{w_n^j} \right) d\theta \tag{7-12}$$

公式(7-12)的一点解释：在这里使用公式(2-10)替换掉公式(7-8)中的 $p(\vec{\theta} \mid \vec{\alpha})$，并使用了式(7-13)和(7-14)的替换（这个过程类似于公式(7-7)）。

$$p(z_n \mid \vec{\theta}) = \prod_{j=1}^{V} (\theta_{z_n})^{w_n^j} \tag{7-13}$$

$$p(w_n \mid z_n,\vec{\beta}) = \prod_{j=1}^{V} (\beta_{z_n,j})^{w_n^j} \tag{7-14}$$

由于公式(7-12)在对潜在主题 z 的求和 sigma 中出现了一对θ和β，所以造成难以求解（intractable）（Dickey, 1983）[1]，由于精确推断（exact inference）不可得，所以 Blei 决定另辟蹊径，采用近似推断（approximate inference）。

## 7.3 推导

### 7.3.1 启蒙

我们先来看看标准的变分贝叶斯（Variational Bayes）求解方法：

变分推断（Variational Inference）的目的是要找出最接近于真实模型的隐变量后验概率分布 p 的那个近似分布 q，然后用 q 来替代 p，那么问题来了，“接近”这个概念如何在具体计算中体现，也就是说如何衡量两个概率分布的相似度，幸运的是，这已经有了现成的工具：通常使用 KL 距离。稍做一点简单的推导，就会发现“优化变分下界”（Evidence lower bound/ELBO）的诀窍。

近似分布 q 到真实分布 p 的 KL 距离的距离公式 $D_{KL}(q \mid\mid p)$，这里使用 q 代表近似变分分布，p 代表真实分布（推导中的 D 是指可观测到的数据 Data，Z 是指不可观测到的隐变量）：

$$D_{KL}(q \mid\mid p) = KL(q(Z) \mid\mid p(Z \mid D)) = \sum_z q(z) \log \frac{q(Z)}{p(Z \mid D)}$$

$$= \sum_z q(z) \log \frac{q(Z)p(D)}{p(Z,D)}$$

$$= \sum_z q(Z)[\log \frac{q(Z)}{p(Z,D)} + \log p(D)]$$

由于 $\sum_z Q(Z) = 1$，因此应用乘法分配律：

$$= \sum_z \left[ q(Z) \log \frac{q(Z)}{p(Z,D)} \right] + \log p(D)$$

式子右边的 $\log P(D)$ 被称为已知/观测到的数据的对数证据(evidence)或对数似然函数(log-likehood)，此时将 $\log p(D)$ 移项到等式左边，即可得到：

$$\log p(D) = D_{KL}(q \mid\mid p) - \underbrace{\sum_z q(Z) \log \frac{q(Z)}{p(Z,D)}}_{E_q\left(\log \frac{q(Z)}{p(Z,D)}\right)} \tag{7-15}$$

(7-15)式中的 $-\sum_z q(Z) \log \frac{q(Z)}{p(Z,D)}$ 便是所谓的变分证据下界（Evidence lower bound/ELBO)，应用函数期望公式(6-8)，所以可以将 ELBO 写作:

$$L(q) = -\sum_z q(Z) \log \frac{q(Z)}{p(Z,D)} = E_q[\log p(Z,D)] - E_q[\log q(Z)] \tag{7-16}$$

我们从另一个角度看下这个 ELBO 的下界公式(7-16)，回忆起 EM 算法推导中出现过的 Jesen 不等式，如果 f 是一个凹函数[37]（concave)，那么根据 Jesen 不等式：

$$f(E[X]) \geq E[f(X)] \tag{7-17}$$

[37] 这里凹函数(concave)和凸函数(convex)与我国中学教材定义相反

如果读者没有学过 Jesen 不等式，请回顾第 6 章 6.1.6 节。

因为对数函数 log 属于凹函数（concave）：

$$\log p(D)=\log\sum\nolimits_{z}p(Z,D)$$

$$=\log\left[\sum_{z}\left(p(Z,D)\frac{q(Z)}{q(Z)}\right)\right]$$

$$=\log\left[E_{q}\left(\frac{p(Z,D)}{q(Z)}\right)\right]$$

$$\geq E_{q}[\log p(Z,D)]-E_{q}[\log q(Z)]$$

因为是不等式的大于等于关系，所以 ELBO 确实是 $\log p(D)$ 的下界(bound)。从式(7-15)就会发现最小化(minimize)KL 距离等价于最大化(maximize) ELBO 式(7-16)，即：

$$\text{minimize KL} \Leftrightarrow \text{maximize ELBO} \tag{7-18}$$

因此推导思路转向研究如何选取 q 函数使得 ELBO 最大，也就是所谓的泛函极值问题。经过漫长复杂的推导，将会得到这个 q 函数。如果做出这个推导并观察规律会发现，在 p 分布函数为指数分布族的情况下，求得的 q 函数也会为属于指数分布族的同样的分布函数。(譬如 p 和 q 都服从 Dirichlet 分布)。

### 7.3.2 变分目标函数

Blei 觉得上文中标准的变分推断（variational inference）比较复杂，他使用了较为简单的方法：convexity-based variational inference。Blei 首先将原有的 LDA 模型图进行了简化，便得到了变分分布模型作为原模型的近似替代。为了方便，这里将原始 LDA 的概率模型图重新画下来，并画出用于替代的变分分布模型图（见图 7-3）。

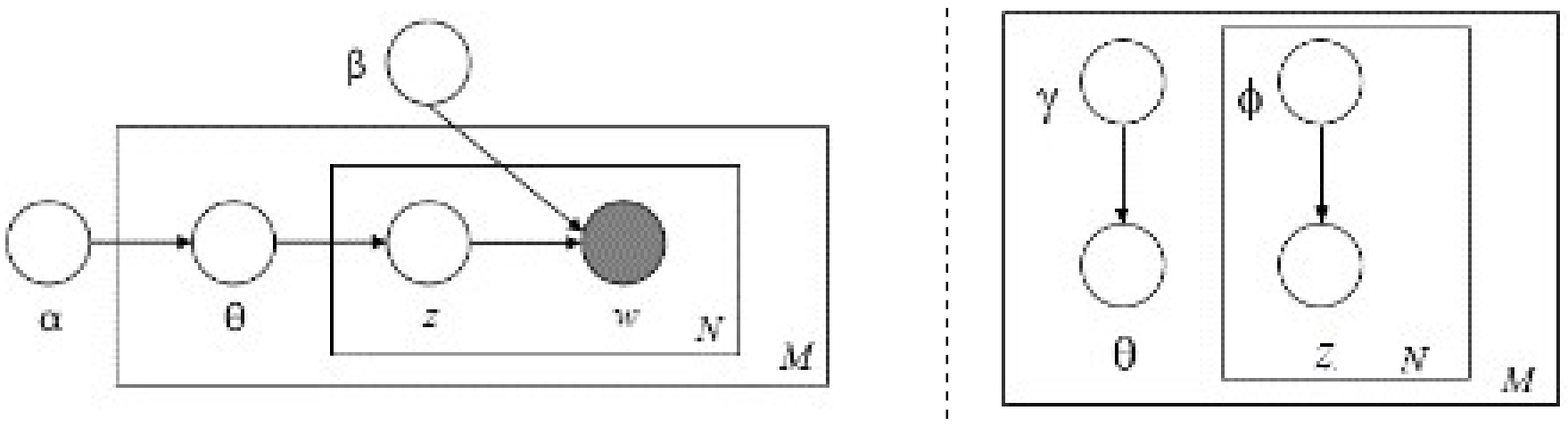

**图 7-3 左图：LDA 概率模型图** **右图：用于近似 LDA 后验分布的变分分布概率模型图**

正是因为θ和β通过节点θ、z 和 w 的连接边耦合（couple）在一起，造成难以求解，又由于原分布采用了指数分布族的 Dirichlet 分布作为θ的先验分布：

$\vec{\theta} \sim Dir(\vec{\alpha})$ 、多项分布（Multinomial Distribution）作为 z 服从的分布：$\vec{z} \sim Mult(\vec{\theta})$，那么就意味着变分分布函数也应该如此。仍然采用 Dirichlet 分布和 Multinomial 分布作为先验分布，绕开标准的变分贝叶斯，直接假设出一个变分分布函数 q 函数，接着我们就回到了我们熟悉的函数求极值方法：求导。

经过上述分析，只要简化原有模型，去掉多余的边和节点（比如节点 w）这个被简化的模型如图 7-3 右图所示。用简化的变分模型的目的是要找出这个最接近于真实分布的变分模型的参数，类似于变分法找出最接近真实极限函数的那个函数。

值得注意的是，图 7-3 右图简化模型中引入了两个新变量：γ和ϕ，用于替代原有的超参数。在这个变分分布中，$\vec{\theta} \sim Dir(\vec{\theta}|\vec{\gamma})$，所以这里γ的维度与θ相同，是文档级别（document level）的参数，每篇文档对应的γ不同，但都为K长度的向量。而$\vec{z} \sim Mult(\vec{z} \mid \vec{\phi})$，所以ϕ的维度与 z 相同，是三维矩阵<d, n, i>。为了简便，后面的公式书写中省略了向量的箭头，用加粗代替。

按照简化后的变分模型图 7-3 右图，变分分布的分布函数如下：

$$q(\theta, \mathbf{z} \mid \gamma, \phi) = q(\theta \mid \gamma)\prod_{n=1}^{N} q(z_n \mid \phi_n) \tag{7-19}$$

式(7-19)中的 N 指的是一篇文章的词数，这里的 Dirichlet 分布参数γ和多项分布（multinomial distribution）参数ϕ均被称为自由变分参数（free variational parameter）。现在工作转向寻找最优化的变分参数γ和ϕ，也就是找到使得变分分布和真实分布 KL 距离最小的最优化的参数：

$$(\gamma^*, \phi^*) = \arg\min_{(\gamma,\phi)} D_{KL}(q(\theta, \mathbf{z} \mid \gamma, \phi) \,||\, p(\theta, \mathbf{z} \mid \mathbf{w}, \alpha, \beta)) \tag{7-20}$$

如同(7-15)的推导，我们可以得到已知数据的对数证据（evidence），如下：

$$\log p(\mathbf{w} \mid \alpha, \beta) = L(\gamma, \phi; \alpha, \beta) + D_{KL}(q(\theta, \mathbf{z} \mid \gamma, \phi) \,||\, p(\theta, \mathbf{z} \mid \mathbf{w}, \alpha, \beta)) \tag{7-21}$$

其中 w 代表当前的一篇文章，回忆起等价关系(7-18)，此时我们可以转向优化变分下界。

### 7.3.3 下界（lower bound）

最基本的思路与标准变分贝叶斯方法一样，Blei 首先利用 Jesen 不等式找到对数似然函数 $\log p(\mathbf{w} \mid \alpha, \beta)$（该对数似然函数也被称为 evidence）的可调节的

ELBO 下界，与公式(7-5)到公式(7-9)推导一样，为求简洁，在 q 函数的书写中中省略了γ和ϕ，将该对数似然公式写出：

$$\log p(\mathbf{w}\mid\alpha,\beta)=\log\int\sum_{\mathbf{z}}p(\theta,\mathbf{z},\mathbf{w}\mid\alpha,\beta)d\theta$$

$$=\log\int\sum_{\mathbf{z}}\frac{p(\theta,\mathbf{z},\mathbf{w}\mid\alpha,\beta)q(\theta,\mathbf{z})}{q(\theta,\mathbf{z})}d\theta$$

可以观察到上式实际就是第 6 章的二维随机变量函数的数学期望公式(6-13)：

$$=\log E_{q(\theta,\mathbf{z})}\left(\frac{p(\theta,\mathbf{z},\mathbf{w}\mid\alpha,\beta)}{q(\theta,\mathbf{z})}\right)$$

利用 Jesen 不等式$f[E(x)]\geq E[f(x)]$，可得下界：

$$\geq E_q\left(\log\frac{p(\theta,\mathbf{z},\mathbf{w}\mid\alpha,\beta)}{q(\theta,\mathbf{z})}\right)=\int\sum_{\mathbf{z}}q(\theta,\mathbf{z})\log\left(\frac{p(\theta,\mathbf{z},\mathbf{w}\mid\alpha,\beta)}{q(\theta,\mathbf{z})}\right)d\theta$$

$$=\int\sum_{\mathbf{z}}q(\theta,\mathbf{z})\log p(\theta,\mathbf{z},\mathbf{w}\mid\alpha,\beta)d\theta-\int\sum_{\mathbf{z}}q(\theta,\mathbf{z})\log q(\theta,\mathbf{z})d\theta$$

$$=E_q[\log p(\theta,\mathbf{z},\mathbf{w}\mid\alpha,\beta)]-E_q[\log q(\theta,\mathbf{z})] \quad (7\text{-}22)$$

(7-22)式便是下界（lower bound）了，现在先来看(7-22)式最右边一项，这里我们做平均场假设，也就是假设可以因式分解，可以得到：

$$q(\theta,\mathbf{z}\mid\gamma,\phi)=q(\theta\mid\gamma)q(\mathbf{z}\mid\phi) \quad (7\text{-}23)$$

值得注意的是，实际情况中，主题的分配 z 的与文章的主题分布θ并不是独立的，这也就造成了一定问题，Blei 管不了这么多了，现在不是讨论这个的时候，他只需要大胆地将他的逻辑向前推进，而将其中的小问题留给了后来人来解决。根据对数函数的性质$\log(A\cdot B)=\log A+\log B$，所以可得：

$$\log q(\theta,\mathbf{z}\mid\gamma,\phi)=\log q(\theta\mid\gamma)+\log q(\mathbf{z}\mid\phi) \quad (7\text{-}24)$$

来看等式(7-22)的左边那一项，这一项的分布为真实分布 p，观察图 7-3 的 LDA 模型图可得到因式分解：

$$\log p(\theta,\mathbf{z},\mathbf{w})=\log p(\theta\mid\alpha)+\log p(\mathbf{z}\mid\theta)+\log p(\mathbf{w}\mid\mathbf{z},\beta) \quad (7\text{-}25)$$

又利用到关于期望的性质公式(6-7)，便可以将下界写为：

$$\begin{aligned}L(\gamma,\phi;\alpha,\beta)=&E_q[\log p(\theta\mid\alpha)]+E_q[\log p(\mathbf{z}\mid\theta)]+E_q[\log p(\mathbf{w}\mid\mathbf{z},\beta)]\\&-E_q[\log q(\theta)]-E_q[\log q(\mathbf{z})]\end{aligned} \quad (7\text{-}26)$$

注意到该式一个鲜明的特点是同时融合了模型参数和变分参数，也即图 7-3 左图的参数α和β，以及右图的参数γ和ϕ同时出现在里面，这就为我们后期的迭代优化提供了契机。

### 7.3.4 下界展开

还记得那句台词吗？“集齐七颗龙珠，可以召唤神龙”。公式(7-26)中一共 5 项，现在需要集齐这 5 项的展开表达式，我们逐一来看。

**第 1 项 ：**

需要注意到 $p(\theta \mid \alpha)$ 是服从 Dirichlet 分布的，因此：

$$p(\theta \mid \alpha) = \frac{\Gamma(\sum_{i=1}^{K}\alpha_i)}{\prod_{i=1}^{K}\Gamma(\alpha_i)}\prod_{i=1}^{K}\theta_i^{\alpha_i - 1}$$

所以：

$$E_q[\log p(\theta \mid \alpha)] = E_q[\sum_{i=1}^{K}(\alpha_i - 1)\log\theta_i + \log\Gamma(\sum_{i=1}^{K}\alpha_i) - \sum_{i=1}^{K}\log\Gamma(\alpha_i)]$$

又因为 q 函数不含α，所以可以将包含α的项从期望中提出：

$$E_q[\log p(\theta \mid \alpha)] = \left(\sum_{i=1}^{K}(\alpha_i - 1)E_q(\log\theta_i)\right) + \log\Gamma(\sum_{i=1}^{K}\alpha_i) - \sum_{i=1}^{K}\log\Gamma(\alpha_i)$$

再利用第 6 章的公式(6-14)可以轻松得到：

$$E_{q(\theta|\gamma)}[\log[\theta_i]] = \Psi(\gamma_i) - \Psi(\sum_{j=1}^{K}\gamma_j)$$

因此，第一项最终得到：

$$E_q[\log p(\theta \mid \alpha)] = \left(\sum_{i=1}^{K}(\alpha_i - 1)(\Psi(\gamma_i) - \Psi(\sum_{j=1}^{K}\gamma_j))\right) + \log\Gamma(\sum_{i=1}^{K}\alpha_i) - \sum_{i=1}^{K}\log\Gamma(\alpha_i)$$

**第 2 项：**

回忆起表 7-1 中对 z 的解释，但是这里的变分推断只针对一篇文档分析，用 $z_{n,i} = 1$ 表示当前文档分配第 n 个词被指定到第 i 号 topic，否则， $z_{n,i} = 0$。并且利用 $\vec{z} \sim Mult(\vec{\theta})$，就可以直接展开第 2 项。

$$E_q[\log p(\mathbf{z} \mid \theta)] = \sum_{n=1}^{N}E_q[\log p(z_n \mid \theta)]$$

$$= \sum_{n=1}^{N}\sum_{i=1}^{K}E_q[\log p(z_{n,i} \mid \theta_i)]$$

$$= \sum_{n=1}^{N}\sum_{i=1}^{K}E_q[\log(\theta_i)^{z_{n,i}}] \qquad // \vec{z} \sim Mult(\vec{\theta})$$

$$= \sum_{n=1}^{N}\sum_{i=1}^{K} E_q[z_{n,i} \log\theta_i]$$

$$= \sum_{n=1}^{N}\sum_{i=1}^{K} \underbrace{E_q(z_{n,i})}_{\int(\phi_{n,i}\cdot z_{n,i})dz} \cdot \underbrace{E_q(\log\theta_i)}_{\Psi(\gamma_i)-\Psi(\sum_{j=1}^{K}\gamma_j)} \qquad // \theta \sim Dir(\theta \mid \gamma)$$

$$= \sum_{n=1}^{N}\sum_{i=1}^{K} \phi_{n,i}\left(\Psi(\gamma_i) - \Psi(\sum_{j=1}^{K}\gamma_j)\right)$$

**第 3 项：**

$$E_q[\log(\mathbf{w} \mid \mathbf{z}, \beta)] = \sum_{n=1}^{N} E_q[\log p(w_n \mid z_n, \beta)]$$

$$= \sum_{n=1}^{N}\sum_{i=1}^{K} E_q[\log(\beta_{i,w_n})^{z_{n,i}}]$$

$$= \sum_{n=1}^{N}\sum_{i=1}^{K} E_q(z_{n,i}) E_q[\log(\beta_{i,w_n})]$$

$$= \sum_{n=1}^{N}\sum_{i=1}^{K} \phi_{n,i} \log(\beta_{i,w_n}) \qquad //\text{q 函数不含}\beta$$

**第 4 项：**

$$-E_q[\log q(\theta \mid \gamma)] = -E_q\left[\log\left(\frac{\Gamma(\sum_{i=1}^{K}\gamma_i)}{\prod_{i=1}^{K}\Gamma(\gamma_i)}\prod_{i=1}^{K}\theta_i^{\gamma_i - 1}\right)\right] \qquad // \theta_i \sim Dir(\theta \mid \gamma)$$

$$= -E_q[\log\Gamma(\sum_{i=1}^{K}\gamma_i)] + E_q[\sum_{i=1}^{K}\log\Gamma(\gamma_i)] - \sum_{i=1}^{K}(\gamma_i - 1)\cdot \underbrace{E_q(\log\theta_i)}_{\Psi(\gamma_i)-\Psi(\sum_{j=1}^{K}\gamma_j)}$$

$$= -\log\Gamma(\sum_{i=1}^{K}\gamma_i) + \sum_{i=1}^{K}\log\Gamma(\gamma_i) - \sum_{i=1}^{K}(\gamma_i - 1)\cdot(\Psi(\gamma_i) - \Psi(\sum_{j=1}^{K}\gamma_j))$$

**第 5 项：**

由于 $z_{n,i} \sim Multinomial(z \mid \phi)$，其中第 n 个词的 topic 指定是 i 号 topic 时 $z_{n,i} = 1$，否则 $z_{n,i} = 0$，可将 $z_{n,i}$ 提到指数，仿照(7-7)式的处理，可得：

$$-E_q[\log q(\mathbf{z} \mid \phi)] = -E_q[\log\left(\prod_{n=1}^{N}\prod_{i=1}^{K}(\phi_{n,i})^{z_{n,i}}\right)]$$

$$= -E_q[\sum_{n=1}^{N}\sum_{i=1}^{K} z_{n,i}\log(\phi_{n,i})]$$

$$= -\sum_{n=1}^{N}\sum_{i=1}^{K} E_q(z_{n,i}\cdot\log\phi_{n,i})$$

$$= -\sum_{n=1}^{N}\sum_{i=1}^{K} E_q(z_{n,i})\cdot E_q(\log\phi_{n,i})$$

$$= -\sum_{n=1}^{N}\sum_{i=1}^{K}\phi_{n,i}\cdot\log\phi_{n,i}$$

**最后，五项齐全，可以写出下界：**

$$\begin{aligned} L(\gamma,\phi;\alpha,\beta) = &\left(\sum_{i=1}^{K}(\alpha_i - 1)(\Psi(\gamma_i) - \Psi(\sum_{j=1}^{K}\gamma_j))\right) + \log\Gamma(\sum_{i=1}^{K}\alpha_i) - \sum_{i=1}^{K}\log\Gamma(\alpha_i) \\ &+\sum_{n=1}^{N}\sum_{i=1}^{K}\phi_{n,i}\left(\Psi(\gamma_i) - \Psi(\sum_{j=1}^{K}\gamma_j)\right) \\ &+\sum_{n=1}^{N}\sum_{i=1}^{K}\phi_{n,i}\log(\beta_{i,w_n}) \\ &-\log\Gamma(\sum_{i=1}^{K}\gamma_i) + \sum_{i=1}^{K}\log\Gamma(\gamma_i) - \sum_{i=1}^{K}(\gamma_i - 1)\cdot(\Psi(\gamma_i) - \Psi(\sum_{j=1}^{K}\gamma_j)) \\ &-\sum_{n=1}^{N}\sum_{i=1}^{K}\phi_{n,i}\cdot\log\phi_{n,i} \end{aligned} \quad (7\text{-}27)$$

你一定认为 Blei 已经迷失在变分下界函数的汪洋大海了，其实得到这一步之后，后面的推导已经非常简单，在下面的小节中，将会展示如何寻找最优变分参数$\phi$和$\gamma$，令其最大化(maximize)变分下界。

### 7.3.5 变分推断

其实 Blei 这个版本的变分推断更像是变分贝叶斯和极大似然估计混合的版本。迭代中的第一个步骤被称为变分推断，第二个步骤被称为参数估计。这个方法其实就是变分EM（Variational EM），不了解普通EM算法的读者请参考 Bishop 写的《Pattern Recognition and Machine Learning》的 9.2~9.4 节（2010），在E步骤，我们使用近似于原始 LDA 后验分布的**变分分布**去寻找变分参数（variational parameter）的最优解。在M步骤，使用E步骤中已获得的变分参数估计去估计

（比如极大似然估计（parameter estimate））原模型参数（model parameter），从而进一步最大化模型的下界（bound）。具体到 LDA 中：

E步骤：目标找到最优的变分参数 $\gamma_d^*$ 和 $\phi_d^*$ 以使得下界函数最大化，见图 7-3 右图所示，这一步的 $\gamma_d^*$ 和 $\phi_d^*$ 是 document level，因此只针对一篇文章优化。获得这些参数之后可以去计算完整数据的 log-likehood 的期望（expectation of the log likehood of the complete data）。

M步骤：是假设已知变分参数ϕ和γ去估计模型参数α和β，这一步的α和β针对整个语料集优化。使用极大似然估计的估计似然函数为：

$$l(\alpha,\beta)=\sum_{d=1}^{M}\log p(\mathbf{w}_d \mid \alpha,\beta)$$

我们先来看 E 步骤，需要注意到在下界公式(7-27)中，注意该公式指的是一篇文章的变分下界。

1. **maximize wrt.** $\phi_{n,i}$

先来研究 $\phi_{n,i}$，也就是第 n 个单词是被第 i 个潜在 topic 产生的概率。应该特别注意到一个词在不同主题的概率之和等于 1，即限制条件：

$$\sum_{i=1}^{K}\phi_{n,i}=1 \tag{7-28}$$

注意该限制条件与 Gibbs 版本的φ归一化，即第 3 章的公式(3-22)不同。

将变分下界公式(7-27)只留下有关 $\phi_{n,i}$ 的项，令 $\beta_{i,v}$ 为令 $p(w_n^v=1\mid z^i=1)$ 的合适的那个单词 v，也就是令 $w_n^v=1$ 的 v。由于存在限制条件(7-28)，利用拉格朗日乘数法求解条件极值，加入拉格朗日乘子 $\lambda_n$：

$$\begin{aligned}L_{[\phi_{n,i}]}=&\sum_{n=1}^{N}\sum_{i=1}^{K}\phi_{n,i}\left(\Psi(\gamma_i)-\Psi(\sum_{j=1}^{K}\gamma_j)\right)+\sum_{n=1}^{N}\sum_{i=1}^{K}\phi_{n,i}\log(\beta_{i,w_n})\\&-\sum_{n=1}^{N}\sum_{i=1}^{K}\phi_{n,i}\cdot\log\phi_{n,i}\\&+\lambda_n(\sum_{j=1}^{K}\phi_{n,j}-1)\end{aligned} \tag{7-29}$$

关于 $\phi_{n,i}$ 求导(其中 $\phi_{n,i}\log(\beta_{i,v})$ 一项用到导数的乘积法则 $(uv)'=u'v+uv'$)：

$$\frac{\partial L_{[\phi_{n,i}]}}{\partial\phi_{n,i}}=\Psi(\gamma_i)-\Psi(\sum_{j=1}^{K}\gamma_j)+\log(\beta_{i,v})-\log\phi_{n,i}-1+\lambda_n$$

令其等于 0，可得：

$$\Psi(\gamma_i) - \Psi(\sum_{j=1}^{K}\gamma_j) + \log(\beta_{i,v}) - \log\phi_{n,i} - 1 + \lambda_n = 0$$

将 $\log\phi_{n,i}$ 移到等式右边，并整理可得：

$$\phi_{n,i} = \exp\{\Psi(\gamma_i) - \Psi(\sum_{j=1}^{K}\gamma_j) + \log(\beta_{i,v}) - 1 + \lambda_n\}$$

$$\propto \beta_{i,w_n} \cdot \exp\{\Psi(\gamma_i) - \Psi(\sum_{j=1}^{K}\gamma_j)\}$$

$$\propto \beta_{i,w_n} \cdot \exp\{\Psi(\gamma_i)\} \quad (7\text{-}30)$$

注意此式中无需考虑 $\lambda_n$，将其忽略，因其对所有的 topic i 而言此数字均相同。

2. **maximize wrt.** $\gamma_i$：

现在轮到 $\gamma_i$ 参数，其中 $i \in \{1,\ldots,K\}$，下界函数只留下 $\gamma_i$ 有关的项：

$$L_{[\gamma_i]} = \sum_{i=1}^{K}(\alpha_i - 1)(\Psi(\gamma_i) - \Psi(\sum_{j=1}^{K}\gamma_j)) + \sum_{n=1}^{N}\sum_{i=1}^{K}\phi_{n,i}\left(\Psi(\gamma_i) - \Psi(\sum_{j=1}^{K}\gamma_j)\right) - \log\Gamma(\sum_{i=1}^{K}\gamma_i) + \sum_{i=1}^{K}\log\Gamma(\gamma_i) - \sum_{i=1}^{K}(\gamma_i - 1)\cdot(\Psi(\gamma_i) - \Psi(\sum_{j=1}^{K}\gamma_j)) \quad (7\text{-}31)$$

公式(7-31)已经得到，如果读者去看 Blei 的《Latent Dirichlet Allocation》论文原文，会发现 Blei 在这里的处理有点问题。

关于 $\gamma_i$ 求导，忽略其他无关参数，这样就消除了一些 sigma 符号。

$$\frac{\partial L_{[\gamma_i]}}{\partial\gamma_i} = (\alpha_i - 1)(\Psi'(\gamma_i) - \Psi'(\sum_{j=1}^{K}\gamma_j)) + \sum_{n=1}^{N}\phi_{n,i}\left(\Psi'(\gamma_i) - \Psi'(\sum_{j=1}^{K}\gamma_j)\right) - \Psi(\sum_{i=1}^{K}\gamma_i) + \Psi(\gamma_i) - \left(\Psi(\gamma_i) - \Psi(\sum_{j=1}^{K}\gamma_j) + (\gamma_i - 1)\cdot(\Psi'(\gamma_i) - \Psi'(\sum_{j=1}^{K}\gamma_j))\right)$$

$$= \left(\Psi'(\gamma_i) - \Psi'(\sum_{j=1}^{K}\gamma_j)\right)\left((\alpha_i - 1) + \sum_{n=1}^{N}\phi_{n,i} - (\gamma_i - 1)\right)$$

令其等于 0：

$$\underbrace{\left(\Psi'(\gamma_i) - \Psi'(\sum_{j=1}^{K}\gamma_j)\right)}_{\neq 0}\underbrace{\left(\alpha_i + \sum_{n=1}^{N}\phi_{n,i} - \gamma_i\right)}_{=0} = 0$$

所以根据这个方程只要令 $\gamma_i$ 取值为下面的(7-32)式，便可以令下界得到最大值：

$$\gamma_i = \alpha_i + \sum_{n=1}^{N} \phi_{n,i} \tag{7-32}$$

可以注意到计算$\phi_{n,i}$的公式 (7-30)的表达式中出现了$\gamma_i$，而计算$\gamma_i$的公式(7-32)的表达式中出现了$\phi_{n,i}$，因此两者交替轮流进行计算，直到下界收敛。

### 7.3.6 参数估计

参数估计(parameter estimation)就是变分 EM 的 M 步骤，在这个步骤中，主要任务是利用 E 步骤已获得的φ和γ来获取模型参数α和β。我们仍然使用变分下界作为真实语料的边缘似然函数（marginal log likelihood）的替代。另外，这一节估计的α和β是 corpus level 的参数，因此似然函数（log-likehood）就是将单个文档的似然函数加起来，而且，总体的变分下界也是单个文档变分下界加起来。在这一节里，总体的变分下界的符号我们仍然“滥用”了上一节的变分下界符号 L。

**1．Maximize Lower Bound wrt. β**

与上一节一样，L 下界函数只留下β有关的项，另外，$\beta_{ij}$表示从$z^i$主题到$w^j$单词的产生概率，而 topic i 下所有词产生概率之和=1，因此有限制条件：

$$\sum_{v=1}^{V} \beta_{i,v} = 1 \tag{7-33}$$

应该注意到这个限制条件与式(7-28)在求和公式上的区别。限制条件意味着需要使用拉格朗日乘数法的公式(6-15)，并使用 d 作为文档编号的下标，写出 L 下界函数：

$$L_{[\beta]} = \sum_{d=1}^{M}\sum_{n=1}^{N_d}\sum_{i=1}^{K} \phi_{d,n,i} \log(\beta_{i,w_n}) + \sum_{i=1}^{K} \lambda_i (\sum_{v=1}^{V} \beta_{i,v} - 1) \tag{7-34}$$

令其关于β求导，令其等于0：

$$\frac{\partial L_{[\beta]}}{\partial \beta_{i,v}} = \sum_{d=1}^{M}\sum_{n=1}^{N_d} \frac{\phi_{d,n,i} \cdot 1(v = w_n)}{\beta_{i,v}} + \lambda_i = 0 \tag{7-35}$$

在式(7-35)中，$w_n$的下标$n \in \{1,\ldots,N_d\}$，$w_n$代表文档 d 的第 n 个词，$1(v = w_n)$是示性函数，表示当满足括号内的条件时返回 1，否则返回 0。

$$\sum_{d=1}^{M}\sum_{n=1}^{N_d}\frac{\phi_{d,n,i}\cdot 1(v=w_n)}{\beta_{i,v}}+\lambda_i=0$$
$$\Rightarrow \beta_{i,v}=-\frac{1}{\lambda_i}\left(\sum_{d=1}^{M}\sum_{n=1}^{N_d}\phi_{d,n,i}\cdot 1(v=w_n)\right)$$

因为$\sum_{v=1}^{V}\beta_{i,v}=1$，因此我们可以忽略拉格朗日乘子$\lambda_i$，直接使用归一化公式：

$$\hat{\beta}_{i,v}\propto\sum_{d=1}^{M}\sum_{n=1}^{N_d}\phi_{d,n,i}\cdot 1(v=w_n) \tag{7-36}$$

**2．Maximize Lower Bound wrt. α**

L 下界函数只留下α有关的项：

$$L_{[\alpha]}=\sum_{d=1}^{M}\left(\left(\sum_{i=1}^{K}(\alpha_i-1)(\Psi(\gamma_{d,i})-\Psi(\sum_{j=1}^{K}\gamma_{d,j}))\right)+\log\Gamma(\sum_{i=1}^{K}\alpha_i)-\sum_{i=1}^{K}\log\Gamma(\alpha_i)\right)$$

关于$\alpha_i$求导，为避免冲突，将上式$L_{[\alpha]}$中的$\log\Gamma(\sum_{i=1}^{K}\alpha_i)$改为$\log\Gamma(\sum_{j=1}^{K}\alpha_j)$：

$$\frac{\partial L_{[\alpha]}}{\partial\alpha_i}=\sum_{d=1}^{M}\left(\Psi(\gamma_{d,i})-\Psi(\sum_{j=1}^{K}\gamma_{d,j})\right)+M\cdot\left(\Psi(\sum_{j=1}^{K}\alpha_j)-\Psi(\alpha_i)\right) \tag{7-37}$$

可以看到这个求导表达式依赖于$\alpha_j$，而$j\neq i$，因此$\frac{\partial L_{[\alpha]}}{\partial\alpha_i}=0$的方程无法直接解出来，必须使用迭代算法来近似估计。Blei 从故纸堆中翻出了牛顿的珍宝：牛顿-拉菲森迭代法。我这里简要讲述下 blei 对α的处理办法，而牛顿方法的详细故事留在另一章讲述。

这里直接给出迭代α的公式（该公式的推导留给牛顿迭代法一章）：

$$\log(\alpha^{t+1})=\log(\alpha^{t})-\frac{\frac{dL}{d\alpha}}{\frac{d^2L}{d\alpha^2}\alpha+\frac{dL}{d\alpha}} \tag{7-38}$$

现在迭代需要的所有的公式都已经齐备，在下一章中，将着重讲解代码实现。

## 7.4 误差讨论

一个优秀的研究者在阅读前人论文时应该带有一些批判性的精神，分析到前人成果的不足，这样才能提出自己新的思想。事实上，如果在推导中提高警觉，便可很轻易地发现推导中存在的一些造成误差的地方，这里简单列举一下。首先，第一个误差出现在公式(7-23)，前文已述。另一个不易发现的误差出现在下界展开的第 2 项和第 3 项、第 5 项的展开式推导中，如果读者的概率论基础足够扎实，会注意到关于期望的一个事实：若干个独立随机变量之积的期望等于各变量的期望之积，即：

$$E(X_1, X_2, ..., X_n) = E(X_1)E(X_2)...E(X_n) \tag{7-39}$$

因此第 2 项的展开推导中的 $E_q[z_{n,i}\log\theta_i]$ 拆开为 $E_q[z_{n,i}]\cdot E_q[\log\theta_i]$ 式存在一定误差的，因为其实际情况中主题的分配 z 的与文章的主题分布θ往往并不是独立的；第3项的展开也是同理，$E_q[z_{n,i}\cdot\log(\beta_{i,w_n})]$ 拆开为 $E_q[z_{n,i}]\cdot E_q[\log(\beta_{i,w_n})]$ 的过程中，β表示从 $z^i$ 主题到 $w^j$ 单词的产生概率，也与主题分配 z 存在一定关系，两者不能完全视为独立；第 5 项的展开推导中 $E_q(z_{n,i}\cdot\log\phi_{n,i})$ 的拆开中，由于 $z_{n,i} \sim Mult(z \mid \phi)$ 更是让这项误差较大。

可以说，Blei 推导 LDA 时做了一些妥协，对某些量进行了近似处理。正如一代物理学家保罗•狄拉克[38]所言：“原先，我只对完全正确的方程感兴趣。然而我所接受的工程训练教导我要容许近似，有时候我能够从这些理论中发现惊人的美，即使它是以近似为基础……。如果没有这些来自工程学的训练，我或许无法在后来的研究作出任何成果……我持续在之后的工作运用这些不完全严谨的工程数学，我相信你们可以从我后来的文章中看出来……那些要求所有计算推导上完全精确的数学家很难在物理上走得很远。”

狄拉克之言机器学习上也是同样适用的，理解了“允许近似”这一点，将会在机器学习之路上走得更远。

[38] 狄拉克在布里斯托大学工程学院学习电机工程。尽管最喜欢的科目是数学，狄拉克后来声称这段工程教育对他影响深远。

# 参考文献

# 第8章 LDA 变分 EM 实现

对于实干家来说，来到了令人兴奋的一章，这一章将系统性地介绍 Blei 的 LDA 的 C 语言源代码的实现。回忆起第一次写程序的情形，运行一段代码，兴奋地看到“hello world”出现在闪闪发光的绿色屏幕，本章也将令人兴奋的 LDA 变分 EM 搬上电脑屏幕，这也算是对前文漫长的推导的一个补偿。

如果初看 Blei 版本的源代码，初学者往往会迷惑于代码中一些精妙的地方。所以这一章的安排是这样的：首先根据前文推导得到的迭代公式写出一个伪代码的框架，后面深入剖析 Blei 的 C 源代码。

## 8.1 伪代码

回忆一下上一章，变分 LDA 重要的步骤就是变分 EM，为了加深记忆，在此处重新写出这些重要的公式。

(1)E-step:

对一篇文档，利用初始化或先前迭代已估计的$\alpha$和$\beta$去估计参数$\gamma$和$\phi$。

$$\phi_{n,i} \propto \beta_{i,w_n} \cdot \exp\{\Psi(\gamma_i)\}$$

$$\gamma_i = \alpha_i + \sum_{n=1}^{N} \phi_{n,i}$$

(2)M-step:

最大化 variational inference 中的下界，求出此时的$\alpha$和$\beta$。

$$\hat{\beta}_{i,v} \propto \sum_{d=1}^{M} \sum_{n=1}^{N_d} \phi_{d,n,i} \cdot 1(v = w_n)$$

$$\log(\alpha^{t+1}) = \log(\alpha^{t}) - \frac{\frac{dL}{d\alpha}}{\frac{d^2L}{d\alpha^2}\alpha + \frac{dL}{d\alpha}}$$

其中$\alpha$在 M 步骤中被反复迭代得到估计值。

不浪费时间，首先直接写出伪代码：

**表 8-1 LDA 的变分 EM 算法步骤**

**输入输出数据：**

输入：主题个数 K

包含 M 篇文章的语料集，每篇文章 d 拥有 $N_d$ 个单词

输出：模型参数：β，θ，z

**算法步骤：**

初始化 $\phi_{n,i}^{0} := 1/K$，其中 i 代表共 K 个主题中的一个主题编号，n 代表 $N_d$ 个单词中其中一个 index。

初始化 $\gamma_i := \alpha_i + N/K$，其中 i 代表共 K 个主题中的一个主题编号。

初始化 $\alpha := 50/K$。

初始化 $\beta_{ij} := 0$，其中 i 代表共 K 个主题中的一个主题编号，j 代表语料词典共 V 个单词中的一个。

```
//E-step（确定 φ 和 γ 以便于计算似然函数）
loglikehood := 0              //似然函数初始化 0
for d = 1 to M{      //遍历语料库的所有文章
    repeat{
        for n=1 to Nd{       //遍历一个文章的词
            for i =1 to K{    //遍历所有主题编号
```

$$\phi_{d,n,i}^{t+1} := \beta_{i,w_n} \exp\left(\Psi(\gamma_{d,i}^{t})\right)$$

```
            }endfor
            将 φ_{d,n,i}^{t+1} 归一化（normalize），使其之和=1
        }endfor
```

$$\gamma^{t+1} := \alpha + \sum_{n=1}^{N} \phi_{d,n}^{t+1}$$

```
    }until  φ_d 和 γ_d 收敛

    loglikehood := loglikelihood + L(γ,φ;α,β)

}endfor


//M-step(极大化变分分布的 log 似然函数)
for d = 1 to M{      //遍历语料库的所有文章
    for i=1 to K{     //遍历所有主题编号
        for j=1 to V{ //遍历语料库中的词典所有单词编号
```

```
            β_{i,j} := β_{i,j} + φ_{d,n,i} w_{d,n,j}
        } endfor
        将 β_i 归一化（normalize），使其之和=1
    } endfor
  } endfor
通过迭代公式(7-38)计算α
if(loglikelihood 收敛) then{
    return parameters
}else{
    Go back to E-step
}endif
```

## 8.2 工程优化分析

可以看到上述迭代算法中反复出现了归一化（normalize）的操作。这在具体工程实践时采用了一个数学上的小技巧，这便是鲜为人知的 log_sum 技巧，给出 log(a)和 log(b)，返回 log(a + b):

$$\begin{aligned}\log(a+b) &= \log\left(a\cdot(1+b\,/\,a)\right)\\ &= \log(a)+\log(1+b\,/\,a)\\ &= \log(a)+\log(1+\exp\{\log(b)-\log(a)\})\end{aligned}$$

有了这个小的技巧，在归一化计算φ的时候，换用对数来归一化就可以加快运算速度：计算出 $\log(\phi_1),\log(\phi_2),\log(\phi_3),\ldots,\log(\phi_K)$之后，利用此技巧就可以不断两两迭代计算出 $\log(\phi_1+\phi_2+\ldots+\phi_K)$，最后要归一化的公式如下：

$$\phi_i := \frac{\phi_i}{\phi_1+\phi_2+\ldots+\phi_K} \Rightarrow \log(\phi_i) := \log(\phi_i) - \log(\phi_1+\phi_2+\ldots+\phi_K) \tag{8-1}$$

采用这种优化后，φ在迭代计算时也换成了 log(φ)，因此迭代公式换成了：

$$\log(\phi_{d,n,i}^{t+1}) := \log(\beta_{i,w_n}) + \Psi(\gamma_{d,i}^{t}) \tag{8-2}$$

另一个工程上的优化是求解γ时的优化，考虑到求解公式(7-32)时已经得到了上一轮迭代的φ信息，因此可以利用这个信息加快速度。来看看 Blei 的代码是如何对γ优化的，Blei 使用了下面的公式：

$$\gamma_i^{t+1} := \gamma_i^t + \sum_{n=1}^{N}\left(\phi_{n,i}^{t+1} - \phi_{n,i}^{t}\right) \tag{8-3}$$

其中上标 t 表示第 t 轮迭代，而下标 i 表示第 i 个 topic，其中

$i \in \{1,2,\ldots,K\}$；N 表示当前文章的单词个数（未去重）。

公式(8-3)是如何得到的呢？细致想来其实不难，首先第一轮初始化时：

$$\gamma_i^0 := \alpha_i + N \cdot \phi_{n,i}^0 = \alpha_i + N / K$$

此后 t+1 轮迭代时必须使用如下迭代公式：

$$\begin{aligned}\gamma_i^{t+1} &:= \alpha_i + \sum_{n=1}^{N} \phi_{n,i}^{t+1} \\ &= \alpha_i + \sum_{n=1}^{N} (\phi_{n,i}^t + \Delta\phi_{n,i}^{t+1}) \\ &= \underbrace{\alpha_i + \sum_{n=1}^{N} (\phi_{n,i}^t)}_{\gamma_i^t} + \sum_{n=1}^{N} \underbrace{\Delta\phi_{n,i}^{t+1}}_{\phi_{n,i}^{t+1} - \phi_{n,i}^t} \\ &= \gamma_i^t + \sum_{n=1}^{N} \left(\phi_{n,i}^{t+1} - \phi_{n,i}^t\right)\end{aligned}$$

公式(8-3)由此得到，公式(8-3)是一个递推公式，由于第一轮初始化时有α信息，后面的递推公式才能成功地去除了α。另外由于只在第 d 篇文档中求解操作，因此忽略ϕ的 d 下标。

在 Blei 的代码中，该段代码出现在 lda-inference.c 的第 70 行左右，出现了下面这句：

```
var_gamma[k] = var_gamma[k] + doc->counts[n]*(phi[n][k] - oldphi[k]);
```

为何出现了 doc->counts[n]这样的乘法因子？这是由于在外层 for 循环中：for (n = 0; n < doc->length; n++)，这里的 doc->length 是 doc 中去重的单词数，所以公式(8-3)变化为：

$$\gamma_i^{t+1} := \gamma_i^t + \sum_{n=1}^{N} \left(\phi_{n,i}^{t+1} - \phi_{n,i}^t\right) = \gamma_i^t + \sum_{n=1}^{doc->length} C_n \left(\phi_{n,i}^{t+1} - \phi_{n,i}^t\right)$$

$C_n$ 即代码中的 doc->counts[n]，指文章中出现的第 n 个单词 word_n 的在该文章内重复出现的计数统计，在针对文章内单词去重后的范围内求和时，需要乘以 $C_n$。

## 8.3 Blei 的变分 LDA（C 语言版）源代码剖析

### 8.3.1 源代码下载

在阅读这一节的源代码分析时，我强烈建议读者去下载一份 Blei 的 C 语言版源代码去阅读，github 上的项目网址为：https://github.com/Blei-

Lab/lda-c

SSH 的 clone 的 url 为：git@github.com:Blei-Lab/lda-c.git

### 8.3.2 使用命令与配置

1.输入格式

得到了伪代码，下面以 Blei 原版 C 代码进行细致分析，首先来看下 Blei 版 LDA 的输入格式：

[M] [term_1]:[count] [term_2]:[count] ... [term_N]:[count]

其中 M 是该文章的去重后的单词数，count 是每个单词出现的次数，term_x 是每个单词的 wordid，这不是 string。

2. 命令行参数

编译后之后，就可以在 shell 调用执行了，训练的调用命令如下：

lda est [alpha] [k] [settings] [data] [random/seeded/*] [directory]

其中[random/seeded/*]代表主题生成方式，“random”随机生成每个主题；“seeded”从一个随机选取的文章中平滑地生成主题分布；或者你可以指定一个以前已经训练好的模型去初始化模型（传入已经训练好的模型名）。

此外，在已经训练好的模型基础上，可以推断(inference)新文章的主题模型，命令如下：

lda inf [settings] [model] [data] [name]

在新数据上执行变分推断时使用已训练好的[model].*文件。

之前训练好的[model].*文件将会包含以下文件：

1. <iteration>.other 包含α;

2. <iteration>.beta 包含主题分布的 log 对数，每一行代表一个主题；在第 k 行，每一个元素是 log(p(w|z=k))

另外有两个文件将会被创建：[name].gamma 是每篇文章的变分狄利克雷参数（variational Dirichlet parameters），也即 doc→topic 分布的矩阵文件，每一行是一个 doc，而每一列是一个 topic 的概率；[name].likelihood 则是每篇文章的似然函数的下界。

3. 配置参数配置

参数配置放置在 settings.txt 文件中，共有以下设置项：

var max iter [integer e.g., 10 or -1]

var convergence [float e.g., 1e-8]

em max iter [integer e.g., 100]

em convergence [float e.g., 1e-5]

alpha [fixed/estimate]

详细解释如下：

[var max iter]

一篇文档的变分分布的坐标上升法（coordinate ascent variational inference）最大的迭代次数。如果设置-1 则为完全执行变分推断，直到变分收敛条件被满足。

[var convergence]

变分推断的收敛条件。(score_old - score) / abs(score_old)如果小于这个数值，则停止迭代（在迭代超过最大迭代次数之后）。请注意这个参数是在一篇特定的文档上的“下界”。

[em max iter]

变分 EM 的是最大的迭代次数。

[em convergence]

变分 EM 的收敛条件。(score_old - score) / abs(score_old)如果小于这个数值，则停止迭代（在迭代超过最大迭代次数之后）。请注意这个参数是整个语料集的似然函数的“下界”。

[alpha]

有两个设置选项，如果设置成[fixed]，则 alpha 在迭代中不更新。如果设置成[estimate]，则 alpha 会与 topic distributions 一起在迭代中被估计。

### 8.3.3 源代码情况一览

表 8-2 源代码中用到的变量一览

| | |
|---|---|
| var_gamma | double 二维数组，存储容量 M X K，doc-topic 分布，每篇文档都会计算其 topic 分布 |
| phi | double 二维数组，存储容量 N(最长的一篇文章的单词数) X K，word-topic 分布，针对每篇文档，计算文档中每个 word 的 topic 分布 |
| alpha | double 类型，只是一个浮点数（由于对称超参数处理） |
| class_word | double 二维数组，存储容量 K X V，β变量 |
| class_total | double 一维数组，存储容量 K，为了将 $\log(\beta_{i,w})$ 归一化而设置 |
| log_prob_w | double 二维数组，存储容量 K X V，$\log(\beta_{i,w})$ 变量 |
| lda_model | lda 的模型参数，里面包括 beta 以及 alpha |
| lda_suffstats | 记录统计信息，比如每个 topic 上每个 word 出现的次数，这是为了计算 lda_model 而存在 |
| corpus | 语料集的全部文档信息 |
| document | 文档的具体信息，包括 word 信息 |

表 8-3 代码文件与解释：

| 文件名 | 典型函数 | 功能解释 |
|---|---|---|
| lda.h | typedef struct lda_suffstats; … | 代码中用到的数据结构 |
| lda-data.c | read_data | 读取训练语料集的文档，返回 corpus 结构体指针 |
| lda-model.c | corpus_initialize_ss/random_initialize_ss | 初始化 LDA 的统计信息 |
| | lda_mle | LDA 变分 EM 的 M 步骤 |
| lda-estimate.c | run_em | 变分 Em 算法框架 |
| | doc_e_step | LDA 变分 EM 的 E 步骤 |
| | read_settings | 读取 setting.txt 配置文件 |
| | infer | 根据已训练好模型去预测 |
| lda-inference.c | lda_inference | 迭代训练计算 γ 以及 ϕ（被 doc_e_step 调用），并且在预测新文档主题分布时被 infer 函数调用 |
| | compute_likelihood | 根据公式(7-27)计算变分下界 |

先来看看代码中使用到一些数据结构体，位于文件 lda.h，如下所示：

```
//文件：lda.h
/* 一篇文档的结构体,word 以 int 指针存储 */
typedef struct
{
```

```
    int* words;
    int* counts;
    int length;
    int total;
} document;

/* 语料集的结构体 */
typedef struct
{
    document* docs;
    //单词总数 V
    int num_terms;
    //文章总数 M
    int num_docs;
} corpus;

typedef struct
{
    double alpha;
     //log(βi,w)
    double** log_prob_w;
    //主题总数 K
    int num_topics;
    int num_terms;
} lda_model;

typedef struct
{
    //topic→word 的统计量，为了未来计算 log(βi,w)所使用的统计量
    double** class_word;
    //为了将 log(βi,w)归一化而设置
    double* class_total;
    double alpha_suffstats;
    int num_docs;
} lda_suffstats;
```

值得一提的是上述代码最后出现的这个 lda_suffstats 结构体，在程序中是一个全局变量，其统计的信息跨越了不同的文档。且来看看它是如何初始化的，根据传入的命令行参数 random/seeded/*不同，有不同的初始化函数相对应。

```
//文件：lda-model.c
/**
功能：传入 seeded。初始化语料集的统计信息
**/
```

```
void corpus_initialize_ss(lda_suffstats* ss, lda_model* model, corpus* c)
{
    int num_topics = model->num_topics;
    int i, k, d, n;
    document* doc;

    for (k = 0; k < num_topics; k++) //遍历所有 topic 编号
    {
        for (i = 0; i < NUM_INIT; i++)
        {
            d = floor(myrand() * c->num_docs);//随机挑选一篇文章
            printf("initialized with document %d\n", d);
            doc = &(c->docs[d]);
            for (n = 0; n < doc->length; n++) //遍历每个单词
            {
                //为每个编号的 topic 的 topic-word 的统计量增加 count
                ss->class_word[k][doc->words[n]] += doc->counts[n];
            }
        }
        //这个 for 循环是为了平滑 smoothed，为每个词的 topic-word 的统计量都初始化+1
        for (n = 0; n < model->num_terms; n++)
        {
            ss->class_word[k][n] += 1.0;
            ss->class_total[k] = ss->class_total[k] + ss->class_word[k][n];
        }
    }
}
```

所谓的 class_word 是 topic→word 的统计量，该变量在E步骤中统计信息，是为了未来估计 $\log(\beta_{i,w})$ 所使用的统计量[39]；class_total 是为了未来计算 $\log(\beta_{i,w})$ 时归一化的目的（代码位于 lda-model.c 文件下的M步骤，即 lda_mle 函数中)。

而当命令行参数传入 random 时，使用了下面的代码进行随机初始化：

```
//文件：lda-model.c
/**
功能：传入 random。初始化语料集的统计信息
**/

void random_initialize_ss(lda_suffstats* ss, lda_model* model)
{
    int num_topics = model->num_topics;
    int num_terms = model->num_terms;
    int k, n;
```

[39] 正如 8.2 节工程优化所述，代码中均使用 log 对数形式保存β值

```
    for (k = 0; k < num_topics; k++)
    {
        for (n = 0; n < num_terms; n++)
        {
            ss->class_word[k][n] += 1.0/num_terms + myrand(); //myrand()是 0~1 的随机小数
            ss->class_total[k] += ss->class_word[k][n];
        }
    }
}
```

### 8.3.4 Variational EM 代码剖析

执行 EM 算法，针对每篇文档，利用其单词以及初始化的β和α信息，更新模型的变分下界，直至收敛。

1. E-step

利用 8.2 节提出的工程优化分析公式(8-1)、(8-2)、(8-3)先写出以下代码迭代求解γ和ϕ（这段代码位于 lda-inference.c)：

```
/*
 * variational inference
 */

double lda_inference(document* doc, lda_model* model, double* var_gamma, double** phi)
{
    double converged = 1;
    double phisum = 0, likelihood = 0;
    double likelihood_old = 0, oldphi[model->num_topics];
    int k, n, var_iter;
    double digamma_gam[model->num_topics];

    // compute posterior dirichlet

    for (k = 0; k < model->num_topics; k++)
    {
        //初始化 γ 以及 ϕ
        var_gamma[k] = model->alpha + (doc->total/((double) model->num_topics));
        digamma_gam[k] = digamma(var_gamma[k]);
        for (n = 0; n < doc->length; n++)
            phi[n][k] = 1.0/model->num_topics;
    }
    var_iter = 0;
```

```
    while ((converged > VAR_CONVERGED) &&
           ((var_iter < VAR_MAX_ITER) || (VAR_MAX_ITER == -1)))
    {
        var_iter++;
        for (n = 0; n < doc->length; n++)
        {
          phisum = 0;
          for (k = 0; k < model->num_topics; k++)
          {
              oldphi[k] = phi[n][k];
              //使用公式(8-2)计算 φ
              phi[n][k] = digamma_gam[k] + model->log_prob_w[k][doc->words[n]];
              if (k > 0)
                  phisum = log_sum(phisum, phi[n][k]);
              else
                  phisum = phi[n][k]; // note, phi is in log space
          }

          for (k = 0; k < model->num_topics; k++)
          {
              //使用公式(8-1)均一化，然后 log(φ) -> φ
              phi[n][k] = exp(phi[n][k] - phisum);
              //使用公式(8-3)计算 γ
              var_gamma[k] =
                  var_gamma[k] + doc->counts[n]*(phi[n][k] - oldphi[k]);
              digamma_gam[k] = digamma(var_gamma[k]);
          }
        }

        likelihood = compute_likelihood(doc, model, phi, var_gamma);
        assert(!isnan(likelihood));
        converged = (likelihood_old - likelihood) / likelihood_old;
        likelihood_old = likelihood;
    }
    return(likelihood);
}
```

2. M-step

后面的代码中，Blei 使用了牛顿迭代法来更新α，我们重新写出这些公式，即：

$$\log(\alpha^{t+1}) = \log(\alpha^{t}) - \frac{\frac{dL}{d\alpha}}{\frac{d^2L}{d\alpha^2}\alpha + \frac{dL}{d\alpha}}$$

可以观察到α迭代公式中需要使用到的因素有$\frac{dL}{d\alpha}$、$\frac{d^2L}{d\alpha^2}$，另外特别提醒的是在具体编程实践中，Blei 对α的处理是使用对称超参数，每个α的值相等，所以只使用一个α，而忽略α的下标。因而第 7 章的求导公式(7-37)就会有所变化，在正式展示代码前需要重新写出这些求导公式：

其中 L 只留下关于α的项变为：

$$L_{[\alpha]} = \sum_{d=1}^{M}\left(\left(\sum_{i=1}^{K}(\alpha-1)(\Psi(\gamma_{d,i})-\Psi(\sum_{j=1}^{K}\gamma_{d,j}))\right)+\log\Gamma(K\cdot\alpha)-K\cdot\log\Gamma(\alpha)\right)$$

求出 L 关于α的一阶导数：

$$\frac{\partial L_{[\alpha]}}{\partial\alpha} = \sum_{d=1}^{M}\left(\sum_{i=1}^{K}\Psi(\gamma_{d,i})-\Psi(\sum_{j=1}^{K}\gamma_{d,j})\right)+M\cdot\left(K\cdot\Psi(K\cdot\alpha)-K\cdot\Psi(\alpha)\right) \tag{8-4}$$

再求出 L 关于α的二阶导数：

$$\frac{\partial^2 L_{[\alpha]}}{\partial\alpha^2} = M\cdot\left(K\cdot K\cdot\Psi'(K\cdot\alpha)-K\cdot\Psi'(\alpha)\right) \tag{8-5}$$

实践中，Blei 首先计算了α的充分统计量 alpha_suffstats（位于 lda-estimate.c 文件下的 doc_e_step 函数中）：

$$\text{alpha_suffstats} = \sum_{d=1}^{M}\left(\sum_{i=1}^{K}\Psi(\gamma_{d,i})-\Psi(\sum_{j=1}^{K}\gamma_{d,j})\right) \tag{8-6}$$

紧接着立即动手实现了三个小函数 alhood、d_alhood、d2_alhood：

```
/*
 * objective function and its derivatives
 */

double alhood(double a, double ss, int D, int K)
{ return(D * (lgamma(K * a) - K * lgamma(a)) + (a - 1) * ss); }

double d_alhood(double a, double ss, int D, int K)
{ return(D * (K * digamma(K * a) - K * digamma(a)) + ss); }

double d2_alhood(double a, int D, int K)
{ return(D * (K * K * trigamma(K * a) - K * trigamma(a))); }
```

函数中的 lgamma 是 C 语言数学库中的 log gamma，而 digamma 在是 log gamma 关于自变量α的导数，trigamma 是 log gamma 关于自变量α的二阶导数，ss 就是前文计算出的 alpha_suffstats。这些函数代码中的变量 D 就是前文推导中所用的整个语料库的文章数 M，double a 就是我们要迭代寻找最优化的α。将上

述三个函数代码翻译成更清晰的数学形式如下：

$$alhood = D\cdot\left(\log\Gamma(K\cdot\alpha) - K\cdot\log\Gamma(\alpha)\right) + (\alpha-1)\cdot alpha_suffstats$$

$$d_alhood = D\cdot\left(K\cdot\Psi(K\cdot\alpha) - K\cdot\Psi(\alpha)\right) + alpha_suffstats$$

$$d2_alhood = D\cdot\left(K\cdot K\cdot\Psi'(K\cdot\alpha) - K\cdot\Psi'(\alpha)\right)$$

之后就是牛顿迭代法循环迭代的主要逻辑函数了：

```
/*
 * newton method
 * @param ss alpha_suffstats
 * @param D num_docs
 * @param K num_topics
 */

double opt_alpha(double ss, int D, int K)
{
    double a, log_a, init_a = 100;
    double f, df, d2f;
    int iter = 0;

    log_a = log(init_a);
    do
    {
        iter++;
        a = exp(log_a);
        if (isnan(a))
        {
            init_a = init_a * 10;
            printf("warning : alpha is nan; new init = %5.5f\n", init_a);
            a = init_a;
            log_a = log(a);
        }
        f = alhood(a, ss, D, K);
        df = d_alhood(a, ss, D, K);
        d2f = d2_alhood(a, D, K);
        log_a = log_a - df/(d2f * a + df);   //公式(7-38)
        printf("alpha maximization : %5.5f   %5.5f\n", f, df);
    }
    //NEWTON_THRESH = 1e-5, MAX_ALPHA_ITER = 1000
    while ((fabs(df) > NEWTON_THRESH) && (iter < MAX_ALPHA_ITER));
    return(exp(log_a));
}
```

轮到β，在实际的代码中β在迭代公式里被冠以 class_word 的名字。并且使用了如下公式：

$$\text{class_word[k][doc->words[n]]}=\beta_{i,v}=\sum_{d=1}^{M}\sum_{n=1}^{N_d}\phi_{d,n,i}\cdot 1(v=w_n)$$

为了将β归一化，代码中又同时计算了 class_total 变量：

$$\text{class_total[k]}=\sum_{v=1}^{V}\beta_{i,v}=\sum_{v=1}^{V}\sum_{d=1}^{M}\sum_{n=1}^{N_d}\phi_{d,n,i}\cdot 1(v=w_n)$$

下面的代码生动地体现了这一点(代码中的 doc->counts[n]是由于程序的输入语料文件格式是[term_N]:[count]，重复的单词只输入 count 次数)。

```
for (n = 0; n < doc->length; n++)
{
    for (k = 0; k < model->num_topics; k++)
    {
        ss->class_word[k][doc->words[n]] += doc->counts[n]*phi[n][k];
        ss->class_total[k] += doc->counts[n]*phi[n][k];
    }
}
```

如果读者亲自去下载阅读 Blei 的源代码，会发现这段代码位于 lda-estimate.c 文件下的 doc_e_step 函数中，这就说明 Blei 的实际代码中将β的迭代提前到了E步骤的函数中（β在前面的理论分析中本属于M步骤计算），这也阐释了变分EM算法的编程实践中E步骤和M步骤有时没有明显的分界线。

由于 E 步骤中ϕ的计算公式(8-2)同样用到了β信息，但是是以$\log(\beta_{i,w})$的形式出现的，因此有了下面的归一化公式：

$$\log(\beta_{i,v}):=\log(\beta_{i,v})-\log(\sum_{v=1}^{V}\beta_{i,v})$$

$\log(\beta_{i,w})$在程序代码中又被冠名为 log_prob_w，换成代码里的变量名：

$$\text{log_prob_w}[k][w]=\log(\text{class_word}[k][w])-\log(\text{class_total}[k])$$

经过前面 E 步骤对于 class_word 和 class_total 统计量的准备，下面的代码就是极大似然估计（M 步骤）了。

```
/*
 * compute MLE lda model from sufficient statistics
 */

void lda_mle(lda_model* model, lda_suffstats* ss, int estimate_alpha)
{
```

```
    int k; int w;
    //compute log(βi,v), then can use it to iteratively compute ϕ
    for (k = 0; k < model->num_topics; k++)
    {
        for (w = 0; w < model->num_terms; w++)
        {
            if (ss->class_word[k][w] > 0)
            {
                model->log_prob_w[k][w] = log(ss->class_word[k][w]) -log(ss->class_total[k]);
            }
            else
                model->log_prob_w[k][w] = -100;
        }
    }
    //setting 配置选项中 alpha=estimate 则 estimate_alpha = 1;否则 estimate_alpha = 0
    if (estimate_alpha == 1)  //newton method iteratively compute α
    {
        model->alpha = opt_alpha(ss->alpha_suffstats,
                                 ss->num_docs,
                                 model->num_topics);

        printf("new alpha = %5.5f\n", model->alpha);
    }
}
```

最后展现给读者的是整个变分 EM 的框架代码：

```
void run_em(char* start, char* directory, corpus* corpus)
{

    int d, n;
    lda_model *model = NULL;
    //var_gamma = M x K
    //phi = N(语料里最长的文章内的单词数) X K
    double **var_gamma, **phi;
    //留给读者去阅读源码:phi 难道在不同文档 d 共用一个?
    // allocate variational parameters

    var_gamma = malloc(sizeof(double*)*(corpus->num_docs));
    for (d = 0; d < corpus->num_docs; d++)
         var_gamma[d] = malloc(sizeof(double) * NTOPICS);

    int max_length = max_corpus_length(corpus);
    phi = malloc(sizeof(double*)*max_length);
    for (n = 0; n < max_length; n++)
         phi[n] = malloc(sizeof(double) * NTOPICS);
```

```
// initialize model

char filename[100];

lda_suffstats* ss = NULL;
//start 是命令行參數
if (strcmp(start, "seeded")==0)
{
    model = new_lda_model(corpus->num_terms, NTOPICS);
    ss = new_lda_suffstats(model);
    corpus_initialize_ss(ss, model, corpus);
    //initial log_prob_w = log(ss->class_word[k][w]) - log(ss->class_total[k])
    lda_mle(model, ss, 0);
    model->alpha = INITIAL_ALPHA;
}
else if (strcmp(start, "random")==0)
{
    model = new_lda_model(corpus->num_terms, NTOPICS);
    ss = new_lda_suffstats(model);
    random_initialize_ss(ss, model);
    //initial log_prob_w = log(ss->class_word[k][w]) - log(ss->class_total[k])
    lda_mle(model, ss, 0);
    model->alpha = INITIAL_ALPHA;
}
else
{
    model = load_lda_model(start);
    ss = new_lda_suffstats(model);
}

sprintf(filename,"%s/000",directory);
save_lda_model(model, filename);

// run expectation maximization

int i = 0;
double likelihood, likelihood_old = 0, converged = 1;
sprintf(filename, "%s/likelihood.dat", directory);
FILE* likelihood_file = fopen(filename, "w");

while (((converged < 0) || (converged > EM_CONVERGED) || (i <= 2)) && (i <= EM_MAX_ITER))
{
    i++; printf("**** em iteration %d ****\n", i);
```

```
        likelihood = 0;
        zero_initialize_ss(ss, model);

        // e-step

        for (d = 0; d < corpus->num_docs; d++)
        {
            if ((d % 1000) == 0) printf("document %d\n",d);
            likelihood += doc_e_step(&(corpus->docs[d]),
                                     var_gamma[d],
                                     phi,
                                     model,
                                     ss);
        }

        // m-step

        lda_mle(model, ss, ESTIMATE_ALPHA);

        // check for convergence

        converged = (likelihood_old - likelihood) / (likelihood_old);
        if (converged < 0) VAR_MAX_ITER = VAR_MAX_ITER * 2;
        likelihood_old = likelihood;

        // output model and likelihood

        fprintf(likelihood_file, "%10.10f\t%5.5e\n", likelihood, converged);
        fflush(likelihood_file);
        if ((i % LAG) == 0)
        {
            sprintf(filename,"%s/%03d",directory, i);
            save_lda_model(model, filename);
            sprintf(filename,"%s/%03d.gamma",directory, i);
            save_gamma(filename, var_gamma, corpus->num_docs, model->num_topics);
        }
    }

    // output the final model

    sprintf(filename,"%s/final",directory);
    save_lda_model(model, filename);
    sprintf(filename,"%s/final.gamma",directory);
    save_gamma(filename, var_gamma, corpus->num_docs, model->num_topics);
```

```
    // output the word assignments (for visualization)

    sprintf(filename, "%s/word-assignments.dat", directory);
    FILE* w_asgn_file = fopen(filename, "w");
    for (d = 0; d < corpus->num_docs; d++)
    {
        if ((d % 100) == 0) printf("final e step document %d\n",d);
        likelihood += lda_inference(&(corpus->docs[d]), model, var_gamma[d], phi);
        write_word_assignment(w_asgn_file, &(corpus->docs[d]), phi, model);
    }
    fclose(w_asgn_file);
    fclose(likelihood_file);
}
```

读者应该注意到判断最终收敛使用了 likehood 来判断，这其实就是在使用下界公式(7-27)，下面的代码严格执行并计算了这个公式。

```
/*
 * compute likelihood bound
 *
 */

double
compute_likelihood(document* doc, lda_model* model, double** phi, double* var_gamma)
{
    double likelihood = 0, digsum = 0, var_gamma_sum = 0, dig[model->num_topics];
    int k, n;

    for (k = 0; k < model->num_topics; k++)
    {
          dig[k] = digamma(var_gamma[k]);
          var_gamma_sum += var_gamma[k];
    }
    digsum = digamma(var_gamma_sum);

    likelihood =
          lgamma(model->alpha * model -> num_topics)
          - model -> num_topics * lgamma(model->alpha)
          - (lgamma(var_gamma_sum));

    for (k = 0; k < model->num_topics; k++)
    {
          likelihood +=
              (model->alpha - 1)*(dig[k] - digsum) + lgamma(var_gamma[k])
              - (var_gamma[k] - 1)*(dig[k] - digsum);
```

```
        for (n = 0; n < doc->length; n++)
        {
          if (phi[n][k] > 0)
          {
              likelihood += doc->counts[n]*
                   (phi[n][k]*((dig[k] - digsum) - log(phi[n][k])
                                 + model->log_prob_w[k][doc->words[n]]));
          }
      }
    }
    return(likelihood);
}
```

### 8.3.5 预测推断新文档

变分 EM 的推断预测功能在 lda_estimate.c 代码文件下的 infer 函数中。可以从下面的这段代码中看到，在为新文档推断主题分布时，var_gamma 二维数组变量(γ)为新分配内存构造出来，可谓白纸一张，而这恰是由于 var_gamma 是 doc-topic 数组，仅与新文档有关，而与历史训练文档无关。而这段程序中调用了前文介绍过的 lda_inference 函数，并传入了 load_model 函数读出的历史训练得到的 log_prob_w 用以迭代计算ϕ（公式(8-2)），皆因 log_prob_w 变量代表 $\log(\beta_{i,w})$，是 topic-word 数组，会受到历史已训练的文档的影响。

```
void infer(char* model_root, char* save, corpus* corpus)
{
    FILE* fileptr;
    char filename[100];
    int i, d, n;
    lda_model *model;
    double **var_gamma, likelihood, **phi;
    document* doc;

    model = load_lda_model(model_root);
    //新分配内存，初始化 var_gamma
    var_gamma = malloc(sizeof(double*)*(corpus->num_docs));
    for (i = 0; i < corpus->num_docs; i++)
          var_gamma[i] = malloc(sizeof(double)*model->num_topics);
    sprintf(filename, "%s-lda-lhood.dat", save);
    fileptr = fopen(filename, "w");
    for (d = 0; d < corpus->num_docs; d++)
    {
```

```
            if (((d % 100) == 0) && (d>0)) printf("document %d\n",d);
            doc = &(corpus->docs[d]);
            phi = (double**) malloc(sizeof(double*) * doc->length);
            for (n = 0; n < doc->length; n++)
                phi[n] = (double*) malloc(sizeof(double) * model->num_topics);
            likelihood = lda_inference(doc, model, var_gamma[d], phi);

            fprintf(fileptr, "%5.5f\n", likelihood);
        }
        fclose(fileptr);
        sprintf(filename, "%s-gamma.dat", save);
        save_gamma(filename, var_gamma, corpus->num_docs, model->num_topics);
}
```

### 8.3.6 load model

如同 8.3.5 节所言，load_model 不读取 gamma 的信息，load_model 读取历史已训练好的模型中的以下信息：

1. 从<iteration>.beta 文件中读取 log_prob_w 信息；
2. 从<iteration>.other 文件中读取出主题个数 K，语料词典单词总数 V，alpha 浮点数值

关于 beta 文件和 other 文件的详细信息请看 8.3.7 节。

### 8.3.7 运行效果与终止条件

如果按照说明 readme 文件运行 blei 版 LDA 程序，屏幕就会出现如下提示：

```
[root@localhost lda]# ./lda est 1.0 100 ../settings.txt lda-input.txt random ./
    reading data from lda-input.txt
    number of docs     : 10
    number of terms    : 3197
    **** em iteration 1 ****
    document 0
    alpha maximization : -12999.24770    -186.45954
    alpha maximization : -1567.95766    -178.24854
    alpha maximization : 2346.00064    -157.78173
    alpha maximization : 3529.08773    -113.75411
```

```
alpha maximization : 3784.24192   -48.77433
alpha maximization : 3807.67408   -6.63685
alpha maximization : 3807.98722   -0.10092
alpha maximization : 3807.98728   -0.00002
alpha maximization : 3807.98728   -0.00000
new alpha = 2.74269
**** em iteration 2 ****
…
```

可见，由于进行了 alpha 迭代估计，所以将 alpha 的估计值也会打印出来。运行结束后会输出以下几个文件：

1. <iteration>.beta

这个文件会在每 n 轮迭代时输出一个 beta 值，比如 000.beta、005.beta 等。而最后一轮迭代后会输出 final.beta。

这个文件的格式与 Gibbs Sampling 版本 LDA 的 phi 文件格式相同，每一行为代表一个 topic，第 k 行的第 v 列=log_prob_w[k][v]，即 $\log\beta_{i,v}$。

2.<iteration>.gamma

这个文件会在每 n 轮迭代时输出一个γ值，比如 000.gamma、005.gamma 等。而最后一轮迭代后会输出 final.gamma。

这个文件的格式与 Gibbs Sampling 版本 LDA 的 theta 文件格式相同，每一行为代表一个 doc，第 m 行代表第 m 篇文档的主题分布，第 m 行的第 k 列=var_gamma[m][k]，即 $\gamma_{m,k}$ 。

有人会发出疑问：Gibbs Sampling 输出文件的 theta(θ)和 phi(φ)文件去哪了？其实在这个版本中使用了超参数 var_gamma 和 log_prob_w 的估计值来代替 theta(θ)和 phi(φ)作为输出。

3.<iteration>.other

这个文件最为简单，就是 alpha，主题个数 K，词典单词数 V 的输出，典型的样例如下:

```
num_topics 100
num_terms 3197
alpha 0.0136578276
```

4. word-assignments.dat

这个文件只会在最后一轮迭代运行结束后输出，这个文件类似 gibbs sampling 版 LDA 的 tassign 文件，该文件的输出结果见图 8-1。

```
617 0000:84 0001:84 0002:84 0003:84 0004:84 0005:84 0006:84 0007:84 0008:84 0009:84 0010:84 0011:84 0012:02 0013:84
1347 0006:95 0009:95 0011:95 0012:95 0013:22 0015:95 0016:95 0020:95 0021:95 0022:95 0024:95 0028:95 0045:95 0052:95
370 2048:71 2049:71 2050:71 2051:71 2052:71 2053:71 0006:71 2055:71 2056:71 2057:71 0010:71 2059:71 2060:71 2061:71
278 2219:55 0009:55 1034:55 0011:55 2060:55 2306:55 0015:55 2065:55 0205:55 1043:55 2268:55 0022:55 0540:55 2309:55
181 0515:93 2421:93 1034:93 0012:93 1368:61 1044:93 0022:61 2078:93 0043:93 0045:93 0052:93 1602:93 0594:93 0089:93
376 0006:56 0007:56 2058:64 2059:64 2065:64 0022:64 2083:64 2085:64 0052:64 2101:64 2105:64 2106:56 2107:64 2108:64
164 0009:19 0011:19 0012:19 0022:19 0024:19 0030:19 0553:19 0052:19 2101:19 0055:19 1084:19 0065:19 0070:19 1121:19
247 2847:07 2816:07 2817:07 0012:07 0015:07 2819:07 2854:07 2083:07 0045:07 1800:07 1076:07 0706:07 2861:07 2849:07
102 1538:69 0009:69 0012:69 0021:69 0022:69 2583:69 1049:69 2588:69 2078:81 2600:81 0553:69 2105:69 1378:69 2915:69
521 2050:11 2062:11 0015:11 2066:11 0022:11 0024:11 0026:11 2078:11 2079:11 2088:11 2091:11 2092:11 0045:11 2105:11
```

**图 8-1 word-assignments.dat 输出样例**

这个文件的每一行是一个 doc 文档的主题指定，每行第一个数字是该文档去重后的单词数，而后续的数字为：word_id:phi[word_id]里概率最大的 topic 编号，注意到这与 Gibbs Sampling 的主题 z 编号的生成方式有很大不同。（想想为什么？）从下面这段代码可以清楚地看到这一点。

```
/*
 * writes the word assignments line for a document to a file
 */
void write_word_assignment(FILE* f, document* doc, double** phi, lda_model* model)
{
    int n;

    fprintf(f, "%03d", doc->length);
    for (n = 0; n < doc->length; n++)
    {
        //argmax(int* array, int K) return max element index from array[0,…,K-1]
        fprintf(f, " %04d:%02d",doc->words[n], argmax(phi[n], model->num_topics));
    }
    fprintf(f, "\n");
    fflush(f);
}
```

5.likelihood.dat

每轮迭代时都会在 EM 步骤之后输出一个似然函数的下界值，输出两列，以 tab 分隔：第一列是当前迭代时，所有文档集总共的下界之和（将每篇文档调用

lda-inference.c 中的 compute_likelihood 加和起来）；第二列是 converged 收敛值，converged = (likelihood_old - likelihood) / (likelihood_old)，这个公式计算得到的 converged 值最终用于终止迭代的条件（converged < 0 或 converged > settings.txt 设置的 EM_CONVERGED）。

文件样例如下：

```
-53850.3207874086      inf
-48731.8030157054    9.50508e-02
-42982.5729925872    1.17977e-01
-41005.7009654659    4.59924e-02
-40497.9670195761    1.23820e-02
-40276.4152025643    5.47069e-03
-40159.6585856079    2.89888e-03
-40089.4897857436    1.74725e-03
```

6. topics.py

另外该程序包中还提供了一个 topics.py 的 python 脚本文件，用来输出人们可以直接理解的 top n 主题词（beta 文件），类似 Gibbs LDA++输出的 twords 文件，操作如下：

[root@localhost lda]# python topics.py 025.beta wordmap.txt 10

如此便输出了每个主题下β概率最大的 10 个特征词：

```
topic 010
    阎王关          254959
    多布 147757
    人間正道是滄桑          276624
    霓虹灯          20385
    查扣 110500
    教导者          273769
    大坪 86792
    西西弗          102334
    舞吧 89873
    軟塌塌          241457
    ……
```

# 值得进一步阅读的参考文献